\documentclass[
 reprint,
 amsmath,
 amssymb,
 aps,
 nofootinbib,
]{revtex4-2}
\usepackage[utf8]{inputenc}
\usepackage{placeins}
\usepackage{graphicx}
\usepackage{physics}
\usepackage{bm}
\usepackage{orcidlink}

\usepackage{dcolumn}

\newcolumntype{d}[1]{D{.}{.}{#1}}
\newcommand\mc[1]{\multicolumn{1}{c}{#1}} 

\newcommand{\alos}{$a_\textrm{los}$ }
\newcommand{\msun}{M$_\odot$}

\newcommand{\apjl}{Astrophys. J.}
\newcommand{\mnras}{Mon. Not. R. Astron. Soc.}
\newcommand{\araa}{Annu. Rev. Astron. Astrophy.}
\newcommand{\aj}{Astron. J.}
\newcommand{\aap}{Astron. Astrophys.}
\newcommand{\sovast}{Sov. Astron.}
\newcommand{\physrep}{Phys. Rep.}
\newcommand{\apjs}{Astrophys. J. Suppl. Ser.}
\newcommand{\pasj}{Publ. Astron. Soc. Jap.}
\newcommand{\bain}{Bull. Astron. Inst. Neth.}
\newcommand{\jcap}{J. Cosm. Astropart. Phys.}
\newcommand{\pasp}{Pub. ASP}
\newcommand{\caa}{Chin. Astron. Astrophys.}
\newcommand{\rmxaa}{Rev. Mex. Astron. Astrophys.}
\newcommand{\apss}{Astrophys. Space Sci.}

\begin{document}

\title{Galactic Structure From Binary Pulsar Accelerations: Beyond Smooth Models}

\author{Thomas Donlon II \orcidlink{0000-0002-7746-8993}}
\affiliation{Department of Physics and Astronomy, University of Alabama in Huntsville, 301 North Sparkman Drive, Huntsville, AL 35816, USA}
\email{thomas.donlon@uah.edu}

\author{Sukanya Chakrabarti \orcidlink{0000-0001-6711-8140}}
\affiliation{Department of Physics and Astronomy, University of Alabama in Huntsville, 301 North Sparkman Drive, Huntsville, AL 35816, USA}

\author{Lawrence M. Widrow \orcidlink{0000-0001-6211-8635}}
\affiliation{Department of Physics, Engineering Physics and Astronomy, Queen’s University, Kingston, ON K7L 3N6, Canada}

\author{Michael T. Lam \orcidlink{0000-0003-0721-651X}}
\affiliation{SETI Institute, 339 N Bernardo Ave Suite 200, Mountain View, CA 94043, USA}
\affiliation{School of Physics and Astronomy, Rochester Institute of Technology, Rochester, NY 14623, USA}
\affiliation{Laboratory for Multiwavelength Astrophysics, Rochester Institute of Technology, Rochester, NY 14623, USA}

\author{Philip Chang \orcidlink{0000-0002-2137-2837}}
\affiliation{Department of Physics, University of Wisconsin-Milwaukee, 3135 N Maryland Ave, Milwaukee, WI 53211, USA}

\author{Alice C. Quillen \orcidlink{0000-0003-1280-2054}}
\affiliation{Department of Physics and Astronomy, University of Rochester, Rochester, NY 14627, USA}


\begin{abstract}
We measure the line-of-sight accelerations of 26 binary pulsars due to the Milky Way's gravitational potential, and produce a 3-dimensional map of the acceleration field of the Galaxy. 
Acceleration measurements directly give us the change in the line-of-sight velocity at present day, without requiring any assumptions inherent to kinematic modeling.
We measure the Oort limit ($\rho_0=0.062\pm0.017$ \msun/pc$^3$) and the dark matter density in the midplane ($\rho_{0,\textrm{DM}}=-0.010\pm0.018$ \msun/pc$^3$); these values are similar to, but have smaller uncertainties than previous pulsar timing measurements of these quantities.  Here, we provide for the first time, values for the Oort constants and the slope of the rotation curve from direct acceleration measurements.  We find that $A=15.4\pm2.6$ km/s/kpc and $B=-13.1\pm2.6$ km/s/kpc (consistent with results from \textit{Gaia}), and the slope of the rotation curve near the Sun is $-2\pm5$ km/s/kpc. We show that the Galactic acceleration field is clearly asymmetric, but due to data limitations it is not yet clear which physical processes drive this asymmetry. We provide updated models of the Galactic potential that account for various sources of disequilibrium; these models are incompatible with commonly used kinematic potentials.  This indicates that use of kinematically  derived Galactic potentials in precision tests (e.g., in tests of general relativity with pulsar timing) may be subject to larger uncertainties than reported.  The acceleration data indicates that the mass of the Galaxy within the Solar circle is $2.3 \times 10^{11}$ M$_\odot$, roughly twice as large as currently accepted models. Additionally, the residuals of the acceleration data compared to existing Galactic models have a dependence on radial position; this trend can be explained if the Sun has an additional acceleration away from the Galactic center. \\\vspace{0.5cm}
\end{abstract}

\maketitle

\section{Introduction} \label{sec:intro}


The Milky Way (MW) is in dynamical disequilibrium due to internal phenomena, including the bar, spiral structure, and external perturbations from satellite galaxies and dark matter substructure. The timescales for these perturbations are comparable to the orbital periods of stars in the disk and stellar halo; as a result, the Galaxy has not achieved relaxation in the statistical sense described by Lynden-Bell (1967) \cite{Lynden-Bell1967}. These disequilibria manifest in the stellar Galactic disk in a variety of ways, such as corrugations, warps, and phase space spirals \citep{Binney1992,Xu2015,Antoja2018}, as well as a north-south asymmetry in the density of stars \citep{Widrow2012,YannyGardner2013,BennettBovy2019}, and moving groups of disk stars \citep{QuillenMinchev2005,Antoja2008,Antoja2011,Hunt2018,Ramos2018,Craig2021}. Disturbances in the HI gas \citep{Levineetal2006,KalberlaDedes2008} have been interpreted as arising from satellite perturbations \citep{Chakrabarti_Blitz2009,Chakrabarti_Blitz2011,Chakrabartietal2019}, and there is apparent bulk motion in the interstellar medium, such as the Radcliffe Wave \citep{Alves2020}.


These disequilibrium features contain valuable information about the MW's structure and history. For example, the phase space spirals in the MW disk may have been generated by the Sagittarius Dwarf Galaxy \citep{Laporte2019} or the gravitational field of the Galactic bar \citep{Khoperskov2019,Hunt2022}.  In general, frequent interactions with substructure leads to a departure from equilibrium in the motions of Milky Way disk stars. 



Historically, studies of the Galaxy's structure have primarily focused on the readily available positions and motions of stars \citep{BlandHawthornGerhard2016}. This method is restrictive, because kinematic information about the gravitational potential is averaged over time and space throughout a given star's orbit. Kinematic techniques, such as Jeans modeling, require simplifying assumptions such as spherical or azimuthal symmetry, and equilibrium \citep{BinneyTremaine2008}; these assumptions restrict the ability to observe spatially complex and transient features, and can therefore only produce an approximation of the Galaxy's true underlying gravitational potential field. 

However, the acceleration of an object is instantaneous in time, and produces a measure of the gravitational potential at a specific location of the Galaxy, as well as a non-local measure of the density field. As a result, direct acceleration measurements allow one to probe the Galaxy's structure in a fine-grained way that is free of kinematic assumptions. Recently, several precision techniques have been developed that enable direct line-of-sight acceleration measurements within a few kpc of the Sun, including extreme precision radial velocity observations \citep{Chakrabarti2020}, pulsar timing  \citep{Chakrabarti2021}, and eclipse timing \citep{Chakrabartietal2022}.

Because pulsars are precise astrophysical clocks, one can directly measure their acceleration and therefore determine the acceleration field of the Galaxy at their locations \citep{Chakrabarti2021,Phillips2021}\footnote{\cite{Bovy2021}'s unpublished pre-print also claims constraints on the Oort constants using the data from C21.  However, we have redone the same analysis using the data gleaned from \cite{Bovy2021}'s figures and C21, and were unable to recover central values nor the uncertainties that are reported in \cite{Bovy2021} In our view, the strength of the claimed constraints in \cite{Bovy2021} is uncertain. } In Phillips et al. \cite{Phillips2021}, accelerations were also inferred from the spin-down rates of solitary pulsars. However, pulsars experience magnetic braking, which is not well-understood, and therefore the relation between the spin-down rates and a pulsar's acceleration is subject to large uncertainties. Since the changes in orbital periods of binary pulsars can be modelled precisely with general relativity \citep{PetersMathews1963,DamourTaylor1991,WeisbergHuang2016}, one can extract Galactic accelerations with far smaller uncertainties.


 Pulsar timing data has been used to test general relativity, via the emission of gravitational waves from binary systems \citep{DamourTaylor1991,WexKramer2020}. Computing this radiation term in the pulsar timing data requires knowing the Galactic potential at the location of the pulsar, which is currently the largest source of uncertainty in these tests \citep{WeisbergHuang2016}.  By obtaining an accurate model of the Galaxy's gravitational potential, one could substantially lower the uncertainty in tests of general relativity.  This can be enabled by direct acceleration measurements that are independent of pulsar timing.  Eclipse timing measurements of precisely timed eclipsing binary stars \citep{Chakrabartietal2022}, which can be employed for this purpose.  Further in the future, extreme-precision radial velocity measurements will also provide direct measurements of the Galactic acceleration \citep{Chakrabarti2020} that can be used to determine the Galactic potential independent of pulsar timing. 

Previously, Chakrabarti et al. (2021) \cite[hereafter C21]{Chakrabarti2021} analyzed line-of-sight acceleration data for 14 binary pulsars; we expand this with updated data for 26 binary pulsars, roughly doubling the size of the dataset. The C21 dataset spanned 2.6 kpc in $R$ and 1.5 kpc in $Z$; our updated dataset extends this to 3.4 kpc in $R$ and 3.6 kpc in $Z$, which allows us to leverage a larger region of the Galaxy when constraining our models. 

In this work, we use the expanded dataset to compute updated values for several different potential models that are fit to the observed accelerations.  In addition to a measurement of the Oort limit and a constraint on the local dark matter density, we are now able to constrain the slope of the rotation curve and  the Oort Constants. We also move beyond smooth potential models, and produce a 3-dimensional, non-parametric map of the MW acceleration field. This map manifests an apparent lack of vertical and azimuthal reflection symmetry about the Sun, which is evidence of local disequilibrium. We demonstrate that a consideration of disequilibrium is required in order to obtain accurate models of the MW's gravitational potential from direct acceleration measurements; this will continue to be important as pulsar timing datasets become larger and more precise. 

\section{Data} \label{sec:data}

We take the distance between the Sun and the Galactic center to be $R_\odot=8.178$ kpc \citep{GravityCollaboration2019}, and the circular speed at the Solar position to be $V_{LSR}=232.8$ km/s \citep{McMillan2017}. Adopting other literature values of $R_\odot$ and $V_{LSR}$ did not substantially change the results of this work. 

This work uses a Galactocentric Cartesian coordinate system, where the Sun is located at positive $x$, $\hat{y}$ points towards the direction of the Sun's motion, and $\hat{z} = -\hat{x} \times \hat{y}$. 

\subsection{Calculating Accelerations}

We follow the methodology in C21 in extracting the Galactic acceleration from measured binary pulsar accelerations.  The observed orbital period of a binary pulsar will be Doppler shifted according to its line-of-sight velocity. If the binary pulsar accelerates along our line of sight, a corresponding change in the Doppler shifted orbital period will be observed. The observed line-of-sight acceleration is expressed as \begin{equation} \label{eq:los_acc}
    a^\textrm{Obs}_{\textrm{los}} = \frac{\dot{P}_b^\textrm{Obs}}{P_b}c,
\end{equation} where $P_b$ is the observed orbital period of the binary and $c$ is the speed of light. 

The observed time rate of change of the binary orbital period ($\dot{P}_b^\textrm{Obs}$) can be decomposed into a sum of different effects: \begin{equation}
    \dot{P}_b^\textrm{Obs} = \dot{P}_b^\textrm{Shk} + \dot{P}_b^\textrm{GR} +  \dot{P}_b^\textrm{Gal}.
\end{equation} 

The quantity $\dot{P}_b^\textrm{Shk}$ is known as the Shklovskii effect \citep{Shklovskii1970}, and is the result of proper motion of a source leading to a change in the distance to that source, producing an apparent acceleration along our line-of-sight. It is computed as \begin{equation}
   \dot{P}_b^\textrm{Shk} = \frac{P_b \mu^2 d}{c},
\end{equation} where $\mu$ is the total proper motion of the source, and $d$ is the distance to that source. 

The quantity $\dot{P}_b^\textrm{GR}$ is the decrease in the orbital period due to radiation of gravitational waves \citep{PetersMathews1963,WeisbergHuang2016}, and is computed as \begin{align*}
    \dot{P}_b^\textrm{GR} = -\frac{192\pi G^{5/3}}{5c^5}\left(\frac{P_b}{2\pi}\right)^{-5/3} \left(1-e^2\right)^{-7/2}
\end{align*}\begin{equation}\label{eq:pdot_gr}
    \times \left(1 + \frac{73}{24}e^2 + \frac{37}{96}e^4\right)\frac{m_p m_c}{\left(m_p + m_c\right)^{1/3}},
\end{equation} where $e$ is the orbital eccentricity of the binary (typically close to 0), $m_p$ is the mass of the pulsar, and $m_c$ is the mass of its companion. 

Because the quantity we are interested in is the acceleration due to the gravitational potential of the MW, we compute the Galactic component of the line-of-sight acceleration as \begin{equation}
    a^\textrm{Gal}_{\textrm{los}} = \frac{\dot{P}_b^\textrm{Gal}}{P_b}c
\end{equation} using Equations \ref{eq:los_acc}-\ref{eq:pdot_gr}. For notational simplicity, we often refer to this quantity as \alos from this point forwards.

\subsection{Pulsar Data}

\begin{figure}
    \centering
    \includegraphics[width=0.45\textwidth]{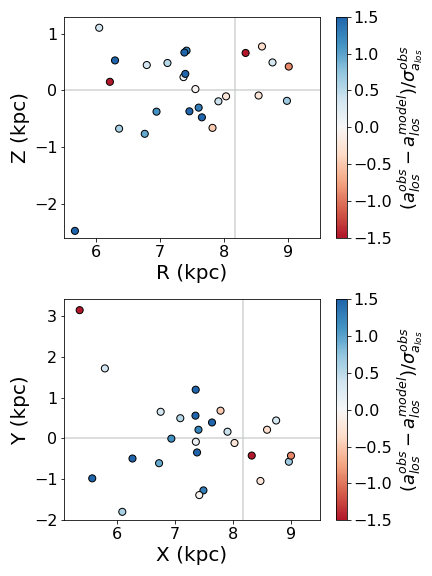}
    \caption{The observed pulsar data. Each source is colored according to the residual of its the observed line-of-sight acceleration and the \textit{Gala MilkyWay2022} potential model, weighted by the uncertainty of each measurement. There is a global trend in the data; points that are further to the left are bluer on average than points that are further to the right. This suggests that the \textit{Gala MilkyWay2022} potential model incorrectly estimates the slope of the rotation curve, or that the Sun is experiencing an additional acceleration away from the Galactic center.}
    \label{fig:resid}
\end{figure}

We selected all available binary pulsars from the Australia Telescope National Facility (ATNF) pulsar catalog \citep{Manchester2005} that satisfied the following criteria: \begin{itemize}
    \item The source had a measured $P_b$ and $\dot{P}_b$;
    \item Had measured parallax $\pi$ and proper motion $\mu$;
    \item Had measured orbital eccentricity $e$, pulsar mass $m_p$, and companion mass $m_c$; or a binary period $P_b>5$ days.
    \item The source cannot be a redback, black widow, etc. or transferring mass with its companion, which can change the orbital period of the system.
    \item The source cannot be in a globular cluster, as the internal accelerations of globular clusters are expected to overpower the Galactic acceleration signal. 
\end{itemize}

\begin{table*}[]
    \centering
    \begin{tabular}{ld{1.7}rrrrr}
        \hline \hline 
        PSR & \mc{Distance} & \mc{$\dot{P}_b^\textrm{Obs}$} & \mc{$\dot{P}_b^\textrm{Shk}$} & \mc{$\dot{P}_b^\textrm{GR}$} & \mc{$a^\textrm{Gal}_{\textrm{los}}$} & Reference \\
         & \mc{(kpc)} & & & & (mm s$^{-1}$ yr$^{-1}$) & \\ \hline
         J0437$-$4715 & 0.157(2) & 3.728(6)E-12 & 3.76(5)E-12 & -3.22(15)E-16 & -0.8(9) & R16 \\
         J0613$-$0200 & 1.14(12) & 3.4(7)E-14 & 3.15(33)E-14 & -8.0(40)E-15 & 1.0(8) & 
         NG \\
         J0737$-$3039A/B & 1.15(22) & -1.252(17)E-12 & 4.4(12)E-16 & -1.2484(8)E-12 & -5.(18) & KMS06, DBT09 \\
         J0740+6620 & 1.0(2) & 1.17(26)E-12 & 1.06(21)E-12 & -6.07(21)E-16 & 2(8) & NG \\
         J0751+1807 & 1.17(5) & -3.50(5)E-14 & 1.19(5)E-14 & -4.4(4)E-14 & -1.2(17) & EPTA, NSK08 \\
         J1012+5307 & 1.0(2) & 6.1(6)E-14 & 8.4(17)E-14 & -6.0(320)E-15 & -3(7) & NG, EPTA \\
         J1017$-$7156 & 1.4(6) & 4.(2)E-13 & 2.0(8)E-13 & -3.9(30)E-16 & 3(4) & PPTA \\
         J1022+1001 & 0.85(6)  & 2.18(9)E-13 & 7.4(20)E-13 & -5.2(8)E-16 & -7.3(28) & EPTA \\
         J1125$-$6014 & 1.5(11) & 7.(1)E-13 & 8.0(60)E-13 & -2.24(30)E-16 & -1(8) & PPTA \\
         J1455$-$3330 & 0.76(6) & 4.5(22)E-12 & 8.0(6)E-13 & - & 5.3(32) & EPTA \\
         B1534+12 & 0.94(7) & -1.366(3)E-13 & 5.3(4)E-14 & -1.93(6)E-13 & 0.8(19) & FST14, DDF21 \\
         J1600$-$3053 & 1.87(3) & 5.(1)E-13 & 2.79(5)E-13 & -1.2(4)E-16 & 1.7(8) & PPTA \\
         J1603$-$7202 & 2.7(13) & 1.9(4)E-13 & 2.2(10)E-13 & - & -0.4(19) & PPTA, WRT22 \\
         J1614$-$2230 & 0.68(4) & 1.33(7)E-12 & 1.32(8)E-12 & -4.23(3)E-16 & 0.1(14) & NG \\
         J1640+2224 & 1.08(28) & 9.5(19)E-12 & 5.3(14)E-12 & -6.(3)E-18 & 2.6(15) & EPTA \\
         J1713+0747 & 1.136(13) & 2.64(73)E-13 & 6.41(7)E-13 & -6.7(2)E-18 & -0.61(12) & EPTA \\
         J1738+0333 & 1.47(10) & -1.70(31)E-14 & 8.2(6)E-15 & -2.9(13)E-14 & 1(4) & FWE12 \\
         J1741+1351 & 3.0(9) & 1.3(4)E-12 & 1.4(4)E-12 & - & -1(4) & NG \\
         J1909$-$3744 & 1.12(3) & 5.09(2)E-13 & 4.94(13)E-13 & -2.90(4)E-15 & 1.2(10) & NG \\
         B1913+16 & 4.1(2) & -2.423(1)E-12 & 1.6(7)E-16 & -2.40263(5)E-12 & -6.96(34) & WH16, DWN18 \\
         J1933$-$6211 & 1.6(3) & 7.(1)E-13 & 6.6(12)E-13 & -1.53(29)E-16 & 0.3(14) & GKF23 \\
         J2043+1711 & 1.4(1) & 1.0(1)E-13 & 6.5(5)E-14 & -2.86(16)E-15 & 2.8(8) & NG \\
         J2129$-$5721 & 3.6(14) & 1.51(9)E-12 & 8.9(35)E-13 & - & 10(6) & PPTA \\
         J2145$-$0750 & 0.71(5) & 1.3(2)E-13 & 1.76(12)E-13 & - & -0.7(4) & PPTA \\
         J2222$-$0137 & 0.2681(12) & 2.554(74)E-13 & 2.796(13)E-13 & -7.98(12)E-15 & -0.76(34) & GFG21 \\
         J2234+0611 & 0.97(4) & 3.1(25)E-12 & 4.78(20)E-12 & -2.57(12)E-17 & -6(9) & SFA19 \\ \hline \hline
    \end{tabular}

    \vspace{1ex}

     {\raggedright Rows without $\dot{P}_b^\textrm{GR}$ data are expected to have negligibly small values of $\dot{P}_b^\textrm{GR}$. \textbf{References}: R16, \cite{Reardon2016}; NG (NANOGrav 15-yr Data Release), \cite{NG15}; KMS06, \cite{Kramer2006}; DBT09, \cite{Deller2009}; EPTA (EPTA-DR2), \cite{EPTADR2}; NSK08, \cite{Nice2008}; FST14, \cite{Fonseca2014}; DDF21, \cite{Ding2021}; PPTA (PPTA-DR2), \cite{PPTADR2}; WRT22, \cite{Walker2022}; FWE12, \cite{Freire2012}; WH16, \cite{WeisbergHuang2016}; DWN18, \cite{Deller2018}; GKF23, \cite{Geyer2023}; GFG21, \cite{Guo2021}; SFA19, \cite{Stovall2019}.  \par}

    \caption{Summary of Pulsar Data.}
    \label{tab:pulsar_data}
\end{table*}

Our final dataset contains 26 binary pulsars; the full list of the pulsars, their distances, their components of $\dot{P}$, and their accelerations are provided in Table \ref{tab:pulsar_data}, and their positions are shown in Figure \ref{fig:resid}. Fourteen of these sources are present in the C21 dataset.  Here, we also include 9 sources from recent work by Moran et al. \cite[][hereafter M23]{Moran2023}.  We do not use the following five sources from the M23 dataset as they do not satisfy our criteria for selection: \begin{itemize}
    \item B1259$-$63: This pulsar has a Be star companion, and experiences regular mass transfer \citep{YiCheng2017}.
    \item J0348$+$0432: This source has no PTA parallax available. M23 used a Bayesian estimate of the parallax taken from a negative \textit{Gaia} DR3 parallax \citep{GaiaDR3,Antoniadis2021,Moran2023b} to calculate the distance to the pulsar, but we choose not to include this source because of its large distance uncertainty. 
    \item J0636$+$5128 (Listed in M23 as J0636+5129): There is evidence for outflows around this source, the pulsar may have an accretion disk, and the pulsar appears to be interacting with its companion in some way \citep{DraghisRomani2018,Chen2021}.
    \item J1756$-$2251: The proper motion in declination of this source is only constrained within an upper limit of 20 mas; M23 use only the proper motion in right ascension for this source (roughly 2 mas). This makes it nontrivial to constrain $\dot{P}_b^\textrm{Shk}$.
    \item J2339$-$0533: This source is a redback \citep{PletschClark2015}.
    
\end{itemize}

Three of these pulsars are new sources, and were not in either the C21 or M23 datasets: J1455$-$3330, J1640$+$2224, and J1933$-$6211; these timing solutions were published after M23. 

C21 listed a parallax of 1.9 mas for J2129$-$5721, given available timing data \citep{Reardonetal2016}, but more recent timing solutions give a parallax of 0.26 mas for this source \citep{PPTADR2}. This erroneous distance measurement, combined with the source's large value of $\dot{P}^\textrm{Obs}_b$, could explain why it had the largest residual of any pulsar considered by C21. 

In the case where $e$, $m_p$ and $m_c$ were not constrained (and therefore $\dot{P}_b^\textrm{GR}$ could not be calculated), the source was still included if its binary period $P_b > 5$ days. Since $\dot{P}_b^\textrm{GR} \propto P_b^{-5/3}$, large orbital periods result in negligibly small values of $\dot{P}_b^\textrm{GR}$ (assuming $m_p$ = $m_c$ = $1.4$ \msun, $e=0$, and $P_b = 5$ days, $\dot{P}_b^\textrm{GR} \sim 10^{-15}$; for comparison, $\dot{P}_b^\textrm{Gal} \sim 10^{-12}$). Our dataset contains 5 sources without $\dot{P}_b^\textrm{GR}$, indicated in Table \ref{tab:pulsar_data} by a dash in the $\dot{P}_b^\textrm{GR}$ column.

\subsection{Line-of-Sight Acceleration Uncertainty}

\begin{figure*}
    \centering
    \includegraphics[width=\textwidth]{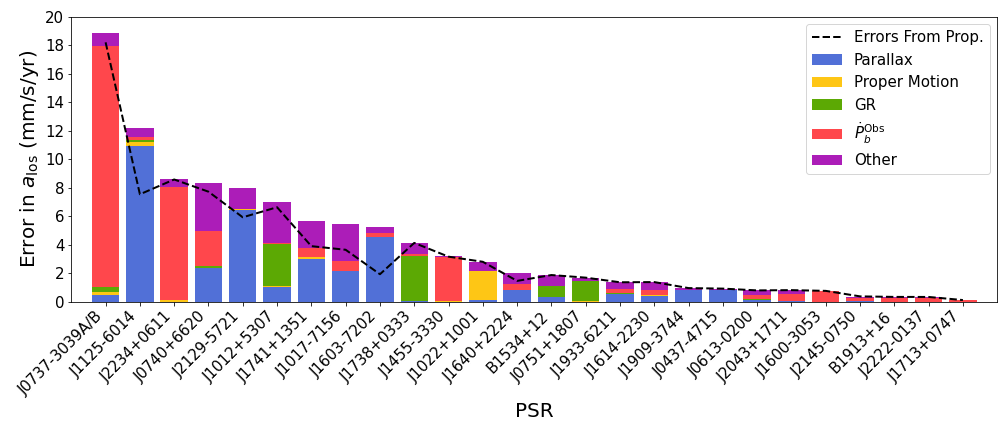}
    \caption{Bootstrapped uncertainties in \alos for each source, broken down by individual component (parallax, proper motion, $\dot{P}^\textrm{GR}_b$, and $\dot{P}^\textrm{Obs}_b$). There is also a contribution to \alos error from covariances between the various components. The dashed black line shows the uncertainty estimated for \alos using error propagation techniques, and is generally in agreement with the bootstrapped uncertainties. }
    \label{fig:rel_errs}
\end{figure*}

We computed the uncertainty in \alos for each source using standard error propagation procedure. This method does not take into account any covariances between the uncertainties in correlated parameters, or the fact that uncertainty in distance contibutes to the total error by both changing the position of the pulsar and also affecting the calculation of $\dot{P}_b^\textrm{Shk}$. As such, it is beneficial to double check that we are not underestimating the uncertanties in $a_\textrm{los}$ by estimating the uncertainties in a second way. 

Figure \ref{fig:rel_errs} shows bootstrapped uncertainties in \alos for each source, broken down into the contribution of the uncertainty of each component to the total \alos uncertainty. The total uncertainty in \alos was estimated by computing the ``true'' \alos for each source assuming its parameters each individually have zero error, and then sampling 10,000 times with the reported errors on each component, assuming that each uncertainty is normally distributed. The contribution of each component to the overall \alos uncertainty was estimated by assuming only that component has zero uncertainty, and then sampling the resulting \alos uncertainty 10,000 times; the difference between the two error estimations gives the contribution of that component. The ``Other'' component in the \alos uncertainties is likely due to additional uncertainty contributed by covariances in the components. 

The dashed black line in Figure \ref{fig:rel_errs} shows the uncertainty that is obtained through error propagation; overall this value appears to be similar to the bootstrap estimates, indicating that the uncertainties obtained via error propagation are reasonable estimates. 

In general, uncertainties in parallax and $\dot{P}^\textrm{Obs}_b$ contribute more to the overall uncertainty in \alos than uncertainties in proper motion and $\dot{P}^\textrm{GR}_b$. As a result, improvements in parallax and $\dot{P}^\textrm{Obs}_b$ measurements should be prioritized in pulsar timing fits in order to obtain more precise \alos measurements in the future.

\section{Radial Dependence of Acceleration Residuals}

\begin{figure}
    \centering
    \includegraphics[width=0.45\textwidth]{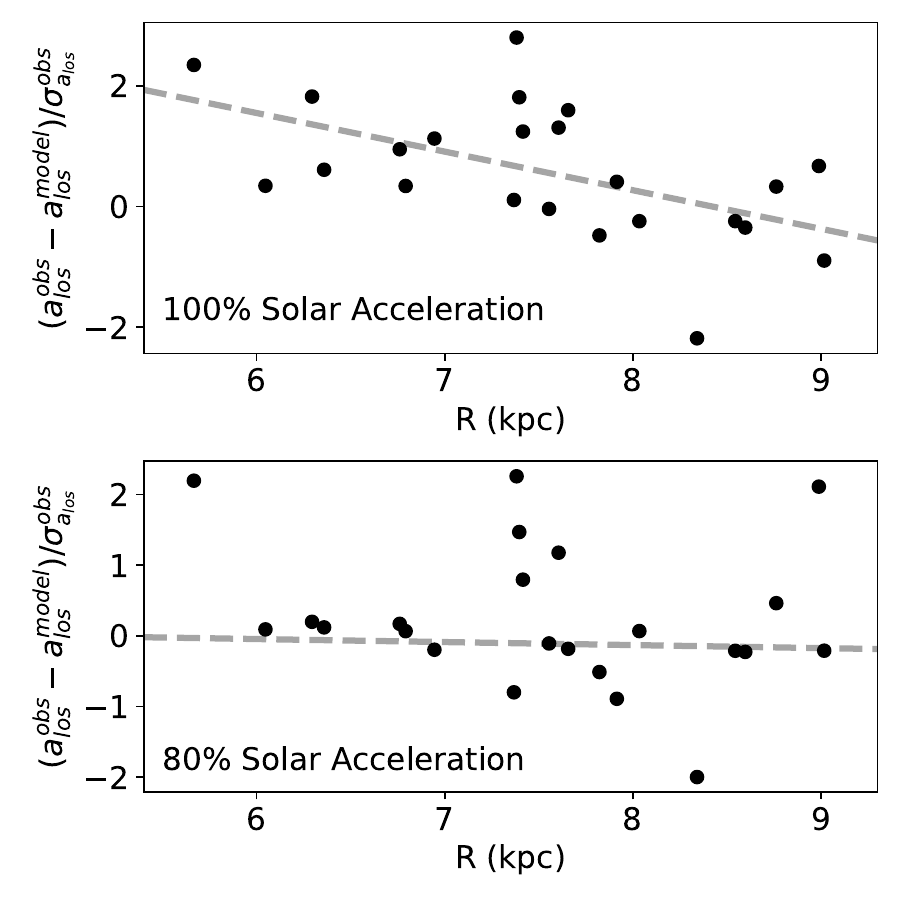}
    \caption{The residuals of the pulsar data and the \textit{Gala MilkyWayPotential2022} model as a function of distance from the Galactic center (black). A linear fit to the data is shown as a dashed gray line. If the model fits the data, one would expect the residuals to be clustered around 0 independent of each pulsar's position; however, there is a radial dependence in the residuals when the Solar acceleration from the \textit{Gala} model is used. If the Sun's acceleration relative to the binary pulsars is actually only 80\% of its presumed value, this trend disappears (although there is still substantial scatter at some radii, particularly at $R$ $\sim$ 7.5 kpc).}
    \label{fig:var_accel}
\end{figure}

The observed acceleration data is shown in Figure \ref{fig:resid} compared to a state-of-the-art theoretical potential model, \textit{MilkyWayPotential2022} from the \verb!Gala! python package \citep{Gala}. There is a global trend in the figure; overall, \textit{MilkyWayPotential2022} underestimates the accelerations of the pulsar sources towards the Galactic center. Because we observe a line-of-sight acceleration relative to the Sun, this results in a positive residual \alos (blue) for sources between the Sun and the Galactic center, and a negative residual \alos (red) for sources with $R>R_\odot$. 

Interestingly, we were not able to remove this trend by varying the total mass of the \textit{Gala MilkyWayPotential2022} model. However, this trend can be explained if the Sun is accelerating relative to the pulsars in the dataset. 

The distribution of quasars, or quasi-stellar objects (QSOs), has been used to measure the acceleration of the Sun relative to a cosmological reference frame \cite{Klioner2021,TitovLambert2013}. These accelerations agree with predictions of the Solar acceleration relative to the Galactic center using existing kinematic potential models of the MW; as a result, the measured acceleration of the Sun relative to the QSO reference frame is typically treated as \textit{identical} to the Sun's acceleration relative to the Galactic center. In other words, it is presumed that the acceleration of the Galaxy relative to the QSO reference frame is negligible. However, there is no reason that the Galactic center is in the same inertial reference frame as the QSOs; if the Galactic center is in-fact accelerating relative to the QSOs, then the acceleration of the Sun relative to the QSOs is \textit{not} a relative acceleration with respect to the Galactic center. 

In this scenario, the Sun could be experiencing an additional acceleration relative to the Galactic center (which cannot be constrained by the QSO measurements). If this is the case, then the observed relative accelerations of each pulsar are actually \begin{equation}
    a_\textrm{los} = \left[\mathbf{a}(\mathbf{x}) - \mathbf{a}(\mathbf{x}_\odot) - \mathbf{a}_\textrm{pec}(\mathbf{x}_\odot) \right]\cdot\hat{\mathbf{d}},
\end{equation} where $\mathbf{a}_\textrm{pec}(\mathbf{x}_\odot)$ is a peculiar acceleration felt by the Sun but not felt by each pulsar (see Equation \ref{eq:alos} for further context) and $\hat{\mathbf{d}}$ is the line-of-sight vector from the Sun to each source. 

In order to remove the observed trend, the Sun would need to feel an additional acceleration of $\mathbf{a}_\textrm{pec}(\mathbf{x}_\odot)=1.1$ mm/s/yr of its predicted radial acceleration away from the center of the Galaxy. The effect of this additional acceleration on the residual of the observed data is shown in Figure \ref{fig:var_accel}. This value was obtained by optimizing a potential model that consisted of \textit{MilkyWayPotential2022} plus an additional acceleration on the Sun relative to the pulsar data (see Section \ref{sec:fits} for a detailed description of the optimization procedure). For this analysis, we removed PSRs J1713+0747, B1913+16, and J2043+1711 because they were outliers that substantially changed the linear fit to the data. 

It is difficult to explain why the Sun would have an additional acceleration relative to the Galactic center, although a relative acceleration could potentially be caused by a nearby massive object, such as a nearby (dim) compact object, or a dark matter subhalo. It is interesting to note that a Jupiter-mass object at a distance of 400 AU from the Sun would produce a Solar acceleration of the correct magnitude to explain the observed global pulsar acceleration trend; constraints on the possible locations of Planet Nine place it roughly in the direction of the Galactic anticenter \citep{delaFuenteMarcos2016,HolmanPayne2016}, potentially making it an explanation for why the Sun would experience an additional acceleration away from the center of the Galaxy that would not be felt by the pulsar sources. However, current constraints on Planet Nine's mass and semimajor axis would only produce a Solar acceleration on the order of $\sim$0.05 mm/s/yr \citep{Batygin2019}; this discrepancy would need to be resolved in order to claim that a ninth planet is the cause of the apparent 1.1 mm/s/yr Solar acceleration. 

Zakamska \& Tremaine (2005) \cite{Zakamska2005} used pulsar and white dwarf timing solutions to constrain the acceleration of the Solar system, and concluded that the Solar system acceleration was consistent with zero. However, their sensitivity was limited to roughly the acceleration of a Jupiter-mass planet at a distance of 200 AU ($\sim 4.5$ mm/s/yr), which is a few times larger than the value we provide here for the Solar acceleration, and is therefore not in tension with our results.

\begin{table*}[]
    \setlength\extrarowheight{9pt}
    \centering
    \begin{tabular}{lllrr}
    \hline \hline
    Local Model & Potential & Parameters & AIC & red.$\chi^2$ \\ \hline
    \multicolumn{2}{l}{Symmetric Models} & & & \\ \hline
    $\alpha$ & $\Phi(R,z) = V^2_{LSR}\ln\left(\frac{R}{R_\odot}\right) + \frac{1}{2}\alpha z^2$ & $\log_{10}(\alpha\cdot\textrm{Gyr}^2)=3.60\pm 0.06$ & 54 & 2.2 \\
    
    $\alpha$-$\beta$ & $\Phi(R,z) = \frac{V^2_{LSR}}{2\beta}\left(\frac{R}{R_\odot}\right)^{2\beta} + \frac{1}{2}\alpha z^2$ & $\log_{10}(\alpha\cdot\textrm{Gyr}^2)=3.51\pm0.13$ & 55 & 2.3 \\
     & & $\beta = 0.14\pm 0.17$ & & \\
    
    Cross ($\alpha$-$\gamma$) & $\Phi(R,z) = \left(V^2_{LSR} + \gamma z^2\right)\ln\left(\frac{R}{R_\odot}\right) + \frac{1}{2}\alpha z^2$ & $\log_{10}(\alpha\cdot\textrm{Gyr}^2)=3.54\pm0.16$ & 55 & 2.3 \\
     & & $\log_{10}(-\gamma\cdot\textrm{Gyr}^2)=3.3\pm0.9$ & & \\

    \textit{MWPotential2014} & \textit{Galpy} Values \citep{Galpy} & - & 66 & 3.0 \\

    \textit{MilkyWayPotential2022} & \textit{Gala} Values \citep{Gala} & - & 70 & 3.5 \\
    
    Damour-Taylor & see Section \ref{sec:dt_pot} & - & 88 & 3.9 \\ \hline \hline

    \multicolumn{2}{l}{Asymmetric Models} & & & \\ \hline
    
    Anharmonic & $\Phi(R,z) = V^2_{LSR}\ln\left(\frac{R}{R_\odot}\right) + \frac{1}{2}\alpha_1 z^2 + \frac{1}{3}\alpha_2 z^3$ & $\log_{10}(\alpha_1 \cdot 
    \textrm{Gyr}^2)=3.51\pm0.08$ & 51 & 2.1 \\
    & & $\log_{10}(\alpha_2 \cdot \textrm{Gyr}^2\cdot\textrm{kpc})=3.4\pm0.2$ & & \\

    $2\alpha$-$\beta$ & $\Phi(R,z) = \frac{V^2_{LSR}}{2\beta}\left(\frac{R}{R_\odot}\right)^{2\beta} + \frac{1}{2}\alpha_1 z^2 + \frac{1}{3}\alpha_2 z^3$ & $\log_{10}(\alpha_1\cdot\textrm{Gyr}^2)=3.56\pm0.11$ & 53 & 2.2 \\
     & & $\log_{10}(\alpha_2 \cdot \textrm{Gyr}^2\cdot\textrm{kpc})=3.5\pm0.2$ & & \\
     & & $\beta = -0.08\pm 0.18$ & & \\

    Local Expansion & see Section \ref{sec:local-expansion-model} & $\log_{10}(-\partial a_R/\partial R\cdot \textrm{Gyr}^2/\textrm{kpc} ) = 2.6\pm0.3$ & 41 & 1.6 \\
    & & $\log_{10}(-\partial a_\phi/\partial \phi \cdot \textrm{Gyr}^2/\textrm{kpc} ) = 3.54\pm0.09$ & & \\
    & & $\log_{10}(\partial a_z/\partial z\cdot \textrm{Gyr}^2/\textrm{kpc} ) = 3.73\pm0.08$ & & \\

    Sinusoidal & $\Phi(R,z) = V^2_{LSR}\ln\left(\frac{R}{R_\odot}\right)$ & $\log_{10}(\alpha\cdot\textrm{Gyr}^2)=3.37\pm 0.08$ & 41 & 1.6 \\
    & \:\:\:\:\:\:\:\:\:\:\:\:\:\:\:\:\:\:\:\:$+ \frac{1}{2}\alpha \left(z + A\sin\left(2\pi R/\lambda + \varphi\right)\right)^2$ & $A = 0.52\pm0.08$ kpc & & \\
    & & $\lambda = 2.46\pm0.15$ kpc & & \\
    & & $\varphi = 3.5\pm1.2$ & & \\
    
    $\alpha$-$\beta$ + 2 Point Mass & $\Phi(R,z) = \frac{V^2_{LSR}}{2\beta}\left(\frac{R}{R_\odot}\right)^{2\beta} + \frac{1}{2}\alpha z^2 + \sum_i^2\frac{GM_i}{|\vec{r}-\vec{r_i}|} $ & $\log_{10}(\alpha\cdot\textrm{Gyr}^2)=3.65\pm 0.09$ & 31 & 0.8 \\
    & & $\beta=-0.43\pm 0.12$ & & \\
    & & $\log_{10}(M_1/M_\odot)=8.3\pm0.3$ & & \\
    & & $x_1=7.4\pm0.2$ kpc & & \\
    & & $y_1=1.2\pm0.1$ kpc & & \\
    & & $z_1=-0.6\pm0.2$ kpc & & \\
    & & $\log_{10}(M_2/M_\odot)=8.0\pm0.5$ & & \\
    & & $x_2=7.5\pm0.1$ kpc & & \\
    & & $y_2=0.5\pm0.1$ kpc & & \\
    & & $z_2=0.9\pm0.2$ kpc & & \\\hline \hline
    \end{tabular}
    \caption{Optimized fits to the local potential models.}
    \label{tab:local_models}
\end{table*}

\section{Model Fits} \label{sec:fits}

Ideally, if one had a complete map of the gravitational field, then its divergence would give a local measure of the density field; however, because we are restricted to one component of the gravitational field with existing data, we must instead rely on models to infer the local density and corresponding features in the Milky Way. 

Given a model for the MW's gravitational potential $\Phi(\vec{x})$, one can compute the acceleration at a given point from \begin{equation}
    \mathbf{a}(\mathbf{x}) = -\nabla\Phi(\mathbf{x}),
\end{equation} which can then be used to obtain a relative line-of-sight acceleration: \begin{equation} \label{eq:alos}
    a_\textrm{los} = \left[\mathbf{a}(\mathbf{x}) - \mathbf{a}(\mathbf{x}_\odot) \right]\cdot\hat{\mathbf{d}}
\end{equation} 

We fit a number of models to the observed \alos data using a maximum likelihood estimation (MLE) technique \citep{Ivezic2014}. We first define a likelihood function, in this case obtained from a $\chi^2$ statistic: \begin{equation}
    \chi^2 = \sum_i^N \frac{\left( a_\textrm{los}^\textrm{obs} - a_\textrm{los}^\textrm{model} \right)^2}{\sigma_{a_\textrm{los}^\textrm{obs}}^2},
\end{equation} where $N$ is the number of sources in the data. The corresponding log-likelihood function is $\ln(L)= -\chi^2/2.$

The log-likelihood function was then maximized using the \verb!differential_evolution! function from the \verb!scipy.optimize! python package \citep{scipy}. Differential evolution was used rather than traditional conjugate gradient descent because some of the models have ``bumpy'' likelihood surfaces, which can cause gradient descent algorithms to get stuck in local minima. 

The covariance matrix ($K$) for the optimized parameters is calculated from the weighted inverse of the Hessian matrix ($H$); \begin{equation}
     K_{ij} = \frac{1}{N}H_{ij}^{-1},
\end{equation} where $N$ is again the number of sources, and \begin{equation}
    H_{ij} = \left[ \frac{\partial^2}{\partial_i\partial_j} \ln (\hat{L}) \right],
\end{equation} where $\hat{L}$ is the likelihood of the optimized parameters given the observed data. The uncertainty in the $i$th optimized parameter ($\sigma_i$) is obtained from the diagonal of the covariance matrix: \begin{equation}
    \sigma_i^2 = \textrm{diag}(K_{ij}).
\end{equation}

The Akaike Information Criterion (AIC) can then be evaluated for each fit model using the formula \begin{equation}
    \textrm{AIC} = 2k - 2\ln(\hat{L}) = 2k + \chi^2,
\end{equation} where $k$ is the number of free parameters in that model. The AIC is a measure of significance of a likelihood score given a model; models with more free parameters typically produce better likelihood scores, so the AIC provides a way of evaluating whether adding additional free parameters to a model actually improves the quality of the fit. A difference of $\Delta$AIC$=6$ between two models is considered to be strong evidence in favor of the model with lower AIC \citep{KassRaftery1995}.

C21 used a Markov Chain Monte Carlo (MCMC) technique to fit models to pulsar acceleration data. We chose to use an MLE technique instead because MCMC codes can quickly become intractable for models with a large number of parameters, such as the ``$\alpha-\beta$ + 2 Point Mass'' model, which has 10 free parameters. This model is feasible in the MLE code, which is much faster and uses much less memory than the corresponding MCMC framework. The C21 MCMC setup is practically identical to our MLE setup, in that the likelihood functions are identical, and the MCMC priors from C21 are (roughly) encoded in the uncertainties in the observed accelerations, which are derived from the usual uncertainty propagation procedure. We compared several of our model fits to fits obtained using C21's MCMC approach, and the optimized parameters agree to well within the reported uncertainties.

The analytic form of each potential model can be found in Tables \ref{tab:local_models} and \ref{tab:global_models}, as well as the best fit parameters for each model and the quality of that fit. Note that the reduced $\chi^2$ value is reported in Tables \ref{tab:local_models} and \ref{tab:global_models}, which is defined as \begin{equation}
    \mathrm{red.}\chi^2 = \frac{\chi^2}{N-k-1}.
\end{equation}

\subsection{Symmetric Local Models} \label{sec:symmetric_local_models}

Our first set of potential models are azimuthally and vertically symmetric models (they are invariant under negation of $\phi$ or $z$). These models were introduced in C21 as local approximations of the potential (Taylor expansions of order $\leq2$), which are only expected to be accurate within a few kpc of the Sun. For this reason, these models are only fit to sources within 3 kpc of the Sun; this removes J2129$-$5721 and B1913+16 from the dataset, and these models are only fit to 24 of the 26 sources. While these local models are useful for evaluating small-scale structure, they are not able to produce estimates of the MW mass or other large-scale quantities; in order to measure those, we later utilize global models in Section \ref{sec:global_models}.

\subsubsection{$\alpha$ Model}

The $\alpha$ model is the simplest model in this work. It is separable in $R$ and $z$, and assumes a flat rotation curve, which is calibrated to the local standard of rest and the Solar position. In the vertical direction, the potential is that of a harmonic oscillator, where $\alpha$ gives the angular frequency of vertical oscillations. The free parameter $\alpha$ also sets the vertical density profile of this model, which is constant and given by $4\pi G \rho = \alpha$. Our measured value of $\alpha$ agrees with the value obtained by C21, and our updated value has smaller uncertainties.

\subsubsection{$\alpha-\beta$ Model}

The $\alpha-\beta$ model is an extension of the $\alpha$ model that considers a sloped rotation curve, which is characterized by $\beta$: \begin{equation} \label{eq:beta}
    \beta = \left[\frac{R}{V_\textrm{circ}}\dv{V_\textrm{circ}}{R}\right]_{R_\odot},
\end{equation} i.e. more positive (negative) values of $\beta$ indicate a more positive (negative) slope of the rotation curve.  Given the smaller extent of the pulsars, C21 could not constrain $\beta$ for this model. Our increased sample of pulsars covers a larger radial extent, which enables us to constrain $\beta$ for the first time. We find that $\beta$ is slightly positive, although it is consistent with zero at the 1$\sigma$ level. Constraining $\beta$ allows us to calculate the Oort constants and an updated estimate of the Oort limit (see Section \ref{sec:oort}).

\subsubsection{``Cross'' Model}

A potential that is separable in $R$ and $z$ (such as the $\alpha$ model) leads to a density that is also separable through the Poisson equation, whereas any realistic potential will not satisfy this condition. In the ``cross'' model, we consider a simple extension of the $\alpha$ model that breaks the separability assumption. The magnitude of this cross term is set by $\gamma$, for which $\gamma<0$ corresponds to oblate isopotential contours, and a larger magnitude of $\gamma$ implies a more oblate potential. \footnote{There is a typo in the C21 values of $\gamma$; what they wrote as $\log_{10}(\gamma/$Gyr$^{-2})=-4.87$ should actually be $\log_{10}(-\gamma/$Gyr$^{-2})=4.87$, etc.}

C21 derive typical values of $\gamma$ to be $\log_{10}(-\gamma\cdot\textrm{Gyr}^2)=2.93$ for a log-spherical MW potential and $\log_{10}(-\gamma\cdot\textrm{Gyr}^2)=3.94$ for a Miyamoto-Nagai Disk potential \citep{MiyamotoNagai1975}.  Both of these values are consistent with our fit value of $\gamma$, although our observed value of $\gamma$ is closer to the spheroid potential than the disk potential. C21 found that their value of $\gamma$ was more disky ($\log_{10}(-\gamma\cdot\textrm{Gyr}^2) = 4.87$) than our best-fit $\gamma$ value ($\log_{10}(-\gamma\cdot\textrm{Gyr}^2) = 3.3$). The difference in these values arises from the updated timing solutions for the pulsars relative to what was used in C21; when fitting the cross model to only the C21 sources, but instead using the more recent values for the accelerations and distances, we obtain a $\gamma$ value that is consistent with our $\gamma$ within 1$\sigma$ uncertainties. 

\begin{figure}
    \centering
    \includegraphics[width=0.45\textwidth]{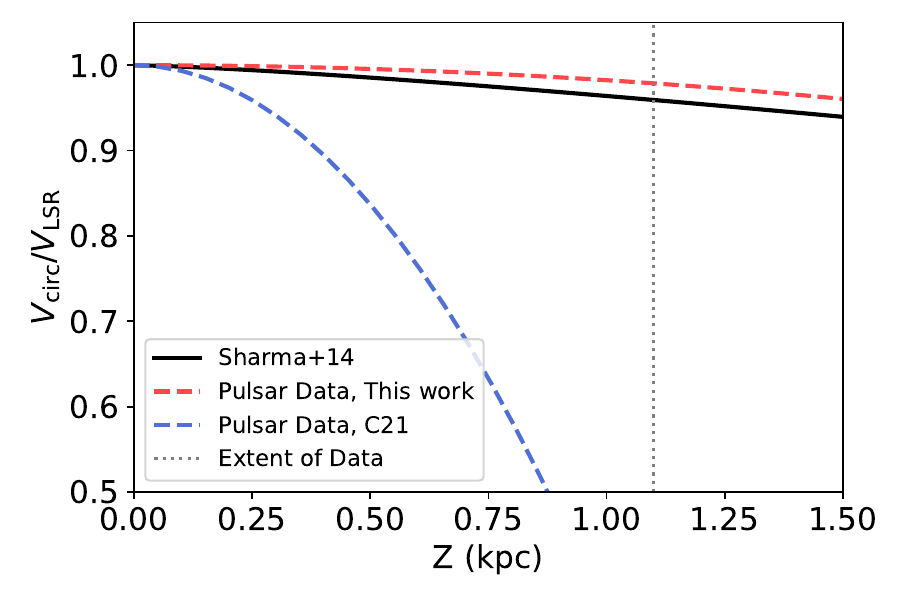}
    \caption{The ratio of $V_\textrm{circ}$ to $V_\textrm{LSR}$ as a function of height from the disk. The dashed red (this work) and blue (C21) lines show the values obtained from the cross model using pulsar acceleration data, and the solid black line shows the values obtained from kinematic data by Sharma et al. (2014) \cite{Sharma2014}. The vertical extent of the pulsar data (which is roughly equal for C21 and our local dataset) is shown as a vertical dotted line. The C21 pulsar data produces a more disky model than our updated model which uses more recent timing solutions, and which is now consistent with the kinematic values from Sharma et al. \cite{Sharma2014} to the vertical extent of the data.}
    \label{fig:vc_vcirc}
\end{figure}

The cross model also gives an approximation for the circular speed as a function of $Z$ at $R=R_\odot$: \begin{equation}
    \frac{V_\textrm{circ}}{V_\textrm{LSR}} \approx 1 - \frac{|\gamma|}{2 V^2_\textrm{LSR}} z^2 .
\end{equation} This expression is plotted in Figure \ref{fig:vc_vcirc} for our best-fit value of $\gamma$, along with the best-fit $\gamma$ value from C21, and a model of $V_\textrm{circ}/V_\textrm{LSR}$ from Sharma et al. \cite{Sharma2014}, who analyzed kinematic data of MW stars. The Sharma et al. \cite{Sharma2014} expression closely agrees with canonical MW potential models, such as the model of Law \& Majewski (2010) \cite{LawMajewski2010}. Overall, $V_\textrm{circ}/V_\textrm{LSR}$ is similar for our pulsar data and the kinematic data; however, the C21 $\gamma$ value produces a very disky potential model that is inconsistent with the kinematic data from Sharma et al. \cite{Sharma2014}. 
As our local dataset has the same vertical extent as the C21 dataset, these differences must be due to improvements in the reported pulsar timing values and the addition of new data.

It is possible to relate $\gamma$ and the scale length of a Miyamoto-Nagai Disk potential with Eq. 15 of C21, by fixing the mass and scale length of the disk to the C21 values, and then numerically solving for the scale height given some value of $\gamma$. This corresponds to a scale height of 1.0 kpc for our fit $\gamma$ value, which is somewhat larger than but generally consistent with MW thick disk populations. The C21 $\gamma$ value corresponds to a scale height of 0.05 kpc, implying a much more condensed disk than we find using the updated pulsar dataset. 

\subsubsection{Damour-Taylor Model} \label{sec:dt_pot}

The orbital decay via the emission of gravitational radiation by binary pulsars can be used
to test general relativity. The Damour-Taylor model \cite{DamourTaylor1991} is an approximation to the MW potential that is commonly used in these studies \citep{DamourTaylor1991,NiceTaylor1995,Lazaridis2009}. While its functional form has changed somewhat since its original definition in Damour \& Taylor \cite{DamourTaylor1991}, the model is essentially a flat rotation curve model with a roughly linear disk acceleration profile, which is similar to our $\alpha$ model. 

We use the version of the model from Lazaridis et al. \cite{Lazaridis2009}, where the acceleration profile is defined as: \begin{align} \label{eq:dtp}
    a_\textrm{los} =& \;\;\;\;\; a_{R,\textrm{los}} + a_{z,\textrm{los}} \notag\\
    =& -\frac{V^2_{LSR}}{R_\odot}\left( \cos l + \frac{\beta}{\beta^2 + \sin^2 l} \right) \cos b \\
    & - \frac{2.27 |z_\textrm{kpc}| + 3.68 (1 - e^{-4.31 |z_\textrm{kpc}|})}{10^{-9} \textrm{ cm s}^{-2}} |\sin b|, \notag
\end{align} where $l$ and $b$ are the Galactic longitude and latitude coordinates of the source, $\beta = (d/R_\odot)\cos b - \cos l$, where $d$ is the distance of the source from the Sun, and $z_\textrm{kpc}$ is the height from the midplane in kpc. The trigonometry of the model is not relevant to the Galactic potential, rather it provides the projection of the Galactic acceleration along our line-of-sight based on the source's position. 

As this potential is separable, we can write the vertical acceleration of this model as \begin{equation}
    a_z = -A|z| - B\left(1 - e^{-C|z|}\right),
\end{equation} which corresponds to a potential of \begin{equation}
    \Phi(z) = \frac{A}{2}z^2 + B\left(|z| + \frac{e^{-C |z|}}{C} \right),
\end{equation} and in the limit of small $z$ we recover the harmonic vertical potential \begin{equation}
    \Phi(z) \approx \frac{1}{2} (A + BC)z^2 = \frac{1}{2}\alpha z^2.
\end{equation} Plugging in the values for $A$, $B$, and $C$ from Equation \ref{eq:dtp}, we can infer a value of $\log_{10}(\alpha) = 3.77$ for this potential, which is somewhat larger than our fit values of $\alpha$ for the $\alpha$, $\alpha-\beta$, and cross potentials. However, note that for several pulsars $Z$ is not small, in which case this expansion is not an accurate approximation of the original potential.

Evaluating this model for our acceleration data produces very large AIC and $\chi^2$ values, indicating that the model is not a good fit to the observed data. This is worrisome, as this model is commonly used in tests of general relativity to obtain $\dot{P}^\textrm{Gal}_b$; if the model is not accurate, this would lead to inconsistencies in the reported values of $\dot{P}^\textrm{GR}_b$ in these studies. The inconsistency between the Damour-Taylor model and the observed data appears to be due to the Damour-Taylor model incorrectly estimating the MW disk density.

A $\Delta$AIC of 2 is positive evidence in favor of the model with the lower AIC, while $\Delta$AIC of 6 provides strong evidence \citep{KassRaftery1995}.  Thus, the $\alpha$, $\alpha-\beta$, and $\alpha-\gamma$ models carry similar statistical confidence. MWPotential2014 \citep{Bovy2017} and MilkyWayPotential2022 \citep{Gala} have lower confidence, and the Damour-Taylor model has even larger AIC and $\chi^2$ values relative to the other kinematic modelsl; these three models are statistically distinct from the first three symmetric models. The extremely large AIC of the Damour-Taylor model indicates that it should not be used to approximate the Galactic acceleration for pulsars near the Sun. 

\begin{table*}[]
    \setlength\extrarowheight{10pt}
    \centering
    \begin{tabular}{llrr}
    \hline \hline
    Global Model & Parameters & AIC & red.$\chi^2$ \\ \hline 
    Hernquist & $\log_{10}(M_\textrm{tot}/M_\odot)=12.7\pm0.2$ & 78 & 3.1 \\ 
     & $r_s = 32\pm 11$ kpc & & \\
    NFW & $\log_{10}(M_\textrm{vir}/M_\odot)=13.0\pm0.2$ & 78 & 3.1 \\ 
     & $r_s = 20\pm 7$ kpc & & \\
    NFW \& Miyamoto-Nagai Disk & $\log_{10}(M_\textrm{vir}/M_\odot)=12.8\pm0.3$ & 84 & 3.5 \\ 
     & $r_s = 16\pm 9$ kpc & & \\
     & $\log_{10}(M_\textrm{disk}/M_\odot)=10\pm2$ & & \\
     & $a = 10\pm 40$ kpc & & \\
     & $b = 0.01\pm 0.5$ kpc & & \\
    \textit{MWPotential2014} & \textit{Galpy} Parameters \citep{Galpy} & 223 & 11.0 \\
    \textit{MilkyWayPotential2022} & \textit{Gala} Parameters \citep{Gala} & 229 & 12.4 \\ \hline \hline
    \end{tabular}
    \caption{Optimized fits to the global potential models.}
    \label{tab:global_models}
\end{table*}

\subsection{Global Models} \label{sec:global_models}

As the local models are only local expansions of the global MW potential field, we cannot use them to estimate the MW mass or the scale radius of the dark matter halo. We now turn to a set of global models, which provide estimates of these quantities below.

The global models, described in Table \ref{tab:global_models}, are combinations of a Hernquist profile \citep{Hernquist1990}, an NFW halo \citep{NFW}, and a Miyamoto-Nagai Disk to fit the observed acceleration data. These fits used all 26 sources. Note that the AIC of these fits cannot be directly compared to the AIC of the local model fits because they use a different set of data. 

The Hernquist and NFW model fits are very similar, with total MW masses of $M_\mathrm{tot}$ = 5-8$\times10^{12}$ M$_\odot$ and scale lengths ($r_s$) in the low tens of kpc. These fits are similar to modern potential models for the MW, except that our model fits place the total MW mass 2-5 times larger than kinematic and dynamical models, which place the total MW mass between 0.7 and $\sim$3$\times10^{12}$ M$_\odot$ \citep{Watkins2010,Newberg2010,Bhattacharjee2014,McMillan2017,Fritz2018,Eilers2019,Gardner2021,Craig2022}. These fits are also an improvement over those of C21, who were not able to constrain a scale radius or mass for their global fits due to the restricted range of available data in that earlier work. 

We also tested the addition of a Miyamoto-Nagai disk to an NFW potential, with mixed results. We were able to constrain the mass of the disk (although it is about an order of magnitude below kinematic estimates -- e.g. Bovy \& Rix \cite{BovyRix2013} and Licquia \& Newman \cite{LicquiaNewman2015} -- and dynamical models, e.g. Newberg et al. \cite{Newberg2010}), but the scale length, $a$, and scale height, $b$, of the disk potential are not well constrained.

The \verb!Galpy! \textit{MWPotential2014} and \verb!Gala! \textit{MilkyWayPotential2022} models are both derived from kinematic data. \textit{MWpotential2014} \cite{Galpy} is fit to a variety of MW kinematic data, including the MW rotation curve, vertical force curve, and total mass measurements; \textit{MilkyWayPotential2022} \cite{Gala} is fit to rotation curve data, the vertical structure of local phase-space, and total mass estimates. The two potential models are ultimately similar and only differ by a few percent in the regions we are interested in.

Both models have a very poor fit to the acceleration data compared to our optimized potential models. It should be noted, however, that the majority of the disagreement between the acceleration data and these kinematic models comes from PSR B1913+16 (the observed acceleration of this source has previously been shown to be inconsistent with the Galactic potential \cite{WeisbergHuang2016,Bovy2021}). If B1913+16 is excluded, \textit{MilkyWayPotential2022} has AIC = 79 and $\chi^2=3.8$ (although the AICs of the other models also drop by $\sim$10 when B1913+16 is removed). The relatively poor fit of \textit{MWPotential2014} and \textit{MilkyWayPotential2022} to the observed acceleration data is in part due to the total mass estimate of the MW, which for \textit{MWPotential2014} is about 9$\times$10$^{11}$ M$_\odot$, and for \textit{MilkyWayPotential2022} is about 1.1$\times10^{12}$ M$_\odot$.

As the Milky Way is believed to be roughly spherically symmetric over large scales, our pulsar acceleration data should be much more sensitive to the amount of mass within the Solar circle than outside it (via the shell theorem). The mass enclosed within 8 kpc of the Galactic center in the NFW global model is $M_\textrm{enc}(8 \textrm{ kpc}) = 2.3 \times 10^{11}$ M$_\odot$, which is $\sim$2.3 times larger than the enclosed mass of the \textit{MilkyWayPotential2022} model ($M_\textrm{enc}(8 \textrm{ kpc}) = 9.8 \times 10^{10}$ M$_\odot$).  Thus, the pulsar acceleration data imply a larger MW mass than kinematic methods. If these larger masses are to be believed, they imply substantial differences between orbits calculated using potentials calibrated to modern kinematic data and the potentials fit to the pulsar data. However, it should be noted that the current global model fits estimate an extremely large circular speed at the location of the Sun (roughly 320 km/s), so it is likely that we are not yet producing precise constraints on the enclosed mass, and it is expected that the addition of future pulsar data will significantly improve these fits.

\section{Inferring Disequilibrium From Accelerations} \label{sec:diseq}

One way that accelerations can be used to probe disequilibrium in the MW disk is by looking at the vertical acceleration profile ($a_z$ vs. $z$). Chakrabarti et al. \cite{Chakrabarti2020} showed that simulations of a MW interacting with dwarf galaxies would have a vertical acceleration profile that is highly asymmetric about the midplane. As a result, showing that $a_z(+z')\neq a_z(-z')$ for some nonzero $z'$, or alternatively that $\dv*{a_z}{z}$ in the midplane is nonzero, indicates disequilibrium in the Galactic disk. 



\begin{figure}
    \centering
    \includegraphics[width=0.45\textwidth]{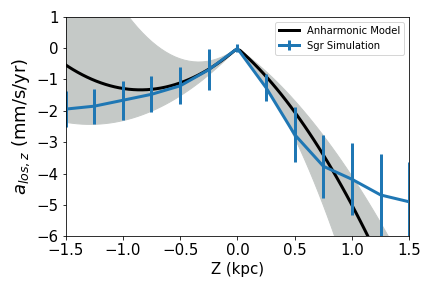}
    \caption{Observed and simulated asymmetry in the vertical acceleration profile of the disk. The fit of the anharmonic potential profile to the observed pulsar data is shown as the black line, and 1$\sigma$ uncertainties are shown as the shaded gray region. The vertical acceleration profile from a simulation of the MW and the Sgr dSph \citep{Chakrabartietal2019} is shown as a blue line with error bars. The shape of the two profiles are roughly the same, suggesting that the observed acceleration asymmetry is consistent with the effects of orbiting dwarf galaxy interactions.}
    \label{fig:anharmonic}
\end{figure}

\begin{table*}[]
    \centering
    \begin{tabular}{lrrrrr} \hline
        Model & A & B & $\dv*{V_\textrm{circ}}{R}$ & $\rho_0$ & $\rho_\textrm{0,DM}$ \\
              & \textit{(km/s/kpc)} & \textit{(km/s/kpc)} & \textit{(km/s/kpc)} & \textit{(M$_\odot$/pc$^3$)} & \textit{(M$_\odot$/pc$^3$)} \\ \hline \hline 
        $\alpha-\beta$ & 12.8$\pm$2.4 & -15.7$\pm$2.4 & 4$\pm$5 & 0.060$\pm$0.018 & -0.012$\pm$0.018 \\
        $2\alpha-\beta$ & 15.4$\pm$2.6 & -13.1$\pm$2.6 & -2$\pm$5 & 0.062$\pm$0.017 & -0.010$\pm$0.018 \\
        $\alpha-\beta$ + 2PM & 20.4$\pm$1.7 & -8.1$\pm$1.7 & -12$\pm$5 &0.066$\pm$0.017 & -0.006$\pm$0.018 \\\hline
    \end{tabular}
    \caption{Oort constants, the slope of the rotation curve, and the (dark matter) density in the midplane for different potential models.}
    \label{tab:oort}
\end{table*}

\subsection{Asymmetric Local Models} \label{sec:local-diseq-models}

We now move on to models that capture disequilibrium features of the Galactic potential, i.e. these potentials are not invariant under negation of $\phi$ and/or $z$. 

In general, the more flexible asymmetric models do a better job of fitting the observed acceleration data. This is strong evidence that the binary pulsar acceleration data contains substantial disequilibrium effects, which must be accounted for in order to obtain an accurate model of the Galactic potential. 

\subsubsection{Anharmonic and 2$\alpha$-$\beta$ Models}

The first disequilibrium model is the anharmonic model, which consists of the $\alpha$ model except the vertical component of the potential is expanded one degree further in a power series expansion for a separable potential:

\begin{equation}\label{eq:pwr}
    \Phi(z) = \sum_{n=2}^3 \frac{\alpha_{n-1}}{n} z^n,
\end{equation} which provides a simple form for $\dv*{a_z}{z}$ in the midplane: \begin{equation}
\eval{\dv{a_z}{z}}_{z=0} = \eval{\dv{z}\left(-\dv{\Phi}{z}\right)}_{z=0} = -\alpha_1.
\end{equation} In essence, $\alpha_1$ can be interpreted as a measure of how much the local disk potential is out of equilibrium.

The additional $z^3$ term allows us to characterize the asymmetry in the vertical acceleration profile, which is shown in Figure \ref{fig:anharmonic}. The observed value of $a_z$ below the disk has a smaller magnitude than $a_Z$ above the disk, which is consistent with simulations of the MW interacting with the Sgr dSph. The methodology of the simulations is described in \cite{Chakrabarti_Blitz2009,Chakrabarti_Blitz2011}, and are the same simulations that are used in \cite{Chakrabartietal2019}. This is consistent with the popular idea that this type of disequilibrium is (at least in part) caused by the disk interacting with orbiting satellites. 

The $2\alpha-\beta$ model is identical to the anharmonic model except that it does not require a flat rotation curve; the interpretation of the $\beta$ parameter is the same as for the $\alpha-\beta$ model. 

We also attempted to fit a model that included a $z^4$ term, although this quartic term could not be constrained with existing data. A model that includes a quartic term would be useful for quantifying the behavior of the phase space spiral that has been observed in the MW disk \citep{Antoja2018}.  

\subsubsection{Local Expansion Model} \label{sec:local-expansion-model}

The local expansion model is a 1st-order Taylor series expansion of the acceleration profile centered at the position of the Sun assuming that the potential is additively separable: \begin{equation}
    a_i(\mathbf{x}) \approx a_i(\mathbf{x}_\odot) +  {\pdv{a_i}{x_i}}\bigg\rvert_\odot (x_i - x_{i,\odot}).
\end{equation} While it is useful to characterize the slope of the rotation curve with $\dv*{a_R}{R}$ and the midplane density of the disk with $\dv*{a_z}{z}$, these terms do not describe disequilibrium features. The nonzero $\dv*{a_\phi}{\phi}$ implies disequilibrium, as it requires that the potential be a function of azimuth, rather than axisymmetric. The AIC of this model is a substantial improvement over the anharmonic and $2\alpha-\beta$ models, which indicates that the MW disk density has substantial azimuthal variation in the Solar neighborhood. 

\subsubsection{Sinusoidal Model}

The MW disk has been shown to contain ripples, which are probably related to the passage of orbiting satellites \citep{Xu2015,BinneySchonrich2018,Laporte2019,McMillan2022}. We explore these features with the sinusoidal model, which follows the same idea as the density model outlined in Xu et al. \cite{Xu2015} of varying the ``midplane'' of the potential as a sine curve, except we use a harmonic oscillator disk potential instead of an exponential density profile. This model outperforms the anharmonic and  $2\alpha-\beta$ models, suggesting that the disk ripples are a prominent feature of the disk potential. We obtain a slightly smaller wavelength and a somewhat larger amplitude for the disk oscillations than was found in the stellar density distribution by Xu et al. \cite{Xu2015}, although these differences could be largely due to the different definitions of the two models; we used a conservative model in order to minimize the number of free parameters that we had to fit, whereas Xu et al. \cite{Xu2015} were able to separately fit the disk density variations above and below the midplane. 

\subsubsection{$\alpha$-$\beta$ + 2 Point Mass Model}

The final model we discuss here is the most flexible model, which consists of the $\alpha-\beta$ model with the addition of an arbitrary number of point masses. The location and mass of each point mass is independently fit to the data simultaneously with the $\alpha$ and $\beta$ parameters. This allows for a flexible non-parametric fit to any various disequilibrium features that may be present in the acceleration data. In order to determine the optimal number of point masses that should be included in the model, we sequentially fit the $\alpha-\beta$ model with $n$ point masses, increasing $n$ until the AIC no longer improved when we added another point mass to the model. The optimal number of point masses (which minimizes AIC) is 2. While the AIC of this model is substantially lower than the other models in this work, the fact that its $\chi^2<1$ could indicate that it is actually overfitting the data.

It is not immediately clear what the physical significance is for the point masses in this model. On the surface, the point masses are simply a flexible model that allows for the generation of a wide range of potential models. It is plausible that this model is producing an approximation of a more complex potential model, and that this presently unknown more complex model is not well represented using our other analytical models. 

One possible interpretation for these point masses is that each point mass corresponds to a large dark matter subhalo in the Solar neighborhood. These dark matter subhalos would have masses of roughly $10^8$ \msun, which is large but not unreasonable for a dark matter subhalo in a MW-mass galaxy \citep[e.g.][]{Garcia-Conde2023}. If this model is actually identifying real dark matter subhalos, then the number density for $M\sim10^{10}$ M$_\odot$ subhalos must be higher than in $\Lambda$-CDM, which predicts that there should only be a few dark matter subhalos with this mass across the entire Galaxy \citep{Reed2005,KunTingEddie2017,GreenVandenbosch2019}. We plan to explore the sensitivity of direct acceleration measurements to dark matter subhalos with a future publication, as it is outside the scope of this work.

\section{Fundamental Galactic Parameters from Pulsar Timing} \label{sec:oort}

\subsection{Oort Constants}

The Galaxy has historically been characterized using the kinematics of stars near the Sun \citep{BlandHawthornGerhard2016}. One way to do this is by empirically estimating the Oort constants \citep{Oort1927}, known as $A$, a measure of azimuthal shear motion in the Solar neighborhood, and $B$, a measure of the rotation curve vorticity, i.e. the scale of epicyclic motion. These constants are related to the rotation curve near the Sun: \begin{align}
    A - B =& \left[\frac{V_\textrm{circ}}{R}\right]_{R_\odot}, \textrm{ and}\\
    A + B =& -\left[\dv{V_\textrm{circ}}{R}\right]_{R_\odot}.
\end{align}

Modern kinematic estimates from \textit{Gaia} data place the Oort constants around $A=15.1$ to $15.3$ km/s/kpc, and $B=-13.4$ to $-11.9$ km/s/kpc \citep{Bovy2017,Li2019}, which are consistent with the most recent estimates from \textit{Gaia} DR3 \citep{Creevey2023}.

From the definition of $\beta$ in the Equation \ref{eq:beta}, we can relate the $\alpha-\beta$ model to the Oort constants:\begin{equation}
    \beta = -\frac{A+B}{A-B},
\end{equation} which implies a value of $\beta \approx -0.1$ from the \textit{Gaia} values for the Oort constants. We can also write the Oort constants as\begin{align}
    A =& \frac{1}{2}\frac{V_{LSR}}{R_\odot}\left( 1 - \beta \right), \textrm{ and} \\
    B =& -\frac{1}{2}\frac{V_{LSR}}{R_\odot}\left( 1 + \beta \right).
\end{align} 

The values of the Oort constants for the three $\alpha-\beta$ models in this work are listed in Table \ref{tab:oort}. It is clear that the different models produce substantially different values for the Oort constants, presumably because they are simultaneously fitting different amounts of local disequilibrium. The $2\alpha-\beta$ model is consistent with the kinematic estimates of the Oort constants, while the other two models are inconsistent with the kinematic estimates. These results suggest that it is not only important to consider disequilibrium in the acceleration models in order to properly extract properties of the Galaxy, but that it is also important \textit{how} this disequilibrium is accounted for. 

\begin{figure}
    \centering
    \includegraphics[width=0.45\textwidth]{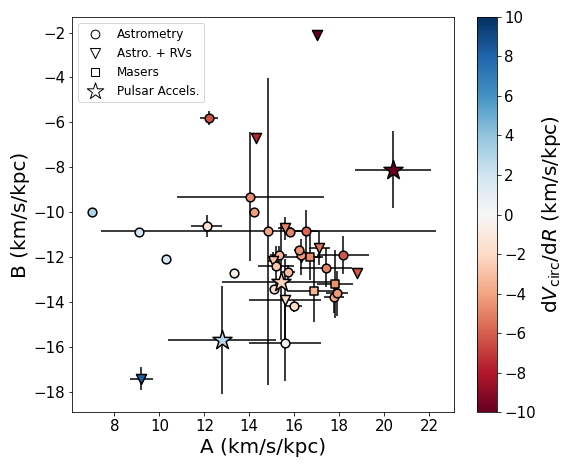}
    \caption{Selected measurements of the Oort constants $A$ and $B$ since 2010. The shape of each point indicates the type of data used to make the measurement, and the color of each point indicates the corresponding slope of the rotation curve given by $\dv*{V_\textrm{circ}}{R} = -(A + B)$. The values we infer from the pulsar timing data are shown as stars; while the three models span a relatively large area on the figure, the $2\alpha-\beta$ model (the pale red star) is in broad agreement with many of the previous measurements. The sources for this data can be found in Table \ref{tab:oort_figure_data}. }
    \label{fig:oort_constants}
\end{figure}

Figure \ref{fig:oort_constants} shows our inferred values of the Oort constants compared with a selection of measurements of the Oort constants since 2010. Our values span a relatively large range, but the values we obtain for the Oort constants from the 2$\alpha-\beta$ model agree well with many previous measurements. 



\subsection{The Slope of the Rotation Curve}

Equation \ref{eq:beta} provides a way to compute the slope of the rotation curve for our $\alpha$-$\beta$ models; the slopes of the rotation curve for each model are provided in Table \ref{tab:oort}. 

While we obtain a range of $\dv*{V_\textrm{circ}}{R}$ values for our different models, the preferred 2$\alpha$-$\beta$ model constrains the slope of the rotation curve to be -2 $\pm$ 5 km/s/kpc at the location of the Sun. This value is consistent with the slope of the rotation curve being flat. However, a slightly declining rotation curve at the Solar location is in good agreement with \textit{Gaia} data, such as Figure 13 of \textit{Gaia} Collaboration et al. \cite{GaiaDR2_maps}. Additionally, P\~oder et al. \cite{Poder2023} contains a compilation of a few different rotation curve models, all but one of which appear to have a value of $\dv*{V_\textrm{circ}}{R} \sim 2$ km/s/kpc at the Solar position. 

\subsection{The Oort Limit and Local Dark Matter Density}

\begin{figure}
    \centering
    \includegraphics[width=0.47\textwidth]{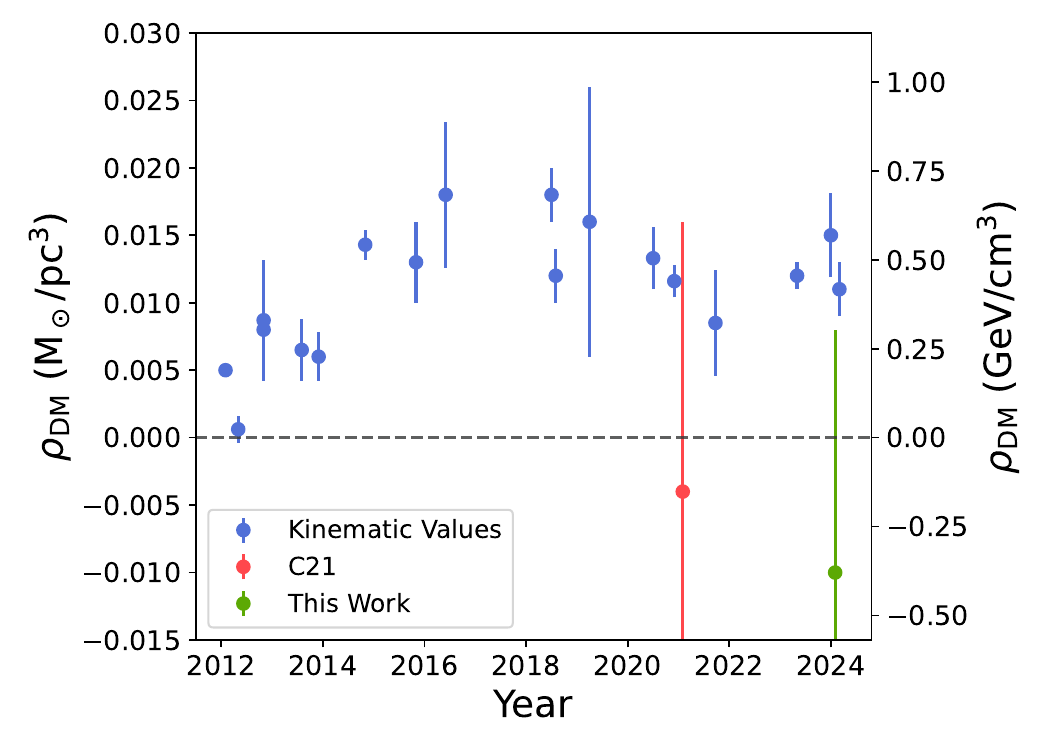}
    \caption{Selected literature measurements of the dark matter density in the Galactic midplane plotted against publication date. The two red points are dark matter midplane density measurements from pulsar acceleration data. Details of the included references are provided in Table \ref{tab:dm_density}.}
    \label{fig:dm_density}
\end{figure}

\begin{figure*}
    \centering
    \includegraphics[width=\textwidth]{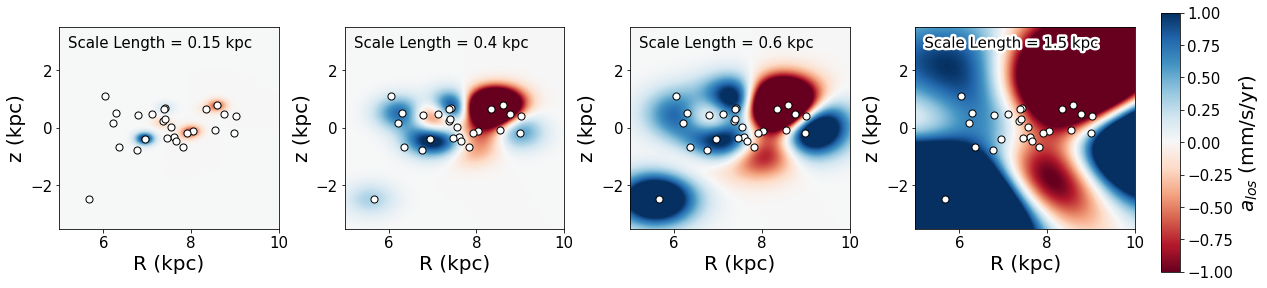}
    \caption{The effect of different choices of scale length on the GPR map. The choice of scale length increases from left to right. The left-most panel has too small of a scale length, as the GPR map only estimates \alos near a source and does not interpolate between sources. The right-most panel has too large a scale length, as it averages over complex substructure and extrapolates \alos far beyond where data is located. The middle two panels show that for a reasonable choice of scale length, the general shapes of the GPR maps are similar, and therefore the exact choice of scale length is not particularly important. }
    \label{fig:goldilocks}
\end{figure*}

\begin{figure*}
    \centering
    \includegraphics[width=\textwidth]{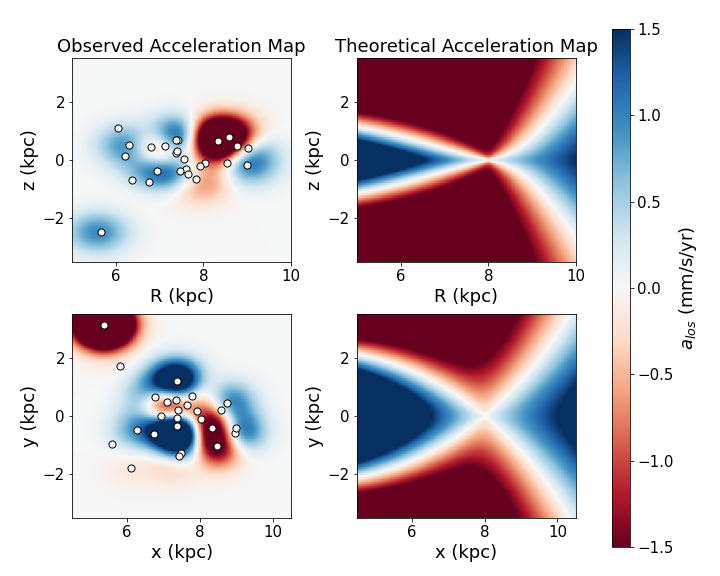}
    \caption{GPR map of the observed line-of-sight acceleration compared to the theoretical line-of-sight accelerations from \textit{Gala MilkyWayPotential2022}. The white points indicate the positions of the binary pulsars. Red (blue) indicates an apparent line-of-sight acceleration towards (away from) the Sun.  The top row shows the $R-Z$ plane at $Y=0$, and the bottom row shows the $X-Y$ plane at $Z=0$. The observed line-of-sight accelerations appear to be very different from the theoretical acceleration maps, implying a substantial departure from an equilibrium acceleration field. }
    \label{fig:gpnr_obs_vs_theo}
\end{figure*}

\begin{figure*}
    \centering
    \includegraphics[width=\textwidth]{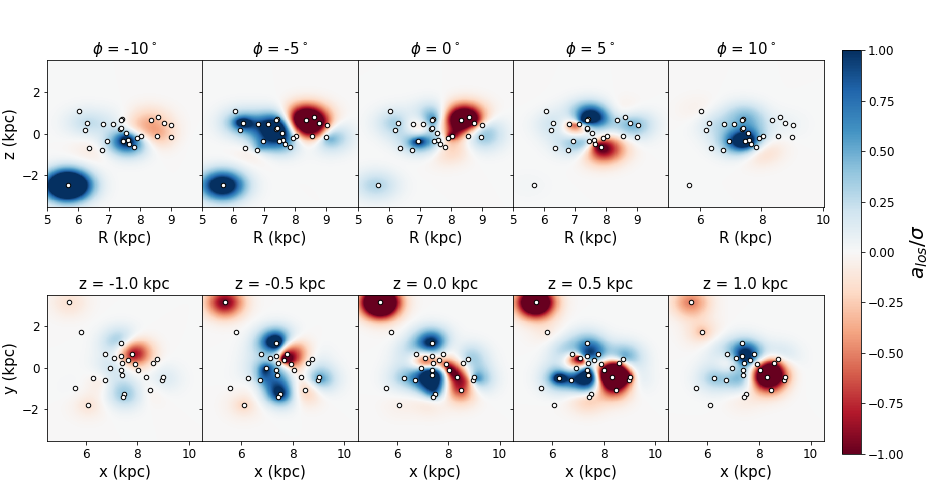}
    \caption{Tomography of the observed line-of-sight acceleration GPR map. Colors are the same as in Figure \ref{fig:gpnr_obs_vs_theo}, except the predicted accelerations have been divided by the corresponding uncertainty in the predicted GPR map in order to emphasize significance of the GPR map values. The top row shows slices of the $R-Z$ plane at different azimuth, where $\phi$ increases in the same direction as the Sun's rotation. The bottom row shows slices of the $X-Y$ plane at different heights below/above the midplane. An equilibrium acceleration field should look identical at $\pm\phi$ and $\pm Z$, but the observed accelerations are substantially different on either side of the Sun. }
    \label{fig:tomography}
\end{figure*}

The Oort limit, or the volumetric mass density in the midplane of the disk, is another characterization of the local Galaxy that is often derived from kinematic data. It can be obtained by solving the Poisson Equation in cylindrical coordinates for the $\alpha-\beta$ models, which leads to \begin{equation}
    4\pi G \rho_0 = \alpha + 2\beta\Omega_\odot^2,
\end{equation} where $\Omega=V_\textrm{circ}/R$ is the angular velocity of the disk at a given radius. Here, $\rho_0 = \rho_\textrm{bary} + \rho_\textrm{DM}$ is the midplane density of the baryonic matter plus the dark matter.

C21 used this prescription to calculate an Oort limit of $\rho_0=0.08^{+0.05}_{-0.02}$ \msun/pc$^3$. When combined with the baryon budget of $\rho_{0,\textrm{bary}}=0.084\pm0.012$ \msun/pc$^3$ from McKee et al. \cite{McKee2015}, this results in a midplane dark matter density of $\rho_{0,\textrm{DM}}=-0.004^{+0.05}_{-0.02}$ \msun/pc$^3$. C21 state that this suggests that values of the Oort limit from Jeans analysis may be an overestimate; their value of $\rho_{0,\textrm{DM}}$ is consistent with there being no dark matter in the midplane, as well as McKee et al. \cite{McKee2015}'s value of $\rho_{0,\textrm{DM}} = 0.013 \pm 0.003~\rm M_{\odot}/\rm pc^{3}$.  

Our results for the Oort limit and the dark matter density in the midplane are given in Table \ref{tab:oort}. These values use a baryon budget built from a literature compilation of the local stellar volume mass density of $\rho_*=0.0468\pm0.0050$ \msun/pc$^3$ \citep{Guo2020} and the gas density from McKee et al. \cite{McKee2015}, $\rho_\textrm{gas}=0.025\pm0.003$ \msun/pc$^3$, which gives a local baryon density of $\rho_\textrm{bary}=0.072\pm0.006$ \msun/pc$^3$. This is similar to, but slightly less than the baryon budget of Bienayme et al. \cite{Bienayme2014}: $\rho_{0,\textrm{bary}}=0.077\pm0.007$ \msun/pc$^3$. 

These updated values are consistent with, and have smaller uncertainty than the measurement of the dark matter density in the midplane from C21. These values are all consistent with no dark matter in the mid-plane. Our measurements of the local dark matter density are lower than recent estimates of the dark matter density obtained from Jeans modeling (between 0.010 and 0.016 \msun/pc$^3$\cite{Bienayme2014,McKee2015,Buch2019,Guo2020,Salomon2020}), which is consistent with the claim of C21 that these kinematic methods may be overestimating the local dark matter density. M23 also use pulsar timing to obtain an Oort limit of $\rho_0=0.036\pm0.021$ \msun/pc$^3$, which is substantially lower than our measurement by about a factor of 2, and does not appear to be consistent with any available baryon budgets. A comparison of these measurements with a compilation of literature values are provided in Figure \ref{fig:dm_density}.

Note that if we instead use the McKee et al. \cite{McKee2015} baryon budget, which has a much larger stellar mass density than the values collated in Guo et al. \cite{Guo2020}, we get $\rho_{0,\textrm{DM}}=-0.024\pm0.021$, $-0.022\pm0.021$, and $-0.018\pm0.021$ \msun/pc$^3$ for our respective models. The McKee et al. \cite{McKee2015} baryon budget is similar to recent estimates from \textit{Gaia} DR2 data, $\rho_{0,\textrm{bary}}\sim0.09$ \msun/pc$^3$ \citep{Widmark2019}. Either of these baryon budgets produce negative values for $\rho_{0,\textrm{DM}}$ that are larger than the error bars in our models, which suggests that these baryon budgets may overestimate the baryon content in the midplane. It has been shown that kinematic modeling can overestimate the MW surface density when the disk is experiencing substantial dynamical disequilibrium \cite{Haines2019}, which could potentially explain why our measurement of $\rho_{0,\textrm{DM}}$ is negative.

Lim et al. \cite{Lim2023} used unsupervised machine learning in the form of normalizing flows to model the phase space distribution of \textit{Gaia} DR3 data. They obtained a value of $\rho_0$ = 0.0617$\pm$ 0.0020 M$_\odot$/pc$^3$ for the Oort limit, which is nearly identical to our pulsar acceleration measurement. Lim et al. \cite{Lim2023} calculated a baryon budget of only 0.0534 $\pm$ 0.0042 M$_\odot$/pc$^3$ from the \textit{Gaia} DR3 data; if we instead use this baryon budget, we obtain a dark matter density in the midplane of $\rho_{0,\textrm{DM}}$ = 0.008 $\pm$ 0.02 M$_\odot$/pc$^3$ for the $2\alpha$-$\beta$ model. This value is still consistent with no dark matter in the Galactic midplane -- however, it is non-negative. 

Some of the differences in measurements of the Oort limit and the corresponding dark matter density may be due to differences in the effective ranges that are probed by the different studies. The ``local'' pulsar dataset lies within $\sim$1 kpc of the MW disk plane, but, for example, the McKee et al. \cite{McKee2015} data is primarily located within a few hundred pc of the midplane. These studies may measure somewhat different densities simply because they average over different vertical extents, so caution should be used when directly comparing these values.

\section{Beyond Smooth Models}

Analytical and parametric models are useful for studying Galactic structure, because they enable one to calculate and compare fundamental Galactic parameters from different datasets. In the regime of a static, equilibrium Galaxy, these parametric models effectively constrain the behavior of the Galaxy. However, we are now aware that the MW is experiencing many disequilibrium effects, which indicates that parametric models are approximations of the Galaxy's behavior. These disequilibrium effects are difficult to quantify analytically, and lend themselves to non-parametric methods with a high degree of flexibility. Here, we develop a non-parametric map of the acceleration field of the Galaxy, and discuss ways to improve this map in the future. 

\subsection{Gaussian Process Regression Maps}

Gaussian Process Regression (GPR) is a non-parametric supervised learning algorithm that trains on measurements of data, and produces an interpolated model of the data and its uncertainties. This interpolated model can be evaluated at any point; in our case, this allows us to build a 3-dimensional map of the line-of-sight acceleration field within several kpc of the Sun. This was done using the \verb!GaussianProcessRegressor! class from \verb!scikit-learn! to generate the GPR model.

GPR requires a characteristic length scale for the interpolated output, which is analogous to the scale of the kernel in a kernel density estimation. Typically, in GPR this scale length is treated as a hyperparameter, which is then optimized over the data in order to produce a GPR model that minimizes the ordinary-least-squares difference between the model and the data, along with a penalty for model complexity, similar to AIC \citep{RasmussenWilliams2005}. However, the ability of the model to successfully optimize the length-scale hyperparameter is dependent on the number of data points available. Ideally, one would have several dozen datapoints per dimension that could be used to obtain a good estimate of the length scale hyperparameter \citep{RasmussenWilliams2005}, but this is not true in our case. By adding or removing a single datapoint, we would obtain dramatically different values for the length scale of the GPR model. 

In order to avoid this problem, we simply set the length scale as 0.5 kpc rather than allow the GPR method to optimize it over the dataset. The rationale for this choice of length scale is shown in Figure \ref{fig:goldilocks}; for too-small choices of the length scale, the GPR model does not interpolate between datapoints, and for too-large choices of the length scale, the GPR model interpolates much further than is reasonable. However, for reasonable choices of length scale (larger than the distance between datapoints but smaller than the overall extent of the data), the GPR maps generally look the same regardless of the exact choice of length scale.

The $R-z$ and $x-y$ planes (2-dimensional slices) from the GPR map are shown in Figure \ref{fig:gpnr_obs_vs_theo}, along with the line-of-sight accelerations that are expected to be observed based on the \textit{MilkyWayPotential2022} potential model. It is immediately clear that the general shapes of the GPR map and the theoretical acceleration maps do not match, which implies that the observed acceleration data is not well-modeled by an equilibrium acceleration profile of the Galaxy. However, note that the theoretical \alos model has structure; this is a map of the tidal field, which is squeezed in the direction of rotation and elongated along the direction towards the Galactic center. 

A useful benchmark for the scale of the observed acceleration perturbations is to consider the radial and vertical components of the acceleration at the location of the Sun. The radial acceleration is roughly \begin{equation}
    a_R \sim \frac{V_\textrm{circ}^2}{R_\odot} = 6.8 \textrm{ mm/s/yr}.
\end{equation} Assuming an infinite density sheet with a surface density of $\Sigma\sim$50 M$_\odot$/pc$^2$, the vertical acceleration is \begin{equation}
    a_z \sim G \Sigma = 0.2 \textrm{ mm/s/yr}.
\end{equation}

The observed variations are on the scale of 0.5-1 mm/s/yr, which is sizeable compared to the vertical acceleration but a fraction of the radial acceleration scale. We are only able to observe variations in \alos, which is a projected combination of $a_R$ and $a_z$, so the variations in \alos we see are on the $\sim10\%$ scale of the total acceleration. 

Further, we provide the tomography of the GPR map in Figure \ref{fig:tomography}. If the Galaxy were in equilibrium, the slices of the map at positive and negative $\phi$ and/or $z$ would be identical, because the Galaxy would be azimuthally symmetric. However, it is clear from the tomography that this is not the case; for example, the slice at $\phi=-5^\circ$ looks substantially different than the slice at $\phi=+5^\circ$. This could indicate a large amount of disequilibrium in the Galaxy's acceleration field near the Sun. 

However, note that the distribution of sources on either side of the Sun are not identical; because the GPR maps interpolate between data points, the differences in the GPR maps on either side of the Sun are likely in part due to differences in where pulsar data is available. However, we point out that the GPR maps in Figure \ref{fig:tomography} are scaled by uncertainty, and regions where there are not many pulsars will not have large values of $a_\mathrm{los}/\sigma$. It is currently unclear exactly how significant these variations are, or how much of the variations are caused by uncertainties on the individual data points or the particular distribution of pulsars. The true significance of these asymmetric features will be revealed as more pulsar data becomes available in the future. 

These GPR maps only consider the propagated uncertainty in \alos for each datapoint. However, each source has its own uncertainty in distance, which will not only change the source's position on the map, but the distance uncertainty is correlated to the uncertainty in \alos in a non-linear way. In the future we plan on creating GPR maps that take this type of correlated uncertainty in the input and output of each datapoint into account. Despite these shortcomings, the proof-of-concept GPR maps presented here are still powerful diagnostics for disequilibrium features and asymmetries in the acceleration profile of the Galaxy.

\section{Conclusions} \label{sec:conclusion}

Here we summarize the main findings of this work: 

\begin{itemize}
    \item We gather direct acceleration measurements for 26 binary pulsars, which span 3.4 kpc in $R$, and 3.6 kpc in $Z$. 
    \item The residual of the pulsar accelerations and accelerations from modern kinematic models appear to have a global gradient in $R$; this trend could not be resolved by varying the mass of the theoretical potential model. However, the trend can be eliminated if the Sun is experiencing a 1.1 mm/s/yr acceleration away from the Galactic center, comparable to a Jupiter-mass object at a distance of 400 AU. 
    \item We fit a collection of symmetric and asymmetric potential models to the observed acceleration data. The more flexible asymmetric models that allow for disequilibrium do a better job of modeling the data compared to the symmetric potentials, indicating that the observed accelerations are consistent with the Galactic disk being in disequilibrium. These asymmetries are consistent with those caused by dwarf galaxies interacting with the MW disk, although it is difficult to determine the exact physical mechanism responsible for the asymmetries given the current limitations of the acceleration data.
    \item The Damour-Taylor Potential \cite{DamourTaylor1991}, which is commonly used to obtain an approximation of the Galactic acceleration in tests of general relativity, is not a good fit to the observed acceleration data. Studies that use this potential to calibrate the observed acceleration of sources due to the gravitational field of the MW will therefore produce systematically inaccurate results, especially for sources with large heights from the Galactic midplane. 
    \item We constrain fundamental Galactic parameters using the acceleration data. Our values of the Oort Constants vary substantially depending on which model we use, but a disequilibrium model that considers the asymmetry of the vertical acceleration profile produces values for the Oort constants of $A=15.4\pm2.6$ km/s/kpc and $B=-13.1\pm2.6$ km/s/kpc. These are $\sim$5$\sigma$ measurements, and are consistent with literature values from kinematic estimates. This model also produces a 3.6$\sigma$ measurement of the Oort limit, or the density in the midplane, of $\rho_0=0.062\pm0.017$ \msun/pc$^3$, and a dark matter density in the midplane of $\rho_{0,\textrm{DM}}=-0.010\pm0.018$ \msun/pc$^3$. However, if we use the baryon budget of Lim et al. \cite{Lim2023}, we obtain a positive dark matter density in the midplane of $\rho_{0,\textrm{DM}}=0.008\pm0.02$ \msun/pc$^3$. The inferred dark matter density is consistent with there being no dark matter density in the Galactic plane. 
    \item We constrain the slope of the rotation curve to be -2 $\pm$ 5 km/s/kpc at the location of the Sun. This is consistent with the slope of the rotation curve being flat, but it is also consistent with the \textit{Gaia} findings that the azimuthal velocity curve in the plane is slightly decreasing at the Solar location.
    \item We measure the oblateness of the potential using a potential model that includes a cross ($\gamma$) term. Our measured value of $\gamma$ relies on updating pulsar timing solutions and produces a similar oblateness and scale height to thick disk kinematic data from Sharma et al. \cite{Sharma2014}.
    \item We provide non-parametric maps of the acceleration field of the Galaxy, which exhibit a substantial amount of variation from the theoretical acceleration maps. The tomography of this map is consistent with a substantial amount of disk disequilibrium, given its asymmetry in $Z$ and $\phi$. 
\end{itemize}

As binary pulsar datasets continue to improve and grow, we will be able to produce progressively better constraints on the MW's gravitational potential and its dark matter content. While more work remains to determine the sources of the currently unexplained variations in accelerations compared to theoretical maps, we have now illustrated the need for flexible disequilibrium models in order to utilize the acceleration data to its full capabilities. 

\begin{acknowledgments}

This work utilizes data from the Australia Telescope National Facility Catalogue \citep{Manchester2005}, which can be found online at \url{http://www.atnf.csiro.au/research/pulsar/psrcat}. MTL graciously acknowledges support received from NSF AAG award number 2009468, and NSF Physics Frontiers Center award number 2020265, which supports the NANOGrav project.  SC gratefully acknowledges support from NSF AST 2009828. 

\textit{Software used:} Numpy \citep{numpy}, Scipy \citep{scipy}, Sci-kit Learn \citep{scikitlearn}, Astropy \citep{astropy}, Matplotlib \citep{matplotlib}, emcee \citep{emcee}, Gala \citep{Gala}, Galpy \citep[][\url{http://github.com/jobovy/galpy}]{Galpy}
\end{acknowledgments}

\bibliographystyle{apsrev4-2}
\bibliography{main.bib}

\begin{thebibliography}{149}%
\makeatletter
\providecommand \@ifxundefined [1]{%
 \@ifx{#1\undefined}
}%
\providecommand \@ifnum [1]{%
 \ifnum #1\expandafter \@firstoftwo
 \else \expandafter \@secondoftwo
 \fi
}%
\providecommand \@ifx [1]{%
 \ifx #1\expandafter \@firstoftwo
 \else \expandafter \@secondoftwo
 \fi
}%
\providecommand \natexlab [1]{#1}%
\providecommand \enquote  [1]{``#1''}%
\providecommand \bibnamefont  [1]{#1}%
\providecommand \bibfnamefont [1]{#1}%
\providecommand \citenamefont [1]{#1}%
\providecommand \href@noop [0]{\@secondoftwo}%
\providecommand \href [0]{\begingroup \@sanitize@url \@href}%
\providecommand \@href[1]{\@@startlink{#1}\@@href}%
\providecommand \@@href[1]{\endgroup#1\@@endlink}%
\providecommand \@sanitize@url [0]{\catcode `\\12\catcode `\$12\catcode `\&12\catcode `\#12\catcode `\^12\catcode `\_12\catcode `\%12\relax}%
\providecommand \@@startlink[1]{}%
\providecommand \@@endlink[0]{}%
\providecommand \url  [0]{\begingroup\@sanitize@url \@url }%
\providecommand \@url [1]{\endgroup\@href {#1}{\urlprefix }}%
\providecommand \urlprefix  [0]{URL }%
\providecommand \Eprint [0]{\href }%
\providecommand \doibase [0]{https://doi.org/}%
\providecommand \selectlanguage [0]{\@gobble}%
\providecommand \bibinfo  [0]{\@secondoftwo}%
\providecommand \bibfield  [0]{\@secondoftwo}%
\providecommand \translation [1]{[#1]}%
\providecommand \BibitemOpen [0]{}%
\providecommand \bibitemStop [0]{}%
\providecommand \bibitemNoStop [0]{.\EOS\space}%
\providecommand \EOS [0]{\spacefactor3000\relax}%
\providecommand \BibitemShut  [1]{\csname bibitem#1\endcsname}%
\let\auto@bib@innerbib\@empty
\bibitem [{\citenamefont {{Lynden-Bell}}(1967)}]{Lynden-Bell1967}%
  \BibitemOpen
  \bibfield  {author} {\bibinfo {author} {\bibfnamefont {D.}~\bibnamefont {{Lynden-Bell}}},\ }\href {https://doi.org/10.1093/mnras/136.1.101} {\bibfield  {journal} {\bibinfo  {journal} {\mnras}\ }\textbf {\bibinfo {volume} {136}},\ \bibinfo {pages} {101} (\bibinfo {year} {1967})}\BibitemShut {NoStop}%
\bibitem [{\citenamefont {{Binney}}(1992)}]{Binney1992}%
  \BibitemOpen
  \bibfield  {author} {\bibinfo {author} {\bibfnamefont {J.}~\bibnamefont {{Binney}}},\ }\href {https://doi.org/10.1146/annurev.aa.30.090192.000411} {\bibfield  {journal} {\bibinfo  {journal} {\araa}\ }\textbf {\bibinfo {volume} {30}},\ \bibinfo {pages} {51} (\bibinfo {year} {1992})}\BibitemShut {NoStop}%
\bibitem [{\citenamefont {{Xu}}\ \emph {et~al.}(2015)\citenamefont {{Xu}}, \citenamefont {{Newberg}}, \citenamefont {{Carlin}}, \citenamefont {{Liu}}, \citenamefont {{Deng}}, \citenamefont {{Li}}, \citenamefont {{Sch{\"o}nrich}},\ and\ \citenamefont {{Yanny}}}]{Xu2015}%
  \BibitemOpen
  \bibfield  {author} {\bibinfo {author} {\bibfnamefont {Y.}~\bibnamefont {{Xu}}}, \bibinfo {author} {\bibfnamefont {H.~J.}\ \bibnamefont {{Newberg}}}, \bibinfo {author} {\bibfnamefont {J.~L.}\ \bibnamefont {{Carlin}}}, \bibinfo {author} {\bibfnamefont {C.}~\bibnamefont {{Liu}}}, \bibinfo {author} {\bibfnamefont {L.}~\bibnamefont {{Deng}}}, \bibinfo {author} {\bibfnamefont {J.}~\bibnamefont {{Li}}}, \bibinfo {author} {\bibfnamefont {R.}~\bibnamefont {{Sch{\"o}nrich}}},\ and\ \bibinfo {author} {\bibfnamefont {B.}~\bibnamefont {{Yanny}}},\ }\href {https://doi.org/10.1088/0004-637X/801/2/105} {\bibfield  {journal} {\bibinfo  {journal} {\apj}\ }\textbf {\bibinfo {volume} {801}},\ \bibinfo {eid} {105} (\bibinfo {year} {2015})},\ \Eprint {https://arxiv.org/abs/1503.00257} {arXiv:1503.00257 [astro-ph.GA]} \BibitemShut {NoStop}%
\bibitem [{\citenamefont {{Antoja}}\ \emph {et~al.}(2018)\citenamefont {{Antoja}}, \citenamefont {{Helmi}}, \citenamefont {{Romero-G{\'o}mez}}, \citenamefont {{Katz}}, \citenamefont {{Babusiaux}}, \citenamefont {{Drimmel}}, \citenamefont {{Evans}}, \citenamefont {{Figueras}}, \citenamefont {{Poggio}}, \citenamefont {{Reyl{\'e}}}, \citenamefont {{Robin}}, \citenamefont {{Seabroke}},\ and\ \citenamefont {{Soubiran}}}]{Antoja2018}%
  \BibitemOpen
  \bibfield  {author} {\bibinfo {author} {\bibfnamefont {T.}~\bibnamefont {{Antoja}}}, \bibinfo {author} {\bibfnamefont {A.}~\bibnamefont {{Helmi}}}, \bibinfo {author} {\bibfnamefont {M.}~\bibnamefont {{Romero-G{\'o}mez}}}, \bibinfo {author} {\bibfnamefont {D.}~\bibnamefont {{Katz}}}, \bibinfo {author} {\bibfnamefont {C.}~\bibnamefont {{Babusiaux}}}, \bibinfo {author} {\bibfnamefont {R.}~\bibnamefont {{Drimmel}}}, \bibinfo {author} {\bibfnamefont {D.~W.}\ \bibnamefont {{Evans}}}, \bibinfo {author} {\bibfnamefont {F.}~\bibnamefont {{Figueras}}}, \bibinfo {author} {\bibfnamefont {E.}~\bibnamefont {{Poggio}}}, \bibinfo {author} {\bibfnamefont {C.}~\bibnamefont {{Reyl{\'e}}}}, \bibinfo {author} {\bibfnamefont {A.~C.}\ \bibnamefont {{Robin}}}, \bibinfo {author} {\bibfnamefont {G.}~\bibnamefont {{Seabroke}}},\ and\ \bibinfo {author} {\bibfnamefont {C.}~\bibnamefont {{Soubiran}}},\ }\href {https://doi.org/10.1038/s41586-018-0510-7} {\bibfield  {journal} {\bibinfo  {journal} {\nat}\ }\textbf {\bibinfo {volume}
  {561}},\ \bibinfo {pages} {360} (\bibinfo {year} {2018})},\ \Eprint {https://arxiv.org/abs/1804.10196} {arXiv:1804.10196 [astro-ph.GA]} \BibitemShut {NoStop}%
\bibitem [{\citenamefont {{Widrow}}\ \emph {et~al.}(2012)\citenamefont {{Widrow}}, \citenamefont {{Gardner}}, \citenamefont {{Yanny}}, \citenamefont {{Dodelson}},\ and\ \citenamefont {{Chen}}}]{Widrow2012}%
  \BibitemOpen
  \bibfield  {author} {\bibinfo {author} {\bibfnamefont {L.~M.}\ \bibnamefont {{Widrow}}}, \bibinfo {author} {\bibfnamefont {S.}~\bibnamefont {{Gardner}}}, \bibinfo {author} {\bibfnamefont {B.}~\bibnamefont {{Yanny}}}, \bibinfo {author} {\bibfnamefont {S.}~\bibnamefont {{Dodelson}}},\ and\ \bibinfo {author} {\bibfnamefont {H.-Y.}\ \bibnamefont {{Chen}}},\ }\href {https://doi.org/10.1088/2041-8205/750/2/L41} {\bibfield  {journal} {\bibinfo  {journal} {\apjl}\ }\textbf {\bibinfo {volume} {750}},\ \bibinfo {eid} {L41} (\bibinfo {year} {2012})},\ \Eprint {https://arxiv.org/abs/1203.6861} {arXiv:1203.6861 [astro-ph.GA]} \BibitemShut {NoStop}%
\bibitem [{\citenamefont {{Yanny}}\ and\ \citenamefont {{Gardner}}(2013)}]{YannyGardner2013}%
  \BibitemOpen
  \bibfield  {author} {\bibinfo {author} {\bibfnamefont {B.}~\bibnamefont {{Yanny}}}\ and\ \bibinfo {author} {\bibfnamefont {S.}~\bibnamefont {{Gardner}}},\ }\href {https://doi.org/10.1088/0004-637X/777/2/91} {\bibfield  {journal} {\bibinfo  {journal} {\apj}\ }\textbf {\bibinfo {volume} {777}},\ \bibinfo {eid} {91} (\bibinfo {year} {2013})},\ \Eprint {https://arxiv.org/abs/1309.2300} {arXiv:1309.2300 [astro-ph.GA]} \BibitemShut {NoStop}%
\bibitem [{\citenamefont {{Bennett}}\ and\ \citenamefont {{Bovy}}(2019)}]{BennettBovy2019}%
  \BibitemOpen
  \bibfield  {author} {\bibinfo {author} {\bibfnamefont {M.}~\bibnamefont {{Bennett}}}\ and\ \bibinfo {author} {\bibfnamefont {J.}~\bibnamefont {{Bovy}}},\ }\href {https://doi.org/10.1093/mnras/sty2813} {\bibfield  {journal} {\bibinfo  {journal} {\mnras}\ }\textbf {\bibinfo {volume} {482}},\ \bibinfo {pages} {1417} (\bibinfo {year} {2019})},\ \Eprint {https://arxiv.org/abs/1809.03507} {arXiv:1809.03507 [astro-ph.GA]} \BibitemShut {NoStop}%
\bibitem [{\citenamefont {{Quillen}}\ and\ \citenamefont {{Minchev}}(2005)}]{QuillenMinchev2005}%
  \BibitemOpen
  \bibfield  {author} {\bibinfo {author} {\bibfnamefont {A.~C.}\ \bibnamefont {{Quillen}}}\ and\ \bibinfo {author} {\bibfnamefont {I.}~\bibnamefont {{Minchev}}},\ }\href {https://doi.org/10.1086/430885} {\bibfield  {journal} {\bibinfo  {journal} {\aj}\ }\textbf {\bibinfo {volume} {130}},\ \bibinfo {pages} {576} (\bibinfo {year} {2005})},\ \Eprint {https://arxiv.org/abs/astro-ph/0502205} {arXiv:astro-ph/0502205 [astro-ph]} \BibitemShut {NoStop}%
\bibitem [{\citenamefont {{Antoja}}\ \emph {et~al.}(2008)\citenamefont {{Antoja}}, \citenamefont {{Figueras}}, \citenamefont {{Fern{\'a}ndez}},\ and\ \citenamefont {{Torra}}}]{Antoja2008}%
  \BibitemOpen
  \bibfield  {author} {\bibinfo {author} {\bibfnamefont {T.}~\bibnamefont {{Antoja}}}, \bibinfo {author} {\bibfnamefont {F.}~\bibnamefont {{Figueras}}}, \bibinfo {author} {\bibfnamefont {D.}~\bibnamefont {{Fern{\'a}ndez}}},\ and\ \bibinfo {author} {\bibfnamefont {J.}~\bibnamefont {{Torra}}},\ }\href {https://doi.org/10.1051/0004-6361:200809519} {\bibfield  {journal} {\bibinfo  {journal} {\aap}\ }\textbf {\bibinfo {volume} {490}},\ \bibinfo {pages} {135} (\bibinfo {year} {2008})},\ \Eprint {https://arxiv.org/abs/0809.0511} {arXiv:0809.0511 [astro-ph]} \BibitemShut {NoStop}%
\bibitem [{\citenamefont {{Antoja}}\ \emph {et~al.}(2011)\citenamefont {{Antoja}}, \citenamefont {{Figueras}}, \citenamefont {{Romero-G{\'o}mez}}, \citenamefont {{Pichardo}}, \citenamefont {{Valenzuela}},\ and\ \citenamefont {{Moreno}}}]{Antoja2011}%
  \BibitemOpen
  \bibfield  {author} {\bibinfo {author} {\bibfnamefont {T.}~\bibnamefont {{Antoja}}}, \bibinfo {author} {\bibfnamefont {F.}~\bibnamefont {{Figueras}}}, \bibinfo {author} {\bibfnamefont {M.}~\bibnamefont {{Romero-G{\'o}mez}}}, \bibinfo {author} {\bibfnamefont {B.}~\bibnamefont {{Pichardo}}}, \bibinfo {author} {\bibfnamefont {O.}~\bibnamefont {{Valenzuela}}},\ and\ \bibinfo {author} {\bibfnamefont {E.}~\bibnamefont {{Moreno}}},\ }\href {https://doi.org/10.1111/j.1365-2966.2011.19190.x} {\bibfield  {journal} {\bibinfo  {journal} {\mnras}\ }\textbf {\bibinfo {volume} {418}},\ \bibinfo {pages} {1423} (\bibinfo {year} {2011})},\ \Eprint {https://arxiv.org/abs/1106.1170} {arXiv:1106.1170 [astro-ph.GA]} \BibitemShut {NoStop}%
\bibitem [{\citenamefont {{Hunt}}\ \emph {et~al.}(2018)\citenamefont {{Hunt}}, \citenamefont {{Hong}}, \citenamefont {{Bovy}}, \citenamefont {{Kawata}},\ and\ \citenamefont {{Grand}}}]{Hunt2018}%
  \BibitemOpen
  \bibfield  {author} {\bibinfo {author} {\bibfnamefont {J.~A.~S.}\ \bibnamefont {{Hunt}}}, \bibinfo {author} {\bibfnamefont {J.}~\bibnamefont {{Hong}}}, \bibinfo {author} {\bibfnamefont {J.}~\bibnamefont {{Bovy}}}, \bibinfo {author} {\bibfnamefont {D.}~\bibnamefont {{Kawata}}},\ and\ \bibinfo {author} {\bibfnamefont {R.~J.~J.}\ \bibnamefont {{Grand}}},\ }\href {https://doi.org/10.1093/mnras/sty2532} {\bibfield  {journal} {\bibinfo  {journal} {\mnras}\ }\textbf {\bibinfo {volume} {481}},\ \bibinfo {pages} {3794} (\bibinfo {year} {2018})},\ \Eprint {https://arxiv.org/abs/1806.02832} {arXiv:1806.02832 [astro-ph.GA]} \BibitemShut {NoStop}%
\bibitem [{\citenamefont {{Ramos}}\ \emph {et~al.}(2018)\citenamefont {{Ramos}}, \citenamefont {{Antoja}},\ and\ \citenamefont {{Figueras}}}]{Ramos2018}%
  \BibitemOpen
  \bibfield  {author} {\bibinfo {author} {\bibfnamefont {P.}~\bibnamefont {{Ramos}}}, \bibinfo {author} {\bibfnamefont {T.}~\bibnamefont {{Antoja}}},\ and\ \bibinfo {author} {\bibfnamefont {F.}~\bibnamefont {{Figueras}}},\ }\href {https://doi.org/10.1051/0004-6361/201833494} {\bibfield  {journal} {\bibinfo  {journal} {\aap}\ }\textbf {\bibinfo {volume} {619}},\ \bibinfo {eid} {A72} (\bibinfo {year} {2018})},\ \Eprint {https://arxiv.org/abs/1805.09790} {arXiv:1805.09790 [astro-ph.GA]} \BibitemShut {NoStop}%
\bibitem [{\citenamefont {{Craig}}\ \emph {et~al.}(2021)\citenamefont {{Craig}}, \citenamefont {{Chakrabarti}}, \citenamefont {{Newberg}},\ and\ \citenamefont {{Quillen}}}]{Craig2021}%
  \BibitemOpen
  \bibfield  {author} {\bibinfo {author} {\bibfnamefont {P.}~\bibnamefont {{Craig}}}, \bibinfo {author} {\bibfnamefont {S.}~\bibnamefont {{Chakrabarti}}}, \bibinfo {author} {\bibfnamefont {H.}~\bibnamefont {{Newberg}}},\ and\ \bibinfo {author} {\bibfnamefont {A.}~\bibnamefont {{Quillen}}},\ }\href {https://doi.org/10.1093/mnras/stab1431} {\bibfield  {journal} {\bibinfo  {journal} {\mnras}\ }\textbf {\bibinfo {volume} {505}},\ \bibinfo {pages} {2561} (\bibinfo {year} {2021})},\ \Eprint {https://arxiv.org/abs/1911.01392} {arXiv:1911.01392 [astro-ph.GA]} \BibitemShut {NoStop}%
\bibitem [{\citenamefont {{Levine}}\ \emph {et~al.}(2006)\citenamefont {{Levine}}, \citenamefont {{Blitz}},\ and\ \citenamefont {{Heiles}}}]{Levineetal2006}%
  \BibitemOpen
  \bibfield  {author} {\bibinfo {author} {\bibfnamefont {E.~S.}\ \bibnamefont {{Levine}}}, \bibinfo {author} {\bibfnamefont {L.}~\bibnamefont {{Blitz}}},\ and\ \bibinfo {author} {\bibfnamefont {C.}~\bibnamefont {{Heiles}}},\ }\href {https://doi.org/10.1126/science.1128455} {\bibfield  {journal} {\bibinfo  {journal} {Science}\ }\textbf {\bibinfo {volume} {312}},\ \bibinfo {pages} {1773} (\bibinfo {year} {2006})},\ \Eprint {https://arxiv.org/abs/astro-ph/0605728} {arXiv:astro-ph/0605728 [astro-ph]} \BibitemShut {NoStop}%
\bibitem [{\citenamefont {{Kalberla}}\ and\ \citenamefont {{Dedes}}(2008)}]{KalberlaDedes2008}%
  \BibitemOpen
  \bibfield  {author} {\bibinfo {author} {\bibfnamefont {P.~M.~W.}\ \bibnamefont {{Kalberla}}}\ and\ \bibinfo {author} {\bibfnamefont {L.}~\bibnamefont {{Dedes}}},\ }\href {https://doi.org/10.1051/0004-6361:20079240} {\bibfield  {journal} {\bibinfo  {journal} {\aap}\ }\textbf {\bibinfo {volume} {487}},\ \bibinfo {pages} {951} (\bibinfo {year} {2008})},\ \Eprint {https://arxiv.org/abs/0804.4831} {arXiv:0804.4831 [astro-ph]} \BibitemShut {NoStop}%
\bibitem [{\citenamefont {{Chakrabarti}}\ and\ \citenamefont {{Blitz}}(2009)}]{Chakrabarti_Blitz2009}%
  \BibitemOpen
  \bibfield  {author} {\bibinfo {author} {\bibfnamefont {S.}~\bibnamefont {{Chakrabarti}}}\ and\ \bibinfo {author} {\bibfnamefont {L.}~\bibnamefont {{Blitz}}},\ }\href {https://doi.org/10.1111/j.1745-3933.2009.00735.x} {\bibfield  {journal} {\bibinfo  {journal} {\mnras}\ }\textbf {\bibinfo {volume} {399}},\ \bibinfo {pages} {L118} (\bibinfo {year} {2009})},\ \Eprint {https://arxiv.org/abs/0812.0821} {arXiv:0812.0821 [astro-ph]} \BibitemShut {NoStop}%
\bibitem [{\citenamefont {{Chakrabarti}}\ and\ \citenamefont {{Blitz}}(2011)}]{Chakrabarti_Blitz2011}%
  \BibitemOpen
  \bibfield  {author} {\bibinfo {author} {\bibfnamefont {S.}~\bibnamefont {{Chakrabarti}}}\ and\ \bibinfo {author} {\bibfnamefont {L.}~\bibnamefont {{Blitz}}},\ }\href {https://doi.org/10.1088/0004-637X/731/1/40} {\bibfield  {journal} {\bibinfo  {journal} {\apj}\ }\textbf {\bibinfo {volume} {731}},\ \bibinfo {eid} {40} (\bibinfo {year} {2011})},\ \Eprint {https://arxiv.org/abs/1007.1982} {arXiv:1007.1982 [astro-ph.GA]} \BibitemShut {NoStop}%
\bibitem [{\citenamefont {{Chakrabarti}}\ \emph {et~al.}(2019)\citenamefont {{Chakrabarti}}, \citenamefont {{Chang}}, \citenamefont {{Price-Whelan}}, \citenamefont {{Read}}, \citenamefont {{Blitz}},\ and\ \citenamefont {{Hernquist}}}]{Chakrabartietal2019}%
  \BibitemOpen
  \bibfield  {author} {\bibinfo {author} {\bibfnamefont {S.}~\bibnamefont {{Chakrabarti}}}, \bibinfo {author} {\bibfnamefont {P.}~\bibnamefont {{Chang}}}, \bibinfo {author} {\bibfnamefont {A.~M.}\ \bibnamefont {{Price-Whelan}}}, \bibinfo {author} {\bibfnamefont {J.}~\bibnamefont {{Read}}}, \bibinfo {author} {\bibfnamefont {L.}~\bibnamefont {{Blitz}}},\ and\ \bibinfo {author} {\bibfnamefont {L.}~\bibnamefont {{Hernquist}}},\ }\href {https://doi.org/10.3847/1538-4357/ab4659} {\bibfield  {journal} {\bibinfo  {journal} {\apj}\ }\textbf {\bibinfo {volume} {886}},\ \bibinfo {eid} {67} (\bibinfo {year} {2019})},\ \Eprint {https://arxiv.org/abs/1906.04203} {arXiv:1906.04203 [astro-ph.GA]} \BibitemShut {NoStop}%
\bibitem [{\citenamefont {{Alves}}\ \emph {et~al.}(2020)\citenamefont {{Alves}}, \citenamefont {{Zucker}}, \citenamefont {{Goodman}}, \citenamefont {{Speagle}}, \citenamefont {{Meingast}}, \citenamefont {{Robitaille}}, \citenamefont {{Finkbeiner}}, \citenamefont {{Schlafly}},\ and\ \citenamefont {{Green}}}]{Alves2020}%
  \BibitemOpen
  \bibfield  {author} {\bibinfo {author} {\bibfnamefont {J.}~\bibnamefont {{Alves}}}, \bibinfo {author} {\bibfnamefont {C.}~\bibnamefont {{Zucker}}}, \bibinfo {author} {\bibfnamefont {A.~A.}\ \bibnamefont {{Goodman}}}, \bibinfo {author} {\bibfnamefont {J.~S.}\ \bibnamefont {{Speagle}}}, \bibinfo {author} {\bibfnamefont {S.}~\bibnamefont {{Meingast}}}, \bibinfo {author} {\bibfnamefont {T.}~\bibnamefont {{Robitaille}}}, \bibinfo {author} {\bibfnamefont {D.~P.}\ \bibnamefont {{Finkbeiner}}}, \bibinfo {author} {\bibfnamefont {E.~F.}\ \bibnamefont {{Schlafly}}},\ and\ \bibinfo {author} {\bibfnamefont {G.~M.}\ \bibnamefont {{Green}}},\ }\href {https://doi.org/10.1038/s41586-019-1874-z} {\bibfield  {journal} {\bibinfo  {journal} {\nat}\ }\textbf {\bibinfo {volume} {578}},\ \bibinfo {pages} {237} (\bibinfo {year} {2020})},\ \Eprint {https://arxiv.org/abs/2001.08748} {arXiv:2001.08748 [astro-ph.GA]} \BibitemShut {NoStop}%
\bibitem [{\citenamefont {{Laporte}}\ \emph {et~al.}(2019)\citenamefont {{Laporte}}, \citenamefont {{Minchev}}, \citenamefont {{Johnston}},\ and\ \citenamefont {{G{\'o}mez}}}]{Laporte2019}%
  \BibitemOpen
  \bibfield  {author} {\bibinfo {author} {\bibfnamefont {C.~F.~P.}\ \bibnamefont {{Laporte}}}, \bibinfo {author} {\bibfnamefont {I.}~\bibnamefont {{Minchev}}}, \bibinfo {author} {\bibfnamefont {K.~V.}\ \bibnamefont {{Johnston}}},\ and\ \bibinfo {author} {\bibfnamefont {F.~A.}\ \bibnamefont {{G{\'o}mez}}},\ }\href {https://doi.org/10.1093/mnras/stz583} {\bibfield  {journal} {\bibinfo  {journal} {\mnras}\ }\textbf {\bibinfo {volume} {485}},\ \bibinfo {pages} {3134} (\bibinfo {year} {2019})},\ \Eprint {https://arxiv.org/abs/1808.00451} {arXiv:1808.00451 [astro-ph.GA]} \BibitemShut {NoStop}%
\bibitem [{\citenamefont {{Khoperskov}}\ \emph {et~al.}(2019)\citenamefont {{Khoperskov}}, \citenamefont {{Di Matteo}}, \citenamefont {{Gerhard}}, \citenamefont {{Katz}}, \citenamefont {{Haywood}}, \citenamefont {{Combes}}, \citenamefont {{Berczik}},\ and\ \citenamefont {{Gomez}}}]{Khoperskov2019}%
  \BibitemOpen
  \bibfield  {author} {\bibinfo {author} {\bibfnamefont {S.}~\bibnamefont {{Khoperskov}}}, \bibinfo {author} {\bibfnamefont {P.}~\bibnamefont {{Di Matteo}}}, \bibinfo {author} {\bibfnamefont {O.}~\bibnamefont {{Gerhard}}}, \bibinfo {author} {\bibfnamefont {D.}~\bibnamefont {{Katz}}}, \bibinfo {author} {\bibfnamefont {M.}~\bibnamefont {{Haywood}}}, \bibinfo {author} {\bibfnamefont {F.}~\bibnamefont {{Combes}}}, \bibinfo {author} {\bibfnamefont {P.}~\bibnamefont {{Berczik}}},\ and\ \bibinfo {author} {\bibfnamefont {A.}~\bibnamefont {{Gomez}}},\ }\href {https://doi.org/10.1051/0004-6361/201834707} {\bibfield  {journal} {\bibinfo  {journal} {\aap}\ }\textbf {\bibinfo {volume} {622}},\ \bibinfo {eid} {L6} (\bibinfo {year} {2019})},\ \Eprint {https://arxiv.org/abs/1811.09205} {arXiv:1811.09205 [astro-ph.GA]} \BibitemShut {NoStop}%
\bibitem [{\citenamefont {{Hunt}}\ \emph {et~al.}(2022)\citenamefont {{Hunt}}, \citenamefont {{Price-Whelan}}, \citenamefont {{Johnston}},\ and\ \citenamefont {{Darragh-Ford}}}]{Hunt2022}%
  \BibitemOpen
  \bibfield  {author} {\bibinfo {author} {\bibfnamefont {J.~A.~S.}\ \bibnamefont {{Hunt}}}, \bibinfo {author} {\bibfnamefont {A.~M.}\ \bibnamefont {{Price-Whelan}}}, \bibinfo {author} {\bibfnamefont {K.~V.}\ \bibnamefont {{Johnston}}},\ and\ \bibinfo {author} {\bibfnamefont {E.}~\bibnamefont {{Darragh-Ford}}},\ }\href {https://doi.org/10.1093/mnrasl/slac082} {\bibfield  {journal} {\bibinfo  {journal} {\mnras}\ }\textbf {\bibinfo {volume} {516}},\ \bibinfo {pages} {L7} (\bibinfo {year} {2022})},\ \Eprint {https://arxiv.org/abs/2206.06125} {arXiv:2206.06125 [astro-ph.GA]} \BibitemShut {NoStop}%
\bibitem [{\citenamefont {{Bland-Hawthorn}}\ and\ \citenamefont {{Gerhard}}(2016)}]{BlandHawthornGerhard2016}%
  \BibitemOpen
  \bibfield  {author} {\bibinfo {author} {\bibfnamefont {J.}~\bibnamefont {{Bland-Hawthorn}}}\ and\ \bibinfo {author} {\bibfnamefont {O.}~\bibnamefont {{Gerhard}}},\ }\href {https://doi.org/10.1146/annurev-astro-081915-023441} {\bibfield  {journal} {\bibinfo  {journal} {\araa}\ }\textbf {\bibinfo {volume} {54}},\ \bibinfo {pages} {529} (\bibinfo {year} {2016})},\ \Eprint {https://arxiv.org/abs/1602.07702} {arXiv:1602.07702 [astro-ph.GA]} \BibitemShut {NoStop}%
\bibitem [{\citenamefont {{Binney}}\ and\ \citenamefont {{Tremaine}}(2008)}]{BinneyTremaine2008}%
  \BibitemOpen
  \bibfield  {author} {\bibinfo {author} {\bibfnamefont {J.}~\bibnamefont {{Binney}}}\ and\ \bibinfo {author} {\bibfnamefont {S.}~\bibnamefont {{Tremaine}}},\ }\href@noop {} {\emph {\bibinfo {title} {{Galactic Dynamics: Second Edition}}}}\ (\bibinfo {year} {2008})\BibitemShut {NoStop}%
\bibitem [{\citenamefont {{Chakrabarti}}\ \emph {et~al.}(2020)\citenamefont {{Chakrabarti}}, \citenamefont {{Wright}}, \citenamefont {{Chang}}, \citenamefont {{Quillen}}, \citenamefont {{Craig}}, \citenamefont {{Territo}}, \citenamefont {{D'Onghia}}, \citenamefont {{Johnston}}, \citenamefont {{De Rosa}}, \citenamefont {{Huber}}, \citenamefont {{Rhode}},\ and\ \citenamefont {{Nielsen}}}]{Chakrabarti2020}%
  \BibitemOpen
  \bibfield  {author} {\bibinfo {author} {\bibfnamefont {S.}~\bibnamefont {{Chakrabarti}}}, \bibinfo {author} {\bibfnamefont {J.}~\bibnamefont {{Wright}}}, \bibinfo {author} {\bibfnamefont {P.}~\bibnamefont {{Chang}}}, \bibinfo {author} {\bibfnamefont {A.}~\bibnamefont {{Quillen}}}, \bibinfo {author} {\bibfnamefont {P.}~\bibnamefont {{Craig}}}, \bibinfo {author} {\bibfnamefont {J.}~\bibnamefont {{Territo}}}, \bibinfo {author} {\bibfnamefont {E.}~\bibnamefont {{D'Onghia}}}, \bibinfo {author} {\bibfnamefont {K.~V.}\ \bibnamefont {{Johnston}}}, \bibinfo {author} {\bibfnamefont {R.~J.}\ \bibnamefont {{De Rosa}}}, \bibinfo {author} {\bibfnamefont {D.}~\bibnamefont {{Huber}}}, \bibinfo {author} {\bibfnamefont {K.~L.}\ \bibnamefont {{Rhode}}},\ and\ \bibinfo {author} {\bibfnamefont {E.}~\bibnamefont {{Nielsen}}},\ }\href {https://doi.org/10.3847/2041-8213/abb9b5} {\bibfield  {journal} {\bibinfo  {journal} {\apjl}\ }\textbf {\bibinfo {volume} {902}},\ \bibinfo {eid} {L28} (\bibinfo {year} {2020})},\ \Eprint
  {https://arxiv.org/abs/2007.15097} {arXiv:2007.15097 [astro-ph.GA]} \BibitemShut {NoStop}%
\bibitem [{\citenamefont {{Chakrabarti}}\ \emph {et~al.}(2021)\citenamefont {{Chakrabarti}}, \citenamefont {{Chang}}, \citenamefont {{Lam}}, \citenamefont {{Vigeland}},\ and\ \citenamefont {{Quillen}}}]{Chakrabarti2021}%
  \BibitemOpen
  \bibfield  {author} {\bibinfo {author} {\bibfnamefont {S.}~\bibnamefont {{Chakrabarti}}}, \bibinfo {author} {\bibfnamefont {P.}~\bibnamefont {{Chang}}}, \bibinfo {author} {\bibfnamefont {M.~T.}\ \bibnamefont {{Lam}}}, \bibinfo {author} {\bibfnamefont {S.~J.}\ \bibnamefont {{Vigeland}}},\ and\ \bibinfo {author} {\bibfnamefont {A.~C.}\ \bibnamefont {{Quillen}}},\ }\href {https://doi.org/10.3847/2041-8213/abd635} {\bibfield  {journal} {\bibinfo  {journal} {\apjl}\ }\textbf {\bibinfo {volume} {907}},\ \bibinfo {eid} {L26} (\bibinfo {year} {2021})},\ \Eprint {https://arxiv.org/abs/2010.04018} {arXiv:2010.04018 [astro-ph.GA]} \BibitemShut {NoStop}%
\bibitem [{\citenamefont {{Chakrabarti}}\ \emph {et~al.}(2022)\citenamefont {{Chakrabarti}}, \citenamefont {{Stevens}}, \citenamefont {{Wright}}, \citenamefont {{Rafikov}}, \citenamefont {{Chang}}, \citenamefont {{Beatty}},\ and\ \citenamefont {{Huber}}}]{Chakrabartietal2022}%
  \BibitemOpen
  \bibfield  {author} {\bibinfo {author} {\bibfnamefont {S.}~\bibnamefont {{Chakrabarti}}}, \bibinfo {author} {\bibfnamefont {D.~J.}\ \bibnamefont {{Stevens}}}, \bibinfo {author} {\bibfnamefont {J.}~\bibnamefont {{Wright}}}, \bibinfo {author} {\bibfnamefont {R.~R.}\ \bibnamefont {{Rafikov}}}, \bibinfo {author} {\bibfnamefont {P.}~\bibnamefont {{Chang}}}, \bibinfo {author} {\bibfnamefont {T.}~\bibnamefont {{Beatty}}},\ and\ \bibinfo {author} {\bibfnamefont {D.}~\bibnamefont {{Huber}}},\ }\href {https://doi.org/10.3847/2041-8213/ac5c43} {\bibfield  {journal} {\bibinfo  {journal} {\apjl}\ }\textbf {\bibinfo {volume} {928}},\ \bibinfo {eid} {L17} (\bibinfo {year} {2022})},\ \Eprint {https://arxiv.org/abs/2112.08231} {arXiv:2112.08231 [astro-ph.GA]} \BibitemShut {NoStop}%
\bibitem [{\citenamefont {{Phillips}}\ \emph {et~al.}(2021)\citenamefont {{Phillips}}, \citenamefont {{Ravi}}, \citenamefont {{Ebadi}},\ and\ \citenamefont {{Walsworth}}}]{Phillips2021}%
  \BibitemOpen
  \bibfield  {author} {\bibinfo {author} {\bibfnamefont {D.~F.}\ \bibnamefont {{Phillips}}}, \bibinfo {author} {\bibfnamefont {A.}~\bibnamefont {{Ravi}}}, \bibinfo {author} {\bibfnamefont {R.}~\bibnamefont {{Ebadi}}},\ and\ \bibinfo {author} {\bibfnamefont {R.~L.}\ \bibnamefont {{Walsworth}}},\ }\href {https://doi.org/10.1103/PhysRevLett.126.141103} {\bibfield  {journal} {\bibinfo  {journal} {\prl}\ }\textbf {\bibinfo {volume} {126}},\ \bibinfo {eid} {141103} (\bibinfo {year} {2021})},\ \Eprint {https://arxiv.org/abs/2008.13052} {arXiv:2008.13052 [astro-ph.GA]} \BibitemShut {NoStop}%
\bibitem [{\citenamefont {{Bovy}}(2020)}]{Bovy2021}%
  \BibitemOpen
  \bibfield  {author} {\bibinfo {author} {\bibfnamefont {J.}~\bibnamefont {{Bovy}}},\ }\href {https://doi.org/10.48550/arXiv.2012.02169} {\bibfield  {journal} {\bibinfo  {journal} {arXiv e-prints}\ ,\ \bibinfo {eid} {arXiv:2012.02169}} (\bibinfo {year} {2020})},\ \Eprint {https://arxiv.org/abs/2012.02169} {arXiv:2012.02169 [astro-ph.GA]} \BibitemShut {NoStop}%
\bibitem [{\citenamefont {{Peters}}\ and\ \citenamefont {{Mathews}}(1963)}]{PetersMathews1963}%
  \BibitemOpen
  \bibfield  {author} {\bibinfo {author} {\bibfnamefont {P.~C.}\ \bibnamefont {{Peters}}}\ and\ \bibinfo {author} {\bibfnamefont {J.}~\bibnamefont {{Mathews}}},\ }\href {https://doi.org/10.1103/PhysRev.131.435} {\bibfield  {journal} {\bibinfo  {journal} {Physical Review}\ }\textbf {\bibinfo {volume} {131}},\ \bibinfo {pages} {435} (\bibinfo {year} {1963})}\BibitemShut {NoStop}%
\bibitem [{\citenamefont {{Damour}}\ and\ \citenamefont {{Taylor}}(1991)}]{DamourTaylor1991}%
  \BibitemOpen
  \bibfield  {author} {\bibinfo {author} {\bibfnamefont {T.}~\bibnamefont {{Damour}}}\ and\ \bibinfo {author} {\bibfnamefont {J.~H.}\ \bibnamefont {{Taylor}}},\ }\href {https://doi.org/10.1086/169585} {\bibfield  {journal} {\bibinfo  {journal} {\apj}\ }\textbf {\bibinfo {volume} {366}},\ \bibinfo {pages} {501} (\bibinfo {year} {1991})}\BibitemShut {NoStop}%
\bibitem [{\citenamefont {{Weisberg}}\ and\ \citenamefont {{Huang}}(2016)}]{WeisbergHuang2016}%
  \BibitemOpen
  \bibfield  {author} {\bibinfo {author} {\bibfnamefont {J.~M.}\ \bibnamefont {{Weisberg}}}\ and\ \bibinfo {author} {\bibfnamefont {Y.}~\bibnamefont {{Huang}}},\ }\href {https://doi.org/10.3847/0004-637X/829/1/55} {\bibfield  {journal} {\bibinfo  {journal} {\apj}\ }\textbf {\bibinfo {volume} {829}},\ \bibinfo {eid} {55} (\bibinfo {year} {2016})},\ \Eprint {https://arxiv.org/abs/1606.02744} {arXiv:1606.02744 [astro-ph.HE]} \BibitemShut {NoStop}%
\bibitem [{\citenamefont {{Wex}}\ and\ \citenamefont {{Kramer}}(2020)}]{WexKramer2020}%
  \BibitemOpen
  \bibfield  {author} {\bibinfo {author} {\bibfnamefont {N.}~\bibnamefont {{Wex}}}\ and\ \bibinfo {author} {\bibfnamefont {M.}~\bibnamefont {{Kramer}}},\ }\href {https://doi.org/10.3390/universe6090156} {\bibfield  {journal} {\bibinfo  {journal} {Universe}\ }\textbf {\bibinfo {volume} {6}},\ \bibinfo {pages} {156} (\bibinfo {year} {2020})}\BibitemShut {NoStop}%
\bibitem [{\citenamefont {{Gravity Collaboration}}\ \emph {et~al.}(2019)\citenamefont {{Gravity Collaboration}}, \citenamefont {{Abuter}}, \citenamefont {{Amorim}}, \citenamefont {{Baub{\"o}ck}}, \citenamefont {{Berger}}, \citenamefont {{Bonnet}}, \citenamefont {{Brandner}}, \citenamefont {{Cl{\'e}net}}, \citenamefont {{Coud{\'e} Du Foresto}}, \citenamefont {{de Zeeuw}}, \citenamefont {{Dexter}}, \citenamefont {{Duvert}}, \citenamefont {{Eckart}}, \citenamefont {{Eisenhauer}}, \citenamefont {{F{\"o}rster Schreiber}}, \citenamefont {{Garcia}}, \citenamefont {{Gao}}, \citenamefont {{Gendron}}, \citenamefont {{Genzel}}, \citenamefont {{Gerhard}}, \citenamefont {{Gillessen}}, \citenamefont {{Habibi}}, \citenamefont {{Haubois}}, \citenamefont {{Henning}}, \citenamefont {{Hippler}}, \citenamefont {{Horrobin}}, \citenamefont {{Jim{\'e}nez-Rosales}}, \citenamefont {{Jocou}}, \citenamefont {{Kervella}}, \citenamefont {{Lacour}}, \citenamefont {{Lapeyr{\`e}re}}, \citenamefont {{Le Bouquin}}, \citenamefont {{L{\'e}na}},
  \citenamefont {{Ott}}, \citenamefont {{Paumard}}, \citenamefont {{Perraut}}, \citenamefont {{Perrin}}, \citenamefont {{Pfuhl}}, \citenamefont {{Rabien}}, \citenamefont {{Rodriguez Coira}}, \citenamefont {{Rousset}}, \citenamefont {{Scheithauer}}, \citenamefont {{Sternberg}}, \citenamefont {{Straub}}, \citenamefont {{Straubmeier}}, \citenamefont {{Sturm}}, \citenamefont {{Tacconi}}, \citenamefont {{Vincent}}, \citenamefont {{von Fellenberg}}, \citenamefont {{Waisberg}}, \citenamefont {{Widmann}}, \citenamefont {{Wieprecht}}, \citenamefont {{Wiezorrek}}, \citenamefont {{Woillez}},\ and\ \citenamefont {{Yazici}}}]{GravityCollaboration2019}%
  \BibitemOpen
  \bibfield  {author} {\bibinfo {author} {\bibnamefont {{Gravity Collaboration}}}, \bibinfo {author} {\bibfnamefont {R.}~\bibnamefont {{Abuter}}}, \bibinfo {author} {\bibfnamefont {A.}~\bibnamefont {{Amorim}}}, \bibinfo {author} {\bibfnamefont {M.}~\bibnamefont {{Baub{\"o}ck}}}, \bibinfo {author} {\bibfnamefont {J.~P.}\ \bibnamefont {{Berger}}}, \bibinfo {author} {\bibfnamefont {H.}~\bibnamefont {{Bonnet}}}, \bibinfo {author} {\bibfnamefont {W.}~\bibnamefont {{Brandner}}}, \bibinfo {author} {\bibfnamefont {Y.}~\bibnamefont {{Cl{\'e}net}}}, \bibinfo {author} {\bibfnamefont {V.}~\bibnamefont {{Coud{\'e} Du Foresto}}}, \bibinfo {author} {\bibfnamefont {P.~T.}\ \bibnamefont {{de Zeeuw}}}, \bibinfo {author} {\bibfnamefont {J.}~\bibnamefont {{Dexter}}}, \bibinfo {author} {\bibfnamefont {G.}~\bibnamefont {{Duvert}}}, \bibinfo {author} {\bibfnamefont {A.}~\bibnamefont {{Eckart}}}, \bibinfo {author} {\bibfnamefont {F.}~\bibnamefont {{Eisenhauer}}}, \bibinfo {author} {\bibfnamefont {N.~M.}\ \bibnamefont {{F{\"o}rster
  Schreiber}}}, \bibinfo {author} {\bibfnamefont {P.}~\bibnamefont {{Garcia}}}, \bibinfo {author} {\bibfnamefont {F.}~\bibnamefont {{Gao}}}, \bibinfo {author} {\bibfnamefont {E.}~\bibnamefont {{Gendron}}}, \bibinfo {author} {\bibfnamefont {R.}~\bibnamefont {{Genzel}}}, \bibinfo {author} {\bibfnamefont {O.}~\bibnamefont {{Gerhard}}}, \bibinfo {author} {\bibfnamefont {S.}~\bibnamefont {{Gillessen}}}, \bibinfo {author} {\bibfnamefont {M.}~\bibnamefont {{Habibi}}}, \bibinfo {author} {\bibfnamefont {X.}~\bibnamefont {{Haubois}}}, \bibinfo {author} {\bibfnamefont {T.}~\bibnamefont {{Henning}}}, \bibinfo {author} {\bibfnamefont {S.}~\bibnamefont {{Hippler}}}, \bibinfo {author} {\bibfnamefont {M.}~\bibnamefont {{Horrobin}}}, \bibinfo {author} {\bibfnamefont {A.}~\bibnamefont {{Jim{\'e}nez-Rosales}}}, \bibinfo {author} {\bibfnamefont {L.}~\bibnamefont {{Jocou}}}, \bibinfo {author} {\bibfnamefont {P.}~\bibnamefont {{Kervella}}}, \bibinfo {author} {\bibfnamefont {S.}~\bibnamefont {{Lacour}}}, \bibinfo {author}
  {\bibfnamefont {V.}~\bibnamefont {{Lapeyr{\`e}re}}}, \bibinfo {author} {\bibfnamefont {J.~B.}\ \bibnamefont {{Le Bouquin}}}, \bibinfo {author} {\bibfnamefont {P.}~\bibnamefont {{L{\'e}na}}}, \bibinfo {author} {\bibfnamefont {T.}~\bibnamefont {{Ott}}}, \bibinfo {author} {\bibfnamefont {T.}~\bibnamefont {{Paumard}}}, \bibinfo {author} {\bibfnamefont {K.}~\bibnamefont {{Perraut}}}, \bibinfo {author} {\bibfnamefont {G.}~\bibnamefont {{Perrin}}}, \bibinfo {author} {\bibfnamefont {O.}~\bibnamefont {{Pfuhl}}}, \bibinfo {author} {\bibfnamefont {S.}~\bibnamefont {{Rabien}}}, \bibinfo {author} {\bibfnamefont {G.}~\bibnamefont {{Rodriguez Coira}}}, \bibinfo {author} {\bibfnamefont {G.}~\bibnamefont {{Rousset}}}, \bibinfo {author} {\bibfnamefont {S.}~\bibnamefont {{Scheithauer}}}, \bibinfo {author} {\bibfnamefont {A.}~\bibnamefont {{Sternberg}}}, \bibinfo {author} {\bibfnamefont {O.}~\bibnamefont {{Straub}}}, \bibinfo {author} {\bibfnamefont {C.}~\bibnamefont {{Straubmeier}}}, \bibinfo {author} {\bibfnamefont
  {E.}~\bibnamefont {{Sturm}}}, \bibinfo {author} {\bibfnamefont {L.~J.}\ \bibnamefont {{Tacconi}}}, \bibinfo {author} {\bibfnamefont {F.}~\bibnamefont {{Vincent}}}, \bibinfo {author} {\bibfnamefont {S.}~\bibnamefont {{von Fellenberg}}}, \bibinfo {author} {\bibfnamefont {I.}~\bibnamefont {{Waisberg}}}, \bibinfo {author} {\bibfnamefont {F.}~\bibnamefont {{Widmann}}}, \bibinfo {author} {\bibfnamefont {E.}~\bibnamefont {{Wieprecht}}}, \bibinfo {author} {\bibfnamefont {E.}~\bibnamefont {{Wiezorrek}}}, \bibinfo {author} {\bibfnamefont {J.}~\bibnamefont {{Woillez}}},\ and\ \bibinfo {author} {\bibfnamefont {S.}~\bibnamefont {{Yazici}}},\ }\href {https://doi.org/10.1051/0004-6361/201935656} {\bibfield  {journal} {\bibinfo  {journal} {\aap}\ }\textbf {\bibinfo {volume} {625}},\ \bibinfo {eid} {L10} (\bibinfo {year} {2019})},\ \Eprint {https://arxiv.org/abs/1904.05721} {arXiv:1904.05721 [astro-ph.GA]} \BibitemShut {NoStop}%
\bibitem [{\citenamefont {{McMillan}}(2017)}]{McMillan2017}%
  \BibitemOpen
  \bibfield  {author} {\bibinfo {author} {\bibfnamefont {P.~J.}\ \bibnamefont {{McMillan}}},\ }\href {https://doi.org/10.1093/mnras/stw2759} {\bibfield  {journal} {\bibinfo  {journal} {\mnras}\ }\textbf {\bibinfo {volume} {465}},\ \bibinfo {pages} {76} (\bibinfo {year} {2017})},\ \Eprint {https://arxiv.org/abs/1608.00971} {arXiv:1608.00971 [astro-ph.GA]} \BibitemShut {NoStop}%
\bibitem [{\citenamefont {{Shklovskii}}(1970)}]{Shklovskii1970}%
  \BibitemOpen
  \bibfield  {author} {\bibinfo {author} {\bibfnamefont {I.~S.}\ \bibnamefont {{Shklovskii}}},\ }\href@noop {} {\bibfield  {journal} {\bibinfo  {journal} {\sovast}\ }\textbf {\bibinfo {volume} {13}},\ \bibinfo {pages} {562} (\bibinfo {year} {1970})}\BibitemShut {NoStop}%
\bibitem [{\citenamefont {{Manchester}}\ \emph {et~al.}(2005)\citenamefont {{Manchester}}, \citenamefont {{Hobbs}}, \citenamefont {{Teoh}},\ and\ \citenamefont {{Hobbs}}}]{Manchester2005}%
  \BibitemOpen
  \bibfield  {author} {\bibinfo {author} {\bibfnamefont {R.~N.}\ \bibnamefont {{Manchester}}}, \bibinfo {author} {\bibfnamefont {G.~B.}\ \bibnamefont {{Hobbs}}}, \bibinfo {author} {\bibfnamefont {A.}~\bibnamefont {{Teoh}}},\ and\ \bibinfo {author} {\bibfnamefont {M.}~\bibnamefont {{Hobbs}}},\ }\href {https://doi.org/10.1086/428488} {\bibfield  {journal} {\bibinfo  {journal} {\aj}\ }\textbf {\bibinfo {volume} {129}},\ \bibinfo {pages} {1993} (\bibinfo {year} {2005})},\ \Eprint {https://arxiv.org/abs/astro-ph/0412641} {arXiv:astro-ph/0412641 [astro-ph]} \BibitemShut {NoStop}%
\bibitem [{\citenamefont {{Reardon}}\ \emph {et~al.}(2016{\natexlab{a}})\citenamefont {{Reardon}}, \citenamefont {{Hobbs}}, \citenamefont {{Coles}}, \citenamefont {{Levin}}, \citenamefont {{Keith}}, \citenamefont {{Bailes}}, \citenamefont {{Bhat}}, \citenamefont {{Burke-Spolaor}}, \citenamefont {{Dai}}, \citenamefont {{Kerr}}, \citenamefont {{Lasky}}, \citenamefont {{Manchester}}, \citenamefont {{Os{\l}owski}}, \citenamefont {{Ravi}}, \citenamefont {{Shannon}}, \citenamefont {{van Straten}}, \citenamefont {{Toomey}}, \citenamefont {{Wang}}, \citenamefont {{Wen}}, \citenamefont {{You}},\ and\ \citenamefont {{Zhu}}}]{Reardon2016}%
  \BibitemOpen
  \bibfield  {author} {\bibinfo {author} {\bibfnamefont {D.~J.}\ \bibnamefont {{Reardon}}}, \bibinfo {author} {\bibfnamefont {G.}~\bibnamefont {{Hobbs}}}, \bibinfo {author} {\bibfnamefont {W.}~\bibnamefont {{Coles}}}, \bibinfo {author} {\bibfnamefont {Y.}~\bibnamefont {{Levin}}}, \bibinfo {author} {\bibfnamefont {M.~J.}\ \bibnamefont {{Keith}}}, \bibinfo {author} {\bibfnamefont {M.}~\bibnamefont {{Bailes}}}, \bibinfo {author} {\bibfnamefont {N.~D.~R.}\ \bibnamefont {{Bhat}}}, \bibinfo {author} {\bibfnamefont {S.}~\bibnamefont {{Burke-Spolaor}}}, \bibinfo {author} {\bibfnamefont {S.}~\bibnamefont {{Dai}}}, \bibinfo {author} {\bibfnamefont {M.}~\bibnamefont {{Kerr}}}, \bibinfo {author} {\bibfnamefont {P.~D.}\ \bibnamefont {{Lasky}}}, \bibinfo {author} {\bibfnamefont {R.~N.}\ \bibnamefont {{Manchester}}}, \bibinfo {author} {\bibfnamefont {S.}~\bibnamefont {{Os{\l}owski}}}, \bibinfo {author} {\bibfnamefont {V.}~\bibnamefont {{Ravi}}}, \bibinfo {author} {\bibfnamefont {R.~M.}\ \bibnamefont {{Shannon}}}, \bibinfo
  {author} {\bibfnamefont {W.}~\bibnamefont {{van Straten}}}, \bibinfo {author} {\bibfnamefont {L.}~\bibnamefont {{Toomey}}}, \bibinfo {author} {\bibfnamefont {J.}~\bibnamefont {{Wang}}}, \bibinfo {author} {\bibfnamefont {L.}~\bibnamefont {{Wen}}}, \bibinfo {author} {\bibfnamefont {X.~P.}\ \bibnamefont {{You}}},\ and\ \bibinfo {author} {\bibfnamefont {X.~J.}\ \bibnamefont {{Zhu}}},\ }\href {https://doi.org/10.1093/mnras/stv2395} {\bibfield  {journal} {\bibinfo  {journal} {\mnras}\ }\textbf {\bibinfo {volume} {455}},\ \bibinfo {pages} {1751} (\bibinfo {year} {2016}{\natexlab{a}})},\ \Eprint {https://arxiv.org/abs/1510.04434} {arXiv:1510.04434 [astro-ph.HE]} \BibitemShut {NoStop}%
\bibitem [{\citenamefont {{Agazie}}\ \emph {et~al.}(2023)\citenamefont {{Agazie}}, \citenamefont {{Alam}}, \citenamefont {{Anumarlapudi}}, \citenamefont {{Archibald}}, \citenamefont {{Arzoumanian}}, \citenamefont {{Baker}}, \citenamefont {{Blecha}}, \citenamefont {{Bonidie}}, \citenamefont {{Brazier}}, \citenamefont {{Brook}}, \citenamefont {{Burke-Spolaor}}, \citenamefont {{B{\'e}csy}}, \citenamefont {{Chapman}}, \citenamefont {{Charisi}}, \citenamefont {{Chatterjee}}, \citenamefont {{Cohen}}, \citenamefont {{Cordes}}, \citenamefont {{Cornish}}, \citenamefont {{Crawford}}, \citenamefont {{Cromartie}}, \citenamefont {{Crowter}}, \citenamefont {{Decesar}}, \citenamefont {{Demorest}}, \citenamefont {{Dolch}}, \citenamefont {{Drachler}}, \citenamefont {{Ferrara}}, \citenamefont {{Fiore}}, \citenamefont {{Fonseca}}, \citenamefont {{Freedman}}, \citenamefont {{Garver-Daniels}}, \citenamefont {{Gentile}}, \citenamefont {{Glaser}}, \citenamefont {{Good}}, \citenamefont {{G{\"u}ltekin}}, \citenamefont {{Hazboun}},
  \citenamefont {{Jennings}}, \citenamefont {{Jessup}}, \citenamefont {{Johnson}}, \citenamefont {{Jones}}, \citenamefont {{Kaiser}}, \citenamefont {{Kaplan}}, \citenamefont {{Kelley}}, \citenamefont {{Kerr}}, \citenamefont {{Key}}, \citenamefont {{Kuske}}, \citenamefont {{Laal}}, \citenamefont {{Lam}}, \citenamefont {{Lamb}}, \citenamefont {{Lazio}}, \citenamefont {{Lewandowska}}, \citenamefont {{Lin}}, \citenamefont {{Liu}}, \citenamefont {{Lorimer}}, \citenamefont {{Luo}}, \citenamefont {{Lynch}}, \citenamefont {{Ma}}, \citenamefont {{Madison}}, \citenamefont {{Maraccini}}, \citenamefont {{McEwen}}, \citenamefont {{McKee}}, \citenamefont {{McLaughlin}}, \citenamefont {{McMann}}, \citenamefont {{Meyers}}, \citenamefont {{Mingarelli}}, \citenamefont {{Mitridate}}, \citenamefont {{Ng}}, \citenamefont {{Nice}}, \citenamefont {{Ocker}}, \citenamefont {{Olum}}, \citenamefont {{Panciu}}, \citenamefont {{Pennucci}}, \citenamefont {{Perera}}, \citenamefont {{Pol}}, \citenamefont {{Radovan}}, \citenamefont
  {{Ransom}}, \citenamefont {{Ray}}, \citenamefont {{Romano}}, \citenamefont {{Salo}}, \citenamefont {{Sardesai}}, \citenamefont {{Schmiedekamp}}, \citenamefont {{Schmiedekamp}}, \citenamefont {{Schmitz}}, \citenamefont {{Shapiro-Albert}}, \citenamefont {{Siemens}}, \citenamefont {{Simon}}, \citenamefont {{Siwek}}, \citenamefont {{Stairs}}, \citenamefont {{Stinebring}}, \citenamefont {{Stovall}}, \citenamefont {{Susobhanan}}, \citenamefont {{Swiggum}}, \citenamefont {{Taylor}}, \citenamefont {{Turner}}, \citenamefont {{Unal}}, \citenamefont {{Vallisneri}}, \citenamefont {{Vigeland}}, \citenamefont {{Wahl}}, \citenamefont {{Wang}}, \citenamefont {{Witt}}, \citenamefont {{Young}},\ and\ \citenamefont {{Nanograv Collaboration}}}]{NG15}%
  \BibitemOpen
  \bibfield  {author} {\bibinfo {author} {\bibfnamefont {G.}~\bibnamefont {{Agazie}}}, \bibinfo {author} {\bibfnamefont {M.~F.}\ \bibnamefont {{Alam}}}, \bibinfo {author} {\bibfnamefont {A.}~\bibnamefont {{Anumarlapudi}}}, \bibinfo {author} {\bibfnamefont {A.~M.}\ \bibnamefont {{Archibald}}}, \bibinfo {author} {\bibfnamefont {Z.}~\bibnamefont {{Arzoumanian}}}, \bibinfo {author} {\bibfnamefont {P.~T.}\ \bibnamefont {{Baker}}}, \bibinfo {author} {\bibfnamefont {L.}~\bibnamefont {{Blecha}}}, \bibinfo {author} {\bibfnamefont {V.}~\bibnamefont {{Bonidie}}}, \bibinfo {author} {\bibfnamefont {A.}~\bibnamefont {{Brazier}}}, \bibinfo {author} {\bibfnamefont {P.~R.}\ \bibnamefont {{Brook}}}, \bibinfo {author} {\bibfnamefont {S.}~\bibnamefont {{Burke-Spolaor}}}, \bibinfo {author} {\bibfnamefont {B.}~\bibnamefont {{B{\'e}csy}}}, \bibinfo {author} {\bibfnamefont {C.}~\bibnamefont {{Chapman}}}, \bibinfo {author} {\bibfnamefont {M.}~\bibnamefont {{Charisi}}}, \bibinfo {author} {\bibfnamefont {S.}~\bibnamefont
  {{Chatterjee}}}, \bibinfo {author} {\bibfnamefont {T.}~\bibnamefont {{Cohen}}}, \bibinfo {author} {\bibfnamefont {J.~M.}\ \bibnamefont {{Cordes}}}, \bibinfo {author} {\bibfnamefont {N.~J.}\ \bibnamefont {{Cornish}}}, \bibinfo {author} {\bibfnamefont {F.}~\bibnamefont {{Crawford}}}, \bibinfo {author} {\bibfnamefont {H.~T.}\ \bibnamefont {{Cromartie}}}, \bibinfo {author} {\bibfnamefont {K.}~\bibnamefont {{Crowter}}}, \bibinfo {author} {\bibfnamefont {M.~E.}\ \bibnamefont {{Decesar}}}, \bibinfo {author} {\bibfnamefont {P.~B.}\ \bibnamefont {{Demorest}}}, \bibinfo {author} {\bibfnamefont {T.}~\bibnamefont {{Dolch}}}, \bibinfo {author} {\bibfnamefont {B.}~\bibnamefont {{Drachler}}}, \bibinfo {author} {\bibfnamefont {E.~C.}\ \bibnamefont {{Ferrara}}}, \bibinfo {author} {\bibfnamefont {W.}~\bibnamefont {{Fiore}}}, \bibinfo {author} {\bibfnamefont {E.}~\bibnamefont {{Fonseca}}}, \bibinfo {author} {\bibfnamefont {G.~E.}\ \bibnamefont {{Freedman}}}, \bibinfo {author} {\bibfnamefont {N.}~\bibnamefont
  {{Garver-Daniels}}}, \bibinfo {author} {\bibfnamefont {P.~A.}\ \bibnamefont {{Gentile}}}, \bibinfo {author} {\bibfnamefont {J.}~\bibnamefont {{Glaser}}}, \bibinfo {author} {\bibfnamefont {D.~C.}\ \bibnamefont {{Good}}}, \bibinfo {author} {\bibfnamefont {K.}~\bibnamefont {{G{\"u}ltekin}}}, \bibinfo {author} {\bibfnamefont {J.~S.}\ \bibnamefont {{Hazboun}}}, \bibinfo {author} {\bibfnamefont {R.~J.}\ \bibnamefont {{Jennings}}}, \bibinfo {author} {\bibfnamefont {C.}~\bibnamefont {{Jessup}}}, \bibinfo {author} {\bibfnamefont {A.~D.}\ \bibnamefont {{Johnson}}}, \bibinfo {author} {\bibfnamefont {M.~L.}\ \bibnamefont {{Jones}}}, \bibinfo {author} {\bibfnamefont {A.~R.}\ \bibnamefont {{Kaiser}}}, \bibinfo {author} {\bibfnamefont {D.~L.}\ \bibnamefont {{Kaplan}}}, \bibinfo {author} {\bibfnamefont {L.~Z.}\ \bibnamefont {{Kelley}}}, \bibinfo {author} {\bibfnamefont {M.}~\bibnamefont {{Kerr}}}, \bibinfo {author} {\bibfnamefont {J.~S.}\ \bibnamefont {{Key}}}, \bibinfo {author} {\bibfnamefont {A.}~\bibnamefont {{Kuske}}},
  \bibinfo {author} {\bibfnamefont {N.}~\bibnamefont {{Laal}}}, \bibinfo {author} {\bibfnamefont {M.~T.}\ \bibnamefont {{Lam}}}, \bibinfo {author} {\bibfnamefont {W.~G.}\ \bibnamefont {{Lamb}}}, \bibinfo {author} {\bibfnamefont {T.~J.~W.}\ \bibnamefont {{Lazio}}}, \bibinfo {author} {\bibfnamefont {N.}~\bibnamefont {{Lewandowska}}}, \bibinfo {author} {\bibfnamefont {Y.}~\bibnamefont {{Lin}}}, \bibinfo {author} {\bibfnamefont {T.}~\bibnamefont {{Liu}}}, \bibinfo {author} {\bibfnamefont {D.~R.}\ \bibnamefont {{Lorimer}}}, \bibinfo {author} {\bibfnamefont {J.}~\bibnamefont {{Luo}}}, \bibinfo {author} {\bibfnamefont {R.~S.}\ \bibnamefont {{Lynch}}}, \bibinfo {author} {\bibfnamefont {C.-P.}\ \bibnamefont {{Ma}}}, \bibinfo {author} {\bibfnamefont {D.~R.}\ \bibnamefont {{Madison}}}, \bibinfo {author} {\bibfnamefont {K.}~\bibnamefont {{Maraccini}}}, \bibinfo {author} {\bibfnamefont {A.}~\bibnamefont {{McEwen}}}, \bibinfo {author} {\bibfnamefont {J.~W.}\ \bibnamefont {{McKee}}}, \bibinfo {author} {\bibfnamefont
  {M.~A.}\ \bibnamefont {{McLaughlin}}}, \bibinfo {author} {\bibfnamefont {N.}~\bibnamefont {{McMann}}}, \bibinfo {author} {\bibfnamefont {B.~W.}\ \bibnamefont {{Meyers}}}, \bibinfo {author} {\bibfnamefont {C.~M.~F.}\ \bibnamefont {{Mingarelli}}}, \bibinfo {author} {\bibfnamefont {A.}~\bibnamefont {{Mitridate}}}, \bibinfo {author} {\bibfnamefont {C.}~\bibnamefont {{Ng}}}, \bibinfo {author} {\bibfnamefont {D.~J.}\ \bibnamefont {{Nice}}}, \bibinfo {author} {\bibfnamefont {S.~K.}\ \bibnamefont {{Ocker}}}, \bibinfo {author} {\bibfnamefont {K.~D.}\ \bibnamefont {{Olum}}}, \bibinfo {author} {\bibfnamefont {E.}~\bibnamefont {{Panciu}}}, \bibinfo {author} {\bibfnamefont {T.~T.}\ \bibnamefont {{Pennucci}}}, \bibinfo {author} {\bibfnamefont {B.~B.~P.}\ \bibnamefont {{Perera}}}, \bibinfo {author} {\bibfnamefont {N.~S.}\ \bibnamefont {{Pol}}}, \bibinfo {author} {\bibfnamefont {H.~A.}\ \bibnamefont {{Radovan}}}, \bibinfo {author} {\bibfnamefont {S.~M.}\ \bibnamefont {{Ransom}}}, \bibinfo {author} {\bibfnamefont {P.~S.}\
  \bibnamefont {{Ray}}}, \bibinfo {author} {\bibfnamefont {J.~D.}\ \bibnamefont {{Romano}}}, \bibinfo {author} {\bibfnamefont {L.}~\bibnamefont {{Salo}}}, \bibinfo {author} {\bibfnamefont {S.~C.}\ \bibnamefont {{Sardesai}}}, \bibinfo {author} {\bibfnamefont {C.}~\bibnamefont {{Schmiedekamp}}}, \bibinfo {author} {\bibfnamefont {A.}~\bibnamefont {{Schmiedekamp}}}, \bibinfo {author} {\bibfnamefont {K.}~\bibnamefont {{Schmitz}}}, \bibinfo {author} {\bibfnamefont {B.~J.}\ \bibnamefont {{Shapiro-Albert}}}, \bibinfo {author} {\bibfnamefont {X.}~\bibnamefont {{Siemens}}}, \bibinfo {author} {\bibfnamefont {J.}~\bibnamefont {{Simon}}}, \bibinfo {author} {\bibfnamefont {M.~S.}\ \bibnamefont {{Siwek}}}, \bibinfo {author} {\bibfnamefont {I.~H.}\ \bibnamefont {{Stairs}}}, \bibinfo {author} {\bibfnamefont {D.~R.}\ \bibnamefont {{Stinebring}}}, \bibinfo {author} {\bibfnamefont {K.}~\bibnamefont {{Stovall}}}, \bibinfo {author} {\bibfnamefont {A.}~\bibnamefont {{Susobhanan}}}, \bibinfo {author} {\bibfnamefont {J.~K.}\
  \bibnamefont {{Swiggum}}}, \bibinfo {author} {\bibfnamefont {S.~R.}\ \bibnamefont {{Taylor}}}, \bibinfo {author} {\bibfnamefont {J.~E.}\ \bibnamefont {{Turner}}}, \bibinfo {author} {\bibfnamefont {C.}~\bibnamefont {{Unal}}}, \bibinfo {author} {\bibfnamefont {M.}~\bibnamefont {{Vallisneri}}}, \bibinfo {author} {\bibfnamefont {S.~J.}\ \bibnamefont {{Vigeland}}}, \bibinfo {author} {\bibfnamefont {H.~M.}\ \bibnamefont {{Wahl}}}, \bibinfo {author} {\bibfnamefont {Q.}~\bibnamefont {{Wang}}}, \bibinfo {author} {\bibfnamefont {C.~A.}\ \bibnamefont {{Witt}}}, \bibinfo {author} {\bibfnamefont {O.}~\bibnamefont {{Young}}},\ and\ \bibinfo {author} {\bibnamefont {{Nanograv Collaboration}}},\ }\href {https://doi.org/10.3847/2041-8213/acda9a} {\bibfield  {journal} {\bibinfo  {journal} {\apjl}\ }\textbf {\bibinfo {volume} {951}},\ \bibinfo {eid} {L9} (\bibinfo {year} {2023})},\ \Eprint {https://arxiv.org/abs/2306.16217} {arXiv:2306.16217 [astro-ph.HE]} \BibitemShut {NoStop}%
\bibitem [{\citenamefont {{Kramer}}\ \emph {et~al.}(2006)\citenamefont {{Kramer}}, \citenamefont {{Stairs}}, \citenamefont {{Manchester}}, \citenamefont {{McLaughlin}}, \citenamefont {{Lyne}}, \citenamefont {{Ferdman}}, \citenamefont {{Burgay}}, \citenamefont {{Lorimer}}, \citenamefont {{Possenti}}, \citenamefont {{D'Amico}}, \citenamefont {{Sarkissian}}, \citenamefont {{Hobbs}}, \citenamefont {{Reynolds}}, \citenamefont {{Freire}},\ and\ \citenamefont {{Camilo}}}]{Kramer2006}%
  \BibitemOpen
  \bibfield  {author} {\bibinfo {author} {\bibfnamefont {M.}~\bibnamefont {{Kramer}}}, \bibinfo {author} {\bibfnamefont {I.~H.}\ \bibnamefont {{Stairs}}}, \bibinfo {author} {\bibfnamefont {R.~N.}\ \bibnamefont {{Manchester}}}, \bibinfo {author} {\bibfnamefont {M.~A.}\ \bibnamefont {{McLaughlin}}}, \bibinfo {author} {\bibfnamefont {A.~G.}\ \bibnamefont {{Lyne}}}, \bibinfo {author} {\bibfnamefont {R.~D.}\ \bibnamefont {{Ferdman}}}, \bibinfo {author} {\bibfnamefont {M.}~\bibnamefont {{Burgay}}}, \bibinfo {author} {\bibfnamefont {D.~R.}\ \bibnamefont {{Lorimer}}}, \bibinfo {author} {\bibfnamefont {A.}~\bibnamefont {{Possenti}}}, \bibinfo {author} {\bibfnamefont {N.}~\bibnamefont {{D'Amico}}}, \bibinfo {author} {\bibfnamefont {J.~M.}\ \bibnamefont {{Sarkissian}}}, \bibinfo {author} {\bibfnamefont {G.~B.}\ \bibnamefont {{Hobbs}}}, \bibinfo {author} {\bibfnamefont {J.~E.}\ \bibnamefont {{Reynolds}}}, \bibinfo {author} {\bibfnamefont {P.~C.~C.}\ \bibnamefont {{Freire}}},\ and\ \bibinfo {author} {\bibfnamefont
  {F.}~\bibnamefont {{Camilo}}},\ }\href {https://doi.org/10.1126/science.1132305} {\bibfield  {journal} {\bibinfo  {journal} {Science}\ }\textbf {\bibinfo {volume} {314}},\ \bibinfo {pages} {97} (\bibinfo {year} {2006})},\ \Eprint {https://arxiv.org/abs/astro-ph/0609417} {arXiv:astro-ph/0609417 [astro-ph]} \BibitemShut {NoStop}%
\bibitem [{\citenamefont {{Deller}}\ \emph {et~al.}(2009)\citenamefont {{Deller}}, \citenamefont {{Bailes}},\ and\ \citenamefont {{Tingay}}}]{Deller2009}%
  \BibitemOpen
  \bibfield  {author} {\bibinfo {author} {\bibfnamefont {A.~T.}\ \bibnamefont {{Deller}}}, \bibinfo {author} {\bibfnamefont {M.}~\bibnamefont {{Bailes}}},\ and\ \bibinfo {author} {\bibfnamefont {S.~J.}\ \bibnamefont {{Tingay}}},\ }\href {https://doi.org/10.1126/science.1167969} {\bibfield  {journal} {\bibinfo  {journal} {Science}\ }\textbf {\bibinfo {volume} {323}},\ \bibinfo {pages} {1327} (\bibinfo {year} {2009})},\ \Eprint {https://arxiv.org/abs/0902.0996} {arXiv:0902.0996 [astro-ph.SR]} \BibitemShut {NoStop}%
\bibitem [{\citenamefont {{EPTA Collaboration}}\ \emph {et~al.}(2023)\citenamefont {{EPTA Collaboration}}, \citenamefont {{Antoniadis}}, \citenamefont {{Babak}}, \citenamefont {{Bak Nielsen}}, \citenamefont {{Bassa}}, \citenamefont {{Berthereau}}, \citenamefont {{Bonetti}}, \citenamefont {{Bortolas}}, \citenamefont {{Brook}}, \citenamefont {{Burgay}}, \citenamefont {{Caballero}}, \citenamefont {{Chalumeau}}, \citenamefont {{Champion}}, \citenamefont {{C{\l}ianlaridis}}, \citenamefont {{Chen}}, \citenamefont {{Cognard}}, \citenamefont {{Desvignes}}, \citenamefont {{Falxa}}, \citenamefont {{Ferdman}}, \citenamefont {{Franchini}}, \citenamefont {{Gair}}, \citenamefont {{Goncharov}}, \citenamefont {{Graikou}}, \citenamefont {{Grie{\ss}meier}}, \citenamefont {{Guillemot}}, \citenamefont {{Guo}}, \citenamefont {{Hu}}, \citenamefont {{Iraci}}, \citenamefont {{Izquierdo-Villalba}}, \citenamefont {{Jang}}, \citenamefont {{Jawor}}, \citenamefont {{Janssen}}, \citenamefont {{Jessner}}, \citenamefont {{Karuppusamy}},
  \citenamefont {{Keane}}, \citenamefont {{Keith}}, \citenamefont {{Kramer}}, \citenamefont {{Krishnakumar}}, \citenamefont {{Lackeos}}, \citenamefont {{Lee}}, \citenamefont {{Liu}}, \citenamefont {{Liu}}, \citenamefont {{Lyne}}, \citenamefont {{McKee}}, \citenamefont {{Main}}, \citenamefont {{Mickaliger}}, \citenamefont {{Ni{\c{t}}u}}, \citenamefont {{Parthasarathy}}, \citenamefont {{Perera}}, \citenamefont {{Perrodin}}, \citenamefont {{Petiteau}}, \citenamefont {{Porayko}}, \citenamefont {{Possenti}}, \citenamefont {{Quelquejay Leclere}}, \citenamefont {{Samajdar}}, \citenamefont {{Sanidas}}, \citenamefont {{Sesana}}, \citenamefont {{Shaifullah}}, \citenamefont {{Speri}}, \citenamefont {{Spiewak}}, \citenamefont {{Stappers}}, \citenamefont {{Susarla}}, \citenamefont {{Theureau}}, \citenamefont {{Tiburzi}}, \citenamefont {{van der Wateren}}, \citenamefont {{Vecchio}}, \citenamefont {{Venkatraman Krishnan}}, \citenamefont {{Verbiest}}, \citenamefont {{Wang}}, \citenamefont {{Wang}},\ and\ \citenamefont
  {{Wu}}}]{EPTADR2}%
  \BibitemOpen
  \bibfield  {author} {\bibinfo {author} {\bibnamefont {{EPTA Collaboration}}}, \bibinfo {author} {\bibfnamefont {J.}~\bibnamefont {{Antoniadis}}}, \bibinfo {author} {\bibfnamefont {S.}~\bibnamefont {{Babak}}}, \bibinfo {author} {\bibfnamefont {A.~S.}\ \bibnamefont {{Bak Nielsen}}}, \bibinfo {author} {\bibfnamefont {C.~G.}\ \bibnamefont {{Bassa}}}, \bibinfo {author} {\bibfnamefont {A.}~\bibnamefont {{Berthereau}}}, \bibinfo {author} {\bibfnamefont {M.}~\bibnamefont {{Bonetti}}}, \bibinfo {author} {\bibfnamefont {E.}~\bibnamefont {{Bortolas}}}, \bibinfo {author} {\bibfnamefont {P.~R.}\ \bibnamefont {{Brook}}}, \bibinfo {author} {\bibfnamefont {M.}~\bibnamefont {{Burgay}}}, \bibinfo {author} {\bibfnamefont {R.~N.}\ \bibnamefont {{Caballero}}}, \bibinfo {author} {\bibfnamefont {A.}~\bibnamefont {{Chalumeau}}}, \bibinfo {author} {\bibfnamefont {D.~J.}\ \bibnamefont {{Champion}}}, \bibinfo {author} {\bibfnamefont {S.}~\bibnamefont {{C{\l}ianlaridis}}}, \bibinfo {author} {\bibfnamefont {S.}~\bibnamefont {{Chen}}},
  \bibinfo {author} {\bibfnamefont {I.}~\bibnamefont {{Cognard}}}, \bibinfo {author} {\bibfnamefont {G.}~\bibnamefont {{Desvignes}}}, \bibinfo {author} {\bibfnamefont {M.}~\bibnamefont {{Falxa}}}, \bibinfo {author} {\bibfnamefont {R.~D.}\ \bibnamefont {{Ferdman}}}, \bibinfo {author} {\bibfnamefont {A.}~\bibnamefont {{Franchini}}}, \bibinfo {author} {\bibfnamefont {J.~R.}\ \bibnamefont {{Gair}}}, \bibinfo {author} {\bibfnamefont {B.}~\bibnamefont {{Goncharov}}}, \bibinfo {author} {\bibfnamefont {E.}~\bibnamefont {{Graikou}}}, \bibinfo {author} {\bibfnamefont {J.~M.}\ \bibnamefont {{Grie{\ss}meier}}}, \bibinfo {author} {\bibfnamefont {L.}~\bibnamefont {{Guillemot}}}, \bibinfo {author} {\bibfnamefont {Y.~J.}\ \bibnamefont {{Guo}}}, \bibinfo {author} {\bibfnamefont {H.}~\bibnamefont {{Hu}}}, \bibinfo {author} {\bibfnamefont {F.}~\bibnamefont {{Iraci}}}, \bibinfo {author} {\bibfnamefont {D.}~\bibnamefont {{Izquierdo-Villalba}}}, \bibinfo {author} {\bibfnamefont {J.}~\bibnamefont {{Jang}}}, \bibinfo {author}
  {\bibfnamefont {J.}~\bibnamefont {{Jawor}}}, \bibinfo {author} {\bibfnamefont {G.~H.}\ \bibnamefont {{Janssen}}}, \bibinfo {author} {\bibfnamefont {A.}~\bibnamefont {{Jessner}}}, \bibinfo {author} {\bibfnamefont {R.}~\bibnamefont {{Karuppusamy}}}, \bibinfo {author} {\bibfnamefont {E.~F.}\ \bibnamefont {{Keane}}}, \bibinfo {author} {\bibfnamefont {M.~J.}\ \bibnamefont {{Keith}}}, \bibinfo {author} {\bibfnamefont {M.}~\bibnamefont {{Kramer}}}, \bibinfo {author} {\bibfnamefont {M.~A.}\ \bibnamefont {{Krishnakumar}}}, \bibinfo {author} {\bibfnamefont {K.}~\bibnamefont {{Lackeos}}}, \bibinfo {author} {\bibfnamefont {K.~J.}\ \bibnamefont {{Lee}}}, \bibinfo {author} {\bibfnamefont {K.}~\bibnamefont {{Liu}}}, \bibinfo {author} {\bibfnamefont {Y.}~\bibnamefont {{Liu}}}, \bibinfo {author} {\bibfnamefont {A.~G.}\ \bibnamefont {{Lyne}}}, \bibinfo {author} {\bibfnamefont {J.~W.}\ \bibnamefont {{McKee}}}, \bibinfo {author} {\bibfnamefont {R.~A.}\ \bibnamefont {{Main}}}, \bibinfo {author} {\bibfnamefont {M.~B.}\
  \bibnamefont {{Mickaliger}}}, \bibinfo {author} {\bibfnamefont {I.~C.}\ \bibnamefont {{Ni{\c{t}}u}}}, \bibinfo {author} {\bibfnamefont {A.}~\bibnamefont {{Parthasarathy}}}, \bibinfo {author} {\bibfnamefont {B.~B.~P.}\ \bibnamefont {{Perera}}}, \bibinfo {author} {\bibfnamefont {D.}~\bibnamefont {{Perrodin}}}, \bibinfo {author} {\bibfnamefont {A.}~\bibnamefont {{Petiteau}}}, \bibinfo {author} {\bibfnamefont {N.~K.}\ \bibnamefont {{Porayko}}}, \bibinfo {author} {\bibfnamefont {A.}~\bibnamefont {{Possenti}}}, \bibinfo {author} {\bibfnamefont {H.}~\bibnamefont {{Quelquejay Leclere}}}, \bibinfo {author} {\bibfnamefont {A.}~\bibnamefont {{Samajdar}}}, \bibinfo {author} {\bibfnamefont {S.~A.}\ \bibnamefont {{Sanidas}}}, \bibinfo {author} {\bibfnamefont {A.}~\bibnamefont {{Sesana}}}, \bibinfo {author} {\bibfnamefont {G.}~\bibnamefont {{Shaifullah}}}, \bibinfo {author} {\bibfnamefont {L.}~\bibnamefont {{Speri}}}, \bibinfo {author} {\bibfnamefont {R.}~\bibnamefont {{Spiewak}}}, \bibinfo {author} {\bibfnamefont
  {B.~W.}\ \bibnamefont {{Stappers}}}, \bibinfo {author} {\bibfnamefont {S.~C.}\ \bibnamefont {{Susarla}}}, \bibinfo {author} {\bibfnamefont {G.}~\bibnamefont {{Theureau}}}, \bibinfo {author} {\bibfnamefont {C.}~\bibnamefont {{Tiburzi}}}, \bibinfo {author} {\bibfnamefont {E.}~\bibnamefont {{van der Wateren}}}, \bibinfo {author} {\bibfnamefont {A.}~\bibnamefont {{Vecchio}}}, \bibinfo {author} {\bibfnamefont {V.}~\bibnamefont {{Venkatraman Krishnan}}}, \bibinfo {author} {\bibfnamefont {J.~P.~W.}\ \bibnamefont {{Verbiest}}}, \bibinfo {author} {\bibfnamefont {J.}~\bibnamefont {{Wang}}}, \bibinfo {author} {\bibfnamefont {L.}~\bibnamefont {{Wang}}},\ and\ \bibinfo {author} {\bibfnamefont {Z.}~\bibnamefont {{Wu}}},\ }\href {https://doi.org/10.1051/0004-6361/202346841} {\bibfield  {journal} {\bibinfo  {journal} {\aap}\ }\textbf {\bibinfo {volume} {678}},\ \bibinfo {eid} {A48} (\bibinfo {year} {2023})},\ \Eprint {https://arxiv.org/abs/2306.16224} {arXiv:2306.16224 [astro-ph.HE]} \BibitemShut {NoStop}%
\bibitem [{\citenamefont {{Nice}}\ \emph {et~al.}(2008)\citenamefont {{Nice}}, \citenamefont {{Stairs}},\ and\ \citenamefont {{Kasian}}}]{Nice2008}%
  \BibitemOpen
  \bibfield  {author} {\bibinfo {author} {\bibfnamefont {D.~J.}\ \bibnamefont {{Nice}}}, \bibinfo {author} {\bibfnamefont {I.~H.}\ \bibnamefont {{Stairs}}},\ and\ \bibinfo {author} {\bibfnamefont {L.~E.}\ \bibnamefont {{Kasian}}},\ }in\ \href {https://doi.org/10.1063/1.2900273} {\emph {\bibinfo {booktitle} {40 Years of Pulsars: Millisecond Pulsars, Magnetars and More}}},\ \bibinfo {series} {American Institute of Physics Conference Series}, Vol.\ \bibinfo {volume} {983},\ \bibinfo {editor} {edited by\ \bibinfo {editor} {\bibfnamefont {C.}~\bibnamefont {{Bassa}}}, \bibinfo {editor} {\bibfnamefont {Z.}~\bibnamefont {{Wang}}}, \bibinfo {editor} {\bibfnamefont {A.}~\bibnamefont {{Cumming}}},\ and\ \bibinfo {editor} {\bibfnamefont {V.~M.}\ \bibnamefont {{Kaspi}}}}\ (\bibinfo {year} {2008})\ pp.\ \bibinfo {pages} {453--458}\BibitemShut {NoStop}%
\bibitem [{\citenamefont {{Fonseca}}\ \emph {et~al.}(2014)\citenamefont {{Fonseca}}, \citenamefont {{Stairs}},\ and\ \citenamefont {{Thorsett}}}]{Fonseca2014}%
  \BibitemOpen
  \bibfield  {author} {\bibinfo {author} {\bibfnamefont {E.}~\bibnamefont {{Fonseca}}}, \bibinfo {author} {\bibfnamefont {I.~H.}\ \bibnamefont {{Stairs}}},\ and\ \bibinfo {author} {\bibfnamefont {S.~E.}\ \bibnamefont {{Thorsett}}},\ }\href {https://doi.org/10.1088/0004-637X/787/1/82} {\bibfield  {journal} {\bibinfo  {journal} {\apj}\ }\textbf {\bibinfo {volume} {787}},\ \bibinfo {eid} {82} (\bibinfo {year} {2014})},\ \Eprint {https://arxiv.org/abs/1402.4836} {arXiv:1402.4836 [astro-ph.HE]} \BibitemShut {NoStop}%
\bibitem [{\citenamefont {{Ding}}\ \emph {et~al.}(2021)\citenamefont {{Ding}}, \citenamefont {{Deller}}, \citenamefont {{Fonseca}}, \citenamefont {{Stairs}}, \citenamefont {{Stappers}},\ and\ \citenamefont {{Lyne}}}]{Ding2021}%
  \BibitemOpen
  \bibfield  {author} {\bibinfo {author} {\bibfnamefont {H.}~\bibnamefont {{Ding}}}, \bibinfo {author} {\bibfnamefont {A.~T.}\ \bibnamefont {{Deller}}}, \bibinfo {author} {\bibfnamefont {E.}~\bibnamefont {{Fonseca}}}, \bibinfo {author} {\bibfnamefont {I.~H.}\ \bibnamefont {{Stairs}}}, \bibinfo {author} {\bibfnamefont {B.}~\bibnamefont {{Stappers}}},\ and\ \bibinfo {author} {\bibfnamefont {A.}~\bibnamefont {{Lyne}}},\ }\href {https://doi.org/10.3847/2041-8213/ac3091} {\bibfield  {journal} {\bibinfo  {journal} {\apjl}\ }\textbf {\bibinfo {volume} {921}},\ \bibinfo {eid} {L19} (\bibinfo {year} {2021})},\ \Eprint {https://arxiv.org/abs/2110.10590} {arXiv:2110.10590 [astro-ph.HE]} \BibitemShut {NoStop}%
\bibitem [{\citenamefont {{Reardon}}\ \emph {et~al.}(2021)\citenamefont {{Reardon}}, \citenamefont {{Shannon}}, \citenamefont {{Cameron}}, \citenamefont {{Goncharov}}, \citenamefont {{Hobbs}}, \citenamefont {{Middleton}}, \citenamefont {{Shamohammadi}}, \citenamefont {{Thyagarajan}}, \citenamefont {{Bailes}}, \citenamefont {{Bhat}}, \citenamefont {{Dai}}, \citenamefont {{Kerr}}, \citenamefont {{Manchester}}, \citenamefont {{Russell}}, \citenamefont {{Spiewak}}, \citenamefont {{Wang}},\ and\ \citenamefont {{Zhu}}}]{PPTADR2}%
  \BibitemOpen
  \bibfield  {author} {\bibinfo {author} {\bibfnamefont {D.~J.}\ \bibnamefont {{Reardon}}}, \bibinfo {author} {\bibfnamefont {R.~M.}\ \bibnamefont {{Shannon}}}, \bibinfo {author} {\bibfnamefont {A.~D.}\ \bibnamefont {{Cameron}}}, \bibinfo {author} {\bibfnamefont {B.}~\bibnamefont {{Goncharov}}}, \bibinfo {author} {\bibfnamefont {G.~B.}\ \bibnamefont {{Hobbs}}}, \bibinfo {author} {\bibfnamefont {H.}~\bibnamefont {{Middleton}}}, \bibinfo {author} {\bibfnamefont {M.}~\bibnamefont {{Shamohammadi}}}, \bibinfo {author} {\bibfnamefont {N.}~\bibnamefont {{Thyagarajan}}}, \bibinfo {author} {\bibfnamefont {M.}~\bibnamefont {{Bailes}}}, \bibinfo {author} {\bibfnamefont {N.~D.~R.}\ \bibnamefont {{Bhat}}}, \bibinfo {author} {\bibfnamefont {S.}~\bibnamefont {{Dai}}}, \bibinfo {author} {\bibfnamefont {M.}~\bibnamefont {{Kerr}}}, \bibinfo {author} {\bibfnamefont {R.~N.}\ \bibnamefont {{Manchester}}}, \bibinfo {author} {\bibfnamefont {C.~J.}\ \bibnamefont {{Russell}}}, \bibinfo {author} {\bibfnamefont {R.}~\bibnamefont
  {{Spiewak}}}, \bibinfo {author} {\bibfnamefont {J.~B.}\ \bibnamefont {{Wang}}},\ and\ \bibinfo {author} {\bibfnamefont {X.~J.}\ \bibnamefont {{Zhu}}},\ }\href {https://doi.org/10.1093/mnras/stab1990} {\bibfield  {journal} {\bibinfo  {journal} {\mnras}\ }\textbf {\bibinfo {volume} {507}},\ \bibinfo {pages} {2137} (\bibinfo {year} {2021})},\ \Eprint {https://arxiv.org/abs/2107.04609} {arXiv:2107.04609 [astro-ph.HE]} \BibitemShut {NoStop}%
\bibitem [{\citenamefont {{Walker}}\ \emph {et~al.}(2022)\citenamefont {{Walker}}, \citenamefont {{Reardon}}, \citenamefont {{Thrane}},\ and\ \citenamefont {{Smith}}}]{Walker2022}%
  \BibitemOpen
  \bibfield  {author} {\bibinfo {author} {\bibfnamefont {K.}~\bibnamefont {{Walker}}}, \bibinfo {author} {\bibfnamefont {D.~J.}\ \bibnamefont {{Reardon}}}, \bibinfo {author} {\bibfnamefont {E.}~\bibnamefont {{Thrane}}},\ and\ \bibinfo {author} {\bibfnamefont {R.}~\bibnamefont {{Smith}}},\ }\href {https://doi.org/10.3847/1538-4357/ac69c6} {\bibfield  {journal} {\bibinfo  {journal} {\apj}\ }\textbf {\bibinfo {volume} {933}},\ \bibinfo {eid} {16} (\bibinfo {year} {2022})},\ \Eprint {https://arxiv.org/abs/2204.11077} {arXiv:2204.11077 [astro-ph.HE]} \BibitemShut {NoStop}%
\bibitem [{\citenamefont {{Freire}}\ \emph {et~al.}(2012)\citenamefont {{Freire}}, \citenamefont {{Wex}}, \citenamefont {{Esposito-Far{\`e}se}}, \citenamefont {{Verbiest}}, \citenamefont {{Bailes}}, \citenamefont {{Jacoby}}, \citenamefont {{Kramer}}, \citenamefont {{Stairs}}, \citenamefont {{Antoniadis}},\ and\ \citenamefont {{Janssen}}}]{Freire2012}%
  \BibitemOpen
  \bibfield  {author} {\bibinfo {author} {\bibfnamefont {P.~C.~C.}\ \bibnamefont {{Freire}}}, \bibinfo {author} {\bibfnamefont {N.}~\bibnamefont {{Wex}}}, \bibinfo {author} {\bibfnamefont {G.}~\bibnamefont {{Esposito-Far{\`e}se}}}, \bibinfo {author} {\bibfnamefont {J.~P.~W.}\ \bibnamefont {{Verbiest}}}, \bibinfo {author} {\bibfnamefont {M.}~\bibnamefont {{Bailes}}}, \bibinfo {author} {\bibfnamefont {B.~A.}\ \bibnamefont {{Jacoby}}}, \bibinfo {author} {\bibfnamefont {M.}~\bibnamefont {{Kramer}}}, \bibinfo {author} {\bibfnamefont {I.~H.}\ \bibnamefont {{Stairs}}}, \bibinfo {author} {\bibfnamefont {J.}~\bibnamefont {{Antoniadis}}},\ and\ \bibinfo {author} {\bibfnamefont {G.~H.}\ \bibnamefont {{Janssen}}},\ }\href {https://doi.org/10.1111/j.1365-2966.2012.21253.x} {\bibfield  {journal} {\bibinfo  {journal} {\mnras}\ }\textbf {\bibinfo {volume} {423}},\ \bibinfo {pages} {3328} (\bibinfo {year} {2012})},\ \Eprint {https://arxiv.org/abs/1205.1450} {arXiv:1205.1450 [astro-ph.GA]} \BibitemShut {NoStop}%
\bibitem [{\citenamefont {{Deller}}\ \emph {et~al.}(2018)\citenamefont {{Deller}}, \citenamefont {{Weisberg}}, \citenamefont {{Nice}},\ and\ \citenamefont {{Chatterjee}}}]{Deller2018}%
  \BibitemOpen
  \bibfield  {author} {\bibinfo {author} {\bibfnamefont {A.~T.}\ \bibnamefont {{Deller}}}, \bibinfo {author} {\bibfnamefont {J.~M.}\ \bibnamefont {{Weisberg}}}, \bibinfo {author} {\bibfnamefont {D.~J.}\ \bibnamefont {{Nice}}},\ and\ \bibinfo {author} {\bibfnamefont {S.}~\bibnamefont {{Chatterjee}}},\ }\href {https://doi.org/10.3847/1538-4357/aacf95} {\bibfield  {journal} {\bibinfo  {journal} {\apj}\ }\textbf {\bibinfo {volume} {862}},\ \bibinfo {eid} {139} (\bibinfo {year} {2018})},\ \Eprint {https://arxiv.org/abs/1806.10265} {arXiv:1806.10265 [astro-ph.SR]} \BibitemShut {NoStop}%
\bibitem [{\citenamefont {{Geyer}}\ \emph {et~al.}(2023)\citenamefont {{Geyer}}, \citenamefont {{Venkatraman Krishnan}}, \citenamefont {{Freire}}, \citenamefont {{Kramer}}, \citenamefont {{Antoniadis}}, \citenamefont {{Bailes}}, \citenamefont {{Bernadich}}, \citenamefont {{Buchner}}, \citenamefont {{Cameron}}, \citenamefont {{Champion}}, \citenamefont {{Karastergiou}}, \citenamefont {{Keith}}, \citenamefont {{Lower}}, \citenamefont {{Os{\l}owski}}, \citenamefont {{Possenti}}, \citenamefont {{Parthasarathy}}, \citenamefont {{Reardon}}, \citenamefont {{Serylak}}, \citenamefont {{Shannon}}, \citenamefont {{Spiewak}}, \citenamefont {{van Straten}},\ and\ \citenamefont {{Verbiest}}}]{Geyer2023}%
  \BibitemOpen
  \bibfield  {author} {\bibinfo {author} {\bibfnamefont {M.}~\bibnamefont {{Geyer}}}, \bibinfo {author} {\bibfnamefont {V.}~\bibnamefont {{Venkatraman Krishnan}}}, \bibinfo {author} {\bibfnamefont {P.~C.~C.}\ \bibnamefont {{Freire}}}, \bibinfo {author} {\bibfnamefont {M.}~\bibnamefont {{Kramer}}}, \bibinfo {author} {\bibfnamefont {J.}~\bibnamefont {{Antoniadis}}}, \bibinfo {author} {\bibfnamefont {M.}~\bibnamefont {{Bailes}}}, \bibinfo {author} {\bibfnamefont {M.~C.~i.}\ \bibnamefont {{Bernadich}}}, \bibinfo {author} {\bibfnamefont {S.}~\bibnamefont {{Buchner}}}, \bibinfo {author} {\bibfnamefont {A.~D.}\ \bibnamefont {{Cameron}}}, \bibinfo {author} {\bibfnamefont {D.~J.}\ \bibnamefont {{Champion}}}, \bibinfo {author} {\bibfnamefont {A.}~\bibnamefont {{Karastergiou}}}, \bibinfo {author} {\bibfnamefont {M.~J.}\ \bibnamefont {{Keith}}}, \bibinfo {author} {\bibfnamefont {M.~E.}\ \bibnamefont {{Lower}}}, \bibinfo {author} {\bibfnamefont {S.}~\bibnamefont {{Os{\l}owski}}}, \bibinfo {author} {\bibfnamefont
  {A.}~\bibnamefont {{Possenti}}}, \bibinfo {author} {\bibfnamefont {A.}~\bibnamefont {{Parthasarathy}}}, \bibinfo {author} {\bibfnamefont {D.~J.}\ \bibnamefont {{Reardon}}}, \bibinfo {author} {\bibfnamefont {M.}~\bibnamefont {{Serylak}}}, \bibinfo {author} {\bibfnamefont {R.~M.}\ \bibnamefont {{Shannon}}}, \bibinfo {author} {\bibfnamefont {R.}~\bibnamefont {{Spiewak}}}, \bibinfo {author} {\bibfnamefont {W.}~\bibnamefont {{van Straten}}},\ and\ \bibinfo {author} {\bibfnamefont {J.~P.~W.}\ \bibnamefont {{Verbiest}}},\ }\href {https://doi.org/10.1051/0004-6361/202244654} {\bibfield  {journal} {\bibinfo  {journal} {\aap}\ }\textbf {\bibinfo {volume} {674}},\ \bibinfo {eid} {A169} (\bibinfo {year} {2023})},\ \Eprint {https://arxiv.org/abs/2304.09060} {arXiv:2304.09060 [astro-ph.HE]} \BibitemShut {NoStop}%
\bibitem [{\citenamefont {{Guo}}\ \emph {et~al.}(2021)\citenamefont {{Guo}}, \citenamefont {{Freire}}, \citenamefont {{Guillemot}}, \citenamefont {{Kramer}}, \citenamefont {{Zhu}}, \citenamefont {{Wex}}, \citenamefont {{McKee}}, \citenamefont {{Deller}}, \citenamefont {{Ding}}, \citenamefont {{Kaplan}}, \citenamefont {{Stappers}}, \citenamefont {{Cognard}}, \citenamefont {{Miao}}, \citenamefont {{Haase}}, \citenamefont {{Keith}}, \citenamefont {{Ransom}},\ and\ \citenamefont {{Theureau}}}]{Guo2021}%
  \BibitemOpen
  \bibfield  {author} {\bibinfo {author} {\bibfnamefont {Y.~J.}\ \bibnamefont {{Guo}}}, \bibinfo {author} {\bibfnamefont {P.~C.~C.}\ \bibnamefont {{Freire}}}, \bibinfo {author} {\bibfnamefont {L.}~\bibnamefont {{Guillemot}}}, \bibinfo {author} {\bibfnamefont {M.}~\bibnamefont {{Kramer}}}, \bibinfo {author} {\bibfnamefont {W.~W.}\ \bibnamefont {{Zhu}}}, \bibinfo {author} {\bibfnamefont {N.}~\bibnamefont {{Wex}}}, \bibinfo {author} {\bibfnamefont {J.~W.}\ \bibnamefont {{McKee}}}, \bibinfo {author} {\bibfnamefont {A.}~\bibnamefont {{Deller}}}, \bibinfo {author} {\bibfnamefont {H.}~\bibnamefont {{Ding}}}, \bibinfo {author} {\bibfnamefont {D.~L.}\ \bibnamefont {{Kaplan}}}, \bibinfo {author} {\bibfnamefont {B.}~\bibnamefont {{Stappers}}}, \bibinfo {author} {\bibfnamefont {I.}~\bibnamefont {{Cognard}}}, \bibinfo {author} {\bibfnamefont {X.}~\bibnamefont {{Miao}}}, \bibinfo {author} {\bibfnamefont {L.}~\bibnamefont {{Haase}}}, \bibinfo {author} {\bibfnamefont {M.}~\bibnamefont {{Keith}}}, \bibinfo {author}
  {\bibfnamefont {S.~M.}\ \bibnamefont {{Ransom}}},\ and\ \bibinfo {author} {\bibfnamefont {G.}~\bibnamefont {{Theureau}}},\ }\href {https://doi.org/10.1051/0004-6361/202141450} {\bibfield  {journal} {\bibinfo  {journal} {\aap}\ }\textbf {\bibinfo {volume} {654}},\ \bibinfo {eid} {A16} (\bibinfo {year} {2021})},\ \Eprint {https://arxiv.org/abs/2107.09474} {arXiv:2107.09474 [astro-ph.HE]} \BibitemShut {NoStop}%
\bibitem [{\citenamefont {{Stovall}}\ \emph {et~al.}(2019)\citenamefont {{Stovall}}, \citenamefont {{Freire}}, \citenamefont {{Antoniadis}}, \citenamefont {{Bagchi}}, \citenamefont {{Deneva}}, \citenamefont {{Garver-Daniels}}, \citenamefont {{Martinez}}, \citenamefont {{McLaughlin}}, \citenamefont {{Arzoumanian}}, \citenamefont {{Blumer}}, \citenamefont {{Brook}}, \citenamefont {{Cromartie}}, \citenamefont {{Demorest}}, \citenamefont {{DeCesar}}, \citenamefont {{Dolch}}, \citenamefont {{Ellis}}, \citenamefont {{Ferdman}}, \citenamefont {{Ferrara}}, \citenamefont {{Fonseca}}, \citenamefont {{Gentile}}, \citenamefont {{Jones}}, \citenamefont {{Lam}}, \citenamefont {{Lorimer}}, \citenamefont {{Lynch}}, \citenamefont {{Ng}}, \citenamefont {{Nice}}, \citenamefont {{Pennucci}}, \citenamefont {{Ransom}}, \citenamefont {{Spiewak}}, \citenamefont {{Stairs}}, \citenamefont {{Swiggum}}, \citenamefont {{Vigeland}},\ and\ \citenamefont {{Zhu}}}]{Stovall2019}%
  \BibitemOpen
  \bibfield  {author} {\bibinfo {author} {\bibfnamefont {K.}~\bibnamefont {{Stovall}}}, \bibinfo {author} {\bibfnamefont {P.~C.~C.}\ \bibnamefont {{Freire}}}, \bibinfo {author} {\bibfnamefont {J.}~\bibnamefont {{Antoniadis}}}, \bibinfo {author} {\bibfnamefont {M.}~\bibnamefont {{Bagchi}}}, \bibinfo {author} {\bibfnamefont {J.~S.}\ \bibnamefont {{Deneva}}}, \bibinfo {author} {\bibfnamefont {N.}~\bibnamefont {{Garver-Daniels}}}, \bibinfo {author} {\bibfnamefont {J.~G.}\ \bibnamefont {{Martinez}}}, \bibinfo {author} {\bibfnamefont {M.~A.}\ \bibnamefont {{McLaughlin}}}, \bibinfo {author} {\bibfnamefont {Z.}~\bibnamefont {{Arzoumanian}}}, \bibinfo {author} {\bibfnamefont {H.}~\bibnamefont {{Blumer}}}, \bibinfo {author} {\bibfnamefont {P.~R.}\ \bibnamefont {{Brook}}}, \bibinfo {author} {\bibfnamefont {H.~T.}\ \bibnamefont {{Cromartie}}}, \bibinfo {author} {\bibfnamefont {P.~B.}\ \bibnamefont {{Demorest}}}, \bibinfo {author} {\bibfnamefont {M.~E.}\ \bibnamefont {{DeCesar}}}, \bibinfo {author} {\bibfnamefont
  {T.}~\bibnamefont {{Dolch}}}, \bibinfo {author} {\bibfnamefont {J.~A.}\ \bibnamefont {{Ellis}}}, \bibinfo {author} {\bibfnamefont {R.~D.}\ \bibnamefont {{Ferdman}}}, \bibinfo {author} {\bibfnamefont {E.~C.}\ \bibnamefont {{Ferrara}}}, \bibinfo {author} {\bibfnamefont {E.}~\bibnamefont {{Fonseca}}}, \bibinfo {author} {\bibfnamefont {P.~A.}\ \bibnamefont {{Gentile}}}, \bibinfo {author} {\bibfnamefont {M.~L.}\ \bibnamefont {{Jones}}}, \bibinfo {author} {\bibfnamefont {M.~T.}\ \bibnamefont {{Lam}}}, \bibinfo {author} {\bibfnamefont {D.~R.}\ \bibnamefont {{Lorimer}}}, \bibinfo {author} {\bibfnamefont {R.~S.}\ \bibnamefont {{Lynch}}}, \bibinfo {author} {\bibfnamefont {C.}~\bibnamefont {{Ng}}}, \bibinfo {author} {\bibfnamefont {D.~J.}\ \bibnamefont {{Nice}}}, \bibinfo {author} {\bibfnamefont {T.~T.}\ \bibnamefont {{Pennucci}}}, \bibinfo {author} {\bibfnamefont {S.~M.}\ \bibnamefont {{Ransom}}}, \bibinfo {author} {\bibfnamefont {R.}~\bibnamefont {{Spiewak}}}, \bibinfo {author} {\bibfnamefont {I.~H.}\ \bibnamefont
  {{Stairs}}}, \bibinfo {author} {\bibfnamefont {J.~K.}\ \bibnamefont {{Swiggum}}}, \bibinfo {author} {\bibfnamefont {S.~J.}\ \bibnamefont {{Vigeland}}},\ and\ \bibinfo {author} {\bibfnamefont {W.~W.}\ \bibnamefont {{Zhu}}},\ }\href {https://doi.org/10.3847/1538-4357/aaf37d} {\bibfield  {journal} {\bibinfo  {journal} {\apj}\ }\textbf {\bibinfo {volume} {870}},\ \bibinfo {eid} {74} (\bibinfo {year} {2019})},\ \Eprint {https://arxiv.org/abs/1809.05064} {arXiv:1809.05064 [astro-ph.HE]} \BibitemShut {NoStop}%
\bibitem [{\citenamefont {{Moran}}\ \emph {et~al.}(2023{\natexlab{a}})\citenamefont {{Moran}}, \citenamefont {{Mingarelli}}, \citenamefont {{Van Tilburg}},\ and\ \citenamefont {{Good}}}]{Moran2023}%
  \BibitemOpen
  \bibfield  {author} {\bibinfo {author} {\bibfnamefont {A.}~\bibnamefont {{Moran}}}, \bibinfo {author} {\bibfnamefont {C.~M.~F.}\ \bibnamefont {{Mingarelli}}}, \bibinfo {author} {\bibfnamefont {K.}~\bibnamefont {{Van Tilburg}}},\ and\ \bibinfo {author} {\bibfnamefont {D.}~\bibnamefont {{Good}}},\ }\href {https://doi.org/10.48550/arXiv.2306.13137} {\bibfield  {journal} {\bibinfo  {journal} {arXiv e-prints}\ ,\ \bibinfo {eid} {arXiv:2306.13137}} (\bibinfo {year} {2023}{\natexlab{a}})},\ \Eprint {https://arxiv.org/abs/2306.13137} {arXiv:2306.13137 [astro-ph.GA]} \BibitemShut {NoStop}%
\bibitem [{\citenamefont {{Yi}}\ and\ \citenamefont {{Cheng}}(2017)}]{YiCheng2017}%
  \BibitemOpen
  \bibfield  {author} {\bibinfo {author} {\bibfnamefont {S.-X.}\ \bibnamefont {{Yi}}}\ and\ \bibinfo {author} {\bibfnamefont {K.~S.}\ \bibnamefont {{Cheng}}},\ }\href {https://doi.org/10.3847/1538-4357/aa7c65} {\bibfield  {journal} {\bibinfo  {journal} {\apj}\ }\textbf {\bibinfo {volume} {844}},\ \bibinfo {eid} {114} (\bibinfo {year} {2017})},\ \Eprint {https://arxiv.org/abs/1706.08715} {arXiv:1706.08715 [astro-ph.HE]} \BibitemShut {NoStop}%
\bibitem [{\citenamefont {{Gaia Collaboration}}\ \emph {et~al.}(2022)\citenamefont {{Gaia Collaboration}}, \citenamefont {{Vallenari}}, \citenamefont {{Brown}}, \citenamefont {{Prusti}}, \citenamefont {{de Bruijne}}, \citenamefont {{Arenou}}, \citenamefont {{Babusiaux}}, \citenamefont {{Biermann}}, \citenamefont {{Creevey}}, \citenamefont {{Ducourant}}, \citenamefont {{Evans}}, \citenamefont {{Eyer}}, \citenamefont {{Guerra}}, \citenamefont {{Hutton}}, \citenamefont {{Jordi}}, \citenamefont {{Klioner}}, \citenamefont {{Lammers}}, \citenamefont {{Lindegren}}, \citenamefont {{Luri}}, \citenamefont {{Mignard}}, \citenamefont {{Panem}}, \citenamefont {{Pourbaix}}, \citenamefont {{Randich}}, \citenamefont {{Sartoretti}}, \citenamefont {{Soubiran}}, \citenamefont {{Tanga}}, \citenamefont {{Walton}}, \citenamefont {{Bailer-Jones}}, \citenamefont {{Bastian}}, \citenamefont {{Drimmel}}, \citenamefont {{Jansen}}, \citenamefont {{Katz}}, \citenamefont {{Lattanzi}}, \citenamefont {{van Leeuwen}}, \citenamefont
  {{Bakker}}, \citenamefont {{Cacciari}}, \citenamefont {{Casta{\~n}eda}}, \citenamefont {{De Angeli}}, \citenamefont {{Fabricius}}, \citenamefont {{Fouesneau}}, \citenamefont {{Fr{\'e}mat}}, \citenamefont {{Galluccio}}, \citenamefont {{Guerrier}}, \citenamefont {{Heiter}}, \citenamefont {{Masana}}, \citenamefont {{Messineo}}, \citenamefont {{Mowlavi}}, \citenamefont {{Nicolas}}, \citenamefont {{Nienartowicz}}, \citenamefont {{Pailler}}, \citenamefont {{Panuzzo}}, \citenamefont {{Riclet}}, \citenamefont {{Roux}}, \citenamefont {{Seabroke}}, \citenamefont {{Sordo{\o}rcit}}, \citenamefont {{Th{\'e}venin}}, \citenamefont {{Gracia-Abril}}, \citenamefont {{Portell}}, \citenamefont {{Teyssier}}, \citenamefont {{Altmann}}, \citenamefont {{Andrae}}, \citenamefont {{Audard}}, \citenamefont {{Bellas-Velidis}}, \citenamefont {{Benson}}, \citenamefont {{Berthier}}, \citenamefont {{Blomme}}, \citenamefont {{Burgess}}, \citenamefont {{Busonero}}, \citenamefont {{Busso}}, \citenamefont {{C{\'a}novas}}, \citenamefont
  {{Carry}}, \citenamefont {{Cellino}}, \citenamefont {{Cheek}}, \citenamefont {{Clementini}}, \citenamefont {{Damerdji}}, \citenamefont {{Davidson}}, \citenamefont {{de Teodoro}}, \citenamefont {{Nu{\~n}ez Campos}}, \citenamefont {{Delchambre}}, \citenamefont {{Dell'Oro}}, \citenamefont {{Esquej}}, \citenamefont {{Fern{\'a}ndez-Hern{\'a}ndez}}, \citenamefont {{Fraile}}, \citenamefont {{Garabato}}, \citenamefont {{Garc{\'\i}a-Lario}}, \citenamefont {{Gosset}}, \citenamefont {{Haigron}}, \citenamefont {{Halbwachs}}, \citenamefont {{Hambly}}, \citenamefont {{Harrison}}, \citenamefont {{Hern{\'a}ndez}}, \citenamefont {{Hestroffer}}, \citenamefont {{Hodgkin}}, \citenamefont {{Holl}}, \citenamefont {{Jan{\ss}en}}, \citenamefont {{Jevardat de Fombelle}}, \citenamefont {{Jordan}}, \citenamefont {{Krone-Martins}}, \citenamefont {{Lanzafame}}, \citenamefont {{L{\"o}ffler}}, \citenamefont {{Marchal}}, \citenamefont {{Marrese}}, \citenamefont {{Moitinho}}, \citenamefont {{Muinonen}}, \citenamefont {{Osborne}},
  \citenamefont {{Pancino}}, \citenamefont {{Pauwels}}, \citenamefont {{Recio-Blanco}}, \citenamefont {{Reyl{\'e}}}, \citenamefont {{Riello}}, \citenamefont {{Rimoldini}}, \citenamefont {{Roegiers}}, \citenamefont {{Rybizki}}, \citenamefont {{Sarro}}, \citenamefont {{Siopis}}, \citenamefont {{Smith}}, \citenamefont {{Sozzetti}}, \citenamefont {{Utrilla}}, \citenamefont {{van Leeuwen}}, \citenamefont {{Abbas}}, \citenamefont {{{\'A}brah{\'a}m}}, \citenamefont {{Abreu Aramburu}}, \citenamefont {{Aerts}}, \citenamefont {{Aguado}}, \citenamefont {{Ajaj}}, \citenamefont {{Aldea-Montero}}, \citenamefont {{Altavilla}}, \citenamefont {{{\'A}lvarez}}, \citenamefont {{Alves}}, \citenamefont {{Anders}}, \citenamefont {{Anderson}}, \citenamefont {{Anglada Varela}}, \citenamefont {{Antoja}}, \citenamefont {{Baines}}, \citenamefont {{Baker}}, \citenamefont {{Balaguer-N{\'u}{\~n}ez}}, \citenamefont {{Balbinot}}, \citenamefont {{Balog}}, \citenamefont {{Barache}}, \citenamefont {{Barbato}}, \citenamefont {{Barros}},
  \citenamefont {{Barstow}}, \citenamefont {{Bartolom{\'e}}}, \citenamefont {{Bassilana}}, \citenamefont {{Bauchet}}, \citenamefont {{Becciani}}, \citenamefont {{Bellazzini}}, \citenamefont {{Berihuete}}, \citenamefont {{Bernet}}, \citenamefont {{Bertone}}, \citenamefont {{Bianchi}}, \citenamefont {{Binnenfeld}}, \citenamefont {{Blanco-Cuaresma}}, \citenamefont {{Blazere}}, \citenamefont {{Boch}}, \citenamefont {{Bombrun}}, \citenamefont {{Bossini}}, \citenamefont {{Bouquillon}}, \citenamefont {{Bragaglia}}, \citenamefont {{Bramante}}, \citenamefont {{Breedt}}, \citenamefont {{Bressan}}, \citenamefont {{Brouillet}}, \citenamefont {{Brugaletta}}, \citenamefont {{Bucciarelli}}, \citenamefont {{Burlacu}}, \citenamefont {{Butkevich}}, \citenamefont {{Buzzi}}, \citenamefont {{Caffau}}, \citenamefont {{Cancelliere}}, \citenamefont {{Cantat-Gaudin}}, \citenamefont {{Carballo}}, \citenamefont {{Carlucci}}, \citenamefont {{Carnerero}}, \citenamefont {{Carrasco}}, \citenamefont {{Casamiquela}}, \citenamefont
  {{Castellani}}, \citenamefont {{Castro-Ginard}}, \citenamefont {{Chaoul}}, \citenamefont {{Charlot}}, \citenamefont {{Chemin}}, \citenamefont {{Chiaramida}}, \citenamefont {{Chiavassa}}, \citenamefont {{Chornay}}, \citenamefont {{Comoretto}}, \citenamefont {{Contursi}}, \citenamefont {{Cooper}}, \citenamefont {{Cornez}}, \citenamefont {{Cowell}}, \citenamefont {{Crifo}}, \citenamefont {{Cropper}}, \citenamefont {{Crosta}}, \citenamefont {{Crowley}}, \citenamefont {{Dafonte}}, \citenamefont {{Dapergolas}}, \citenamefont {{David}}, \citenamefont {{David}}, \citenamefont {{de Laverny}}, \citenamefont {{De Luise}}, \citenamefont {{De March}}, \citenamefont {{De Ridder}}, \citenamefont {{de Souza}}, \citenamefont {{de Torres}}, \citenamefont {{del Peloso}}, \citenamefont {{del Pozo}}, \citenamefont {{Delbo}}, \citenamefont {{Delgado}}, \citenamefont {{Delisle}}, \citenamefont {{Demouchy}}, \citenamefont {{Dharmawardena}}, \citenamefont {{Di Matteo}}, \citenamefont {{Diakite}}, \citenamefont {{Diener}},
  \citenamefont {{Distefano}}, \citenamefont {{Dolding}}, \citenamefont {{Edvardsson}}, \citenamefont {{Enke}}, \citenamefont {{Fabre}}, \citenamefont {{Fabrizio}}, \citenamefont {{Faigler}}, \citenamefont {{Fedorets}}, \citenamefont {{Fernique}}, \citenamefont {{Fienga}}, \citenamefont {{Figueras}}, \citenamefont {{Fournier}}, \citenamefont {{Fouron}}, \citenamefont {{Fragkoudi}}, \citenamefont {{Gai}}, \citenamefont {{Garcia-Gutierrez}}, \citenamefont {{Garcia-Reinaldos}}, \citenamefont {{Garc{\'\i}a-Torres}}, \citenamefont {{Garofalo}}, \citenamefont {{Gavel}}, \citenamefont {{Gavras}}, \citenamefont {{Gerlach}}, \citenamefont {{Geyer}}, \citenamefont {{Giacobbe}}, \citenamefont {{Gilmore}}, \citenamefont {{Girona}}, \citenamefont {{Giuffrida}}, \citenamefont {{Gomel}}, \citenamefont {{Gomez}}, \citenamefont {{Gonz{\'a}lez-N{\'u}{\~n}ez}}, \citenamefont {{Gonz{\'a}lez-Santamar{\'\i}a}}, \citenamefont {{Gonz{\'a}lez-Vidal}}, \citenamefont {{Granvik}}, \citenamefont {{Guillout}}, \citenamefont {{Guiraud}},
  \citenamefont {{Guti{\'e}rrez-S{\'a}nchez}}, \citenamefont {{Guy}}, \citenamefont {{Hatzidimitriou}}, \citenamefont {{Hauser}}, \citenamefont {{Haywood}}, \citenamefont {{Helmer}}, \citenamefont {{Helmi}}, \citenamefont {{Sarmiento}}, \citenamefont {{Hidalgo}}, \citenamefont {{Hilger}}, \citenamefont {{H{\l}adczuk}}, \citenamefont {{Hobbs}}, \citenamefont {{Holland}}, \citenamefont {{Huckle}}, \citenamefont {{Jardine}}, \citenamefont {{Jasniewicz}}, \citenamefont {{Jean-Antoine Piccolo}}, \citenamefont {{Jim{\'e}nez-Arranz}}, \citenamefont {{Jorissen}}, \citenamefont {{Juaristi Campillo}}, \citenamefont {{Julbe}}, \citenamefont {{Karbevska}}, \citenamefont {{Kervella}}, \citenamefont {{Khanna}}, \citenamefont {{Kontizas}}, \citenamefont {{Kordopatis}}, \citenamefont {{Korn}}, \citenamefont {{K{\'o}sp{\'a}l}}, \citenamefont {{Kostrzewa-Rutkowska}}, \citenamefont {{Kruszy{\'n}ska}}, \citenamefont {{Kun}}, \citenamefont {{Laizeau}}, \citenamefont {{Lambert}}, \citenamefont {{Lanza}}, \citenamefont {{Lasne}},
  \citenamefont {{Le Campion}}, \citenamefont {{Lebreton}}, \citenamefont {{Lebzelter}}, \citenamefont {{Leccia}}, \citenamefont {{Leclerc}}, \citenamefont {{Lecoeur-Taibi}}, \citenamefont {{Liao}}, \citenamefont {{Licata}}, \citenamefont {{Lindstr{\o}m}}, \citenamefont {{Lister}}, \citenamefont {{Livanou}}, \citenamefont {{Lobel}}, \citenamefont {{Lorca}}, \citenamefont {{Loup}}, \citenamefont {{Madrero Pardo}}, \citenamefont {{Magdaleno Romeo}}, \citenamefont {{Managau}}, \citenamefont {{Mann}}, \citenamefont {{Manteiga}}, \citenamefont {{Marchant}}, \citenamefont {{Marconi}}, \citenamefont {{Marcos}}, \citenamefont {{Marcos Santos}}, \citenamefont {{Mar{\'\i}n Pina}}, \citenamefont {{Marinoni}}, \citenamefont {{Marocco}}, \citenamefont {{Marshall}}, \citenamefont {{Polo}}, \citenamefont {{Mart{\'\i}n-Fleitas}}, \citenamefont {{Marton}}, \citenamefont {{Mary}}, \citenamefont {{Masip}}, \citenamefont {{Massari}}, \citenamefont {{Mastrobuono-Battisti}}, \citenamefont {{Mazeh}}, \citenamefont {{McMillan}},
  \citenamefont {{Messina}}, \citenamefont {{Michalik}}, \citenamefont {{Millar}}, \citenamefont {{Mints}}, \citenamefont {{Molina}}, \citenamefont {{Molinaro}}, \citenamefont {{Moln{\'a}r}}, \citenamefont {{Monari}}, \citenamefont {{Mongui{\'o}}}, \citenamefont {{Montegriffo}}, \citenamefont {{Montero}}, \citenamefont {{Mor}}, \citenamefont {{Mora}}, \citenamefont {{Morbidelli}}, \citenamefont {{Morel}}, \citenamefont {{Morris}}, \citenamefont {{Muraveva}}, \citenamefont {{Murphy}}, \citenamefont {{Musella}}, \citenamefont {{Nagy}}, \citenamefont {{Noval}}, \citenamefont {{Oca{\~n}a}}, \citenamefont {{Ogden}}, \citenamefont {{Ordenovic}}, \citenamefont {{Osinde}}, \citenamefont {{Pagani}}, \citenamefont {{Pagano}}, \citenamefont {{Palaversa}}, \citenamefont {{Palicio}}, \citenamefont {{Pallas-Quintela}}, \citenamefont {{Panahi}}, \citenamefont {{Payne-Wardenaar}}, \citenamefont {{Pe{\~n}alosa Esteller}}, \citenamefont {{Penttil{\"a}}}, \citenamefont {{Pichon}}, \citenamefont {{Piersimoni}}, \citenamefont
  {{Pineau}}, \citenamefont {{Plachy}}, \citenamefont {{Plum}}, \citenamefont {{Poggio}}, \citenamefont {{Pr{\v{s}}a}}, \citenamefont {{Pulone}}, \citenamefont {{Racero}}, \citenamefont {{Ragaini}}, \citenamefont {{Rainer}}, \citenamefont {{Raiteri}}, \citenamefont {{Rambaux}}, \citenamefont {{Ramos}}, \citenamefont {{Ramos-Lerate}}, \citenamefont {{Re Fiorentin}}, \citenamefont {{Regibo}}, \citenamefont {{Richards}}, \citenamefont {{Rios Diaz}}, \citenamefont {{Ripepi}}, \citenamefont {{Riva}}, \citenamefont {{Rix}}, \citenamefont {{Rixon}}, \citenamefont {{Robichon}}, \citenamefont {{Robin}}, \citenamefont {{Robin}}, \citenamefont {{Roelens}}, \citenamefont {{Rogues}}, \citenamefont {{Rohrbasser}}, \citenamefont {{Romero-G{\'o}mez}}, \citenamefont {{Rowell}}, \citenamefont {{Royer}}, \citenamefont {{Ruz Mieres}}, \citenamefont {{Rybicki}}, \citenamefont {{Sadowski}}, \citenamefont {{S{\'a}ez N{\'u}{\~n}ez}}, \citenamefont {{Sagrist{\`a} Sell{\'e}s}}, \citenamefont {{Sahlmann}}, \citenamefont {{Salguero}},
  \citenamefont {{Samaras}}, \citenamefont {{Sanchez Gimenez}}, \citenamefont {{Sanna}}, \citenamefont {{Santove{\~n}a}}, \citenamefont {{Sarasso}}, \citenamefont {{Schultheis}}, \citenamefont {{Sciacca}}, \citenamefont {{Segol}}, \citenamefont {{Segovia}}, \citenamefont {{S{\'e}gransan}}, \citenamefont {{Semeux}}, \citenamefont {{Shahaf}}, \citenamefont {{Siddiqui}}, \citenamefont {{Siebert}}, \citenamefont {{Siltala}}, \citenamefont {{Silvelo}}, \citenamefont {{Slezak}}, \citenamefont {{Slezak}}, \citenamefont {{Smart}}, \citenamefont {{Snaith}}, \citenamefont {{Solano}}, \citenamefont {{Solitro}}, \citenamefont {{Souami}}, \citenamefont {{Souchay}}, \citenamefont {{Spagna}}, \citenamefont {{Spina}}, \citenamefont {{Spoto}}, \citenamefont {{Steele}}, \citenamefont {{Steidelm{\"u}ller}}, \citenamefont {{Stephenson}}, \citenamefont {{S{\"u}veges}}, \citenamefont {{Surdej}}, \citenamefont {{Szabados}}, \citenamefont {{Szegedi-Elek}}, \citenamefont {{Taris}}, \citenamefont {{Taylo}}, \citenamefont {{Teixeira}},
  \citenamefont {{Tolomei}}, \citenamefont {{Tonello}}, \citenamefont {{Torra}}, \citenamefont {{Torra}}, \citenamefont {{Torralba Elipe}}, \citenamefont {{Trabucchi}}, \citenamefont {{Tsounis}}, \citenamefont {{Turon}}, \citenamefont {{Ulla}}, \citenamefont {{Unger}}, \citenamefont {{Vaillant}}, \citenamefont {{van Dillen}}, \citenamefont {{van Reeven}}, \citenamefont {{Vanel}}, \citenamefont {{Vecchiato}}, \citenamefont {{Viala}}, \citenamefont {{Vicente}}, \citenamefont {{Voutsinas}}, \citenamefont {{Weiler}}, \citenamefont {{Wevers}}, \citenamefont {{Wyrzykowski}}, \citenamefont {{Yoldas}}, \citenamefont {{Yvard}}, \citenamefont {{Zhao}}, \citenamefont {{Zorec}}, \citenamefont {{Zucker}},\ and\ \citenamefont {{Zwitter}}}]{GaiaDR3}%
  \BibitemOpen
  \bibfield  {author} {\bibinfo {author} {\bibnamefont {{Gaia Collaboration}}}, \bibinfo {author} {\bibfnamefont {A.}~\bibnamefont {{Vallenari}}}, \bibinfo {author} {\bibfnamefont {A.~G.~A.}\ \bibnamefont {{Brown}}}, \bibinfo {author} {\bibfnamefont {T.}~\bibnamefont {{Prusti}}}, \bibinfo {author} {\bibfnamefont {J.~H.~J.}\ \bibnamefont {{de Bruijne}}}, \bibinfo {author} {\bibfnamefont {F.}~\bibnamefont {{Arenou}}}, \bibinfo {author} {\bibfnamefont {C.}~\bibnamefont {{Babusiaux}}}, \bibinfo {author} {\bibfnamefont {M.}~\bibnamefont {{Biermann}}}, \bibinfo {author} {\bibfnamefont {O.~L.}\ \bibnamefont {{Creevey}}}, \bibinfo {author} {\bibfnamefont {C.}~\bibnamefont {{Ducourant}}}, \bibinfo {author} {\bibfnamefont {D.~W.}\ \bibnamefont {{Evans}}}, \bibinfo {author} {\bibfnamefont {L.}~\bibnamefont {{Eyer}}}, \bibinfo {author} {\bibfnamefont {R.}~\bibnamefont {{Guerra}}}, \bibinfo {author} {\bibfnamefont {A.}~\bibnamefont {{Hutton}}}, \bibinfo {author} {\bibfnamefont {C.}~\bibnamefont {{Jordi}}}, \bibinfo
  {author} {\bibfnamefont {S.~A.}\ \bibnamefont {{Klioner}}}, \bibinfo {author} {\bibfnamefont {U.~L.}\ \bibnamefont {{Lammers}}}, \bibinfo {author} {\bibfnamefont {L.}~\bibnamefont {{Lindegren}}}, \bibinfo {author} {\bibfnamefont {X.}~\bibnamefont {{Luri}}}, \bibinfo {author} {\bibfnamefont {F.}~\bibnamefont {{Mignard}}}, \bibinfo {author} {\bibfnamefont {C.}~\bibnamefont {{Panem}}}, \bibinfo {author} {\bibfnamefont {D.}~\bibnamefont {{Pourbaix}}}, \bibinfo {author} {\bibfnamefont {S.}~\bibnamefont {{Randich}}}, \bibinfo {author} {\bibfnamefont {P.}~\bibnamefont {{Sartoretti}}}, \bibinfo {author} {\bibfnamefont {C.}~\bibnamefont {{Soubiran}}}, \bibinfo {author} {\bibfnamefont {P.}~\bibnamefont {{Tanga}}}, \bibinfo {author} {\bibfnamefont {N.~A.}\ \bibnamefont {{Walton}}}, \bibinfo {author} {\bibfnamefont {C.~A.~L.}\ \bibnamefont {{Bailer-Jones}}}, \bibinfo {author} {\bibfnamefont {U.}~\bibnamefont {{Bastian}}}, \bibinfo {author} {\bibfnamefont {R.}~\bibnamefont {{Drimmel}}}, \bibinfo {author} {\bibfnamefont
  {F.}~\bibnamefont {{Jansen}}}, \bibinfo {author} {\bibfnamefont {D.}~\bibnamefont {{Katz}}}, \bibinfo {author} {\bibfnamefont {M.~G.}\ \bibnamefont {{Lattanzi}}}, \bibinfo {author} {\bibfnamefont {F.}~\bibnamefont {{van Leeuwen}}}, \bibinfo {author} {\bibfnamefont {J.}~\bibnamefont {{Bakker}}}, \bibinfo {author} {\bibfnamefont {C.}~\bibnamefont {{Cacciari}}}, \bibinfo {author} {\bibfnamefont {J.}~\bibnamefont {{Casta{\~n}eda}}}, \bibinfo {author} {\bibfnamefont {F.}~\bibnamefont {{De Angeli}}}, \bibinfo {author} {\bibfnamefont {C.}~\bibnamefont {{Fabricius}}}, \bibinfo {author} {\bibfnamefont {M.}~\bibnamefont {{Fouesneau}}}, \bibinfo {author} {\bibfnamefont {Y.}~\bibnamefont {{Fr{\'e}mat}}}, \bibinfo {author} {\bibfnamefont {L.}~\bibnamefont {{Galluccio}}}, \bibinfo {author} {\bibfnamefont {A.}~\bibnamefont {{Guerrier}}}, \bibinfo {author} {\bibfnamefont {U.}~\bibnamefont {{Heiter}}}, \bibinfo {author} {\bibfnamefont {E.}~\bibnamefont {{Masana}}}, \bibinfo {author} {\bibfnamefont {R.}~\bibnamefont
  {{Messineo}}}, \bibinfo {author} {\bibfnamefont {N.}~\bibnamefont {{Mowlavi}}}, \bibinfo {author} {\bibfnamefont {C.}~\bibnamefont {{Nicolas}}}, \bibinfo {author} {\bibfnamefont {K.}~\bibnamefont {{Nienartowicz}}}, \bibinfo {author} {\bibfnamefont {F.}~\bibnamefont {{Pailler}}}, \bibinfo {author} {\bibfnamefont {P.}~\bibnamefont {{Panuzzo}}}, \bibinfo {author} {\bibfnamefont {F.}~\bibnamefont {{Riclet}}}, \bibinfo {author} {\bibfnamefont {W.}~\bibnamefont {{Roux}}}, \bibinfo {author} {\bibfnamefont {G.~M.}\ \bibnamefont {{Seabroke}}}, \bibinfo {author} {\bibfnamefont {R.}~\bibnamefont {{Sordo{\o}rcit}}}, \bibinfo {author} {\bibfnamefont {F.}~\bibnamefont {{Th{\'e}venin}}}, \bibinfo {author} {\bibfnamefont {G.}~\bibnamefont {{Gracia-Abril}}}, \bibinfo {author} {\bibfnamefont {J.}~\bibnamefont {{Portell}}}, \bibinfo {author} {\bibfnamefont {D.}~\bibnamefont {{Teyssier}}}, \bibinfo {author} {\bibfnamefont {M.}~\bibnamefont {{Altmann}}}, \bibinfo {author} {\bibfnamefont {R.}~\bibnamefont {{Andrae}}}, \bibinfo
  {author} {\bibfnamefont {M.}~\bibnamefont {{Audard}}}, \bibinfo {author} {\bibfnamefont {I.}~\bibnamefont {{Bellas-Velidis}}}, \bibinfo {author} {\bibfnamefont {K.}~\bibnamefont {{Benson}}}, \bibinfo {author} {\bibfnamefont {J.}~\bibnamefont {{Berthier}}}, \bibinfo {author} {\bibfnamefont {R.}~\bibnamefont {{Blomme}}}, \bibinfo {author} {\bibfnamefont {P.~W.}\ \bibnamefont {{Burgess}}}, \bibinfo {author} {\bibfnamefont {D.}~\bibnamefont {{Busonero}}}, \bibinfo {author} {\bibfnamefont {G.}~\bibnamefont {{Busso}}}, \bibinfo {author} {\bibfnamefont {H.}~\bibnamefont {{C{\'a}novas}}}, \bibinfo {author} {\bibfnamefont {B.}~\bibnamefont {{Carry}}}, \bibinfo {author} {\bibfnamefont {A.}~\bibnamefont {{Cellino}}}, \bibinfo {author} {\bibfnamefont {N.}~\bibnamefont {{Cheek}}}, \bibinfo {author} {\bibfnamefont {G.}~\bibnamefont {{Clementini}}}, \bibinfo {author} {\bibfnamefont {Y.}~\bibnamefont {{Damerdji}}}, \bibinfo {author} {\bibfnamefont {M.}~\bibnamefont {{Davidson}}}, \bibinfo {author} {\bibfnamefont
  {P.}~\bibnamefont {{de Teodoro}}}, \bibinfo {author} {\bibfnamefont {M.}~\bibnamefont {{Nu{\~n}ez Campos}}}, \bibinfo {author} {\bibfnamefont {L.}~\bibnamefont {{Delchambre}}}, \bibinfo {author} {\bibfnamefont {A.}~\bibnamefont {{Dell'Oro}}}, \bibinfo {author} {\bibfnamefont {P.}~\bibnamefont {{Esquej}}}, \bibinfo {author} {\bibfnamefont {J.}~\bibnamefont {{Fern{\'a}ndez-Hern{\'a}ndez}}}, \bibinfo {author} {\bibfnamefont {E.}~\bibnamefont {{Fraile}}}, \bibinfo {author} {\bibfnamefont {D.}~\bibnamefont {{Garabato}}}, \bibinfo {author} {\bibfnamefont {P.}~\bibnamefont {{Garc{\'\i}a-Lario}}}, \bibinfo {author} {\bibfnamefont {E.}~\bibnamefont {{Gosset}}}, \bibinfo {author} {\bibfnamefont {R.}~\bibnamefont {{Haigron}}}, \bibinfo {author} {\bibfnamefont {J.~L.}\ \bibnamefont {{Halbwachs}}}, \bibinfo {author} {\bibfnamefont {N.~C.}\ \bibnamefont {{Hambly}}}, \bibinfo {author} {\bibfnamefont {D.~L.}\ \bibnamefont {{Harrison}}}, \bibinfo {author} {\bibfnamefont {J.}~\bibnamefont {{Hern{\'a}ndez}}}, \bibinfo
  {author} {\bibfnamefont {D.}~\bibnamefont {{Hestroffer}}}, \bibinfo {author} {\bibfnamefont {S.~T.}\ \bibnamefont {{Hodgkin}}}, \bibinfo {author} {\bibfnamefont {B.}~\bibnamefont {{Holl}}}, \bibinfo {author} {\bibfnamefont {K.}~\bibnamefont {{Jan{\ss}en}}}, \bibinfo {author} {\bibfnamefont {G.}~\bibnamefont {{Jevardat de Fombelle}}}, \bibinfo {author} {\bibfnamefont {S.}~\bibnamefont {{Jordan}}}, \bibinfo {author} {\bibfnamefont {A.}~\bibnamefont {{Krone-Martins}}}, \bibinfo {author} {\bibfnamefont {A.~C.}\ \bibnamefont {{Lanzafame}}}, \bibinfo {author} {\bibfnamefont {W.}~\bibnamefont {{L{\"o}ffler}}}, \bibinfo {author} {\bibfnamefont {O.}~\bibnamefont {{Marchal}}}, \bibinfo {author} {\bibfnamefont {P.~M.}\ \bibnamefont {{Marrese}}}, \bibinfo {author} {\bibfnamefont {A.}~\bibnamefont {{Moitinho}}}, \bibinfo {author} {\bibfnamefont {K.}~\bibnamefont {{Muinonen}}}, \bibinfo {author} {\bibfnamefont {P.}~\bibnamefont {{Osborne}}}, \bibinfo {author} {\bibfnamefont {E.}~\bibnamefont {{Pancino}}}, \bibinfo
  {author} {\bibfnamefont {T.}~\bibnamefont {{Pauwels}}}, \bibinfo {author} {\bibfnamefont {A.}~\bibnamefont {{Recio-Blanco}}}, \bibinfo {author} {\bibfnamefont {C.}~\bibnamefont {{Reyl{\'e}}}}, \bibinfo {author} {\bibfnamefont {M.}~\bibnamefont {{Riello}}}, \bibinfo {author} {\bibfnamefont {L.}~\bibnamefont {{Rimoldini}}}, \bibinfo {author} {\bibfnamefont {T.}~\bibnamefont {{Roegiers}}}, \bibinfo {author} {\bibfnamefont {J.}~\bibnamefont {{Rybizki}}}, \bibinfo {author} {\bibfnamefont {L.~M.}\ \bibnamefont {{Sarro}}}, \bibinfo {author} {\bibfnamefont {C.}~\bibnamefont {{Siopis}}}, \bibinfo {author} {\bibfnamefont {M.}~\bibnamefont {{Smith}}}, \bibinfo {author} {\bibfnamefont {A.}~\bibnamefont {{Sozzetti}}}, \bibinfo {author} {\bibfnamefont {E.}~\bibnamefont {{Utrilla}}}, \bibinfo {author} {\bibfnamefont {M.}~\bibnamefont {{van Leeuwen}}}, \bibinfo {author} {\bibfnamefont {U.}~\bibnamefont {{Abbas}}}, \bibinfo {author} {\bibfnamefont {P.}~\bibnamefont {{{\'A}brah{\'a}m}}}, \bibinfo {author} {\bibfnamefont
  {A.}~\bibnamefont {{Abreu Aramburu}}}, \bibinfo {author} {\bibfnamefont {C.}~\bibnamefont {{Aerts}}}, \bibinfo {author} {\bibfnamefont {J.~J.}\ \bibnamefont {{Aguado}}}, \bibinfo {author} {\bibfnamefont {M.}~\bibnamefont {{Ajaj}}}, \bibinfo {author} {\bibfnamefont {F.}~\bibnamefont {{Aldea-Montero}}}, \bibinfo {author} {\bibfnamefont {G.}~\bibnamefont {{Altavilla}}}, \bibinfo {author} {\bibfnamefont {M.~A.}\ \bibnamefont {{{\'A}lvarez}}}, \bibinfo {author} {\bibfnamefont {J.}~\bibnamefont {{Alves}}}, \bibinfo {author} {\bibfnamefont {F.}~\bibnamefont {{Anders}}}, \bibinfo {author} {\bibfnamefont {R.~I.}\ \bibnamefont {{Anderson}}}, \bibinfo {author} {\bibfnamefont {E.}~\bibnamefont {{Anglada Varela}}}, \bibinfo {author} {\bibfnamefont {T.}~\bibnamefont {{Antoja}}}, \bibinfo {author} {\bibfnamefont {D.}~\bibnamefont {{Baines}}}, \bibinfo {author} {\bibfnamefont {S.~G.}\ \bibnamefont {{Baker}}}, \bibinfo {author} {\bibfnamefont {L.}~\bibnamefont {{Balaguer-N{\'u}{\~n}ez}}}, \bibinfo {author} {\bibfnamefont
  {E.}~\bibnamefont {{Balbinot}}}, \bibinfo {author} {\bibfnamefont {Z.}~\bibnamefont {{Balog}}}, \bibinfo {author} {\bibfnamefont {C.}~\bibnamefont {{Barache}}}, \bibinfo {author} {\bibfnamefont {D.}~\bibnamefont {{Barbato}}}, \bibinfo {author} {\bibfnamefont {M.}~\bibnamefont {{Barros}}}, \bibinfo {author} {\bibfnamefont {M.~A.}\ \bibnamefont {{Barstow}}}, \bibinfo {author} {\bibfnamefont {S.}~\bibnamefont {{Bartolom{\'e}}}}, \bibinfo {author} {\bibfnamefont {J.~L.}\ \bibnamefont {{Bassilana}}}, \bibinfo {author} {\bibfnamefont {N.}~\bibnamefont {{Bauchet}}}, \bibinfo {author} {\bibfnamefont {U.}~\bibnamefont {{Becciani}}}, \bibinfo {author} {\bibfnamefont {M.}~\bibnamefont {{Bellazzini}}}, \bibinfo {author} {\bibfnamefont {A.}~\bibnamefont {{Berihuete}}}, \bibinfo {author} {\bibfnamefont {M.}~\bibnamefont {{Bernet}}}, \bibinfo {author} {\bibfnamefont {S.}~\bibnamefont {{Bertone}}}, \bibinfo {author} {\bibfnamefont {L.}~\bibnamefont {{Bianchi}}}, \bibinfo {author} {\bibfnamefont {A.}~\bibnamefont
  {{Binnenfeld}}}, \bibinfo {author} {\bibfnamefont {S.}~\bibnamefont {{Blanco-Cuaresma}}}, \bibinfo {author} {\bibfnamefont {A.}~\bibnamefont {{Blazere}}}, \bibinfo {author} {\bibfnamefont {T.}~\bibnamefont {{Boch}}}, \bibinfo {author} {\bibfnamefont {A.}~\bibnamefont {{Bombrun}}}, \bibinfo {author} {\bibfnamefont {D.}~\bibnamefont {{Bossini}}}, \bibinfo {author} {\bibfnamefont {S.}~\bibnamefont {{Bouquillon}}}, \bibinfo {author} {\bibfnamefont {A.}~\bibnamefont {{Bragaglia}}}, \bibinfo {author} {\bibfnamefont {L.}~\bibnamefont {{Bramante}}}, \bibinfo {author} {\bibfnamefont {E.}~\bibnamefont {{Breedt}}}, \bibinfo {author} {\bibfnamefont {A.}~\bibnamefont {{Bressan}}}, \bibinfo {author} {\bibfnamefont {N.}~\bibnamefont {{Brouillet}}}, \bibinfo {author} {\bibfnamefont {E.}~\bibnamefont {{Brugaletta}}}, \bibinfo {author} {\bibfnamefont {B.}~\bibnamefont {{Bucciarelli}}}, \bibinfo {author} {\bibfnamefont {A.}~\bibnamefont {{Burlacu}}}, \bibinfo {author} {\bibfnamefont {A.~G.}\ \bibnamefont {{Butkevich}}},
  \bibinfo {author} {\bibfnamefont {R.}~\bibnamefont {{Buzzi}}}, \bibinfo {author} {\bibfnamefont {E.}~\bibnamefont {{Caffau}}}, \bibinfo {author} {\bibfnamefont {R.}~\bibnamefont {{Cancelliere}}}, \bibinfo {author} {\bibfnamefont {T.}~\bibnamefont {{Cantat-Gaudin}}}, \bibinfo {author} {\bibfnamefont {R.}~\bibnamefont {{Carballo}}}, \bibinfo {author} {\bibfnamefont {T.}~\bibnamefont {{Carlucci}}}, \bibinfo {author} {\bibfnamefont {M.~I.}\ \bibnamefont {{Carnerero}}}, \bibinfo {author} {\bibfnamefont {J.~M.}\ \bibnamefont {{Carrasco}}}, \bibinfo {author} {\bibfnamefont {L.}~\bibnamefont {{Casamiquela}}}, \bibinfo {author} {\bibfnamefont {M.}~\bibnamefont {{Castellani}}}, \bibinfo {author} {\bibfnamefont {A.}~\bibnamefont {{Castro-Ginard}}}, \bibinfo {author} {\bibfnamefont {L.}~\bibnamefont {{Chaoul}}}, \bibinfo {author} {\bibfnamefont {P.}~\bibnamefont {{Charlot}}}, \bibinfo {author} {\bibfnamefont {L.}~\bibnamefont {{Chemin}}}, \bibinfo {author} {\bibfnamefont {V.}~\bibnamefont {{Chiaramida}}}, \bibinfo
  {author} {\bibfnamefont {A.}~\bibnamefont {{Chiavassa}}}, \bibinfo {author} {\bibfnamefont {N.}~\bibnamefont {{Chornay}}}, \bibinfo {author} {\bibfnamefont {G.}~\bibnamefont {{Comoretto}}}, \bibinfo {author} {\bibfnamefont {G.}~\bibnamefont {{Contursi}}}, \bibinfo {author} {\bibfnamefont {W.~J.}\ \bibnamefont {{Cooper}}}, \bibinfo {author} {\bibfnamefont {T.}~\bibnamefont {{Cornez}}}, \bibinfo {author} {\bibfnamefont {S.}~\bibnamefont {{Cowell}}}, \bibinfo {author} {\bibfnamefont {F.}~\bibnamefont {{Crifo}}}, \bibinfo {author} {\bibfnamefont {M.}~\bibnamefont {{Cropper}}}, \bibinfo {author} {\bibfnamefont {M.}~\bibnamefont {{Crosta}}}, \bibinfo {author} {\bibfnamefont {C.}~\bibnamefont {{Crowley}}}, \bibinfo {author} {\bibfnamefont {C.}~\bibnamefont {{Dafonte}}}, \bibinfo {author} {\bibfnamefont {A.}~\bibnamefont {{Dapergolas}}}, \bibinfo {author} {\bibfnamefont {M.}~\bibnamefont {{David}}}, \bibinfo {author} {\bibfnamefont {P.}~\bibnamefont {{David}}}, \bibinfo {author} {\bibfnamefont {P.}~\bibnamefont
  {{de Laverny}}}, \bibinfo {author} {\bibfnamefont {F.}~\bibnamefont {{De Luise}}}, \bibinfo {author} {\bibfnamefont {R.}~\bibnamefont {{De March}}}, \bibinfo {author} {\bibfnamefont {J.}~\bibnamefont {{De Ridder}}}, \bibinfo {author} {\bibfnamefont {R.}~\bibnamefont {{de Souza}}}, \bibinfo {author} {\bibfnamefont {A.}~\bibnamefont {{de Torres}}}, \bibinfo {author} {\bibfnamefont {E.~F.}\ \bibnamefont {{del Peloso}}}, \bibinfo {author} {\bibfnamefont {E.}~\bibnamefont {{del Pozo}}}, \bibinfo {author} {\bibfnamefont {M.}~\bibnamefont {{Delbo}}}, \bibinfo {author} {\bibfnamefont {A.}~\bibnamefont {{Delgado}}}, \bibinfo {author} {\bibfnamefont {J.~B.}\ \bibnamefont {{Delisle}}}, \bibinfo {author} {\bibfnamefont {C.}~\bibnamefont {{Demouchy}}}, \bibinfo {author} {\bibfnamefont {T.~E.}\ \bibnamefont {{Dharmawardena}}}, \bibinfo {author} {\bibfnamefont {P.}~\bibnamefont {{Di Matteo}}}, \bibinfo {author} {\bibfnamefont {S.}~\bibnamefont {{Diakite}}}, \bibinfo {author} {\bibfnamefont {C.}~\bibnamefont {{Diener}}},
  \bibinfo {author} {\bibfnamefont {E.}~\bibnamefont {{Distefano}}}, \bibinfo {author} {\bibfnamefont {C.}~\bibnamefont {{Dolding}}}, \bibinfo {author} {\bibfnamefont {B.}~\bibnamefont {{Edvardsson}}}, \bibinfo {author} {\bibfnamefont {H.}~\bibnamefont {{Enke}}}, \bibinfo {author} {\bibfnamefont {C.}~\bibnamefont {{Fabre}}}, \bibinfo {author} {\bibfnamefont {M.}~\bibnamefont {{Fabrizio}}}, \bibinfo {author} {\bibfnamefont {S.}~\bibnamefont {{Faigler}}}, \bibinfo {author} {\bibfnamefont {G.}~\bibnamefont {{Fedorets}}}, \bibinfo {author} {\bibfnamefont {P.}~\bibnamefont {{Fernique}}}, \bibinfo {author} {\bibfnamefont {A.}~\bibnamefont {{Fienga}}}, \bibinfo {author} {\bibfnamefont {F.}~\bibnamefont {{Figueras}}}, \bibinfo {author} {\bibfnamefont {Y.}~\bibnamefont {{Fournier}}}, \bibinfo {author} {\bibfnamefont {C.}~\bibnamefont {{Fouron}}}, \bibinfo {author} {\bibfnamefont {F.}~\bibnamefont {{Fragkoudi}}}, \bibinfo {author} {\bibfnamefont {M.}~\bibnamefont {{Gai}}}, \bibinfo {author} {\bibfnamefont
  {A.}~\bibnamefont {{Garcia-Gutierrez}}}, \bibinfo {author} {\bibfnamefont {M.}~\bibnamefont {{Garcia-Reinaldos}}}, \bibinfo {author} {\bibfnamefont {M.}~\bibnamefont {{Garc{\'\i}a-Torres}}}, \bibinfo {author} {\bibfnamefont {A.}~\bibnamefont {{Garofalo}}}, \bibinfo {author} {\bibfnamefont {A.}~\bibnamefont {{Gavel}}}, \bibinfo {author} {\bibfnamefont {P.}~\bibnamefont {{Gavras}}}, \bibinfo {author} {\bibfnamefont {E.}~\bibnamefont {{Gerlach}}}, \bibinfo {author} {\bibfnamefont {R.}~\bibnamefont {{Geyer}}}, \bibinfo {author} {\bibfnamefont {P.}~\bibnamefont {{Giacobbe}}}, \bibinfo {author} {\bibfnamefont {G.}~\bibnamefont {{Gilmore}}}, \bibinfo {author} {\bibfnamefont {S.}~\bibnamefont {{Girona}}}, \bibinfo {author} {\bibfnamefont {G.}~\bibnamefont {{Giuffrida}}}, \bibinfo {author} {\bibfnamefont {R.}~\bibnamefont {{Gomel}}}, \bibinfo {author} {\bibfnamefont {A.}~\bibnamefont {{Gomez}}}, \bibinfo {author} {\bibfnamefont {J.}~\bibnamefont {{Gonz{\'a}lez-N{\'u}{\~n}ez}}}, \bibinfo {author} {\bibfnamefont
  {I.}~\bibnamefont {{Gonz{\'a}lez-Santamar{\'\i}a}}}, \bibinfo {author} {\bibfnamefont {J.~J.}\ \bibnamefont {{Gonz{\'a}lez-Vidal}}}, \bibinfo {author} {\bibfnamefont {M.}~\bibnamefont {{Granvik}}}, \bibinfo {author} {\bibfnamefont {P.}~\bibnamefont {{Guillout}}}, \bibinfo {author} {\bibfnamefont {J.}~\bibnamefont {{Guiraud}}}, \bibinfo {author} {\bibfnamefont {R.}~\bibnamefont {{Guti{\'e}rrez-S{\'a}nchez}}}, \bibinfo {author} {\bibfnamefont {L.~P.}\ \bibnamefont {{Guy}}}, \bibinfo {author} {\bibfnamefont {D.}~\bibnamefont {{Hatzidimitriou}}}, \bibinfo {author} {\bibfnamefont {M.}~\bibnamefont {{Hauser}}}, \bibinfo {author} {\bibfnamefont {M.}~\bibnamefont {{Haywood}}}, \bibinfo {author} {\bibfnamefont {A.}~\bibnamefont {{Helmer}}}, \bibinfo {author} {\bibfnamefont {A.}~\bibnamefont {{Helmi}}}, \bibinfo {author} {\bibfnamefont {M.~H.}\ \bibnamefont {{Sarmiento}}}, \bibinfo {author} {\bibfnamefont {S.~L.}\ \bibnamefont {{Hidalgo}}}, \bibinfo {author} {\bibfnamefont {T.}~\bibnamefont {{Hilger}}}, \bibinfo
  {author} {\bibfnamefont {N.}~\bibnamefont {{H{\l}adczuk}}}, \bibinfo {author} {\bibfnamefont {D.}~\bibnamefont {{Hobbs}}}, \bibinfo {author} {\bibfnamefont {G.}~\bibnamefont {{Holland}}}, \bibinfo {author} {\bibfnamefont {H.~E.}\ \bibnamefont {{Huckle}}}, \bibinfo {author} {\bibfnamefont {K.}~\bibnamefont {{Jardine}}}, \bibinfo {author} {\bibfnamefont {G.}~\bibnamefont {{Jasniewicz}}}, \bibinfo {author} {\bibfnamefont {A.}~\bibnamefont {{Jean-Antoine Piccolo}}}, \bibinfo {author} {\bibfnamefont {{\'O}.}~\bibnamefont {{Jim{\'e}nez-Arranz}}}, \bibinfo {author} {\bibfnamefont {A.}~\bibnamefont {{Jorissen}}}, \bibinfo {author} {\bibfnamefont {J.}~\bibnamefont {{Juaristi Campillo}}}, \bibinfo {author} {\bibfnamefont {F.}~\bibnamefont {{Julbe}}}, \bibinfo {author} {\bibfnamefont {L.}~\bibnamefont {{Karbevska}}}, \bibinfo {author} {\bibfnamefont {P.}~\bibnamefont {{Kervella}}}, \bibinfo {author} {\bibfnamefont {S.}~\bibnamefont {{Khanna}}}, \bibinfo {author} {\bibfnamefont {M.}~\bibnamefont {{Kontizas}}}, \bibinfo
  {author} {\bibfnamefont {G.}~\bibnamefont {{Kordopatis}}}, \bibinfo {author} {\bibfnamefont {A.~J.}\ \bibnamefont {{Korn}}}, \bibinfo {author} {\bibfnamefont {{\'A}.}~\bibnamefont {{K{\'o}sp{\'a}l}}}, \bibinfo {author} {\bibfnamefont {Z.}~\bibnamefont {{Kostrzewa-Rutkowska}}}, \bibinfo {author} {\bibfnamefont {K.}~\bibnamefont {{Kruszy{\'n}ska}}}, \bibinfo {author} {\bibfnamefont {M.}~\bibnamefont {{Kun}}}, \bibinfo {author} {\bibfnamefont {P.}~\bibnamefont {{Laizeau}}}, \bibinfo {author} {\bibfnamefont {S.}~\bibnamefont {{Lambert}}}, \bibinfo {author} {\bibfnamefont {A.~F.}\ \bibnamefont {{Lanza}}}, \bibinfo {author} {\bibfnamefont {Y.}~\bibnamefont {{Lasne}}}, \bibinfo {author} {\bibfnamefont {J.~F.}\ \bibnamefont {{Le Campion}}}, \bibinfo {author} {\bibfnamefont {Y.}~\bibnamefont {{Lebreton}}}, \bibinfo {author} {\bibfnamefont {T.}~\bibnamefont {{Lebzelter}}}, \bibinfo {author} {\bibfnamefont {S.}~\bibnamefont {{Leccia}}}, \bibinfo {author} {\bibfnamefont {N.}~\bibnamefont {{Leclerc}}}, \bibinfo {author}
  {\bibfnamefont {I.}~\bibnamefont {{Lecoeur-Taibi}}}, \bibinfo {author} {\bibfnamefont {S.}~\bibnamefont {{Liao}}}, \bibinfo {author} {\bibfnamefont {E.~L.}\ \bibnamefont {{Licata}}}, \bibinfo {author} {\bibfnamefont {H.~E.~P.}\ \bibnamefont {{Lindstr{\o}m}}}, \bibinfo {author} {\bibfnamefont {T.~A.}\ \bibnamefont {{Lister}}}, \bibinfo {author} {\bibfnamefont {E.}~\bibnamefont {{Livanou}}}, \bibinfo {author} {\bibfnamefont {A.}~\bibnamefont {{Lobel}}}, \bibinfo {author} {\bibfnamefont {A.}~\bibnamefont {{Lorca}}}, \bibinfo {author} {\bibfnamefont {C.}~\bibnamefont {{Loup}}}, \bibinfo {author} {\bibfnamefont {P.}~\bibnamefont {{Madrero Pardo}}}, \bibinfo {author} {\bibfnamefont {A.}~\bibnamefont {{Magdaleno Romeo}}}, \bibinfo {author} {\bibfnamefont {S.}~\bibnamefont {{Managau}}}, \bibinfo {author} {\bibfnamefont {R.~G.}\ \bibnamefont {{Mann}}}, \bibinfo {author} {\bibfnamefont {M.}~\bibnamefont {{Manteiga}}}, \bibinfo {author} {\bibfnamefont {J.~M.}\ \bibnamefont {{Marchant}}}, \bibinfo {author}
  {\bibfnamefont {M.}~\bibnamefont {{Marconi}}}, \bibinfo {author} {\bibfnamefont {J.}~\bibnamefont {{Marcos}}}, \bibinfo {author} {\bibfnamefont {M.~M.~S.}\ \bibnamefont {{Marcos Santos}}}, \bibinfo {author} {\bibfnamefont {D.}~\bibnamefont {{Mar{\'\i}n Pina}}}, \bibinfo {author} {\bibfnamefont {S.}~\bibnamefont {{Marinoni}}}, \bibinfo {author} {\bibfnamefont {F.}~\bibnamefont {{Marocco}}}, \bibinfo {author} {\bibfnamefont {D.~J.}\ \bibnamefont {{Marshall}}}, \bibinfo {author} {\bibfnamefont {L.~M.}\ \bibnamefont {{Polo}}}, \bibinfo {author} {\bibfnamefont {J.~M.}\ \bibnamefont {{Mart{\'\i}n-Fleitas}}}, \bibinfo {author} {\bibfnamefont {G.}~\bibnamefont {{Marton}}}, \bibinfo {author} {\bibfnamefont {N.}~\bibnamefont {{Mary}}}, \bibinfo {author} {\bibfnamefont {A.}~\bibnamefont {{Masip}}}, \bibinfo {author} {\bibfnamefont {D.}~\bibnamefont {{Massari}}}, \bibinfo {author} {\bibfnamefont {A.}~\bibnamefont {{Mastrobuono-Battisti}}}, \bibinfo {author} {\bibfnamefont {T.}~\bibnamefont {{Mazeh}}}, \bibinfo {author}
  {\bibfnamefont {P.~J.}\ \bibnamefont {{McMillan}}}, \bibinfo {author} {\bibfnamefont {S.}~\bibnamefont {{Messina}}}, \bibinfo {author} {\bibfnamefont {D.}~\bibnamefont {{Michalik}}}, \bibinfo {author} {\bibfnamefont {N.~R.}\ \bibnamefont {{Millar}}}, \bibinfo {author} {\bibfnamefont {A.}~\bibnamefont {{Mints}}}, \bibinfo {author} {\bibfnamefont {D.}~\bibnamefont {{Molina}}}, \bibinfo {author} {\bibfnamefont {R.}~\bibnamefont {{Molinaro}}}, \bibinfo {author} {\bibfnamefont {L.}~\bibnamefont {{Moln{\'a}r}}}, \bibinfo {author} {\bibfnamefont {G.}~\bibnamefont {{Monari}}}, \bibinfo {author} {\bibfnamefont {M.}~\bibnamefont {{Mongui{\'o}}}}, \bibinfo {author} {\bibfnamefont {P.}~\bibnamefont {{Montegriffo}}}, \bibinfo {author} {\bibfnamefont {A.}~\bibnamefont {{Montero}}}, \bibinfo {author} {\bibfnamefont {R.}~\bibnamefont {{Mor}}}, \bibinfo {author} {\bibfnamefont {A.}~\bibnamefont {{Mora}}}, \bibinfo {author} {\bibfnamefont {R.}~\bibnamefont {{Morbidelli}}}, \bibinfo {author} {\bibfnamefont {T.}~\bibnamefont
  {{Morel}}}, \bibinfo {author} {\bibfnamefont {D.}~\bibnamefont {{Morris}}}, \bibinfo {author} {\bibfnamefont {T.}~\bibnamefont {{Muraveva}}}, \bibinfo {author} {\bibfnamefont {C.~P.}\ \bibnamefont {{Murphy}}}, \bibinfo {author} {\bibfnamefont {I.}~\bibnamefont {{Musella}}}, \bibinfo {author} {\bibfnamefont {Z.}~\bibnamefont {{Nagy}}}, \bibinfo {author} {\bibfnamefont {L.}~\bibnamefont {{Noval}}}, \bibinfo {author} {\bibfnamefont {F.}~\bibnamefont {{Oca{\~n}a}}}, \bibinfo {author} {\bibfnamefont {A.}~\bibnamefont {{Ogden}}}, \bibinfo {author} {\bibfnamefont {C.}~\bibnamefont {{Ordenovic}}}, \bibinfo {author} {\bibfnamefont {J.~O.}\ \bibnamefont {{Osinde}}}, \bibinfo {author} {\bibfnamefont {C.}~\bibnamefont {{Pagani}}}, \bibinfo {author} {\bibfnamefont {I.}~\bibnamefont {{Pagano}}}, \bibinfo {author} {\bibfnamefont {L.}~\bibnamefont {{Palaversa}}}, \bibinfo {author} {\bibfnamefont {P.~A.}\ \bibnamefont {{Palicio}}}, \bibinfo {author} {\bibfnamefont {L.}~\bibnamefont {{Pallas-Quintela}}}, \bibinfo {author}
  {\bibfnamefont {A.}~\bibnamefont {{Panahi}}}, \bibinfo {author} {\bibfnamefont {S.}~\bibnamefont {{Payne-Wardenaar}}}, \bibinfo {author} {\bibfnamefont {X.}~\bibnamefont {{Pe{\~n}alosa Esteller}}}, \bibinfo {author} {\bibfnamefont {A.}~\bibnamefont {{Penttil{\"a}}}}, \bibinfo {author} {\bibfnamefont {B.}~\bibnamefont {{Pichon}}}, \bibinfo {author} {\bibfnamefont {A.~M.}\ \bibnamefont {{Piersimoni}}}, \bibinfo {author} {\bibfnamefont {F.~X.}\ \bibnamefont {{Pineau}}}, \bibinfo {author} {\bibfnamefont {E.}~\bibnamefont {{Plachy}}}, \bibinfo {author} {\bibfnamefont {G.}~\bibnamefont {{Plum}}}, \bibinfo {author} {\bibfnamefont {E.}~\bibnamefont {{Poggio}}}, \bibinfo {author} {\bibfnamefont {A.}~\bibnamefont {{Pr{\v{s}}a}}}, \bibinfo {author} {\bibfnamefont {L.}~\bibnamefont {{Pulone}}}, \bibinfo {author} {\bibfnamefont {E.}~\bibnamefont {{Racero}}}, \bibinfo {author} {\bibfnamefont {S.}~\bibnamefont {{Ragaini}}}, \bibinfo {author} {\bibfnamefont {M.}~\bibnamefont {{Rainer}}}, \bibinfo {author} {\bibfnamefont
  {C.~M.}\ \bibnamefont {{Raiteri}}}, \bibinfo {author} {\bibfnamefont {N.}~\bibnamefont {{Rambaux}}}, \bibinfo {author} {\bibfnamefont {P.}~\bibnamefont {{Ramos}}}, \bibinfo {author} {\bibfnamefont {M.}~\bibnamefont {{Ramos-Lerate}}}, \bibinfo {author} {\bibfnamefont {P.}~\bibnamefont {{Re Fiorentin}}}, \bibinfo {author} {\bibfnamefont {S.}~\bibnamefont {{Regibo}}}, \bibinfo {author} {\bibfnamefont {P.~J.}\ \bibnamefont {{Richards}}}, \bibinfo {author} {\bibfnamefont {C.}~\bibnamefont {{Rios Diaz}}}, \bibinfo {author} {\bibfnamefont {V.}~\bibnamefont {{Ripepi}}}, \bibinfo {author} {\bibfnamefont {A.}~\bibnamefont {{Riva}}}, \bibinfo {author} {\bibfnamefont {H.~W.}\ \bibnamefont {{Rix}}}, \bibinfo {author} {\bibfnamefont {G.}~\bibnamefont {{Rixon}}}, \bibinfo {author} {\bibfnamefont {N.}~\bibnamefont {{Robichon}}}, \bibinfo {author} {\bibfnamefont {A.~C.}\ \bibnamefont {{Robin}}}, \bibinfo {author} {\bibfnamefont {C.}~\bibnamefont {{Robin}}}, \bibinfo {author} {\bibfnamefont {M.}~\bibnamefont {{Roelens}}},
  \bibinfo {author} {\bibfnamefont {H.~R.~O.}\ \bibnamefont {{Rogues}}}, \bibinfo {author} {\bibfnamefont {L.}~\bibnamefont {{Rohrbasser}}}, \bibinfo {author} {\bibfnamefont {M.}~\bibnamefont {{Romero-G{\'o}mez}}}, \bibinfo {author} {\bibfnamefont {N.}~\bibnamefont {{Rowell}}}, \bibinfo {author} {\bibfnamefont {F.}~\bibnamefont {{Royer}}}, \bibinfo {author} {\bibfnamefont {D.}~\bibnamefont {{Ruz Mieres}}}, \bibinfo {author} {\bibfnamefont {K.~A.}\ \bibnamefont {{Rybicki}}}, \bibinfo {author} {\bibfnamefont {G.}~\bibnamefont {{Sadowski}}}, \bibinfo {author} {\bibfnamefont {A.}~\bibnamefont {{S{\'a}ez N{\'u}{\~n}ez}}}, \bibinfo {author} {\bibfnamefont {A.}~\bibnamefont {{Sagrist{\`a} Sell{\'e}s}}}, \bibinfo {author} {\bibfnamefont {J.}~\bibnamefont {{Sahlmann}}}, \bibinfo {author} {\bibfnamefont {E.}~\bibnamefont {{Salguero}}}, \bibinfo {author} {\bibfnamefont {N.}~\bibnamefont {{Samaras}}}, \bibinfo {author} {\bibfnamefont {V.}~\bibnamefont {{Sanchez Gimenez}}}, \bibinfo {author} {\bibfnamefont
  {N.}~\bibnamefont {{Sanna}}}, \bibinfo {author} {\bibfnamefont {R.}~\bibnamefont {{Santove{\~n}a}}}, \bibinfo {author} {\bibfnamefont {M.}~\bibnamefont {{Sarasso}}}, \bibinfo {author} {\bibfnamefont {M.}~\bibnamefont {{Schultheis}}}, \bibinfo {author} {\bibfnamefont {E.}~\bibnamefont {{Sciacca}}}, \bibinfo {author} {\bibfnamefont {M.}~\bibnamefont {{Segol}}}, \bibinfo {author} {\bibfnamefont {J.~C.}\ \bibnamefont {{Segovia}}}, \bibinfo {author} {\bibfnamefont {D.}~\bibnamefont {{S{\'e}gransan}}}, \bibinfo {author} {\bibfnamefont {D.}~\bibnamefont {{Semeux}}}, \bibinfo {author} {\bibfnamefont {S.}~\bibnamefont {{Shahaf}}}, \bibinfo {author} {\bibfnamefont {H.~I.}\ \bibnamefont {{Siddiqui}}}, \bibinfo {author} {\bibfnamefont {A.}~\bibnamefont {{Siebert}}}, \bibinfo {author} {\bibfnamefont {L.}~\bibnamefont {{Siltala}}}, \bibinfo {author} {\bibfnamefont {A.}~\bibnamefont {{Silvelo}}}, \bibinfo {author} {\bibfnamefont {E.}~\bibnamefont {{Slezak}}}, \bibinfo {author} {\bibfnamefont {I.}~\bibnamefont {{Slezak}}},
  \bibinfo {author} {\bibfnamefont {R.~L.}\ \bibnamefont {{Smart}}}, \bibinfo {author} {\bibfnamefont {O.~N.}\ \bibnamefont {{Snaith}}}, \bibinfo {author} {\bibfnamefont {E.}~\bibnamefont {{Solano}}}, \bibinfo {author} {\bibfnamefont {F.}~\bibnamefont {{Solitro}}}, \bibinfo {author} {\bibfnamefont {D.}~\bibnamefont {{Souami}}}, \bibinfo {author} {\bibfnamefont {J.}~\bibnamefont {{Souchay}}}, \bibinfo {author} {\bibfnamefont {A.}~\bibnamefont {{Spagna}}}, \bibinfo {author} {\bibfnamefont {L.}~\bibnamefont {{Spina}}}, \bibinfo {author} {\bibfnamefont {F.}~\bibnamefont {{Spoto}}}, \bibinfo {author} {\bibfnamefont {I.~A.}\ \bibnamefont {{Steele}}}, \bibinfo {author} {\bibfnamefont {H.}~\bibnamefont {{Steidelm{\"u}ller}}}, \bibinfo {author} {\bibfnamefont {C.~A.}\ \bibnamefont {{Stephenson}}}, \bibinfo {author} {\bibfnamefont {M.}~\bibnamefont {{S{\"u}veges}}}, \bibinfo {author} {\bibfnamefont {J.}~\bibnamefont {{Surdej}}}, \bibinfo {author} {\bibfnamefont {L.}~\bibnamefont {{Szabados}}}, \bibinfo {author}
  {\bibfnamefont {E.}~\bibnamefont {{Szegedi-Elek}}}, \bibinfo {author} {\bibfnamefont {F.}~\bibnamefont {{Taris}}}, \bibinfo {author} {\bibfnamefont {M.~B.}\ \bibnamefont {{Taylo}}}, \bibinfo {author} {\bibfnamefont {R.}~\bibnamefont {{Teixeira}}}, \bibinfo {author} {\bibfnamefont {L.}~\bibnamefont {{Tolomei}}}, \bibinfo {author} {\bibfnamefont {N.}~\bibnamefont {{Tonello}}}, \bibinfo {author} {\bibfnamefont {F.}~\bibnamefont {{Torra}}}, \bibinfo {author} {\bibfnamefont {J.}~\bibnamefont {{Torra}}}, \bibinfo {author} {\bibfnamefont {G.}~\bibnamefont {{Torralba Elipe}}}, \bibinfo {author} {\bibfnamefont {M.}~\bibnamefont {{Trabucchi}}}, \bibinfo {author} {\bibfnamefont {A.~T.}\ \bibnamefont {{Tsounis}}}, \bibinfo {author} {\bibfnamefont {C.}~\bibnamefont {{Turon}}}, \bibinfo {author} {\bibfnamefont {A.}~\bibnamefont {{Ulla}}}, \bibinfo {author} {\bibfnamefont {N.}~\bibnamefont {{Unger}}}, \bibinfo {author} {\bibfnamefont {M.~V.}\ \bibnamefont {{Vaillant}}}, \bibinfo {author} {\bibfnamefont {E.}~\bibnamefont
  {{van Dillen}}}, \bibinfo {author} {\bibfnamefont {W.}~\bibnamefont {{van Reeven}}}, \bibinfo {author} {\bibfnamefont {O.}~\bibnamefont {{Vanel}}}, \bibinfo {author} {\bibfnamefont {A.}~\bibnamefont {{Vecchiato}}}, \bibinfo {author} {\bibfnamefont {Y.}~\bibnamefont {{Viala}}}, \bibinfo {author} {\bibfnamefont {D.}~\bibnamefont {{Vicente}}}, \bibinfo {author} {\bibfnamefont {S.}~\bibnamefont {{Voutsinas}}}, \bibinfo {author} {\bibfnamefont {M.}~\bibnamefont {{Weiler}}}, \bibinfo {author} {\bibfnamefont {T.}~\bibnamefont {{Wevers}}}, \bibinfo {author} {\bibfnamefont {L.}~\bibnamefont {{Wyrzykowski}}}, \bibinfo {author} {\bibfnamefont {A.}~\bibnamefont {{Yoldas}}}, \bibinfo {author} {\bibfnamefont {P.}~\bibnamefont {{Yvard}}}, \bibinfo {author} {\bibfnamefont {H.}~\bibnamefont {{Zhao}}}, \bibinfo {author} {\bibfnamefont {J.}~\bibnamefont {{Zorec}}}, \bibinfo {author} {\bibfnamefont {S.}~\bibnamefont {{Zucker}}},\ and\ \bibinfo {author} {\bibfnamefont {T.}~\bibnamefont {{Zwitter}}},\ }\href
  {https://doi.org/10.48550/arXiv.2208.00211} {\bibfield  {journal} {\bibinfo  {journal} {arXiv e-prints}\ ,\ \bibinfo {eid} {arXiv:2208.00211}} (\bibinfo {year} {2022})},\ \Eprint {https://arxiv.org/abs/2208.00211} {arXiv:2208.00211 [astro-ph.GA]} \BibitemShut {NoStop}%
\bibitem [{\citenamefont {{Antoniadis}}(2021)}]{Antoniadis2021}%
  \BibitemOpen
  \bibfield  {author} {\bibinfo {author} {\bibfnamefont {J.}~\bibnamefont {{Antoniadis}}},\ }\href {https://doi.org/10.1093/mnras/staa3595} {\bibfield  {journal} {\bibinfo  {journal} {\mnras}\ }\textbf {\bibinfo {volume} {501}},\ \bibinfo {pages} {1116} (\bibinfo {year} {2021})},\ \Eprint {https://arxiv.org/abs/2011.08075} {arXiv:2011.08075 [astro-ph.HE]} \BibitemShut {NoStop}%
\bibitem [{\citenamefont {{Moran}}\ \emph {et~al.}(2023{\natexlab{b}})\citenamefont {{Moran}}, \citenamefont {{Mingarelli}}, \citenamefont {{Bedell}}, \citenamefont {{Good}},\ and\ \citenamefont {{Spergel}}}]{Moran2023b}%
  \BibitemOpen
  \bibfield  {author} {\bibinfo {author} {\bibfnamefont {A.}~\bibnamefont {{Moran}}}, \bibinfo {author} {\bibfnamefont {C.~M.~F.}\ \bibnamefont {{Mingarelli}}}, \bibinfo {author} {\bibfnamefont {M.}~\bibnamefont {{Bedell}}}, \bibinfo {author} {\bibfnamefont {D.}~\bibnamefont {{Good}}},\ and\ \bibinfo {author} {\bibfnamefont {D.~N.}\ \bibnamefont {{Spergel}}},\ }\href {https://doi.org/10.3847/1538-4357/acec75} {\bibfield  {journal} {\bibinfo  {journal} {\apj}\ }\textbf {\bibinfo {volume} {954}},\ \bibinfo {eid} {89} (\bibinfo {year} {2023}{\natexlab{b}})},\ \Eprint {https://arxiv.org/abs/2210.10816} {arXiv:2210.10816 [astro-ph.IM]} \BibitemShut {NoStop}%
\bibitem [{\citenamefont {{Draghis}}\ and\ \citenamefont {{Romani}}(2018)}]{DraghisRomani2018}%
  \BibitemOpen
  \bibfield  {author} {\bibinfo {author} {\bibfnamefont {P.}~\bibnamefont {{Draghis}}}\ and\ \bibinfo {author} {\bibfnamefont {R.~W.}\ \bibnamefont {{Romani}}},\ }\href {https://doi.org/10.3847/2041-8213/aad2db} {\bibfield  {journal} {\bibinfo  {journal} {\apjl}\ }\textbf {\bibinfo {volume} {862}},\ \bibinfo {eid} {L6} (\bibinfo {year} {2018})},\ \Eprint {https://arxiv.org/abs/1807.04249} {arXiv:1807.04249 [astro-ph.HE]} \BibitemShut {NoStop}%
\bibitem [{\citenamefont {{Chen}}(2021)}]{Chen2021}%
  \BibitemOpen
  \bibfield  {author} {\bibinfo {author} {\bibfnamefont {W.-C.}\ \bibnamefont {{Chen}}},\ }\href {https://doi.org/10.1093/mnras/staa3701} {\bibfield  {journal} {\bibinfo  {journal} {\mnras}\ }\textbf {\bibinfo {volume} {501}},\ \bibinfo {pages} {2327} (\bibinfo {year} {2021})},\ \Eprint {https://arxiv.org/abs/2011.13923} {arXiv:2011.13923 [astro-ph.HE]} \BibitemShut {NoStop}%
\bibitem [{\citenamefont {{Pletsch}}\ and\ \citenamefont {{Clark}}(2015)}]{PletschClark2015}%
  \BibitemOpen
  \bibfield  {author} {\bibinfo {author} {\bibfnamefont {H.~J.}\ \bibnamefont {{Pletsch}}}\ and\ \bibinfo {author} {\bibfnamefont {C.~J.}\ \bibnamefont {{Clark}}},\ }\href {https://doi.org/10.1088/0004-637X/807/1/18} {\bibfield  {journal} {\bibinfo  {journal} {\apj}\ }\textbf {\bibinfo {volume} {807}},\ \bibinfo {eid} {18} (\bibinfo {year} {2015})},\ \Eprint {https://arxiv.org/abs/1504.07466} {arXiv:1504.07466 [astro-ph.HE]} \BibitemShut {NoStop}%
\bibitem [{\citenamefont {{Reardon}}\ \emph {et~al.}(2016{\natexlab{b}})\citenamefont {{Reardon}}, \citenamefont {{Hobbs}}, \citenamefont {{Coles}}, \citenamefont {{Levin}}, \citenamefont {{Keith}}, \citenamefont {{Bailes}}, \citenamefont {{Bhat}}, \citenamefont {{Burke-Spolaor}}, \citenamefont {{Dai}}, \citenamefont {{Kerr}}, \citenamefont {{Lasky}}, \citenamefont {{Manchester}}, \citenamefont {{Os{\l}owski}}, \citenamefont {{Ravi}}, \citenamefont {{Shannon}}, \citenamefont {{van Straten}}, \citenamefont {{Toomey}}, \citenamefont {{Wang}}, \citenamefont {{Wen}}, \citenamefont {{You}},\ and\ \citenamefont {{Zhu}}}]{Reardonetal2016}%
  \BibitemOpen
  \bibfield  {author} {\bibinfo {author} {\bibfnamefont {D.~J.}\ \bibnamefont {{Reardon}}}, \bibinfo {author} {\bibfnamefont {G.}~\bibnamefont {{Hobbs}}}, \bibinfo {author} {\bibfnamefont {W.}~\bibnamefont {{Coles}}}, \bibinfo {author} {\bibfnamefont {Y.}~\bibnamefont {{Levin}}}, \bibinfo {author} {\bibfnamefont {M.~J.}\ \bibnamefont {{Keith}}}, \bibinfo {author} {\bibfnamefont {M.}~\bibnamefont {{Bailes}}}, \bibinfo {author} {\bibfnamefont {N.~D.~R.}\ \bibnamefont {{Bhat}}}, \bibinfo {author} {\bibfnamefont {S.}~\bibnamefont {{Burke-Spolaor}}}, \bibinfo {author} {\bibfnamefont {S.}~\bibnamefont {{Dai}}}, \bibinfo {author} {\bibfnamefont {M.}~\bibnamefont {{Kerr}}}, \bibinfo {author} {\bibfnamefont {P.~D.}\ \bibnamefont {{Lasky}}}, \bibinfo {author} {\bibfnamefont {R.~N.}\ \bibnamefont {{Manchester}}}, \bibinfo {author} {\bibfnamefont {S.}~\bibnamefont {{Os{\l}owski}}}, \bibinfo {author} {\bibfnamefont {V.}~\bibnamefont {{Ravi}}}, \bibinfo {author} {\bibfnamefont {R.~M.}\ \bibnamefont {{Shannon}}}, \bibinfo
  {author} {\bibfnamefont {W.}~\bibnamefont {{van Straten}}}, \bibinfo {author} {\bibfnamefont {L.}~\bibnamefont {{Toomey}}}, \bibinfo {author} {\bibfnamefont {J.}~\bibnamefont {{Wang}}}, \bibinfo {author} {\bibfnamefont {L.}~\bibnamefont {{Wen}}}, \bibinfo {author} {\bibfnamefont {X.~P.}\ \bibnamefont {{You}}},\ and\ \bibinfo {author} {\bibfnamefont {X.~J.}\ \bibnamefont {{Zhu}}},\ }\href {https://doi.org/10.1093/mnras/stv2395} {\bibfield  {journal} {\bibinfo  {journal} {\mnras}\ }\textbf {\bibinfo {volume} {455}},\ \bibinfo {pages} {1751} (\bibinfo {year} {2016}{\natexlab{b}})},\ \Eprint {https://arxiv.org/abs/1510.04434} {arXiv:1510.04434 [astro-ph.HE]} \BibitemShut {NoStop}%
\bibitem [{\citenamefont {Price-Whelan}(2017)}]{Gala}%
  \BibitemOpen
  \bibfield  {author} {\bibinfo {author} {\bibfnamefont {A.~M.}\ \bibnamefont {Price-Whelan}},\ }\bibfield  {journal} {\bibinfo  {journal} {The Journal of Open Source Software}\ }\textbf {\bibinfo {volume} {2}},\ \href {https://doi.org/10.21105/joss.00388} {10.21105/joss.00388} (\bibinfo {year} {2017})\BibitemShut {NoStop}%
\bibitem [{\citenamefont {{Gaia Collaboration}}\ \emph {et~al.}(2021)\citenamefont {{Gaia Collaboration}}, \citenamefont {{Klioner}}, \citenamefont {{Mignard}}, \citenamefont {{Lindegren}}, \citenamefont {{Bastian}}, \citenamefont {{McMillan}}, \citenamefont {{Hern{\'a}ndez}}, \citenamefont {{Hobbs}}, \citenamefont {{Ramos-Lerate}}, \citenamefont {{Biermann}}, \citenamefont {{Bombrun}}, \citenamefont {{de Torres}}, \citenamefont {{Gerlach}}, \citenamefont {{Geyer}}, \citenamefont {{Hilger}}, \citenamefont {{Lammers}}, \citenamefont {{Steidelm{\"u}ller}}, \citenamefont {{Stephenson}}, \citenamefont {{Brown}}, \citenamefont {{Vallenari}}, \citenamefont {{Prusti}}, \citenamefont {{de Bruijne}}, \citenamefont {{Babusiaux}}, \citenamefont {{Creevey}}, \citenamefont {{Evans}}, \citenamefont {{Eyer}}, \citenamefont {{Hutton}}, \citenamefont {{Jansen}}, \citenamefont {{Jordi}}, \citenamefont {{Luri}}, \citenamefont {{Panem}}, \citenamefont {{Pourbaix}}, \citenamefont {{Randich}}, \citenamefont {{Sartoretti}},
  \citenamefont {{Soubiran}}, \citenamefont {{Walton}}, \citenamefont {{Arenou}}, \citenamefont {{Bailer-Jones}}, \citenamefont {{Cropper}}, \citenamefont {{Drimmel}}, \citenamefont {{Katz}}, \citenamefont {{Lattanzi}}, \citenamefont {{van Leeuwen}}, \citenamefont {{Bakker}}, \citenamefont {{Casta{\~n}eda}}, \citenamefont {{De Angeli}}, \citenamefont {{Ducourant}}, \citenamefont {{Fabricius}}, \citenamefont {{Fouesneau}}, \citenamefont {{Fr{\'e}mat}}, \citenamefont {{Guerra}}, \citenamefont {{Guerrier}}, \citenamefont {{Guiraud}}, \citenamefont {{Jean-Antoine Piccolo}}, \citenamefont {{Masana}}, \citenamefont {{Messineo}}, \citenamefont {{Mowlavi}}, \citenamefont {{Nicolas}}, \citenamefont {{Nienartowicz}}, \citenamefont {{Pailler}}, \citenamefont {{Panuzzo}}, \citenamefont {{Riclet}}, \citenamefont {{Roux}}, \citenamefont {{Seabroke}}, \citenamefont {{Sordo}}, \citenamefont {{Tanga}}, \citenamefont {{Th{\'e}venin}}, \citenamefont {{Gracia-Abril}}, \citenamefont {{Portell}}, \citenamefont {{Teyssier}},
  \citenamefont {{Altmann}}, \citenamefont {{Andrae}}, \citenamefont {{Bellas-Velidis}}, \citenamefont {{Benson}}, \citenamefont {{Berthier}}, \citenamefont {{Blomme}}, \citenamefont {{Brugaletta}}, \citenamefont {{Burgess}}, \citenamefont {{Busso}}, \citenamefont {{Carry}}, \citenamefont {{Cellino}}, \citenamefont {{Cheek}}, \citenamefont {{Clementini}}, \citenamefont {{Damerdji}}, \citenamefont {{Davidson}}, \citenamefont {{Delchambre}}, \citenamefont {{Dell'Oro}}, \citenamefont {{Fern{\'a}ndez-Hern{\'a}ndez}}, \citenamefont {{Galluccio}}, \citenamefont {{Garc{\'\i}a-Lario}}, \citenamefont {{Garcia-Reinaldos}}, \citenamefont {{Gonz{\'a}lez-N{\'u}{\~n}ez}}, \citenamefont {{Gosset}}, \citenamefont {{Haigron}}, \citenamefont {{Halbwachs}}, \citenamefont {{Hambly}}, \citenamefont {{Harrison}}, \citenamefont {{Hatzidimitriou}}, \citenamefont {{Heiter}}, \citenamefont {{Hestroffer}}, \citenamefont {{Hodgkin}}, \citenamefont {{Holl}}, \citenamefont {{Jan{\ss}en}}, \citenamefont {{Jevardat de Fombelle}},
  \citenamefont {{Jordan}}, \citenamefont {{Krone-Martins}}, \citenamefont {{Lanzafame}}, \citenamefont {{L{\"o}ffler}}, \citenamefont {{Lorca}}, \citenamefont {{Manteiga}}, \citenamefont {{Marchal}}, \citenamefont {{Marrese}}, \citenamefont {{Moitinho}}, \citenamefont {{Mora}}, \citenamefont {{Muinonen}}, \citenamefont {{Osborne}}, \citenamefont {{Pancino}}, \citenamefont {{Pauwels}}, \citenamefont {{Recio-Blanco}}, \citenamefont {{Richards}}, \citenamefont {{Riello}}, \citenamefont {{Rimoldini}}, \citenamefont {{Robin}}, \citenamefont {{Roegiers}}, \citenamefont {{Rybizki}}, \citenamefont {{Sarro}}, \citenamefont {{Siopis}}, \citenamefont {{Smith}}, \citenamefont {{Sozzetti}}, \citenamefont {{Ulla}}, \citenamefont {{Utrilla}}, \citenamefont {{van Leeuwen}}, \citenamefont {{van Reeven}}, \citenamefont {{Abbas}}, \citenamefont {{Abreu Aramburu}}, \citenamefont {{Accart}}, \citenamefont {{Aerts}}, \citenamefont {{Aguado}}, \citenamefont {{Ajaj}}, \citenamefont {{Altavilla}}, \citenamefont {{{\'A}lvarez}},
  \citenamefont {{{\'A}lvarez Cid-Fuentes}}, \citenamefont {{Alves}}, \citenamefont {{Anderson}}, \citenamefont {{Anglada Varela}}, \citenamefont {{Antoja}}, \citenamefont {{Audard}}, \citenamefont {{Baines}}, \citenamefont {{Baker}}, \citenamefont {{Balaguer-N{\'u}{\~n}ez}}, \citenamefont {{Balbinot}}, \citenamefont {{Balog}}, \citenamefont {{Barache}}, \citenamefont {{Barbato}}, \citenamefont {{Barros}}, \citenamefont {{Barstow}}, \citenamefont {{Bartolom{\'e}}}, \citenamefont {{Bassilana}}, \citenamefont {{Bauchet}}, \citenamefont {{Baudesson-Stella}}, \citenamefont {{Becciani}}, \citenamefont {{Bellazzini}}, \citenamefont {{Bernet}}, \citenamefont {{Bertone}}, \citenamefont {{Bianchi}}, \citenamefont {{Blanco-Cuaresma}}, \citenamefont {{Boch}}, \citenamefont {{Bossini}}, \citenamefont {{Bouquillon}}, \citenamefont {{Bramante}}, \citenamefont {{Breedt}}, \citenamefont {{Bressan}}, \citenamefont {{Brouillet}}, \citenamefont {{Bucciarelli}}, \citenamefont {{Burlacu}}, \citenamefont {{Busonero}},
  \citenamefont {{Butkevich}}, \citenamefont {{Buzzi}}, \citenamefont {{Caffau}}, \citenamefont {{Cancelliere}}, \citenamefont {{C{\'a}novas}}, \citenamefont {{Cantat-Gaudin}}, \citenamefont {{Carballo}}, \citenamefont {{Carlucci}}, \citenamefont {{Carnerero}}, \citenamefont {{Carrasco}}, \citenamefont {{Casamiquela}}, \citenamefont {{Castellani}}, \citenamefont {{Castro-Ginard}}, \citenamefont {{Castro Sampol}}, \citenamefont {{Chaoul}}, \citenamefont {{Charlot}}, \citenamefont {{Chemin}}, \citenamefont {{Chiavassa}}, \citenamefont {{Comoretto}}, \citenamefont {{Cooper}}, \citenamefont {{Cornez}}, \citenamefont {{Cowell}}, \citenamefont {{Crifo}}, \citenamefont {{Crosta}}, \citenamefont {{Crowley}}, \citenamefont {{Dafonte}}, \citenamefont {{Dapergolas}}, \citenamefont {{David}}, \citenamefont {{David}}, \citenamefont {{de Laverny}}, \citenamefont {{De Luise}}, \citenamefont {{De March}}, \citenamefont {{De Ridder}}, \citenamefont {{de Souza}}, \citenamefont {{de Teodoro}}, \citenamefont {{del Peloso}},
  \citenamefont {{del Pozo}}, \citenamefont {{Delgado}}, \citenamefont {{Delgado}}, \citenamefont {{Delisle}}, \citenamefont {{Di Matteo}}, \citenamefont {{Diakite}}, \citenamefont {{Diener}}, \citenamefont {{Distefano}}, \citenamefont {{Dolding}}, \citenamefont {{Eappachen}}, \citenamefont {{Enke}}, \citenamefont {{Esquej}}, \citenamefont {{Fabre}}, \citenamefont {{Fabrizio}}, \citenamefont {{Faigler}}, \citenamefont {{Fedorets}}, \citenamefont {{Fernique}}, \citenamefont {{Fienga}}, \citenamefont {{Figueras}}, \citenamefont {{Fouron}}, \citenamefont {{Fragkoudi}}, \citenamefont {{Fraile}}, \citenamefont {{Franke}}, \citenamefont {{Gai}}, \citenamefont {{Garabato}}, \citenamefont {{Garcia-Gutierrez}}, \citenamefont {{Garc{\'\i}a-Torres}}, \citenamefont {{Garofalo}}, \citenamefont {{Gavras}}, \citenamefont {{Giacobbe}}, \citenamefont {{Gilmore}}, \citenamefont {{Girona}}, \citenamefont {{Giuffrida}}, \citenamefont {{Gomez}}, \citenamefont {{Gonzalez-Santamaria}}, \citenamefont {{Gonz{\'a}lez-Vidal}},
  \citenamefont {{Granvik}}, \citenamefont {{Guti{\'e}rrez-S{\'a}nchez}}, \citenamefont {{Guy}}, \citenamefont {{Hauser}}, \citenamefont {{Haywood}}, \citenamefont {{Helmi}}, \citenamefont {{Hidalgo}}, \citenamefont {{H{\l}adczuk}}, \citenamefont {{Holland}}, \citenamefont {{Huckle}}, \citenamefont {{Jasniewicz}}, \citenamefont {{Jonker}}, \citenamefont {{Juaristi Campillo}}, \citenamefont {{Julbe}}, \citenamefont {{Karbevska}}, \citenamefont {{Kervella}}, \citenamefont {{Khanna}}, \citenamefont {{Kochoska}}, \citenamefont {{Kordopatis}}, \citenamefont {{Korn}}, \citenamefont {{Kostrzewa-Rutkowska}}, \citenamefont {{Kruszy{\'n}ska}}, \citenamefont {{Lambert}}, \citenamefont {{Lanza}}, \citenamefont {{Lasne}}, \citenamefont {{Le Campion}}, \citenamefont {{Le Fustec}}, \citenamefont {{Lebreton}}, \citenamefont {{Lebzelter}}, \citenamefont {{Leccia}}, \citenamefont {{Leclerc}}, \citenamefont {{Lecoeur-Taibi}}, \citenamefont {{Liao}}, \citenamefont {{Licata}}, \citenamefont {{Lindstr{\o}m}}, \citenamefont
  {{Lister}}, \citenamefont {{Livanou}}, \citenamefont {{Lobel}}, \citenamefont {{Madrero Pardo}}, \citenamefont {{Managau}}, \citenamefont {{Mann}}, \citenamefont {{Marchant}}, \citenamefont {{Marconi}}, \citenamefont {{Marcos Santos}}, \citenamefont {{Marinoni}}, \citenamefont {{Marocco}}, \citenamefont {{Marshall}}, \citenamefont {{Martin Polo}}, \citenamefont {{Mart{\'\i}n-Fleitas}}, \citenamefont {{Masip}}, \citenamefont {{Massari}}, \citenamefont {{Mastrobuono-Battisti}}, \citenamefont {{Mazeh}}, \citenamefont {{Messina}}, \citenamefont {{Michalik}}, \citenamefont {{Millar}}, \citenamefont {{Mints}}, \citenamefont {{Molina}}, \citenamefont {{Molinaro}}, \citenamefont {{Moln{\'a}r}}, \citenamefont {{Montegriffo}}, \citenamefont {{Mor}}, \citenamefont {{Morbidelli}}, \citenamefont {{Morel}}, \citenamefont {{Morris}}, \citenamefont {{Mulone}}, \citenamefont {{Munoz}}, \citenamefont {{Muraveva}}, \citenamefont {{Murphy}}, \citenamefont {{Musella}}, \citenamefont {{Noval}}, \citenamefont {{Ord{\'e}novic}},
  \citenamefont {{Orr{\`u}}}, \citenamefont {{Osinde}}, \citenamefont {{Pagani}}, \citenamefont {{Pagano}}, \citenamefont {{Palaversa}}, \citenamefont {{Palicio}}, \citenamefont {{Panahi}}, \citenamefont {{Pawlak}}, \citenamefont {{Pe{\~n}alosa Esteller}}, \citenamefont {{Penttil{\"a}}}, \citenamefont {{Piersimoni}}, \citenamefont {{Pineau}}, \citenamefont {{Plachy}}, \citenamefont {{Plum}}, \citenamefont {{Poggio}}, \citenamefont {{Poretti}}, \citenamefont {{Poujoulet}}, \citenamefont {{Pr{\v{s}}a}}, \citenamefont {{Pulone}}, \citenamefont {{Racero}}, \citenamefont {{Ragaini}}, \citenamefont {{Rainer}}, \citenamefont {{Raiteri}}, \citenamefont {{Rambaux}}, \citenamefont {{Ramos}}, \citenamefont {{Re Fiorentin}}, \citenamefont {{Regibo}}, \citenamefont {{Reyl{\'e}}}, \citenamefont {{Ripepi}}, \citenamefont {{Riva}}, \citenamefont {{Rixon}}, \citenamefont {{Robichon}}, \citenamefont {{Robin}}, \citenamefont {{Roelens}}, \citenamefont {{Rohrbasser}}, \citenamefont {{Romero-G{\'o}mez}}, \citenamefont {{Rowell}},
  \citenamefont {{Royer}}, \citenamefont {{Rybicki}}, \citenamefont {{Sadowski}}, \citenamefont {{Sagrist{\`a} Sell{\'e}s}}, \citenamefont {{Sahlmann}}, \citenamefont {{Salgado}}, \citenamefont {{Salguero}}, \citenamefont {{Samaras}}, \citenamefont {{Sanchez Gimenez}}, \citenamefont {{Sanna}}, \citenamefont {{Santove{\~n}a}}, \citenamefont {{Sarasso}}, \citenamefont {{Schultheis}}, \citenamefont {{Sciacca}}, \citenamefont {{Segol}}, \citenamefont {{Segovia}}, \citenamefont {{S{\'e}gransan}}, \citenamefont {{Semeux}}, \citenamefont {{Siddiqui}}, \citenamefont {{Siebert}}, \citenamefont {{Siltala}}, \citenamefont {{Slezak}}, \citenamefont {{Smart}}, \citenamefont {{Solano}}, \citenamefont {{Solitro}}, \citenamefont {{Souami}}, \citenamefont {{Souchay}}, \citenamefont {{Spagna}}, \citenamefont {{Spoto}}, \citenamefont {{Steele}}, \citenamefont {{S{\"u}veges}}, \citenamefont {{Szabados}}, \citenamefont {{Szegedi-Elek}}, \citenamefont {{Taris}}, \citenamefont {{Tauran}}, \citenamefont {{Taylor}}, \citenamefont
  {{Teixeira}}, \citenamefont {{Thuillot}}, \citenamefont {{Tonello}}, \citenamefont {{Torra}}, \citenamefont {{Torra}}, \citenamefont {{Turon}}, \citenamefont {{Unger}}, \citenamefont {{Vaillant}}, \citenamefont {{van Dillen}}, \citenamefont {{Vanel}}, \citenamefont {{Vecchiato}}, \citenamefont {{Viala}}, \citenamefont {{Vicente}}, \citenamefont {{Voutsinas}}, \citenamefont {{Weiler}}, \citenamefont {{Wevers}}, \citenamefont {{Wyrzykowski}}, \citenamefont {{Yoldas}}, \citenamefont {{Yvard}}, \citenamefont {{Zhao}}, \citenamefont {{Zorec}}, \citenamefont {{Zucker}}, \citenamefont {{Zurbach}},\ and\ \citenamefont {{Zwitter}}}]{Klioner2021}%
  \BibitemOpen
  \bibfield  {author} {\bibinfo {author} {\bibnamefont {{Gaia Collaboration}}}, \bibinfo {author} {\bibfnamefont {S.~A.}\ \bibnamefont {{Klioner}}}, \bibinfo {author} {\bibfnamefont {F.}~\bibnamefont {{Mignard}}}, \bibinfo {author} {\bibfnamefont {L.}~\bibnamefont {{Lindegren}}}, \bibinfo {author} {\bibfnamefont {U.}~\bibnamefont {{Bastian}}}, \bibinfo {author} {\bibfnamefont {P.~J.}\ \bibnamefont {{McMillan}}}, \bibinfo {author} {\bibfnamefont {J.}~\bibnamefont {{Hern{\'a}ndez}}}, \bibinfo {author} {\bibfnamefont {D.}~\bibnamefont {{Hobbs}}}, \bibinfo {author} {\bibfnamefont {M.}~\bibnamefont {{Ramos-Lerate}}}, \bibinfo {author} {\bibfnamefont {M.}~\bibnamefont {{Biermann}}}, \bibinfo {author} {\bibfnamefont {A.}~\bibnamefont {{Bombrun}}}, \bibinfo {author} {\bibfnamefont {A.}~\bibnamefont {{de Torres}}}, \bibinfo {author} {\bibfnamefont {E.}~\bibnamefont {{Gerlach}}}, \bibinfo {author} {\bibfnamefont {R.}~\bibnamefont {{Geyer}}}, \bibinfo {author} {\bibfnamefont {T.}~\bibnamefont {{Hilger}}}, \bibinfo
  {author} {\bibfnamefont {U.}~\bibnamefont {{Lammers}}}, \bibinfo {author} {\bibfnamefont {H.}~\bibnamefont {{Steidelm{\"u}ller}}}, \bibinfo {author} {\bibfnamefont {C.~A.}\ \bibnamefont {{Stephenson}}}, \bibinfo {author} {\bibfnamefont {A.~G.~A.}\ \bibnamefont {{Brown}}}, \bibinfo {author} {\bibfnamefont {A.}~\bibnamefont {{Vallenari}}}, \bibinfo {author} {\bibfnamefont {T.}~\bibnamefont {{Prusti}}}, \bibinfo {author} {\bibfnamefont {J.~H.~J.}\ \bibnamefont {{de Bruijne}}}, \bibinfo {author} {\bibfnamefont {C.}~\bibnamefont {{Babusiaux}}}, \bibinfo {author} {\bibfnamefont {O.~L.}\ \bibnamefont {{Creevey}}}, \bibinfo {author} {\bibfnamefont {D.~W.}\ \bibnamefont {{Evans}}}, \bibinfo {author} {\bibfnamefont {L.}~\bibnamefont {{Eyer}}}, \bibinfo {author} {\bibfnamefont {A.}~\bibnamefont {{Hutton}}}, \bibinfo {author} {\bibfnamefont {F.}~\bibnamefont {{Jansen}}}, \bibinfo {author} {\bibfnamefont {C.}~\bibnamefont {{Jordi}}}, \bibinfo {author} {\bibfnamefont {X.}~\bibnamefont {{Luri}}}, \bibinfo {author}
  {\bibfnamefont {C.}~\bibnamefont {{Panem}}}, \bibinfo {author} {\bibfnamefont {D.}~\bibnamefont {{Pourbaix}}}, \bibinfo {author} {\bibfnamefont {S.}~\bibnamefont {{Randich}}}, \bibinfo {author} {\bibfnamefont {P.}~\bibnamefont {{Sartoretti}}}, \bibinfo {author} {\bibfnamefont {C.}~\bibnamefont {{Soubiran}}}, \bibinfo {author} {\bibfnamefont {N.~A.}\ \bibnamefont {{Walton}}}, \bibinfo {author} {\bibfnamefont {F.}~\bibnamefont {{Arenou}}}, \bibinfo {author} {\bibfnamefont {C.~A.~L.}\ \bibnamefont {{Bailer-Jones}}}, \bibinfo {author} {\bibfnamefont {M.}~\bibnamefont {{Cropper}}}, \bibinfo {author} {\bibfnamefont {R.}~\bibnamefont {{Drimmel}}}, \bibinfo {author} {\bibfnamefont {D.}~\bibnamefont {{Katz}}}, \bibinfo {author} {\bibfnamefont {M.~G.}\ \bibnamefont {{Lattanzi}}}, \bibinfo {author} {\bibfnamefont {F.}~\bibnamefont {{van Leeuwen}}}, \bibinfo {author} {\bibfnamefont {J.}~\bibnamefont {{Bakker}}}, \bibinfo {author} {\bibfnamefont {J.}~\bibnamefont {{Casta{\~n}eda}}}, \bibinfo {author} {\bibfnamefont
  {F.}~\bibnamefont {{De Angeli}}}, \bibinfo {author} {\bibfnamefont {C.}~\bibnamefont {{Ducourant}}}, \bibinfo {author} {\bibfnamefont {C.}~\bibnamefont {{Fabricius}}}, \bibinfo {author} {\bibfnamefont {M.}~\bibnamefont {{Fouesneau}}}, \bibinfo {author} {\bibfnamefont {Y.}~\bibnamefont {{Fr{\'e}mat}}}, \bibinfo {author} {\bibfnamefont {R.}~\bibnamefont {{Guerra}}}, \bibinfo {author} {\bibfnamefont {A.}~\bibnamefont {{Guerrier}}}, \bibinfo {author} {\bibfnamefont {J.}~\bibnamefont {{Guiraud}}}, \bibinfo {author} {\bibfnamefont {A.}~\bibnamefont {{Jean-Antoine Piccolo}}}, \bibinfo {author} {\bibfnamefont {E.}~\bibnamefont {{Masana}}}, \bibinfo {author} {\bibfnamefont {R.}~\bibnamefont {{Messineo}}}, \bibinfo {author} {\bibfnamefont {N.}~\bibnamefont {{Mowlavi}}}, \bibinfo {author} {\bibfnamefont {C.}~\bibnamefont {{Nicolas}}}, \bibinfo {author} {\bibfnamefont {K.}~\bibnamefont {{Nienartowicz}}}, \bibinfo {author} {\bibfnamefont {F.}~\bibnamefont {{Pailler}}}, \bibinfo {author} {\bibfnamefont {P.}~\bibnamefont
  {{Panuzzo}}}, \bibinfo {author} {\bibfnamefont {F.}~\bibnamefont {{Riclet}}}, \bibinfo {author} {\bibfnamefont {W.}~\bibnamefont {{Roux}}}, \bibinfo {author} {\bibfnamefont {G.~M.}\ \bibnamefont {{Seabroke}}}, \bibinfo {author} {\bibfnamefont {R.}~\bibnamefont {{Sordo}}}, \bibinfo {author} {\bibfnamefont {P.}~\bibnamefont {{Tanga}}}, \bibinfo {author} {\bibfnamefont {F.}~\bibnamefont {{Th{\'e}venin}}}, \bibinfo {author} {\bibfnamefont {G.}~\bibnamefont {{Gracia-Abril}}}, \bibinfo {author} {\bibfnamefont {J.}~\bibnamefont {{Portell}}}, \bibinfo {author} {\bibfnamefont {D.}~\bibnamefont {{Teyssier}}}, \bibinfo {author} {\bibfnamefont {M.}~\bibnamefont {{Altmann}}}, \bibinfo {author} {\bibfnamefont {R.}~\bibnamefont {{Andrae}}}, \bibinfo {author} {\bibfnamefont {I.}~\bibnamefont {{Bellas-Velidis}}}, \bibinfo {author} {\bibfnamefont {K.}~\bibnamefont {{Benson}}}, \bibinfo {author} {\bibfnamefont {J.}~\bibnamefont {{Berthier}}}, \bibinfo {author} {\bibfnamefont {R.}~\bibnamefont {{Blomme}}}, \bibinfo {author}
  {\bibfnamefont {E.}~\bibnamefont {{Brugaletta}}}, \bibinfo {author} {\bibfnamefont {P.~W.}\ \bibnamefont {{Burgess}}}, \bibinfo {author} {\bibfnamefont {G.}~\bibnamefont {{Busso}}}, \bibinfo {author} {\bibfnamefont {B.}~\bibnamefont {{Carry}}}, \bibinfo {author} {\bibfnamefont {A.}~\bibnamefont {{Cellino}}}, \bibinfo {author} {\bibfnamefont {N.}~\bibnamefont {{Cheek}}}, \bibinfo {author} {\bibfnamefont {G.}~\bibnamefont {{Clementini}}}, \bibinfo {author} {\bibfnamefont {Y.}~\bibnamefont {{Damerdji}}}, \bibinfo {author} {\bibfnamefont {M.}~\bibnamefont {{Davidson}}}, \bibinfo {author} {\bibfnamefont {L.}~\bibnamefont {{Delchambre}}}, \bibinfo {author} {\bibfnamefont {A.}~\bibnamefont {{Dell'Oro}}}, \bibinfo {author} {\bibfnamefont {J.}~\bibnamefont {{Fern{\'a}ndez-Hern{\'a}ndez}}}, \bibinfo {author} {\bibfnamefont {L.}~\bibnamefont {{Galluccio}}}, \bibinfo {author} {\bibfnamefont {P.}~\bibnamefont {{Garc{\'\i}a-Lario}}}, \bibinfo {author} {\bibfnamefont {M.}~\bibnamefont {{Garcia-Reinaldos}}}, \bibinfo
  {author} {\bibfnamefont {J.}~\bibnamefont {{Gonz{\'a}lez-N{\'u}{\~n}ez}}}, \bibinfo {author} {\bibfnamefont {E.}~\bibnamefont {{Gosset}}}, \bibinfo {author} {\bibfnamefont {R.}~\bibnamefont {{Haigron}}}, \bibinfo {author} {\bibfnamefont {J.~L.}\ \bibnamefont {{Halbwachs}}}, \bibinfo {author} {\bibfnamefont {N.~C.}\ \bibnamefont {{Hambly}}}, \bibinfo {author} {\bibfnamefont {D.~L.}\ \bibnamefont {{Harrison}}}, \bibinfo {author} {\bibfnamefont {D.}~\bibnamefont {{Hatzidimitriou}}}, \bibinfo {author} {\bibfnamefont {U.}~\bibnamefont {{Heiter}}}, \bibinfo {author} {\bibfnamefont {D.}~\bibnamefont {{Hestroffer}}}, \bibinfo {author} {\bibfnamefont {S.~T.}\ \bibnamefont {{Hodgkin}}}, \bibinfo {author} {\bibfnamefont {B.}~\bibnamefont {{Holl}}}, \bibinfo {author} {\bibfnamefont {K.}~\bibnamefont {{Jan{\ss}en}}}, \bibinfo {author} {\bibfnamefont {G.}~\bibnamefont {{Jevardat de Fombelle}}}, \bibinfo {author} {\bibfnamefont {S.}~\bibnamefont {{Jordan}}}, \bibinfo {author} {\bibfnamefont {A.}~\bibnamefont
  {{Krone-Martins}}}, \bibinfo {author} {\bibfnamefont {A.~C.}\ \bibnamefont {{Lanzafame}}}, \bibinfo {author} {\bibfnamefont {W.}~\bibnamefont {{L{\"o}ffler}}}, \bibinfo {author} {\bibfnamefont {A.}~\bibnamefont {{Lorca}}}, \bibinfo {author} {\bibfnamefont {M.}~\bibnamefont {{Manteiga}}}, \bibinfo {author} {\bibfnamefont {O.}~\bibnamefont {{Marchal}}}, \bibinfo {author} {\bibfnamefont {P.~M.}\ \bibnamefont {{Marrese}}}, \bibinfo {author} {\bibfnamefont {A.}~\bibnamefont {{Moitinho}}}, \bibinfo {author} {\bibfnamefont {A.}~\bibnamefont {{Mora}}}, \bibinfo {author} {\bibfnamefont {K.}~\bibnamefont {{Muinonen}}}, \bibinfo {author} {\bibfnamefont {P.}~\bibnamefont {{Osborne}}}, \bibinfo {author} {\bibfnamefont {E.}~\bibnamefont {{Pancino}}}, \bibinfo {author} {\bibfnamefont {T.}~\bibnamefont {{Pauwels}}}, \bibinfo {author} {\bibfnamefont {A.}~\bibnamefont {{Recio-Blanco}}}, \bibinfo {author} {\bibfnamefont {P.~J.}\ \bibnamefont {{Richards}}}, \bibinfo {author} {\bibfnamefont {M.}~\bibnamefont {{Riello}}},
  \bibinfo {author} {\bibfnamefont {L.}~\bibnamefont {{Rimoldini}}}, \bibinfo {author} {\bibfnamefont {A.~C.}\ \bibnamefont {{Robin}}}, \bibinfo {author} {\bibfnamefont {T.}~\bibnamefont {{Roegiers}}}, \bibinfo {author} {\bibfnamefont {J.}~\bibnamefont {{Rybizki}}}, \bibinfo {author} {\bibfnamefont {L.~M.}\ \bibnamefont {{Sarro}}}, \bibinfo {author} {\bibfnamefont {C.}~\bibnamefont {{Siopis}}}, \bibinfo {author} {\bibfnamefont {M.}~\bibnamefont {{Smith}}}, \bibinfo {author} {\bibfnamefont {A.}~\bibnamefont {{Sozzetti}}}, \bibinfo {author} {\bibfnamefont {A.}~\bibnamefont {{Ulla}}}, \bibinfo {author} {\bibfnamefont {E.}~\bibnamefont {{Utrilla}}}, \bibinfo {author} {\bibfnamefont {M.}~\bibnamefont {{van Leeuwen}}}, \bibinfo {author} {\bibfnamefont {W.}~\bibnamefont {{van Reeven}}}, \bibinfo {author} {\bibfnamefont {U.}~\bibnamefont {{Abbas}}}, \bibinfo {author} {\bibfnamefont {A.}~\bibnamefont {{Abreu Aramburu}}}, \bibinfo {author} {\bibfnamefont {S.}~\bibnamefont {{Accart}}}, \bibinfo {author} {\bibfnamefont
  {C.}~\bibnamefont {{Aerts}}}, \bibinfo {author} {\bibfnamefont {J.~J.}\ \bibnamefont {{Aguado}}}, \bibinfo {author} {\bibfnamefont {M.}~\bibnamefont {{Ajaj}}}, \bibinfo {author} {\bibfnamefont {G.}~\bibnamefont {{Altavilla}}}, \bibinfo {author} {\bibfnamefont {M.~A.}\ \bibnamefont {{{\'A}lvarez}}}, \bibinfo {author} {\bibfnamefont {J.}~\bibnamefont {{{\'A}lvarez Cid-Fuentes}}}, \bibinfo {author} {\bibfnamefont {J.}~\bibnamefont {{Alves}}}, \bibinfo {author} {\bibfnamefont {R.~I.}\ \bibnamefont {{Anderson}}}, \bibinfo {author} {\bibfnamefont {E.}~\bibnamefont {{Anglada Varela}}}, \bibinfo {author} {\bibfnamefont {T.}~\bibnamefont {{Antoja}}}, \bibinfo {author} {\bibfnamefont {M.}~\bibnamefont {{Audard}}}, \bibinfo {author} {\bibfnamefont {D.}~\bibnamefont {{Baines}}}, \bibinfo {author} {\bibfnamefont {S.~G.}\ \bibnamefont {{Baker}}}, \bibinfo {author} {\bibfnamefont {L.}~\bibnamefont {{Balaguer-N{\'u}{\~n}ez}}}, \bibinfo {author} {\bibfnamefont {E.}~\bibnamefont {{Balbinot}}}, \bibinfo {author}
  {\bibfnamefont {Z.}~\bibnamefont {{Balog}}}, \bibinfo {author} {\bibfnamefont {C.}~\bibnamefont {{Barache}}}, \bibinfo {author} {\bibfnamefont {D.}~\bibnamefont {{Barbato}}}, \bibinfo {author} {\bibfnamefont {M.}~\bibnamefont {{Barros}}}, \bibinfo {author} {\bibfnamefont {M.~A.}\ \bibnamefont {{Barstow}}}, \bibinfo {author} {\bibfnamefont {S.}~\bibnamefont {{Bartolom{\'e}}}}, \bibinfo {author} {\bibfnamefont {J.~L.}\ \bibnamefont {{Bassilana}}}, \bibinfo {author} {\bibfnamefont {N.}~\bibnamefont {{Bauchet}}}, \bibinfo {author} {\bibfnamefont {A.}~\bibnamefont {{Baudesson-Stella}}}, \bibinfo {author} {\bibfnamefont {U.}~\bibnamefont {{Becciani}}}, \bibinfo {author} {\bibfnamefont {M.}~\bibnamefont {{Bellazzini}}}, \bibinfo {author} {\bibfnamefont {M.}~\bibnamefont {{Bernet}}}, \bibinfo {author} {\bibfnamefont {S.}~\bibnamefont {{Bertone}}}, \bibinfo {author} {\bibfnamefont {L.}~\bibnamefont {{Bianchi}}}, \bibinfo {author} {\bibfnamefont {S.}~\bibnamefont {{Blanco-Cuaresma}}}, \bibinfo {author} {\bibfnamefont
  {T.}~\bibnamefont {{Boch}}}, \bibinfo {author} {\bibfnamefont {D.}~\bibnamefont {{Bossini}}}, \bibinfo {author} {\bibfnamefont {S.}~\bibnamefont {{Bouquillon}}}, \bibinfo {author} {\bibfnamefont {L.}~\bibnamefont {{Bramante}}}, \bibinfo {author} {\bibfnamefont {E.}~\bibnamefont {{Breedt}}}, \bibinfo {author} {\bibfnamefont {A.}~\bibnamefont {{Bressan}}}, \bibinfo {author} {\bibfnamefont {N.}~\bibnamefont {{Brouillet}}}, \bibinfo {author} {\bibfnamefont {B.}~\bibnamefont {{Bucciarelli}}}, \bibinfo {author} {\bibfnamefont {A.}~\bibnamefont {{Burlacu}}}, \bibinfo {author} {\bibfnamefont {D.}~\bibnamefont {{Busonero}}}, \bibinfo {author} {\bibfnamefont {A.~G.}\ \bibnamefont {{Butkevich}}}, \bibinfo {author} {\bibfnamefont {R.}~\bibnamefont {{Buzzi}}}, \bibinfo {author} {\bibfnamefont {E.}~\bibnamefont {{Caffau}}}, \bibinfo {author} {\bibfnamefont {R.}~\bibnamefont {{Cancelliere}}}, \bibinfo {author} {\bibfnamefont {H.}~\bibnamefont {{C{\'a}novas}}}, \bibinfo {author} {\bibfnamefont {T.}~\bibnamefont
  {{Cantat-Gaudin}}}, \bibinfo {author} {\bibfnamefont {R.}~\bibnamefont {{Carballo}}}, \bibinfo {author} {\bibfnamefont {T.}~\bibnamefont {{Carlucci}}}, \bibinfo {author} {\bibfnamefont {M.~I.}\ \bibnamefont {{Carnerero}}}, \bibinfo {author} {\bibfnamefont {J.~M.}\ \bibnamefont {{Carrasco}}}, \bibinfo {author} {\bibfnamefont {L.}~\bibnamefont {{Casamiquela}}}, \bibinfo {author} {\bibfnamefont {M.}~\bibnamefont {{Castellani}}}, \bibinfo {author} {\bibfnamefont {A.}~\bibnamefont {{Castro-Ginard}}}, \bibinfo {author} {\bibfnamefont {P.}~\bibnamefont {{Castro Sampol}}}, \bibinfo {author} {\bibfnamefont {L.}~\bibnamefont {{Chaoul}}}, \bibinfo {author} {\bibfnamefont {P.}~\bibnamefont {{Charlot}}}, \bibinfo {author} {\bibfnamefont {L.}~\bibnamefont {{Chemin}}}, \bibinfo {author} {\bibfnamefont {A.}~\bibnamefont {{Chiavassa}}}, \bibinfo {author} {\bibfnamefont {G.}~\bibnamefont {{Comoretto}}}, \bibinfo {author} {\bibfnamefont {W.~J.}\ \bibnamefont {{Cooper}}}, \bibinfo {author} {\bibfnamefont {T.}~\bibnamefont
  {{Cornez}}}, \bibinfo {author} {\bibfnamefont {S.}~\bibnamefont {{Cowell}}}, \bibinfo {author} {\bibfnamefont {F.}~\bibnamefont {{Crifo}}}, \bibinfo {author} {\bibfnamefont {M.}~\bibnamefont {{Crosta}}}, \bibinfo {author} {\bibfnamefont {C.}~\bibnamefont {{Crowley}}}, \bibinfo {author} {\bibfnamefont {C.}~\bibnamefont {{Dafonte}}}, \bibinfo {author} {\bibfnamefont {A.}~\bibnamefont {{Dapergolas}}}, \bibinfo {author} {\bibfnamefont {M.}~\bibnamefont {{David}}}, \bibinfo {author} {\bibfnamefont {P.}~\bibnamefont {{David}}}, \bibinfo {author} {\bibfnamefont {P.}~\bibnamefont {{de Laverny}}}, \bibinfo {author} {\bibfnamefont {F.}~\bibnamefont {{De Luise}}}, \bibinfo {author} {\bibfnamefont {R.}~\bibnamefont {{De March}}}, \bibinfo {author} {\bibfnamefont {J.}~\bibnamefont {{De Ridder}}}, \bibinfo {author} {\bibfnamefont {R.}~\bibnamefont {{de Souza}}}, \bibinfo {author} {\bibfnamefont {P.}~\bibnamefont {{de Teodoro}}}, \bibinfo {author} {\bibfnamefont {E.~F.}\ \bibnamefont {{del Peloso}}}, \bibinfo {author}
  {\bibfnamefont {E.}~\bibnamefont {{del Pozo}}}, \bibinfo {author} {\bibfnamefont {A.}~\bibnamefont {{Delgado}}}, \bibinfo {author} {\bibfnamefont {H.~E.}\ \bibnamefont {{Delgado}}}, \bibinfo {author} {\bibfnamefont {J.~B.}\ \bibnamefont {{Delisle}}}, \bibinfo {author} {\bibfnamefont {P.}~\bibnamefont {{Di Matteo}}}, \bibinfo {author} {\bibfnamefont {S.}~\bibnamefont {{Diakite}}}, \bibinfo {author} {\bibfnamefont {C.}~\bibnamefont {{Diener}}}, \bibinfo {author} {\bibfnamefont {E.}~\bibnamefont {{Distefano}}}, \bibinfo {author} {\bibfnamefont {C.}~\bibnamefont {{Dolding}}}, \bibinfo {author} {\bibfnamefont {D.}~\bibnamefont {{Eappachen}}}, \bibinfo {author} {\bibfnamefont {H.}~\bibnamefont {{Enke}}}, \bibinfo {author} {\bibfnamefont {P.}~\bibnamefont {{Esquej}}}, \bibinfo {author} {\bibfnamefont {C.}~\bibnamefont {{Fabre}}}, \bibinfo {author} {\bibfnamefont {M.}~\bibnamefont {{Fabrizio}}}, \bibinfo {author} {\bibfnamefont {S.}~\bibnamefont {{Faigler}}}, \bibinfo {author} {\bibfnamefont {G.}~\bibnamefont
  {{Fedorets}}}, \bibinfo {author} {\bibfnamefont {P.}~\bibnamefont {{Fernique}}}, \bibinfo {author} {\bibfnamefont {A.}~\bibnamefont {{Fienga}}}, \bibinfo {author} {\bibfnamefont {F.}~\bibnamefont {{Figueras}}}, \bibinfo {author} {\bibfnamefont {C.}~\bibnamefont {{Fouron}}}, \bibinfo {author} {\bibfnamefont {F.}~\bibnamefont {{Fragkoudi}}}, \bibinfo {author} {\bibfnamefont {E.}~\bibnamefont {{Fraile}}}, \bibinfo {author} {\bibfnamefont {F.}~\bibnamefont {{Franke}}}, \bibinfo {author} {\bibfnamefont {M.}~\bibnamefont {{Gai}}}, \bibinfo {author} {\bibfnamefont {D.}~\bibnamefont {{Garabato}}}, \bibinfo {author} {\bibfnamefont {A.}~\bibnamefont {{Garcia-Gutierrez}}}, \bibinfo {author} {\bibfnamefont {M.}~\bibnamefont {{Garc{\'\i}a-Torres}}}, \bibinfo {author} {\bibfnamefont {A.}~\bibnamefont {{Garofalo}}}, \bibinfo {author} {\bibfnamefont {P.}~\bibnamefont {{Gavras}}}, \bibinfo {author} {\bibfnamefont {P.}~\bibnamefont {{Giacobbe}}}, \bibinfo {author} {\bibfnamefont {G.}~\bibnamefont {{Gilmore}}}, \bibinfo
  {author} {\bibfnamefont {S.}~\bibnamefont {{Girona}}}, \bibinfo {author} {\bibfnamefont {G.}~\bibnamefont {{Giuffrida}}}, \bibinfo {author} {\bibfnamefont {A.}~\bibnamefont {{Gomez}}}, \bibinfo {author} {\bibfnamefont {I.}~\bibnamefont {{Gonzalez-Santamaria}}}, \bibinfo {author} {\bibfnamefont {J.~J.}\ \bibnamefont {{Gonz{\'a}lez-Vidal}}}, \bibinfo {author} {\bibfnamefont {M.}~\bibnamefont {{Granvik}}}, \bibinfo {author} {\bibfnamefont {R.}~\bibnamefont {{Guti{\'e}rrez-S{\'a}nchez}}}, \bibinfo {author} {\bibfnamefont {L.~P.}\ \bibnamefont {{Guy}}}, \bibinfo {author} {\bibfnamefont {M.}~\bibnamefont {{Hauser}}}, \bibinfo {author} {\bibfnamefont {M.}~\bibnamefont {{Haywood}}}, \bibinfo {author} {\bibfnamefont {A.}~\bibnamefont {{Helmi}}}, \bibinfo {author} {\bibfnamefont {S.~L.}\ \bibnamefont {{Hidalgo}}}, \bibinfo {author} {\bibfnamefont {N.}~\bibnamefont {{H{\l}adczuk}}}, \bibinfo {author} {\bibfnamefont {G.}~\bibnamefont {{Holland}}}, \bibinfo {author} {\bibfnamefont {H.~E.}\ \bibnamefont {{Huckle}}},
  \bibinfo {author} {\bibfnamefont {G.}~\bibnamefont {{Jasniewicz}}}, \bibinfo {author} {\bibfnamefont {P.~G.}\ \bibnamefont {{Jonker}}}, \bibinfo {author} {\bibfnamefont {J.}~\bibnamefont {{Juaristi Campillo}}}, \bibinfo {author} {\bibfnamefont {F.}~\bibnamefont {{Julbe}}}, \bibinfo {author} {\bibfnamefont {L.}~\bibnamefont {{Karbevska}}}, \bibinfo {author} {\bibfnamefont {P.}~\bibnamefont {{Kervella}}}, \bibinfo {author} {\bibfnamefont {S.}~\bibnamefont {{Khanna}}}, \bibinfo {author} {\bibfnamefont {A.}~\bibnamefont {{Kochoska}}}, \bibinfo {author} {\bibfnamefont {G.}~\bibnamefont {{Kordopatis}}}, \bibinfo {author} {\bibfnamefont {A.~J.}\ \bibnamefont {{Korn}}}, \bibinfo {author} {\bibfnamefont {Z.}~\bibnamefont {{Kostrzewa-Rutkowska}}}, \bibinfo {author} {\bibfnamefont {K.}~\bibnamefont {{Kruszy{\'n}ska}}}, \bibinfo {author} {\bibfnamefont {S.}~\bibnamefont {{Lambert}}}, \bibinfo {author} {\bibfnamefont {A.~F.}\ \bibnamefont {{Lanza}}}, \bibinfo {author} {\bibfnamefont {Y.}~\bibnamefont {{Lasne}}},
  \bibinfo {author} {\bibfnamefont {J.~F.}\ \bibnamefont {{Le Campion}}}, \bibinfo {author} {\bibfnamefont {Y.}~\bibnamefont {{Le Fustec}}}, \bibinfo {author} {\bibfnamefont {Y.}~\bibnamefont {{Lebreton}}}, \bibinfo {author} {\bibfnamefont {T.}~\bibnamefont {{Lebzelter}}}, \bibinfo {author} {\bibfnamefont {S.}~\bibnamefont {{Leccia}}}, \bibinfo {author} {\bibfnamefont {N.}~\bibnamefont {{Leclerc}}}, \bibinfo {author} {\bibfnamefont {I.}~\bibnamefont {{Lecoeur-Taibi}}}, \bibinfo {author} {\bibfnamefont {S.}~\bibnamefont {{Liao}}}, \bibinfo {author} {\bibfnamefont {E.}~\bibnamefont {{Licata}}}, \bibinfo {author} {\bibfnamefont {H.~E.~P.}\ \bibnamefont {{Lindstr{\o}m}}}, \bibinfo {author} {\bibfnamefont {T.~A.}\ \bibnamefont {{Lister}}}, \bibinfo {author} {\bibfnamefont {E.}~\bibnamefont {{Livanou}}}, \bibinfo {author} {\bibfnamefont {A.}~\bibnamefont {{Lobel}}}, \bibinfo {author} {\bibfnamefont {P.}~\bibnamefont {{Madrero Pardo}}}, \bibinfo {author} {\bibfnamefont {S.}~\bibnamefont {{Managau}}}, \bibinfo
  {author} {\bibfnamefont {R.~G.}\ \bibnamefont {{Mann}}}, \bibinfo {author} {\bibfnamefont {J.~M.}\ \bibnamefont {{Marchant}}}, \bibinfo {author} {\bibfnamefont {M.}~\bibnamefont {{Marconi}}}, \bibinfo {author} {\bibfnamefont {M.~M.~S.}\ \bibnamefont {{Marcos Santos}}}, \bibinfo {author} {\bibfnamefont {S.}~\bibnamefont {{Marinoni}}}, \bibinfo {author} {\bibfnamefont {F.}~\bibnamefont {{Marocco}}}, \bibinfo {author} {\bibfnamefont {D.~J.}\ \bibnamefont {{Marshall}}}, \bibinfo {author} {\bibfnamefont {L.}~\bibnamefont {{Martin Polo}}}, \bibinfo {author} {\bibfnamefont {J.~M.}\ \bibnamefont {{Mart{\'\i}n-Fleitas}}}, \bibinfo {author} {\bibfnamefont {A.}~\bibnamefont {{Masip}}}, \bibinfo {author} {\bibfnamefont {D.}~\bibnamefont {{Massari}}}, \bibinfo {author} {\bibfnamefont {A.}~\bibnamefont {{Mastrobuono-Battisti}}}, \bibinfo {author} {\bibfnamefont {T.}~\bibnamefont {{Mazeh}}}, \bibinfo {author} {\bibfnamefont {S.}~\bibnamefont {{Messina}}}, \bibinfo {author} {\bibfnamefont {D.}~\bibnamefont {{Michalik}}},
  \bibinfo {author} {\bibfnamefont {N.~R.}\ \bibnamefont {{Millar}}}, \bibinfo {author} {\bibfnamefont {A.}~\bibnamefont {{Mints}}}, \bibinfo {author} {\bibfnamefont {D.}~\bibnamefont {{Molina}}}, \bibinfo {author} {\bibfnamefont {R.}~\bibnamefont {{Molinaro}}}, \bibinfo {author} {\bibfnamefont {L.}~\bibnamefont {{Moln{\'a}r}}}, \bibinfo {author} {\bibfnamefont {P.}~\bibnamefont {{Montegriffo}}}, \bibinfo {author} {\bibfnamefont {R.}~\bibnamefont {{Mor}}}, \bibinfo {author} {\bibfnamefont {R.}~\bibnamefont {{Morbidelli}}}, \bibinfo {author} {\bibfnamefont {T.}~\bibnamefont {{Morel}}}, \bibinfo {author} {\bibfnamefont {D.}~\bibnamefont {{Morris}}}, \bibinfo {author} {\bibfnamefont {A.~F.}\ \bibnamefont {{Mulone}}}, \bibinfo {author} {\bibfnamefont {D.}~\bibnamefont {{Munoz}}}, \bibinfo {author} {\bibfnamefont {T.}~\bibnamefont {{Muraveva}}}, \bibinfo {author} {\bibfnamefont {C.~P.}\ \bibnamefont {{Murphy}}}, \bibinfo {author} {\bibfnamefont {I.}~\bibnamefont {{Musella}}}, \bibinfo {author} {\bibfnamefont
  {L.}~\bibnamefont {{Noval}}}, \bibinfo {author} {\bibfnamefont {C.}~\bibnamefont {{Ord{\'e}novic}}}, \bibinfo {author} {\bibfnamefont {G.}~\bibnamefont {{Orr{\`u}}}}, \bibinfo {author} {\bibfnamefont {J.}~\bibnamefont {{Osinde}}}, \bibinfo {author} {\bibfnamefont {C.}~\bibnamefont {{Pagani}}}, \bibinfo {author} {\bibfnamefont {I.}~\bibnamefont {{Pagano}}}, \bibinfo {author} {\bibfnamefont {L.}~\bibnamefont {{Palaversa}}}, \bibinfo {author} {\bibfnamefont {P.~A.}\ \bibnamefont {{Palicio}}}, \bibinfo {author} {\bibfnamefont {A.}~\bibnamefont {{Panahi}}}, \bibinfo {author} {\bibfnamefont {M.}~\bibnamefont {{Pawlak}}}, \bibinfo {author} {\bibfnamefont {X.}~\bibnamefont {{Pe{\~n}alosa Esteller}}}, \bibinfo {author} {\bibfnamefont {A.}~\bibnamefont {{Penttil{\"a}}}}, \bibinfo {author} {\bibfnamefont {A.~M.}\ \bibnamefont {{Piersimoni}}}, \bibinfo {author} {\bibfnamefont {F.~X.}\ \bibnamefont {{Pineau}}}, \bibinfo {author} {\bibfnamefont {E.}~\bibnamefont {{Plachy}}}, \bibinfo {author} {\bibfnamefont
  {G.}~\bibnamefont {{Plum}}}, \bibinfo {author} {\bibfnamefont {E.}~\bibnamefont {{Poggio}}}, \bibinfo {author} {\bibfnamefont {E.}~\bibnamefont {{Poretti}}}, \bibinfo {author} {\bibfnamefont {E.}~\bibnamefont {{Poujoulet}}}, \bibinfo {author} {\bibfnamefont {A.}~\bibnamefont {{Pr{\v{s}}a}}}, \bibinfo {author} {\bibfnamefont {L.}~\bibnamefont {{Pulone}}}, \bibinfo {author} {\bibfnamefont {E.}~\bibnamefont {{Racero}}}, \bibinfo {author} {\bibfnamefont {S.}~\bibnamefont {{Ragaini}}}, \bibinfo {author} {\bibfnamefont {M.}~\bibnamefont {{Rainer}}}, \bibinfo {author} {\bibfnamefont {C.~M.}\ \bibnamefont {{Raiteri}}}, \bibinfo {author} {\bibfnamefont {N.}~\bibnamefont {{Rambaux}}}, \bibinfo {author} {\bibfnamefont {P.}~\bibnamefont {{Ramos}}}, \bibinfo {author} {\bibfnamefont {P.}~\bibnamefont {{Re Fiorentin}}}, \bibinfo {author} {\bibfnamefont {S.}~\bibnamefont {{Regibo}}}, \bibinfo {author} {\bibfnamefont {C.}~\bibnamefont {{Reyl{\'e}}}}, \bibinfo {author} {\bibfnamefont {V.}~\bibnamefont {{Ripepi}}}, \bibinfo
  {author} {\bibfnamefont {A.}~\bibnamefont {{Riva}}}, \bibinfo {author} {\bibfnamefont {G.}~\bibnamefont {{Rixon}}}, \bibinfo {author} {\bibfnamefont {N.}~\bibnamefont {{Robichon}}}, \bibinfo {author} {\bibfnamefont {C.}~\bibnamefont {{Robin}}}, \bibinfo {author} {\bibfnamefont {M.}~\bibnamefont {{Roelens}}}, \bibinfo {author} {\bibfnamefont {L.}~\bibnamefont {{Rohrbasser}}}, \bibinfo {author} {\bibfnamefont {M.}~\bibnamefont {{Romero-G{\'o}mez}}}, \bibinfo {author} {\bibfnamefont {N.}~\bibnamefont {{Rowell}}}, \bibinfo {author} {\bibfnamefont {F.}~\bibnamefont {{Royer}}}, \bibinfo {author} {\bibfnamefont {K.~A.}\ \bibnamefont {{Rybicki}}}, \bibinfo {author} {\bibfnamefont {G.}~\bibnamefont {{Sadowski}}}, \bibinfo {author} {\bibfnamefont {A.}~\bibnamefont {{Sagrist{\`a} Sell{\'e}s}}}, \bibinfo {author} {\bibfnamefont {J.}~\bibnamefont {{Sahlmann}}}, \bibinfo {author} {\bibfnamefont {J.}~\bibnamefont {{Salgado}}}, \bibinfo {author} {\bibfnamefont {E.}~\bibnamefont {{Salguero}}}, \bibinfo {author}
  {\bibfnamefont {N.}~\bibnamefont {{Samaras}}}, \bibinfo {author} {\bibfnamefont {V.}~\bibnamefont {{Sanchez Gimenez}}}, \bibinfo {author} {\bibfnamefont {N.}~\bibnamefont {{Sanna}}}, \bibinfo {author} {\bibfnamefont {R.}~\bibnamefont {{Santove{\~n}a}}}, \bibinfo {author} {\bibfnamefont {M.}~\bibnamefont {{Sarasso}}}, \bibinfo {author} {\bibfnamefont {M.}~\bibnamefont {{Schultheis}}}, \bibinfo {author} {\bibfnamefont {E.}~\bibnamefont {{Sciacca}}}, \bibinfo {author} {\bibfnamefont {M.}~\bibnamefont {{Segol}}}, \bibinfo {author} {\bibfnamefont {J.~C.}\ \bibnamefont {{Segovia}}}, \bibinfo {author} {\bibfnamefont {D.}~\bibnamefont {{S{\'e}gransan}}}, \bibinfo {author} {\bibfnamefont {D.}~\bibnamefont {{Semeux}}}, \bibinfo {author} {\bibfnamefont {H.~I.}\ \bibnamefont {{Siddiqui}}}, \bibinfo {author} {\bibfnamefont {A.}~\bibnamefont {{Siebert}}}, \bibinfo {author} {\bibfnamefont {L.}~\bibnamefont {{Siltala}}}, \bibinfo {author} {\bibfnamefont {E.}~\bibnamefont {{Slezak}}}, \bibinfo {author} {\bibfnamefont
  {R.~L.}\ \bibnamefont {{Smart}}}, \bibinfo {author} {\bibfnamefont {E.}~\bibnamefont {{Solano}}}, \bibinfo {author} {\bibfnamefont {F.}~\bibnamefont {{Solitro}}}, \bibinfo {author} {\bibfnamefont {D.}~\bibnamefont {{Souami}}}, \bibinfo {author} {\bibfnamefont {J.}~\bibnamefont {{Souchay}}}, \bibinfo {author} {\bibfnamefont {A.}~\bibnamefont {{Spagna}}}, \bibinfo {author} {\bibfnamefont {F.}~\bibnamefont {{Spoto}}}, \bibinfo {author} {\bibfnamefont {I.~A.}\ \bibnamefont {{Steele}}}, \bibinfo {author} {\bibfnamefont {M.}~\bibnamefont {{S{\"u}veges}}}, \bibinfo {author} {\bibfnamefont {L.}~\bibnamefont {{Szabados}}}, \bibinfo {author} {\bibfnamefont {E.}~\bibnamefont {{Szegedi-Elek}}}, \bibinfo {author} {\bibfnamefont {F.}~\bibnamefont {{Taris}}}, \bibinfo {author} {\bibfnamefont {G.}~\bibnamefont {{Tauran}}}, \bibinfo {author} {\bibfnamefont {M.~B.}\ \bibnamefont {{Taylor}}}, \bibinfo {author} {\bibfnamefont {R.}~\bibnamefont {{Teixeira}}}, \bibinfo {author} {\bibfnamefont {W.}~\bibnamefont {{Thuillot}}},
  \bibinfo {author} {\bibfnamefont {N.}~\bibnamefont {{Tonello}}}, \bibinfo {author} {\bibfnamefont {F.}~\bibnamefont {{Torra}}}, \bibinfo {author} {\bibfnamefont {J.}~\bibnamefont {{Torra}}}, \bibinfo {author} {\bibfnamefont {C.}~\bibnamefont {{Turon}}}, \bibinfo {author} {\bibfnamefont {N.}~\bibnamefont {{Unger}}}, \bibinfo {author} {\bibfnamefont {M.}~\bibnamefont {{Vaillant}}}, \bibinfo {author} {\bibfnamefont {E.}~\bibnamefont {{van Dillen}}}, \bibinfo {author} {\bibfnamefont {O.}~\bibnamefont {{Vanel}}}, \bibinfo {author} {\bibfnamefont {A.}~\bibnamefont {{Vecchiato}}}, \bibinfo {author} {\bibfnamefont {Y.}~\bibnamefont {{Viala}}}, \bibinfo {author} {\bibfnamefont {D.}~\bibnamefont {{Vicente}}}, \bibinfo {author} {\bibfnamefont {S.}~\bibnamefont {{Voutsinas}}}, \bibinfo {author} {\bibfnamefont {M.}~\bibnamefont {{Weiler}}}, \bibinfo {author} {\bibfnamefont {T.}~\bibnamefont {{Wevers}}}, \bibinfo {author} {\bibfnamefont {{\L}.}~\bibnamefont {{Wyrzykowski}}}, \bibinfo {author} {\bibfnamefont
  {A.}~\bibnamefont {{Yoldas}}}, \bibinfo {author} {\bibfnamefont {P.}~\bibnamefont {{Yvard}}}, \bibinfo {author} {\bibfnamefont {H.}~\bibnamefont {{Zhao}}}, \bibinfo {author} {\bibfnamefont {J.}~\bibnamefont {{Zorec}}}, \bibinfo {author} {\bibfnamefont {S.}~\bibnamefont {{Zucker}}}, \bibinfo {author} {\bibfnamefont {C.}~\bibnamefont {{Zurbach}}},\ and\ \bibinfo {author} {\bibfnamefont {T.}~\bibnamefont {{Zwitter}}},\ }\href {https://doi.org/10.1051/0004-6361/202039734} {\bibfield  {journal} {\bibinfo  {journal} {\aap}\ }\textbf {\bibinfo {volume} {649}},\ \bibinfo {eid} {A9} (\bibinfo {year} {2021})},\ \Eprint {https://arxiv.org/abs/2012.02036} {arXiv:2012.02036 [astro-ph.GA]} \BibitemShut {NoStop}%
\bibitem [{\citenamefont {{Titov}}\ and\ \citenamefont {{Lambert}}(2013)}]{TitovLambert2013}%
  \BibitemOpen
  \bibfield  {author} {\bibinfo {author} {\bibfnamefont {O.}~\bibnamefont {{Titov}}}\ and\ \bibinfo {author} {\bibfnamefont {S.}~\bibnamefont {{Lambert}}},\ }\href {https://doi.org/10.1051/0004-6361/201321806} {\bibfield  {journal} {\bibinfo  {journal} {\aap}\ }\textbf {\bibinfo {volume} {559}},\ \bibinfo {eid} {A95} (\bibinfo {year} {2013})},\ \Eprint {https://arxiv.org/abs/1310.2723} {arXiv:1310.2723 [astro-ph.IM]} \BibitemShut {NoStop}%
\bibitem [{\citenamefont {{de la Fuente Marcos}}\ and\ \citenamefont {{de la Fuente Marcos}}(2016)}]{delaFuenteMarcos2016}%
  \BibitemOpen
  \bibfield  {author} {\bibinfo {author} {\bibfnamefont {C.}~\bibnamefont {{de la Fuente Marcos}}}\ and\ \bibinfo {author} {\bibfnamefont {R.}~\bibnamefont {{de la Fuente Marcos}}},\ }\href {https://doi.org/10.1093/mnras/stw1778} {\bibfield  {journal} {\bibinfo  {journal} {\mnras}\ }\textbf {\bibinfo {volume} {462}},\ \bibinfo {pages} {1972} (\bibinfo {year} {2016})},\ \Eprint {https://arxiv.org/abs/1607.05633} {arXiv:1607.05633 [astro-ph.EP]} \BibitemShut {NoStop}%
\bibitem [{\citenamefont {{Holman}}\ and\ \citenamefont {{Payne}}(2016)}]{HolmanPayne2016}%
  \BibitemOpen
  \bibfield  {author} {\bibinfo {author} {\bibfnamefont {M.~J.}\ \bibnamefont {{Holman}}}\ and\ \bibinfo {author} {\bibfnamefont {M.~J.}\ \bibnamefont {{Payne}}},\ }\href {https://doi.org/10.3847/0004-6256/152/4/94} {\bibfield  {journal} {\bibinfo  {journal} {\aj}\ }\textbf {\bibinfo {volume} {152}},\ \bibinfo {eid} {94} (\bibinfo {year} {2016})},\ \Eprint {https://arxiv.org/abs/1604.03180} {arXiv:1604.03180 [astro-ph.EP]} \BibitemShut {NoStop}%
\bibitem [{\citenamefont {{Batygin}}\ \emph {et~al.}(2019)\citenamefont {{Batygin}}, \citenamefont {{Adams}}, \citenamefont {{Brown}},\ and\ \citenamefont {{Becker}}}]{Batygin2019}%
  \BibitemOpen
  \bibfield  {author} {\bibinfo {author} {\bibfnamefont {K.}~\bibnamefont {{Batygin}}}, \bibinfo {author} {\bibfnamefont {F.~C.}\ \bibnamefont {{Adams}}}, \bibinfo {author} {\bibfnamefont {M.~E.}\ \bibnamefont {{Brown}}},\ and\ \bibinfo {author} {\bibfnamefont {J.~C.}\ \bibnamefont {{Becker}}},\ }\href {https://doi.org/10.1016/j.physrep.2019.01.009} {\bibfield  {journal} {\bibinfo  {journal} {\physrep}\ }\textbf {\bibinfo {volume} {805}},\ \bibinfo {pages} {1} (\bibinfo {year} {2019})},\ \Eprint {https://arxiv.org/abs/1902.10103} {arXiv:1902.10103 [astro-ph.EP]} \BibitemShut {NoStop}%
\bibitem [{\citenamefont {{Zakamska}}\ and\ \citenamefont {{Tremaine}}(2005)}]{Zakamska2005}%
  \BibitemOpen
  \bibfield  {author} {\bibinfo {author} {\bibfnamefont {N.~L.}\ \bibnamefont {{Zakamska}}}\ and\ \bibinfo {author} {\bibfnamefont {S.}~\bibnamefont {{Tremaine}}},\ }\href {https://doi.org/10.1086/444476} {\bibfield  {journal} {\bibinfo  {journal} {\aj}\ }\textbf {\bibinfo {volume} {130}},\ \bibinfo {pages} {1939} (\bibinfo {year} {2005})},\ \Eprint {https://arxiv.org/abs/astro-ph/0506548} {arXiv:astro-ph/0506548 [astro-ph]} \BibitemShut {NoStop}%
\bibitem [{\citenamefont {{Bovy}}(2015)}]{Galpy}%
  \BibitemOpen
  \bibfield  {author} {\bibinfo {author} {\bibfnamefont {J.}~\bibnamefont {{Bovy}}},\ }\href {https://doi.org/10.1088/0067-0049/216/2/29} {\bibfield  {journal} {\bibinfo  {journal} {\apjs}\ }\textbf {\bibinfo {volume} {216}},\ \bibinfo {eid} {29} (\bibinfo {year} {2015})},\ \Eprint {https://arxiv.org/abs/1412.3451} {arXiv:1412.3451 [astro-ph.GA]} \BibitemShut {NoStop}%
\bibitem [{\citenamefont {{Ivezi{\'c}}}\ \emph {et~al.}(2014)\citenamefont {{Ivezi{\'c}}}, \citenamefont {{Connolly}}, \citenamefont {{VanderPlas}},\ and\ \citenamefont {{Gray}}}]{Ivezic2014}%
  \BibitemOpen
  \bibfield  {author} {\bibinfo {author} {\bibfnamefont {{\v{Z}}.}~\bibnamefont {{Ivezi{\'c}}}}, \bibinfo {author} {\bibfnamefont {A.~J.}\ \bibnamefont {{Connolly}}}, \bibinfo {author} {\bibfnamefont {J.~T.}\ \bibnamefont {{VanderPlas}}},\ and\ \bibinfo {author} {\bibfnamefont {A.}~\bibnamefont {{Gray}}},\ }\href {https://doi.org/10.1515/9781400848911} {\emph {\bibinfo {title} {{Statistics, Data Mining, and Machine Learning in Astronomy: A Practical Python Guide for the Analysis of Survey Data}}}}\ (\bibinfo {year} {2014})\BibitemShut {NoStop}%
\bibitem [{\citenamefont {Virtanen}\ \emph {et~al.}(2020)\citenamefont {Virtanen}, \citenamefont {Gommers}, \citenamefont {Oliphant}, \citenamefont {Haberland}, \citenamefont {Reddy}, \citenamefont {Cournapeau}, \citenamefont {Burovski}, \citenamefont {Peterson}, \citenamefont {Weckesser}, \citenamefont {Bright}, \citenamefont {{van der Walt}}, \citenamefont {Brett}, \citenamefont {Wilson}, \citenamefont {Millman}, \citenamefont {Mayorov}, \citenamefont {Nelson}, \citenamefont {Jones}, \citenamefont {Kern}, \citenamefont {Larson}, \citenamefont {Carey}, \citenamefont {Polat}, \citenamefont {Feng}, \citenamefont {Moore}, \citenamefont {{VanderPlas}}, \citenamefont {Laxalde}, \citenamefont {Perktold}, \citenamefont {Cimrman}, \citenamefont {Henriksen}, \citenamefont {Quintero}, \citenamefont {Harris}, \citenamefont {Archibald}, \citenamefont {Ribeiro}, \citenamefont {Pedregosa}, \citenamefont {{van Mulbregt}},\ and\ \citenamefont {{SciPy 1.0 Contributors}}}]{scipy}%
  \BibitemOpen
  \bibfield  {author} {\bibinfo {author} {\bibfnamefont {P.}~\bibnamefont {Virtanen}}, \bibinfo {author} {\bibfnamefont {R.}~\bibnamefont {Gommers}}, \bibinfo {author} {\bibfnamefont {T.~E.}\ \bibnamefont {Oliphant}}, \bibinfo {author} {\bibfnamefont {M.}~\bibnamefont {Haberland}}, \bibinfo {author} {\bibfnamefont {T.}~\bibnamefont {Reddy}}, \bibinfo {author} {\bibfnamefont {D.}~\bibnamefont {Cournapeau}}, \bibinfo {author} {\bibfnamefont {E.}~\bibnamefont {Burovski}}, \bibinfo {author} {\bibfnamefont {P.}~\bibnamefont {Peterson}}, \bibinfo {author} {\bibfnamefont {W.}~\bibnamefont {Weckesser}}, \bibinfo {author} {\bibfnamefont {J.}~\bibnamefont {Bright}}, \bibinfo {author} {\bibfnamefont {S.~J.}\ \bibnamefont {{van der Walt}}}, \bibinfo {author} {\bibfnamefont {M.}~\bibnamefont {Brett}}, \bibinfo {author} {\bibfnamefont {J.}~\bibnamefont {Wilson}}, \bibinfo {author} {\bibfnamefont {K.~J.}\ \bibnamefont {Millman}}, \bibinfo {author} {\bibfnamefont {N.}~\bibnamefont {Mayorov}}, \bibinfo {author} {\bibfnamefont
  {A.~R.~J.}\ \bibnamefont {Nelson}}, \bibinfo {author} {\bibfnamefont {E.}~\bibnamefont {Jones}}, \bibinfo {author} {\bibfnamefont {R.}~\bibnamefont {Kern}}, \bibinfo {author} {\bibfnamefont {E.}~\bibnamefont {Larson}}, \bibinfo {author} {\bibfnamefont {C.~J.}\ \bibnamefont {Carey}}, \bibinfo {author} {\bibfnamefont {{\.I}.}~\bibnamefont {Polat}}, \bibinfo {author} {\bibfnamefont {Y.}~\bibnamefont {Feng}}, \bibinfo {author} {\bibfnamefont {E.~W.}\ \bibnamefont {Moore}}, \bibinfo {author} {\bibfnamefont {J.}~\bibnamefont {{VanderPlas}}}, \bibinfo {author} {\bibfnamefont {D.}~\bibnamefont {Laxalde}}, \bibinfo {author} {\bibfnamefont {J.}~\bibnamefont {Perktold}}, \bibinfo {author} {\bibfnamefont {R.}~\bibnamefont {Cimrman}}, \bibinfo {author} {\bibfnamefont {I.}~\bibnamefont {Henriksen}}, \bibinfo {author} {\bibfnamefont {E.~A.}\ \bibnamefont {Quintero}}, \bibinfo {author} {\bibfnamefont {C.~R.}\ \bibnamefont {Harris}}, \bibinfo {author} {\bibfnamefont {A.~M.}\ \bibnamefont {Archibald}}, \bibinfo {author}
  {\bibfnamefont {A.~H.}\ \bibnamefont {Ribeiro}}, \bibinfo {author} {\bibfnamefont {F.}~\bibnamefont {Pedregosa}}, \bibinfo {author} {\bibfnamefont {P.}~\bibnamefont {{van Mulbregt}}},\ and\ \bibinfo {author} {\bibnamefont {{SciPy 1.0 Contributors}}},\ }\href {https://doi.org/10.1038/s41592-019-0686-2} {\bibfield  {journal} {\bibinfo  {journal} {Nature Methods}\ }\textbf {\bibinfo {volume} {17}},\ \bibinfo {pages} {261} (\bibinfo {year} {2020})}\BibitemShut {NoStop}%
\bibitem [{\citenamefont {Kass}\ and\ \citenamefont {Raftery}(1995)}]{KassRaftery1995}%
  \BibitemOpen
  \bibfield  {author} {\bibinfo {author} {\bibfnamefont {R.~E.}\ \bibnamefont {Kass}}\ and\ \bibinfo {author} {\bibfnamefont {A.~E.}\ \bibnamefont {Raftery}},\ }\href {https://doi.org/10.1080/01621459.1995.10476572} {\bibfield  {journal} {\bibinfo  {journal} {Journal of the American Statistical Association}\ }\textbf {\bibinfo {volume} {90}},\ \bibinfo {pages} {773} (\bibinfo {year} {1995})},\ \Eprint {https://arxiv.org/abs/https://www.tandfonline.com/doi/pdf/10.1080/01621459.1995.10476572} {https://www.tandfonline.com/doi/pdf/10.1080/01621459.1995.10476572} \BibitemShut {NoStop}%
\bibitem [{\citenamefont {{Miyamoto}}\ and\ \citenamefont {{Nagai}}(1975)}]{MiyamotoNagai1975}%
  \BibitemOpen
  \bibfield  {author} {\bibinfo {author} {\bibfnamefont {M.}~\bibnamefont {{Miyamoto}}}\ and\ \bibinfo {author} {\bibfnamefont {R.}~\bibnamefont {{Nagai}}},\ }\href@noop {} {\bibfield  {journal} {\bibinfo  {journal} {\pasj}\ }\textbf {\bibinfo {volume} {27}},\ \bibinfo {pages} {533} (\bibinfo {year} {1975})}\BibitemShut {NoStop}%
\bibitem [{\citenamefont {{Sharma}}\ \emph {et~al.}(2014)\citenamefont {{Sharma}}, \citenamefont {{Bland-Hawthorn}}, \citenamefont {{Binney}}, \citenamefont {{Freeman}}, \citenamefont {{Steinmetz}}, \citenamefont {{Boeche}}, \citenamefont {{Bienaym{\'e}}}, \citenamefont {{Gibson}}, \citenamefont {{Gilmore}}, \citenamefont {{Grebel}}, \citenamefont {{Helmi}}, \citenamefont {{Kordopatis}}, \citenamefont {{Munari}}, \citenamefont {{Navarro}}, \citenamefont {{Parker}}, \citenamefont {{Reid}}, \citenamefont {{Seabroke}}, \citenamefont {{Siebert}}, \citenamefont {{Watson}}, \citenamefont {{Williams}}, \citenamefont {{Wyse}},\ and\ \citenamefont {{Zwitter}}}]{Sharma2014}%
  \BibitemOpen
  \bibfield  {author} {\bibinfo {author} {\bibfnamefont {S.}~\bibnamefont {{Sharma}}}, \bibinfo {author} {\bibfnamefont {J.}~\bibnamefont {{Bland-Hawthorn}}}, \bibinfo {author} {\bibfnamefont {J.}~\bibnamefont {{Binney}}}, \bibinfo {author} {\bibfnamefont {K.~C.}\ \bibnamefont {{Freeman}}}, \bibinfo {author} {\bibfnamefont {M.}~\bibnamefont {{Steinmetz}}}, \bibinfo {author} {\bibfnamefont {C.}~\bibnamefont {{Boeche}}}, \bibinfo {author} {\bibfnamefont {O.}~\bibnamefont {{Bienaym{\'e}}}}, \bibinfo {author} {\bibfnamefont {B.~K.}\ \bibnamefont {{Gibson}}}, \bibinfo {author} {\bibfnamefont {G.~F.}\ \bibnamefont {{Gilmore}}}, \bibinfo {author} {\bibfnamefont {E.~K.}\ \bibnamefont {{Grebel}}}, \bibinfo {author} {\bibfnamefont {A.}~\bibnamefont {{Helmi}}}, \bibinfo {author} {\bibfnamefont {G.}~\bibnamefont {{Kordopatis}}}, \bibinfo {author} {\bibfnamefont {U.}~\bibnamefont {{Munari}}}, \bibinfo {author} {\bibfnamefont {J.~F.}\ \bibnamefont {{Navarro}}}, \bibinfo {author} {\bibfnamefont {Q.~A.}\ \bibnamefont
  {{Parker}}}, \bibinfo {author} {\bibfnamefont {W.~A.}\ \bibnamefont {{Reid}}}, \bibinfo {author} {\bibfnamefont {G.~M.}\ \bibnamefont {{Seabroke}}}, \bibinfo {author} {\bibfnamefont {A.}~\bibnamefont {{Siebert}}}, \bibinfo {author} {\bibfnamefont {F.}~\bibnamefont {{Watson}}}, \bibinfo {author} {\bibfnamefont {M.~E.~K.}\ \bibnamefont {{Williams}}}, \bibinfo {author} {\bibfnamefont {R.~F.~G.}\ \bibnamefont {{Wyse}}},\ and\ \bibinfo {author} {\bibfnamefont {T.}~\bibnamefont {{Zwitter}}},\ }\href {https://doi.org/10.1088/0004-637X/793/1/51} {\bibfield  {journal} {\bibinfo  {journal} {\apj}\ }\textbf {\bibinfo {volume} {793}},\ \bibinfo {eid} {51} (\bibinfo {year} {2014})},\ \Eprint {https://arxiv.org/abs/1405.7435} {arXiv:1405.7435 [astro-ph.GA]} \BibitemShut {NoStop}%
\bibitem [{\citenamefont {{Law}}\ and\ \citenamefont {{Majewski}}(2010)}]{LawMajewski2010}%
  \BibitemOpen
  \bibfield  {author} {\bibinfo {author} {\bibfnamefont {D.~R.}\ \bibnamefont {{Law}}}\ and\ \bibinfo {author} {\bibfnamefont {S.~R.}\ \bibnamefont {{Majewski}}},\ }\href {https://doi.org/10.1088/0004-637X/714/1/229} {\bibfield  {journal} {\bibinfo  {journal} {\apj}\ }\textbf {\bibinfo {volume} {714}},\ \bibinfo {pages} {229} (\bibinfo {year} {2010})},\ \Eprint {https://arxiv.org/abs/1003.1132} {arXiv:1003.1132 [astro-ph.GA]} \BibitemShut {NoStop}%
\bibitem [{\citenamefont {{Nice}}\ and\ \citenamefont {{Taylor}}(1995)}]{NiceTaylor1995}%
  \BibitemOpen
  \bibfield  {author} {\bibinfo {author} {\bibfnamefont {D.~J.}\ \bibnamefont {{Nice}}}\ and\ \bibinfo {author} {\bibfnamefont {J.~H.}\ \bibnamefont {{Taylor}}},\ }\href {https://doi.org/10.1086/175367} {\bibfield  {journal} {\bibinfo  {journal} {\apj}\ }\textbf {\bibinfo {volume} {441}},\ \bibinfo {pages} {429} (\bibinfo {year} {1995})}\BibitemShut {NoStop}%
\bibitem [{\citenamefont {{Lazaridis}}\ \emph {et~al.}(2009)\citenamefont {{Lazaridis}}, \citenamefont {{Wex}}, \citenamefont {{Jessner}}, \citenamefont {{Kramer}}, \citenamefont {{Stappers}}, \citenamefont {{Janssen}}, \citenamefont {{Desvignes}}, \citenamefont {{Purver}}, \citenamefont {{Cognard}}, \citenamefont {{Theureau}}, \citenamefont {{Lyne}}, \citenamefont {{Jordan}},\ and\ \citenamefont {{Zensus}}}]{Lazaridis2009}%
  \BibitemOpen
  \bibfield  {author} {\bibinfo {author} {\bibfnamefont {K.}~\bibnamefont {{Lazaridis}}}, \bibinfo {author} {\bibfnamefont {N.}~\bibnamefont {{Wex}}}, \bibinfo {author} {\bibfnamefont {A.}~\bibnamefont {{Jessner}}}, \bibinfo {author} {\bibfnamefont {M.}~\bibnamefont {{Kramer}}}, \bibinfo {author} {\bibfnamefont {B.~W.}\ \bibnamefont {{Stappers}}}, \bibinfo {author} {\bibfnamefont {G.~H.}\ \bibnamefont {{Janssen}}}, \bibinfo {author} {\bibfnamefont {G.}~\bibnamefont {{Desvignes}}}, \bibinfo {author} {\bibfnamefont {M.~B.}\ \bibnamefont {{Purver}}}, \bibinfo {author} {\bibfnamefont {I.}~\bibnamefont {{Cognard}}}, \bibinfo {author} {\bibfnamefont {G.}~\bibnamefont {{Theureau}}}, \bibinfo {author} {\bibfnamefont {A.~G.}\ \bibnamefont {{Lyne}}}, \bibinfo {author} {\bibfnamefont {C.~A.}\ \bibnamefont {{Jordan}}},\ and\ \bibinfo {author} {\bibfnamefont {J.~A.}\ \bibnamefont {{Zensus}}},\ }\href {https://doi.org/10.1111/j.1365-2966.2009.15481.x} {\bibfield  {journal} {\bibinfo  {journal} {\mnras}\ }\textbf {\bibinfo
  {volume} {400}},\ \bibinfo {pages} {805} (\bibinfo {year} {2009})},\ \Eprint {https://arxiv.org/abs/0908.0285} {arXiv:0908.0285 [astro-ph.GA]} \BibitemShut {NoStop}%
\bibitem [{\citenamefont {{Bovy}}(2017)}]{Bovy2017}%
  \BibitemOpen
  \bibfield  {author} {\bibinfo {author} {\bibfnamefont {J.}~\bibnamefont {{Bovy}}},\ }\href {https://doi.org/10.1093/mnrasl/slx027} {\bibfield  {journal} {\bibinfo  {journal} {\mnras}\ }\textbf {\bibinfo {volume} {468}},\ \bibinfo {pages} {L63} (\bibinfo {year} {2017})},\ \Eprint {https://arxiv.org/abs/1610.07610} {arXiv:1610.07610 [astro-ph.GA]} \BibitemShut {NoStop}%
\bibitem [{\citenamefont {{Hernquist}}(1990)}]{Hernquist1990}%
  \BibitemOpen
  \bibfield  {author} {\bibinfo {author} {\bibfnamefont {L.}~\bibnamefont {{Hernquist}}},\ }\href {https://doi.org/10.1086/168845} {\bibfield  {journal} {\bibinfo  {journal} {\apj}\ }\textbf {\bibinfo {volume} {356}},\ \bibinfo {pages} {359} (\bibinfo {year} {1990})}\BibitemShut {NoStop}%
\bibitem [{\citenamefont {{Navarro}}\ \emph {et~al.}(1997)\citenamefont {{Navarro}}, \citenamefont {{Frenk}},\ and\ \citenamefont {{White}}}]{NFW}%
  \BibitemOpen
  \bibfield  {author} {\bibinfo {author} {\bibfnamefont {J.~F.}\ \bibnamefont {{Navarro}}}, \bibinfo {author} {\bibfnamefont {C.~S.}\ \bibnamefont {{Frenk}}},\ and\ \bibinfo {author} {\bibfnamefont {S.~D.~M.}\ \bibnamefont {{White}}},\ }\href {https://doi.org/10.1086/304888} {\bibfield  {journal} {\bibinfo  {journal} {\apj}\ }\textbf {\bibinfo {volume} {490}},\ \bibinfo {pages} {493} (\bibinfo {year} {1997})},\ \Eprint {https://arxiv.org/abs/astro-ph/9611107} {arXiv:astro-ph/9611107 [astro-ph]} \BibitemShut {NoStop}%
\bibitem [{\citenamefont {{Watkins}}\ \emph {et~al.}(2010)\citenamefont {{Watkins}}, \citenamefont {{Evans}},\ and\ \citenamefont {{An}}}]{Watkins2010}%
  \BibitemOpen
  \bibfield  {author} {\bibinfo {author} {\bibfnamefont {L.~L.}\ \bibnamefont {{Watkins}}}, \bibinfo {author} {\bibfnamefont {N.~W.}\ \bibnamefont {{Evans}}},\ and\ \bibinfo {author} {\bibfnamefont {J.~H.}\ \bibnamefont {{An}}},\ }\href {https://doi.org/10.1111/j.1365-2966.2010.16708.x} {\bibfield  {journal} {\bibinfo  {journal} {\mnras}\ }\textbf {\bibinfo {volume} {406}},\ \bibinfo {pages} {264} (\bibinfo {year} {2010})},\ \Eprint {https://arxiv.org/abs/1002.4565} {arXiv:1002.4565 [astro-ph.GA]} \BibitemShut {NoStop}%
\bibitem [{\citenamefont {{Newberg}}\ \emph {et~al.}(2010)\citenamefont {{Newberg}}, \citenamefont {{Willett}}, \citenamefont {{Yanny}},\ and\ \citenamefont {{Xu}}}]{Newberg2010}%
  \BibitemOpen
  \bibfield  {author} {\bibinfo {author} {\bibfnamefont {H.~J.}\ \bibnamefont {{Newberg}}}, \bibinfo {author} {\bibfnamefont {B.~A.}\ \bibnamefont {{Willett}}}, \bibinfo {author} {\bibfnamefont {B.}~\bibnamefont {{Yanny}}},\ and\ \bibinfo {author} {\bibfnamefont {Y.}~\bibnamefont {{Xu}}},\ }\href {https://doi.org/10.1088/0004-637X/711/1/32} {\bibfield  {journal} {\bibinfo  {journal} {\apj}\ }\textbf {\bibinfo {volume} {711}},\ \bibinfo {pages} {32} (\bibinfo {year} {2010})},\ \Eprint {https://arxiv.org/abs/1001.0576} {arXiv:1001.0576 [astro-ph.GA]} \BibitemShut {NoStop}%
\bibitem [{\citenamefont {{Bhattacharjee}}\ \emph {et~al.}(2014)\citenamefont {{Bhattacharjee}}, \citenamefont {{Chaudhury}},\ and\ \citenamefont {{Kundu}}}]{Bhattacharjee2014}%
  \BibitemOpen
  \bibfield  {author} {\bibinfo {author} {\bibfnamefont {P.}~\bibnamefont {{Bhattacharjee}}}, \bibinfo {author} {\bibfnamefont {S.}~\bibnamefont {{Chaudhury}}},\ and\ \bibinfo {author} {\bibfnamefont {S.}~\bibnamefont {{Kundu}}},\ }\href {https://doi.org/10.1088/0004-637X/785/1/63} {\bibfield  {journal} {\bibinfo  {journal} {\apj}\ }\textbf {\bibinfo {volume} {785}},\ \bibinfo {eid} {63} (\bibinfo {year} {2014})},\ \Eprint {https://arxiv.org/abs/1310.2659} {arXiv:1310.2659 [astro-ph.GA]} \BibitemShut {NoStop}%
\bibitem [{\citenamefont {{Fritz}}\ \emph {et~al.}(2018)\citenamefont {{Fritz}}, \citenamefont {{Battaglia}}, \citenamefont {{Pawlowski}}, \citenamefont {{Kallivayalil}}, \citenamefont {{van der Marel}}, \citenamefont {{Sohn}}, \citenamefont {{Brook}},\ and\ \citenamefont {{Besla}}}]{Fritz2018}%
  \BibitemOpen
  \bibfield  {author} {\bibinfo {author} {\bibfnamefont {T.~K.}\ \bibnamefont {{Fritz}}}, \bibinfo {author} {\bibfnamefont {G.}~\bibnamefont {{Battaglia}}}, \bibinfo {author} {\bibfnamefont {M.~S.}\ \bibnamefont {{Pawlowski}}}, \bibinfo {author} {\bibfnamefont {N.}~\bibnamefont {{Kallivayalil}}}, \bibinfo {author} {\bibfnamefont {R.}~\bibnamefont {{van der Marel}}}, \bibinfo {author} {\bibfnamefont {S.~T.}\ \bibnamefont {{Sohn}}}, \bibinfo {author} {\bibfnamefont {C.}~\bibnamefont {{Brook}}},\ and\ \bibinfo {author} {\bibfnamefont {G.}~\bibnamefont {{Besla}}},\ }\href {https://doi.org/10.1051/0004-6361/201833343} {\bibfield  {journal} {\bibinfo  {journal} {\aap}\ }\textbf {\bibinfo {volume} {619}},\ \bibinfo {eid} {A103} (\bibinfo {year} {2018})},\ \Eprint {https://arxiv.org/abs/1805.00908} {arXiv:1805.00908 [astro-ph.GA]} \BibitemShut {NoStop}%
\bibitem [{\citenamefont {{Eilers}}\ \emph {et~al.}(2019)\citenamefont {{Eilers}}, \citenamefont {{Hogg}}, \citenamefont {{Rix}},\ and\ \citenamefont {{Ness}}}]{Eilers2019}%
  \BibitemOpen
  \bibfield  {author} {\bibinfo {author} {\bibfnamefont {A.-C.}\ \bibnamefont {{Eilers}}}, \bibinfo {author} {\bibfnamefont {D.~W.}\ \bibnamefont {{Hogg}}}, \bibinfo {author} {\bibfnamefont {H.-W.}\ \bibnamefont {{Rix}}},\ and\ \bibinfo {author} {\bibfnamefont {M.~K.}\ \bibnamefont {{Ness}}},\ }\href {https://doi.org/10.3847/1538-4357/aaf648} {\bibfield  {journal} {\bibinfo  {journal} {\apj}\ }\textbf {\bibinfo {volume} {871}},\ \bibinfo {eid} {120} (\bibinfo {year} {2019})},\ \Eprint {https://arxiv.org/abs/1810.09466} {arXiv:1810.09466 [astro-ph.GA]} \BibitemShut {NoStop}%
\bibitem [{\citenamefont {{Gardner}}\ \emph {et~al.}(2021)\citenamefont {{Gardner}}, \citenamefont {{McDermott}},\ and\ \citenamefont {{Yanny}}}]{Gardner2021}%
  \BibitemOpen
  \bibfield  {author} {\bibinfo {author} {\bibfnamefont {S.}~\bibnamefont {{Gardner}}}, \bibinfo {author} {\bibfnamefont {S.~D.}\ \bibnamefont {{McDermott}}},\ and\ \bibinfo {author} {\bibfnamefont {B.}~\bibnamefont {{Yanny}}},\ }\href {https://doi.org/10.1016/j.ppnp.2021.103904} {\bibfield  {journal} {\bibinfo  {journal} {Progress in Particle and Nuclear Physics}\ }\textbf {\bibinfo {volume} {121}},\ \bibinfo {eid} {103904} (\bibinfo {year} {2021})},\ \Eprint {https://arxiv.org/abs/2106.13284} {arXiv:2106.13284 [astro-ph.GA]} \BibitemShut {NoStop}%
\bibitem [{\citenamefont {{Craig}}\ \emph {et~al.}(2022)\citenamefont {{Craig}}, \citenamefont {{Chakrabarti}}, \citenamefont {{Baum}},\ and\ \citenamefont {{Lewis}}}]{Craig2022}%
  \BibitemOpen
  \bibfield  {author} {\bibinfo {author} {\bibfnamefont {P.~A.}\ \bibnamefont {{Craig}}}, \bibinfo {author} {\bibfnamefont {S.}~\bibnamefont {{Chakrabarti}}}, \bibinfo {author} {\bibfnamefont {S.}~\bibnamefont {{Baum}}},\ and\ \bibinfo {author} {\bibfnamefont {B.~T.}\ \bibnamefont {{Lewis}}},\ }\href {https://doi.org/10.1093/mnras/stac2308} {\bibfield  {journal} {\bibinfo  {journal} {\mnras}\ }\textbf {\bibinfo {volume} {517}},\ \bibinfo {pages} {1737} (\bibinfo {year} {2022})}\BibitemShut {NoStop}%
\bibitem [{\citenamefont {{Bovy}}\ and\ \citenamefont {{Rix}}(2013)}]{BovyRix2013}%
  \BibitemOpen
  \bibfield  {author} {\bibinfo {author} {\bibfnamefont {J.}~\bibnamefont {{Bovy}}}\ and\ \bibinfo {author} {\bibfnamefont {H.-W.}\ \bibnamefont {{Rix}}},\ }\href {https://doi.org/10.1088/0004-637X/779/2/115} {\bibfield  {journal} {\bibinfo  {journal} {\apj}\ }\textbf {\bibinfo {volume} {779}},\ \bibinfo {eid} {115} (\bibinfo {year} {2013})},\ \Eprint {https://arxiv.org/abs/1309.0809} {arXiv:1309.0809 [astro-ph.GA]} \BibitemShut {NoStop}%
\bibitem [{\citenamefont {{Licquia}}\ and\ \citenamefont {{Newman}}(2015)}]{LicquiaNewman2015}%
  \BibitemOpen
  \bibfield  {author} {\bibinfo {author} {\bibfnamefont {T.~C.}\ \bibnamefont {{Licquia}}}\ and\ \bibinfo {author} {\bibfnamefont {J.~A.}\ \bibnamefont {{Newman}}},\ }\href {https://doi.org/10.1088/0004-637X/806/1/96} {\bibfield  {journal} {\bibinfo  {journal} {\apj}\ }\textbf {\bibinfo {volume} {806}},\ \bibinfo {eid} {96} (\bibinfo {year} {2015})},\ \Eprint {https://arxiv.org/abs/1407.1078} {arXiv:1407.1078 [astro-ph.GA]} \BibitemShut {NoStop}%
\bibitem [{\citenamefont {{Binney}}\ and\ \citenamefont {{Sch{\"o}nrich}}(2018)}]{BinneySchonrich2018}%
  \BibitemOpen
  \bibfield  {author} {\bibinfo {author} {\bibfnamefont {J.}~\bibnamefont {{Binney}}}\ and\ \bibinfo {author} {\bibfnamefont {R.}~\bibnamefont {{Sch{\"o}nrich}}},\ }\href {https://doi.org/10.1093/mnras/sty2378} {\bibfield  {journal} {\bibinfo  {journal} {\mnras}\ }\textbf {\bibinfo {volume} {481}},\ \bibinfo {pages} {1501} (\bibinfo {year} {2018})},\ \Eprint {https://arxiv.org/abs/1807.09819} {arXiv:1807.09819 [astro-ph.GA]} \BibitemShut {NoStop}%
\bibitem [{\citenamefont {{McMillan}}\ \emph {et~al.}(2022)\citenamefont {{McMillan}}, \citenamefont {{Petersson}}, \citenamefont {{Tepper-Garcia}}, \citenamefont {{Bland-Hawthorn}}, \citenamefont {{Antoja}}, \citenamefont {{Chemin}}, \citenamefont {{Figueras}}, \citenamefont {{Khanna}}, \citenamefont {{Kordopatis}}, \citenamefont {{Ramos}}, \citenamefont {{Romero-G{\'o}mez}},\ and\ \citenamefont {{Seabroke}}}]{McMillan2022}%
  \BibitemOpen
  \bibfield  {author} {\bibinfo {author} {\bibfnamefont {P.~J.}\ \bibnamefont {{McMillan}}}, \bibinfo {author} {\bibfnamefont {J.}~\bibnamefont {{Petersson}}}, \bibinfo {author} {\bibfnamefont {T.}~\bibnamefont {{Tepper-Garcia}}}, \bibinfo {author} {\bibfnamefont {J.}~\bibnamefont {{Bland-Hawthorn}}}, \bibinfo {author} {\bibfnamefont {T.}~\bibnamefont {{Antoja}}}, \bibinfo {author} {\bibfnamefont {L.}~\bibnamefont {{Chemin}}}, \bibinfo {author} {\bibfnamefont {F.}~\bibnamefont {{Figueras}}}, \bibinfo {author} {\bibfnamefont {S.}~\bibnamefont {{Khanna}}}, \bibinfo {author} {\bibfnamefont {G.}~\bibnamefont {{Kordopatis}}}, \bibinfo {author} {\bibfnamefont {P.}~\bibnamefont {{Ramos}}}, \bibinfo {author} {\bibfnamefont {M.}~\bibnamefont {{Romero-G{\'o}mez}}},\ and\ \bibinfo {author} {\bibfnamefont {G.}~\bibnamefont {{Seabroke}}},\ }\href {https://doi.org/10.1093/mnras/stac2571} {\bibfield  {journal} {\bibinfo  {journal} {\mnras}\ }\textbf {\bibinfo {volume} {516}},\ \bibinfo {pages} {4988} (\bibinfo {year}
  {2022})},\ \Eprint {https://arxiv.org/abs/2206.04059} {arXiv:2206.04059 [astro-ph.GA]} \BibitemShut {NoStop}%
\bibitem [{\citenamefont {{Garc{\'\i}a-Conde}}\ \emph {et~al.}(2023)\citenamefont {{Garc{\'\i}a-Conde}}, \citenamefont {{Antoja}}, \citenamefont {{Roca-F{\`a}brega}}, \citenamefont {{G{\'o}mez}}, \citenamefont {{Ramos}}, \citenamefont {{Garavito-Camargo}},\ and\ \citenamefont {{G{\'o}mez-Flechoso}}}]{Garcia-Conde2023}%
  \BibitemOpen
  \bibfield  {author} {\bibinfo {author} {\bibfnamefont {B.}~\bibnamefont {{Garc{\'\i}a-Conde}}}, \bibinfo {author} {\bibfnamefont {T.}~\bibnamefont {{Antoja}}}, \bibinfo {author} {\bibfnamefont {S.}~\bibnamefont {{Roca-F{\`a}brega}}}, \bibinfo {author} {\bibfnamefont {F.}~\bibnamefont {{G{\'o}mez}}}, \bibinfo {author} {\bibfnamefont {P.}~\bibnamefont {{Ramos}}}, \bibinfo {author} {\bibfnamefont {N.}~\bibnamefont {{Garavito-Camargo}}},\ and\ \bibinfo {author} {\bibfnamefont {M.}~\bibnamefont {{G{\'o}mez-Flechoso}}},\ }\href {https://doi.org/10.48550/arXiv.2311.07137} {\bibfield  {journal} {\bibinfo  {journal} {arXiv e-prints}\ ,\ \bibinfo {eid} {arXiv:2311.07137}} (\bibinfo {year} {2023})},\ \Eprint {https://arxiv.org/abs/2311.07137} {arXiv:2311.07137 [astro-ph.GA]} \BibitemShut {NoStop}%
\bibitem [{\citenamefont {{Reed}}\ \emph {et~al.}(2005)\citenamefont {{Reed}}, \citenamefont {{Governato}}, \citenamefont {{Quinn}}, \citenamefont {{Gardner}}, \citenamefont {{Stadel}},\ and\ \citenamefont {{Lake}}}]{Reed2005}%
  \BibitemOpen
  \bibfield  {author} {\bibinfo {author} {\bibfnamefont {D.}~\bibnamefont {{Reed}}}, \bibinfo {author} {\bibfnamefont {F.}~\bibnamefont {{Governato}}}, \bibinfo {author} {\bibfnamefont {T.}~\bibnamefont {{Quinn}}}, \bibinfo {author} {\bibfnamefont {J.}~\bibnamefont {{Gardner}}}, \bibinfo {author} {\bibfnamefont {J.}~\bibnamefont {{Stadel}}},\ and\ \bibinfo {author} {\bibfnamefont {G.}~\bibnamefont {{Lake}}},\ }\href {https://doi.org/10.1111/j.1365-2966.2005.09020.x} {\bibfield  {journal} {\bibinfo  {journal} {\mnras}\ }\textbf {\bibinfo {volume} {359}},\ \bibinfo {pages} {1537} (\bibinfo {year} {2005})},\ \Eprint {https://arxiv.org/abs/astro-ph/0406034} {arXiv:astro-ph/0406034 [astro-ph]} \BibitemShut {NoStop}%
\bibitem [{\citenamefont {{Chua}}\ \emph {et~al.}(2017)\citenamefont {{Chua}}, \citenamefont {{Pillepich}}, \citenamefont {{Rodriguez-Gomez}}, \citenamefont {{Vogelsberger}}, \citenamefont {{Bird}},\ and\ \citenamefont {{Hernquist}}}]{KunTingEddie2017}%
  \BibitemOpen
  \bibfield  {author} {\bibinfo {author} {\bibfnamefont {K.~T.~E.}\ \bibnamefont {{Chua}}}, \bibinfo {author} {\bibfnamefont {A.}~\bibnamefont {{Pillepich}}}, \bibinfo {author} {\bibfnamefont {V.}~\bibnamefont {{Rodriguez-Gomez}}}, \bibinfo {author} {\bibfnamefont {M.}~\bibnamefont {{Vogelsberger}}}, \bibinfo {author} {\bibfnamefont {S.}~\bibnamefont {{Bird}}},\ and\ \bibinfo {author} {\bibfnamefont {L.}~\bibnamefont {{Hernquist}}},\ }\href {https://doi.org/10.1093/mnras/stx2238} {\bibfield  {journal} {\bibinfo  {journal} {\mnras}\ }\textbf {\bibinfo {volume} {472}},\ \bibinfo {pages} {4343} (\bibinfo {year} {2017})},\ \Eprint {https://arxiv.org/abs/1611.07991} {arXiv:1611.07991 [astro-ph.GA]} \BibitemShut {NoStop}%
\bibitem [{\citenamefont {{Green}}\ and\ \citenamefont {{van den Bosch}}(2019)}]{GreenVandenbosch2019}%
  \BibitemOpen
  \bibfield  {author} {\bibinfo {author} {\bibfnamefont {S.~B.}\ \bibnamefont {{Green}}}\ and\ \bibinfo {author} {\bibfnamefont {F.~C.}\ \bibnamefont {{van den Bosch}}},\ }\href {https://doi.org/10.1093/mnras/stz2767} {\bibfield  {journal} {\bibinfo  {journal} {\mnras}\ }\textbf {\bibinfo {volume} {490}},\ \bibinfo {pages} {2091} (\bibinfo {year} {2019})},\ \Eprint {https://arxiv.org/abs/1908.08537} {arXiv:1908.08537 [astro-ph.GA]} \BibitemShut {NoStop}%
\bibitem [{\citenamefont {{Oort}}(1927)}]{Oort1927}%
  \BibitemOpen
  \bibfield  {author} {\bibinfo {author} {\bibfnamefont {J.~H.}\ \bibnamefont {{Oort}}},\ }\href@noop {} {\bibfield  {journal} {\bibinfo  {journal} {\bain}\ }\textbf {\bibinfo {volume} {3}},\ \bibinfo {pages} {275} (\bibinfo {year} {1927})}\BibitemShut {NoStop}%
\bibitem [{\citenamefont {{Li}}\ \emph {et~al.}(2019)\citenamefont {{Li}}, \citenamefont {{Zhao}},\ and\ \citenamefont {{Yang}}}]{Li2019}%
  \BibitemOpen
  \bibfield  {author} {\bibinfo {author} {\bibfnamefont {C.}~\bibnamefont {{Li}}}, \bibinfo {author} {\bibfnamefont {G.}~\bibnamefont {{Zhao}}},\ and\ \bibinfo {author} {\bibfnamefont {C.}~\bibnamefont {{Yang}}},\ }\href {https://doi.org/10.3847/1538-4357/ab0104} {\bibfield  {journal} {\bibinfo  {journal} {\apj}\ }\textbf {\bibinfo {volume} {872}},\ \bibinfo {eid} {205} (\bibinfo {year} {2019})}\BibitemShut {NoStop}%
\bibitem [{\citenamefont {{Gaia Collaboration}}\ \emph {et~al.}(2023)\citenamefont {{Gaia Collaboration}}, \citenamefont {{Creevey}}, \citenamefont {{Sarro}}, \citenamefont {{Lobel}}, \citenamefont {{Pancino}}, \citenamefont {{Andrae}}, \citenamefont {{Smart}}, \citenamefont {{Clementini}}, \citenamefont {{Heiter}}, \citenamefont {{Korn}}, \citenamefont {{Fouesneau}}, \citenamefont {{Fr{\'e}mat}}, \citenamefont {{De Angeli}}, \citenamefont {{Vallenari}}, \citenamefont {{Harrison}}, \citenamefont {{Th{\'e}venin}}, \citenamefont {{Reyl{\'e}}}, \citenamefont {{Sordo}}, \citenamefont {{Garofalo}}, \citenamefont {{Brown}}, \citenamefont {{Eyer}}, \citenamefont {{Prusti}}, \citenamefont {{de Bruijne}}, \citenamefont {{Arenou}}, \citenamefont {{Babusiaux}}, \citenamefont {{Biermann}}, \citenamefont {{Ducourant}}, \citenamefont {{Evans}}, \citenamefont {{Guerra}}, \citenamefont {{Hutton}}, \citenamefont {{Jordi}}, \citenamefont {{Klioner}}, \citenamefont {{Lammers}}, \citenamefont {{Lindegren}}, \citenamefont
  {{Luri}}, \citenamefont {{Mignard}}, \citenamefont {{Panem}}, \citenamefont {{Pourbaix}}, \citenamefont {{Randich}}, \citenamefont {{Sartoretti}}, \citenamefont {{Soubiran}}, \citenamefont {{Tanga}}, \citenamefont {{Walton}}, \citenamefont {{Bailer-Jones}}, \citenamefont {{Bastian}}, \citenamefont {{Drimmel}}, \citenamefont {{Jansen}}, \citenamefont {{Katz}}, \citenamefont {{Lattanzi}}, \citenamefont {{van Leeuwen}}, \citenamefont {{Bakker}}, \citenamefont {{Cacciari}}, \citenamefont {{Casta{\~n}eda}}, \citenamefont {{Fabricius}}, \citenamefont {{Galluccio}}, \citenamefont {{Guerrier}}, \citenamefont {{Masana}}, \citenamefont {{Messineo}}, \citenamefont {{Mowlavi}}, \citenamefont {{Nicolas}}, \citenamefont {{Nienartowicz}}, \citenamefont {{Pailler}}, \citenamefont {{Panuzzo}}, \citenamefont {{Riclet}}, \citenamefont {{Roux}}, \citenamefont {{Seabroke}}, \citenamefont {{Gracia-Abril}}, \citenamefont {{Portell}}, \citenamefont {{Teyssier}}, \citenamefont {{Altmann}}, \citenamefont {{Audard}}, \citenamefont
  {{Bellas-Velidis}}, \citenamefont {{Benson}}, \citenamefont {{Berthier}}, \citenamefont {{Blomme}}, \citenamefont {{Burgess}}, \citenamefont {{Busonero}}, \citenamefont {{Busso}}, \citenamefont {{C{\'a}novas}}, \citenamefont {{Carry}}, \citenamefont {{Cellino}}, \citenamefont {{Cheek}}, \citenamefont {{Damerdji}}, \citenamefont {{Davidson}}, \citenamefont {{de Teodoro}}, \citenamefont {{Nu{\~n}ez Campos}}, \citenamefont {{Delchambre}}, \citenamefont {{Dell'Oro}}, \citenamefont {{Esquej}}, \citenamefont {{Fern{\'a}ndez-Hern{\'a}ndez}}, \citenamefont {{Fraile}}, \citenamefont {{Garabato}}, \citenamefont {{Garc{\'\i}a-Lario}}, \citenamefont {{Gosset}}, \citenamefont {{Haigron}}, \citenamefont {{Halbwachs}}, \citenamefont {{Hambly}}, \citenamefont {{Hern{\'a}ndez}}, \citenamefont {{Hestroffer}}, \citenamefont {{Hodgkin}}, \citenamefont {{Holl}}, \citenamefont {{Jan{\ss}en}}, \citenamefont {{Jevardat de Fombelle}}, \citenamefont {{Jordan}}, \citenamefont {{Krone-Martins}}, \citenamefont {{Lanzafame}},
  \citenamefont {{L{\"o}ffler}}, \citenamefont {{Marchal}}, \citenamefont {{Marrese}}, \citenamefont {{Moitinho}}, \citenamefont {{Muinonen}}, \citenamefont {{Osborne}}, \citenamefont {{Pauwels}}, \citenamefont {{Recio-Blanco}}, \citenamefont {{Riello}}, \citenamefont {{Rimoldini}}, \citenamefont {{Roegiers}}, \citenamefont {{Rybizki}}, \citenamefont {{Siopis}}, \citenamefont {{Smith}}, \citenamefont {{Sozzetti}}, \citenamefont {{Utrilla}}, \citenamefont {{van Leeuwen}}, \citenamefont {{Abbas}}, \citenamefont {{{\'A}brah{\'a}m}}, \citenamefont {{Abreu Aramburu}}, \citenamefont {{Aerts}}, \citenamefont {{Aguado}}, \citenamefont {{Ajaj}}, \citenamefont {{Aldea-Montero}}, \citenamefont {{Altavilla}}, \citenamefont {{{\'A}lvarez}}, \citenamefont {{Alves}}, \citenamefont {{Anders}}, \citenamefont {{Anderson}}, \citenamefont {{Anglada Varela}}, \citenamefont {{Antoja}}, \citenamefont {{Baines}}, \citenamefont {{Baker}}, \citenamefont {{Balaguer-N{\'u}{\~n}ez}}, \citenamefont {{Balbinot}}, \citenamefont {{Balog}},
  \citenamefont {{Barache}}, \citenamefont {{Barbato}}, \citenamefont {{Barros}}, \citenamefont {{Barstow}}, \citenamefont {{Bartolom{\'e}}}, \citenamefont {{Bassilana}}, \citenamefont {{Bauchet}}, \citenamefont {{Becciani}}, \citenamefont {{Bellazzini}}, \citenamefont {{Berihuete}}, \citenamefont {{Bernet}}, \citenamefont {{Bertone}}, \citenamefont {{Bianchi}}, \citenamefont {{Binnenfeld}}, \citenamefont {{Blanco-Cuaresma}}, \citenamefont {{Boch}}, \citenamefont {{Bombrun}}, \citenamefont {{Bossini}}, \citenamefont {{Bouquillon}}, \citenamefont {{Bragaglia}}, \citenamefont {{Bramante}}, \citenamefont {{Breedt}}, \citenamefont {{Bressan}}, \citenamefont {{Brouillet}}, \citenamefont {{Brugaletta}}, \citenamefont {{Bucciarelli}}, \citenamefont {{Burlacu}}, \citenamefont {{Butkevich}}, \citenamefont {{Buzzi}}, \citenamefont {{Caffau}}, \citenamefont {{Cancelliere}}, \citenamefont {{Cantat-Gaudin}}, \citenamefont {{Carballo}}, \citenamefont {{Carlucci}}, \citenamefont {{Carnerero}}, \citenamefont {{Carrasco}},
  \citenamefont {{Casamiquela}}, \citenamefont {{Castellani}}, \citenamefont {{Castro-Ginard}}, \citenamefont {{Chaoul}}, \citenamefont {{Charlot}}, \citenamefont {{Chemin}}, \citenamefont {{Chiaramida}}, \citenamefont {{Chiavassa}}, \citenamefont {{Chornay}}, \citenamefont {{Comoretto}}, \citenamefont {{Contursi}}, \citenamefont {{Cooper}}, \citenamefont {{Cornez}}, \citenamefont {{Cowell}}, \citenamefont {{Crifo}}, \citenamefont {{Cropper}}, \citenamefont {{Crosta}}, \citenamefont {{Crowley}}, \citenamefont {{Dafonte}}, \citenamefont {{Dapergolas}}, \citenamefont {{David}}, \citenamefont {{de Laverny}}, \citenamefont {{De Luise}}, \citenamefont {{De March}}, \citenamefont {{De Ridder}}, \citenamefont {{de Souza}}, \citenamefont {{de Torres}}, \citenamefont {{del Peloso}}, \citenamefont {{del Pozo}}, \citenamefont {{Delbo}}, \citenamefont {{Delgado}}, \citenamefont {{Delisle}}, \citenamefont {{Demouchy}}, \citenamefont {{Dharmawardena}}, \citenamefont {{Di Matteo}}, \citenamefont {{Diakite}}, \citenamefont
  {{Diener}}, \citenamefont {{Distefano}}, \citenamefont {{Dolding}}, \citenamefont {{Enke}}, \citenamefont {{Fabre}}, \citenamefont {{Fabrizio}}, \citenamefont {{Faigler}}, \citenamefont {{Fedorets}}, \citenamefont {{Fernique}}, \citenamefont {{Figueras}}, \citenamefont {{Fournier}}, \citenamefont {{Fouron}}, \citenamefont {{Fragkoudi}}, \citenamefont {{Gai}}, \citenamefont {{Garcia-Gutierrez}}, \citenamefont {{Garcia-Reinaldos}}, \citenamefont {{Garc{\'\i}a-Torres}}, \citenamefont {{Gavel}}, \citenamefont {{Gavras}}, \citenamefont {{Gerlach}}, \citenamefont {{Geyer}}, \citenamefont {{Giacobbe}}, \citenamefont {{Gilmore}}, \citenamefont {{Girona}}, \citenamefont {{Giuffrida}}, \citenamefont {{Gomel}}, \citenamefont {{Gomez}}, \citenamefont {{Gonz{\'a}lez-N{\'u}{\~n}ez}}, \citenamefont {{Gonz{\'a}lez-Santamar{\'\i}a}}, \citenamefont {{Gonz{\'a}lez-Vidal}}, \citenamefont {{Granvik}}, \citenamefont {{Guillout}}, \citenamefont {{Guiraud}}, \citenamefont {{Guti{\'e}rrez-S{\'a}nchez}}, \citenamefont {{Guy}},
  \citenamefont {{Hatzidimitriou}}, \citenamefont {{Hauser}}, \citenamefont {{Haywood}}, \citenamefont {{Helmer}}, \citenamefont {{Helmi}}, \citenamefont {{Hilger}}, \citenamefont {{Sarmiento}}, \citenamefont {{Hidalgo}}, \citenamefont {{H{\l}adczuk}}, \citenamefont {{Hobbs}}, \citenamefont {{Holland}}, \citenamefont {{Huckle}}, \citenamefont {{Jardine}}, \citenamefont {{Jasniewicz}}, \citenamefont {{Jean-Antoine Piccolo}}, \citenamefont {{Jim{\'e}nez-Arranz}}, \citenamefont {{Juaristi Campillo}}, \citenamefont {{Julbe}}, \citenamefont {{Karbevska}}, \citenamefont {{Kervella}}, \citenamefont {{Khanna}}, \citenamefont {{Kordopatis}}, \citenamefont {{K{\'o}sp{\'a}l}}, \citenamefont {{Kostrzewa-Rutkowska}}, \citenamefont {{Kruszy{\'n}ska}}, \citenamefont {{Kun}}, \citenamefont {{Laizeau}}, \citenamefont {{Lambert}}, \citenamefont {{Lanza}}, \citenamefont {{Lasne}}, \citenamefont {{Le Campion}}, \citenamefont {{Lebreton}}, \citenamefont {{Lebzelter}}, \citenamefont {{Leccia}}, \citenamefont {{Leclerc}},
  \citenamefont {{Lecoeur-Taibi}}, \citenamefont {{Liao}}, \citenamefont {{Licata}}, \citenamefont {{Lindstr{\o}m}}, \citenamefont {{Lister}}, \citenamefont {{Livanou}}, \citenamefont {{Lorca}}, \citenamefont {{Loup}}, \citenamefont {{Madrero Pardo}}, \citenamefont {{Magdaleno Romeo}}, \citenamefont {{Managau}}, \citenamefont {{Mann}}, \citenamefont {{Manteiga}}, \citenamefont {{Marchant}}, \citenamefont {{Marconi}}, \citenamefont {{Marcos}}, \citenamefont {{Marcos Santos}}, \citenamefont {{Mar{\'\i}n Pina}}, \citenamefont {{Marinoni}}, \citenamefont {{Marocco}}, \citenamefont {{Marshall}}, \citenamefont {{Martin Polo}}, \citenamefont {{Mart{\'\i}n-Fleitas}}, \citenamefont {{Marton}}, \citenamefont {{Mary}}, \citenamefont {{Masip}}, \citenamefont {{Massari}}, \citenamefont {{Mastrobuono-Battisti}}, \citenamefont {{Mazeh}}, \citenamefont {{McMillan}}, \citenamefont {{Messina}}, \citenamefont {{Michalik}}, \citenamefont {{Millar}}, \citenamefont {{Mints}}, \citenamefont {{Molina}}, \citenamefont {{Molinaro}},
  \citenamefont {{Moln{\'a}r}}, \citenamefont {{Monari}}, \citenamefont {{Mongui{\'o}}}, \citenamefont {{Montegriffo}}, \citenamefont {{Montero}}, \citenamefont {{Mor}}, \citenamefont {{Mora}}, \citenamefont {{Morbidelli}}, \citenamefont {{Morel}}, \citenamefont {{Morris}}, \citenamefont {{Muraveva}}, \citenamefont {{Murphy}}, \citenamefont {{Musella}}, \citenamefont {{Nagy}}, \citenamefont {{Noval}}, \citenamefont {{Oca{\~n}a}}, \citenamefont {{Ogden}}, \citenamefont {{Ordenovic}}, \citenamefont {{Osinde}}, \citenamefont {{Pagani}}, \citenamefont {{Pagano}}, \citenamefont {{Palaversa}}, \citenamefont {{Palicio}}, \citenamefont {{Pallas-Quintela}}, \citenamefont {{Panahi}}, \citenamefont {{Payne-Wardenaar}}, \citenamefont {{Pe{\~n}alosa Esteller}}, \citenamefont {{Penttil{\"a}}}, \citenamefont {{Pichon}}, \citenamefont {{Piersimoni}}, \citenamefont {{Pineau}}, \citenamefont {{Plachy}}, \citenamefont {{Plum}}, \citenamefont {{Poggio}}, \citenamefont {{Pr{\v{s}}a}}, \citenamefont {{Pulone}}, \citenamefont
  {{Racero}}, \citenamefont {{Ragaini}}, \citenamefont {{Rainer}}, \citenamefont {{Raiteri}}, \citenamefont {{Ramos}}, \citenamefont {{Ramos-Lerate}}, \citenamefont {{Re Fiorentin}}, \citenamefont {{Regibo}}, \citenamefont {{Richards}}, \citenamefont {{Rios Diaz}}, \citenamefont {{Ripepi}}, \citenamefont {{Riva}}, \citenamefont {{Rix}}, \citenamefont {{Rixon}}, \citenamefont {{Robichon}}, \citenamefont {{Robin}}, \citenamefont {{Robin}}, \citenamefont {{Roelens}}, \citenamefont {{Rogues}}, \citenamefont {{Rohrbasser}}, \citenamefont {{Romero-G{\'o}mez}}, \citenamefont {{Rowell}}, \citenamefont {{Royer}}, \citenamefont {{Ruz Mieres}}, \citenamefont {{Rybicki}}, \citenamefont {{Sadowski}}, \citenamefont {{S{\'a}ez N{\'u}{\~n}ez}}, \citenamefont {{Sagrist{\`a} Sell{\'e}s}}, \citenamefont {{Sahlmann}}, \citenamefont {{Salguero}}, \citenamefont {{Samaras}}, \citenamefont {{Sanchez Gimenez}}, \citenamefont {{Sanna}}, \citenamefont {{Santove{\~n}a}}, \citenamefont {{Sarasso}}, \citenamefont {{Schultheis}},
  \citenamefont {{Sciacca}}, \citenamefont {{Segol}}, \citenamefont {{Segovia}}, \citenamefont {{S{\'e}gransan}}, \citenamefont {{Semeux}}, \citenamefont {{Shahaf}}, \citenamefont {{Siddiqui}}, \citenamefont {{Siebert}}, \citenamefont {{Siltala}}, \citenamefont {{Silvelo}}, \citenamefont {{Slezak}}, \citenamefont {{Slezak}}, \citenamefont {{Snaith}}, \citenamefont {{Solano}}, \citenamefont {{Solitro}}, \citenamefont {{Souami}}, \citenamefont {{Souchay}}, \citenamefont {{Spagna}}, \citenamefont {{Spina}}, \citenamefont {{Spoto}}, \citenamefont {{Steele}}, \citenamefont {{Steidelm{\"u}ller}}, \citenamefont {{Stephenson}}, \citenamefont {{S{\"u}veges}}, \citenamefont {{Surdej}}, \citenamefont {{Szabados}}, \citenamefont {{Szegedi-Elek}}, \citenamefont {{Taris}}, \citenamefont {{Taylor}}, \citenamefont {{Teixeira}}, \citenamefont {{Tolomei}}, \citenamefont {{Tonello}}, \citenamefont {{Torra}}, \citenamefont {{Torra}}, \citenamefont {{Torralba Elipe}}, \citenamefont {{Trabucchi}}, \citenamefont {{Tsounis}},
  \citenamefont {{Turon}}, \citenamefont {{Ulla}}, \citenamefont {{Unger}}, \citenamefont {{Vaillant}}, \citenamefont {{van Dillen}}, \citenamefont {{van Reeven}}, \citenamefont {{Vanel}}, \citenamefont {{Vecchiato}}, \citenamefont {{Viala}}, \citenamefont {{Vicente}}, \citenamefont {{Voutsinas}}, \citenamefont {{Weiler}}, \citenamefont {{Wevers}}, \citenamefont {{Wyrzykowski}}, \citenamefont {{Yoldas}}, \citenamefont {{Yvard}}, \citenamefont {{Zhao}}, \citenamefont {{Zorec}}, \citenamefont {{Zucker}},\ and\ \citenamefont {{Zwitter}}}]{Creevey2023}%
  \BibitemOpen
  \bibfield  {author} {\bibinfo {author} {\bibnamefont {{Gaia Collaboration}}}, \bibinfo {author} {\bibfnamefont {O.~L.}\ \bibnamefont {{Creevey}}}, \bibinfo {author} {\bibfnamefont {L.~M.}\ \bibnamefont {{Sarro}}}, \bibinfo {author} {\bibfnamefont {A.}~\bibnamefont {{Lobel}}}, \bibinfo {author} {\bibfnamefont {E.}~\bibnamefont {{Pancino}}}, \bibinfo {author} {\bibfnamefont {R.}~\bibnamefont {{Andrae}}}, \bibinfo {author} {\bibfnamefont {R.~L.}\ \bibnamefont {{Smart}}}, \bibinfo {author} {\bibfnamefont {G.}~\bibnamefont {{Clementini}}}, \bibinfo {author} {\bibfnamefont {U.}~\bibnamefont {{Heiter}}}, \bibinfo {author} {\bibfnamefont {A.~J.}\ \bibnamefont {{Korn}}}, \bibinfo {author} {\bibfnamefont {M.}~\bibnamefont {{Fouesneau}}}, \bibinfo {author} {\bibfnamefont {Y.}~\bibnamefont {{Fr{\'e}mat}}}, \bibinfo {author} {\bibfnamefont {F.}~\bibnamefont {{De Angeli}}}, \bibinfo {author} {\bibfnamefont {A.}~\bibnamefont {{Vallenari}}}, \bibinfo {author} {\bibfnamefont {D.~L.}\ \bibnamefont {{Harrison}}}, \bibinfo
  {author} {\bibfnamefont {F.}~\bibnamefont {{Th{\'e}venin}}}, \bibinfo {author} {\bibfnamefont {C.}~\bibnamefont {{Reyl{\'e}}}}, \bibinfo {author} {\bibfnamefont {R.}~\bibnamefont {{Sordo}}}, \bibinfo {author} {\bibfnamefont {A.}~\bibnamefont {{Garofalo}}}, \bibinfo {author} {\bibfnamefont {A.~G.~A.}\ \bibnamefont {{Brown}}}, \bibinfo {author} {\bibfnamefont {L.}~\bibnamefont {{Eyer}}}, \bibinfo {author} {\bibfnamefont {T.}~\bibnamefont {{Prusti}}}, \bibinfo {author} {\bibfnamefont {J.~H.~J.}\ \bibnamefont {{de Bruijne}}}, \bibinfo {author} {\bibfnamefont {F.}~\bibnamefont {{Arenou}}}, \bibinfo {author} {\bibfnamefont {C.}~\bibnamefont {{Babusiaux}}}, \bibinfo {author} {\bibfnamefont {M.}~\bibnamefont {{Biermann}}}, \bibinfo {author} {\bibfnamefont {C.}~\bibnamefont {{Ducourant}}}, \bibinfo {author} {\bibfnamefont {D.~W.}\ \bibnamefont {{Evans}}}, \bibinfo {author} {\bibfnamefont {R.}~\bibnamefont {{Guerra}}}, \bibinfo {author} {\bibfnamefont {A.}~\bibnamefont {{Hutton}}}, \bibinfo {author} {\bibfnamefont
  {C.}~\bibnamefont {{Jordi}}}, \bibinfo {author} {\bibfnamefont {S.~A.}\ \bibnamefont {{Klioner}}}, \bibinfo {author} {\bibfnamefont {U.~L.}\ \bibnamefont {{Lammers}}}, \bibinfo {author} {\bibfnamefont {L.}~\bibnamefont {{Lindegren}}}, \bibinfo {author} {\bibfnamefont {X.}~\bibnamefont {{Luri}}}, \bibinfo {author} {\bibfnamefont {F.}~\bibnamefont {{Mignard}}}, \bibinfo {author} {\bibfnamefont {C.}~\bibnamefont {{Panem}}}, \bibinfo {author} {\bibfnamefont {D.}~\bibnamefont {{Pourbaix}}}, \bibinfo {author} {\bibfnamefont {S.}~\bibnamefont {{Randich}}}, \bibinfo {author} {\bibfnamefont {P.}~\bibnamefont {{Sartoretti}}}, \bibinfo {author} {\bibfnamefont {C.}~\bibnamefont {{Soubiran}}}, \bibinfo {author} {\bibfnamefont {P.}~\bibnamefont {{Tanga}}}, \bibinfo {author} {\bibfnamefont {N.~A.}\ \bibnamefont {{Walton}}}, \bibinfo {author} {\bibfnamefont {C.~A.~L.}\ \bibnamefont {{Bailer-Jones}}}, \bibinfo {author} {\bibfnamefont {U.}~\bibnamefont {{Bastian}}}, \bibinfo {author} {\bibfnamefont {R.}~\bibnamefont
  {{Drimmel}}}, \bibinfo {author} {\bibfnamefont {F.}~\bibnamefont {{Jansen}}}, \bibinfo {author} {\bibfnamefont {D.}~\bibnamefont {{Katz}}}, \bibinfo {author} {\bibfnamefont {M.~G.}\ \bibnamefont {{Lattanzi}}}, \bibinfo {author} {\bibfnamefont {F.}~\bibnamefont {{van Leeuwen}}}, \bibinfo {author} {\bibfnamefont {J.}~\bibnamefont {{Bakker}}}, \bibinfo {author} {\bibfnamefont {C.}~\bibnamefont {{Cacciari}}}, \bibinfo {author} {\bibfnamefont {J.}~\bibnamefont {{Casta{\~n}eda}}}, \bibinfo {author} {\bibfnamefont {C.}~\bibnamefont {{Fabricius}}}, \bibinfo {author} {\bibfnamefont {L.}~\bibnamefont {{Galluccio}}}, \bibinfo {author} {\bibfnamefont {A.}~\bibnamefont {{Guerrier}}}, \bibinfo {author} {\bibfnamefont {E.}~\bibnamefont {{Masana}}}, \bibinfo {author} {\bibfnamefont {R.}~\bibnamefont {{Messineo}}}, \bibinfo {author} {\bibfnamefont {N.}~\bibnamefont {{Mowlavi}}}, \bibinfo {author} {\bibfnamefont {C.}~\bibnamefont {{Nicolas}}}, \bibinfo {author} {\bibfnamefont {K.}~\bibnamefont {{Nienartowicz}}}, \bibinfo
  {author} {\bibfnamefont {F.}~\bibnamefont {{Pailler}}}, \bibinfo {author} {\bibfnamefont {P.}~\bibnamefont {{Panuzzo}}}, \bibinfo {author} {\bibfnamefont {F.}~\bibnamefont {{Riclet}}}, \bibinfo {author} {\bibfnamefont {W.}~\bibnamefont {{Roux}}}, \bibinfo {author} {\bibfnamefont {G.~M.}\ \bibnamefont {{Seabroke}}}, \bibinfo {author} {\bibfnamefont {G.}~\bibnamefont {{Gracia-Abril}}}, \bibinfo {author} {\bibfnamefont {J.}~\bibnamefont {{Portell}}}, \bibinfo {author} {\bibfnamefont {D.}~\bibnamefont {{Teyssier}}}, \bibinfo {author} {\bibfnamefont {M.}~\bibnamefont {{Altmann}}}, \bibinfo {author} {\bibfnamefont {M.}~\bibnamefont {{Audard}}}, \bibinfo {author} {\bibfnamefont {I.}~\bibnamefont {{Bellas-Velidis}}}, \bibinfo {author} {\bibfnamefont {K.}~\bibnamefont {{Benson}}}, \bibinfo {author} {\bibfnamefont {J.}~\bibnamefont {{Berthier}}}, \bibinfo {author} {\bibfnamefont {R.}~\bibnamefont {{Blomme}}}, \bibinfo {author} {\bibfnamefont {P.~W.}\ \bibnamefont {{Burgess}}}, \bibinfo {author} {\bibfnamefont
  {D.}~\bibnamefont {{Busonero}}}, \bibinfo {author} {\bibfnamefont {G.}~\bibnamefont {{Busso}}}, \bibinfo {author} {\bibfnamefont {H.}~\bibnamefont {{C{\'a}novas}}}, \bibinfo {author} {\bibfnamefont {B.}~\bibnamefont {{Carry}}}, \bibinfo {author} {\bibfnamefont {A.}~\bibnamefont {{Cellino}}}, \bibinfo {author} {\bibfnamefont {N.}~\bibnamefont {{Cheek}}}, \bibinfo {author} {\bibfnamefont {Y.}~\bibnamefont {{Damerdji}}}, \bibinfo {author} {\bibfnamefont {M.}~\bibnamefont {{Davidson}}}, \bibinfo {author} {\bibfnamefont {P.}~\bibnamefont {{de Teodoro}}}, \bibinfo {author} {\bibfnamefont {M.}~\bibnamefont {{Nu{\~n}ez Campos}}}, \bibinfo {author} {\bibfnamefont {L.}~\bibnamefont {{Delchambre}}}, \bibinfo {author} {\bibfnamefont {A.}~\bibnamefont {{Dell'Oro}}}, \bibinfo {author} {\bibfnamefont {P.}~\bibnamefont {{Esquej}}}, \bibinfo {author} {\bibfnamefont {J.}~\bibnamefont {{Fern{\'a}ndez-Hern{\'a}ndez}}}, \bibinfo {author} {\bibfnamefont {E.}~\bibnamefont {{Fraile}}}, \bibinfo {author} {\bibfnamefont
  {D.}~\bibnamefont {{Garabato}}}, \bibinfo {author} {\bibfnamefont {P.}~\bibnamefont {{Garc{\'\i}a-Lario}}}, \bibinfo {author} {\bibfnamefont {E.}~\bibnamefont {{Gosset}}}, \bibinfo {author} {\bibfnamefont {R.}~\bibnamefont {{Haigron}}}, \bibinfo {author} {\bibfnamefont {J.~L.}\ \bibnamefont {{Halbwachs}}}, \bibinfo {author} {\bibfnamefont {N.~C.}\ \bibnamefont {{Hambly}}}, \bibinfo {author} {\bibfnamefont {J.}~\bibnamefont {{Hern{\'a}ndez}}}, \bibinfo {author} {\bibfnamefont {D.}~\bibnamefont {{Hestroffer}}}, \bibinfo {author} {\bibfnamefont {S.~T.}\ \bibnamefont {{Hodgkin}}}, \bibinfo {author} {\bibfnamefont {B.}~\bibnamefont {{Holl}}}, \bibinfo {author} {\bibfnamefont {K.}~\bibnamefont {{Jan{\ss}en}}}, \bibinfo {author} {\bibfnamefont {G.}~\bibnamefont {{Jevardat de Fombelle}}}, \bibinfo {author} {\bibfnamefont {S.}~\bibnamefont {{Jordan}}}, \bibinfo {author} {\bibfnamefont {A.}~\bibnamefont {{Krone-Martins}}}, \bibinfo {author} {\bibfnamefont {A.~C.}\ \bibnamefont {{Lanzafame}}}, \bibinfo {author}
  {\bibfnamefont {W.}~\bibnamefont {{L{\"o}ffler}}}, \bibinfo {author} {\bibfnamefont {O.}~\bibnamefont {{Marchal}}}, \bibinfo {author} {\bibfnamefont {P.~M.}\ \bibnamefont {{Marrese}}}, \bibinfo {author} {\bibfnamefont {A.}~\bibnamefont {{Moitinho}}}, \bibinfo {author} {\bibfnamefont {K.}~\bibnamefont {{Muinonen}}}, \bibinfo {author} {\bibfnamefont {P.}~\bibnamefont {{Osborne}}}, \bibinfo {author} {\bibfnamefont {T.}~\bibnamefont {{Pauwels}}}, \bibinfo {author} {\bibfnamefont {A.}~\bibnamefont {{Recio-Blanco}}}, \bibinfo {author} {\bibfnamefont {M.}~\bibnamefont {{Riello}}}, \bibinfo {author} {\bibfnamefont {L.}~\bibnamefont {{Rimoldini}}}, \bibinfo {author} {\bibfnamefont {T.}~\bibnamefont {{Roegiers}}}, \bibinfo {author} {\bibfnamefont {J.}~\bibnamefont {{Rybizki}}}, \bibinfo {author} {\bibfnamefont {C.}~\bibnamefont {{Siopis}}}, \bibinfo {author} {\bibfnamefont {M.}~\bibnamefont {{Smith}}}, \bibinfo {author} {\bibfnamefont {A.}~\bibnamefont {{Sozzetti}}}, \bibinfo {author} {\bibfnamefont {E.}~\bibnamefont
  {{Utrilla}}}, \bibinfo {author} {\bibfnamefont {M.}~\bibnamefont {{van Leeuwen}}}, \bibinfo {author} {\bibfnamefont {U.}~\bibnamefont {{Abbas}}}, \bibinfo {author} {\bibfnamefont {P.}~\bibnamefont {{{\'A}brah{\'a}m}}}, \bibinfo {author} {\bibfnamefont {A.}~\bibnamefont {{Abreu Aramburu}}}, \bibinfo {author} {\bibfnamefont {C.}~\bibnamefont {{Aerts}}}, \bibinfo {author} {\bibfnamefont {J.~J.}\ \bibnamefont {{Aguado}}}, \bibinfo {author} {\bibfnamefont {M.}~\bibnamefont {{Ajaj}}}, \bibinfo {author} {\bibfnamefont {F.}~\bibnamefont {{Aldea-Montero}}}, \bibinfo {author} {\bibfnamefont {G.}~\bibnamefont {{Altavilla}}}, \bibinfo {author} {\bibfnamefont {M.~A.}\ \bibnamefont {{{\'A}lvarez}}}, \bibinfo {author} {\bibfnamefont {J.}~\bibnamefont {{Alves}}}, \bibinfo {author} {\bibfnamefont {F.}~\bibnamefont {{Anders}}}, \bibinfo {author} {\bibfnamefont {R.~I.}\ \bibnamefont {{Anderson}}}, \bibinfo {author} {\bibfnamefont {E.}~\bibnamefont {{Anglada Varela}}}, \bibinfo {author} {\bibfnamefont {T.}~\bibnamefont
  {{Antoja}}}, \bibinfo {author} {\bibfnamefont {D.}~\bibnamefont {{Baines}}}, \bibinfo {author} {\bibfnamefont {S.~G.}\ \bibnamefont {{Baker}}}, \bibinfo {author} {\bibfnamefont {L.}~\bibnamefont {{Balaguer-N{\'u}{\~n}ez}}}, \bibinfo {author} {\bibfnamefont {E.}~\bibnamefont {{Balbinot}}}, \bibinfo {author} {\bibfnamefont {Z.}~\bibnamefont {{Balog}}}, \bibinfo {author} {\bibfnamefont {C.}~\bibnamefont {{Barache}}}, \bibinfo {author} {\bibfnamefont {D.}~\bibnamefont {{Barbato}}}, \bibinfo {author} {\bibfnamefont {M.}~\bibnamefont {{Barros}}}, \bibinfo {author} {\bibfnamefont {M.~A.}\ \bibnamefont {{Barstow}}}, \bibinfo {author} {\bibfnamefont {S.}~\bibnamefont {{Bartolom{\'e}}}}, \bibinfo {author} {\bibfnamefont {J.~L.}\ \bibnamefont {{Bassilana}}}, \bibinfo {author} {\bibfnamefont {N.}~\bibnamefont {{Bauchet}}}, \bibinfo {author} {\bibfnamefont {U.}~\bibnamefont {{Becciani}}}, \bibinfo {author} {\bibfnamefont {M.}~\bibnamefont {{Bellazzini}}}, \bibinfo {author} {\bibfnamefont {A.}~\bibnamefont
  {{Berihuete}}}, \bibinfo {author} {\bibfnamefont {M.}~\bibnamefont {{Bernet}}}, \bibinfo {author} {\bibfnamefont {S.}~\bibnamefont {{Bertone}}}, \bibinfo {author} {\bibfnamefont {L.}~\bibnamefont {{Bianchi}}}, \bibinfo {author} {\bibfnamefont {A.}~\bibnamefont {{Binnenfeld}}}, \bibinfo {author} {\bibfnamefont {S.}~\bibnamefont {{Blanco-Cuaresma}}}, \bibinfo {author} {\bibfnamefont {T.}~\bibnamefont {{Boch}}}, \bibinfo {author} {\bibfnamefont {A.}~\bibnamefont {{Bombrun}}}, \bibinfo {author} {\bibfnamefont {D.}~\bibnamefont {{Bossini}}}, \bibinfo {author} {\bibfnamefont {S.}~\bibnamefont {{Bouquillon}}}, \bibinfo {author} {\bibfnamefont {A.}~\bibnamefont {{Bragaglia}}}, \bibinfo {author} {\bibfnamefont {L.}~\bibnamefont {{Bramante}}}, \bibinfo {author} {\bibfnamefont {E.}~\bibnamefont {{Breedt}}}, \bibinfo {author} {\bibfnamefont {A.}~\bibnamefont {{Bressan}}}, \bibinfo {author} {\bibfnamefont {N.}~\bibnamefont {{Brouillet}}}, \bibinfo {author} {\bibfnamefont {E.}~\bibnamefont {{Brugaletta}}}, \bibinfo
  {author} {\bibfnamefont {B.}~\bibnamefont {{Bucciarelli}}}, \bibinfo {author} {\bibfnamefont {A.}~\bibnamefont {{Burlacu}}}, \bibinfo {author} {\bibfnamefont {A.~G.}\ \bibnamefont {{Butkevich}}}, \bibinfo {author} {\bibfnamefont {R.}~\bibnamefont {{Buzzi}}}, \bibinfo {author} {\bibfnamefont {E.}~\bibnamefont {{Caffau}}}, \bibinfo {author} {\bibfnamefont {R.}~\bibnamefont {{Cancelliere}}}, \bibinfo {author} {\bibfnamefont {T.}~\bibnamefont {{Cantat-Gaudin}}}, \bibinfo {author} {\bibfnamefont {R.}~\bibnamefont {{Carballo}}}, \bibinfo {author} {\bibfnamefont {T.}~\bibnamefont {{Carlucci}}}, \bibinfo {author} {\bibfnamefont {M.~I.}\ \bibnamefont {{Carnerero}}}, \bibinfo {author} {\bibfnamefont {J.~M.}\ \bibnamefont {{Carrasco}}}, \bibinfo {author} {\bibfnamefont {L.}~\bibnamefont {{Casamiquela}}}, \bibinfo {author} {\bibfnamefont {M.}~\bibnamefont {{Castellani}}}, \bibinfo {author} {\bibfnamefont {A.}~\bibnamefont {{Castro-Ginard}}}, \bibinfo {author} {\bibfnamefont {L.}~\bibnamefont {{Chaoul}}}, \bibinfo
  {author} {\bibfnamefont {P.}~\bibnamefont {{Charlot}}}, \bibinfo {author} {\bibfnamefont {L.}~\bibnamefont {{Chemin}}}, \bibinfo {author} {\bibfnamefont {V.}~\bibnamefont {{Chiaramida}}}, \bibinfo {author} {\bibfnamefont {A.}~\bibnamefont {{Chiavassa}}}, \bibinfo {author} {\bibfnamefont {N.}~\bibnamefont {{Chornay}}}, \bibinfo {author} {\bibfnamefont {G.}~\bibnamefont {{Comoretto}}}, \bibinfo {author} {\bibfnamefont {G.}~\bibnamefont {{Contursi}}}, \bibinfo {author} {\bibfnamefont {W.~J.}\ \bibnamefont {{Cooper}}}, \bibinfo {author} {\bibfnamefont {T.}~\bibnamefont {{Cornez}}}, \bibinfo {author} {\bibfnamefont {S.}~\bibnamefont {{Cowell}}}, \bibinfo {author} {\bibfnamefont {F.}~\bibnamefont {{Crifo}}}, \bibinfo {author} {\bibfnamefont {M.}~\bibnamefont {{Cropper}}}, \bibinfo {author} {\bibfnamefont {M.}~\bibnamefont {{Crosta}}}, \bibinfo {author} {\bibfnamefont {C.}~\bibnamefont {{Crowley}}}, \bibinfo {author} {\bibfnamefont {C.}~\bibnamefont {{Dafonte}}}, \bibinfo {author} {\bibfnamefont {A.}~\bibnamefont
  {{Dapergolas}}}, \bibinfo {author} {\bibfnamefont {P.}~\bibnamefont {{David}}}, \bibinfo {author} {\bibfnamefont {P.}~\bibnamefont {{de Laverny}}}, \bibinfo {author} {\bibfnamefont {F.}~\bibnamefont {{De Luise}}}, \bibinfo {author} {\bibfnamefont {R.}~\bibnamefont {{De March}}}, \bibinfo {author} {\bibfnamefont {J.}~\bibnamefont {{De Ridder}}}, \bibinfo {author} {\bibfnamefont {R.}~\bibnamefont {{de Souza}}}, \bibinfo {author} {\bibfnamefont {A.}~\bibnamefont {{de Torres}}}, \bibinfo {author} {\bibfnamefont {E.~F.}\ \bibnamefont {{del Peloso}}}, \bibinfo {author} {\bibfnamefont {E.}~\bibnamefont {{del Pozo}}}, \bibinfo {author} {\bibfnamefont {M.}~\bibnamefont {{Delbo}}}, \bibinfo {author} {\bibfnamefont {A.}~\bibnamefont {{Delgado}}}, \bibinfo {author} {\bibfnamefont {J.~B.}\ \bibnamefont {{Delisle}}}, \bibinfo {author} {\bibfnamefont {C.}~\bibnamefont {{Demouchy}}}, \bibinfo {author} {\bibfnamefont {T.~E.}\ \bibnamefont {{Dharmawardena}}}, \bibinfo {author} {\bibfnamefont {P.}~\bibnamefont {{Di Matteo}}},
  \bibinfo {author} {\bibfnamefont {S.}~\bibnamefont {{Diakite}}}, \bibinfo {author} {\bibfnamefont {C.}~\bibnamefont {{Diener}}}, \bibinfo {author} {\bibfnamefont {E.}~\bibnamefont {{Distefano}}}, \bibinfo {author} {\bibfnamefont {C.}~\bibnamefont {{Dolding}}}, \bibinfo {author} {\bibfnamefont {H.}~\bibnamefont {{Enke}}}, \bibinfo {author} {\bibfnamefont {C.}~\bibnamefont {{Fabre}}}, \bibinfo {author} {\bibfnamefont {M.}~\bibnamefont {{Fabrizio}}}, \bibinfo {author} {\bibfnamefont {S.}~\bibnamefont {{Faigler}}}, \bibinfo {author} {\bibfnamefont {G.}~\bibnamefont {{Fedorets}}}, \bibinfo {author} {\bibfnamefont {P.}~\bibnamefont {{Fernique}}}, \bibinfo {author} {\bibfnamefont {F.}~\bibnamefont {{Figueras}}}, \bibinfo {author} {\bibfnamefont {Y.}~\bibnamefont {{Fournier}}}, \bibinfo {author} {\bibfnamefont {C.}~\bibnamefont {{Fouron}}}, \bibinfo {author} {\bibfnamefont {F.}~\bibnamefont {{Fragkoudi}}}, \bibinfo {author} {\bibfnamefont {M.}~\bibnamefont {{Gai}}}, \bibinfo {author} {\bibfnamefont
  {A.}~\bibnamefont {{Garcia-Gutierrez}}}, \bibinfo {author} {\bibfnamefont {M.}~\bibnamefont {{Garcia-Reinaldos}}}, \bibinfo {author} {\bibfnamefont {M.}~\bibnamefont {{Garc{\'\i}a-Torres}}}, \bibinfo {author} {\bibfnamefont {A.}~\bibnamefont {{Gavel}}}, \bibinfo {author} {\bibfnamefont {P.}~\bibnamefont {{Gavras}}}, \bibinfo {author} {\bibfnamefont {E.}~\bibnamefont {{Gerlach}}}, \bibinfo {author} {\bibfnamefont {R.}~\bibnamefont {{Geyer}}}, \bibinfo {author} {\bibfnamefont {P.}~\bibnamefont {{Giacobbe}}}, \bibinfo {author} {\bibfnamefont {G.}~\bibnamefont {{Gilmore}}}, \bibinfo {author} {\bibfnamefont {S.}~\bibnamefont {{Girona}}}, \bibinfo {author} {\bibfnamefont {G.}~\bibnamefont {{Giuffrida}}}, \bibinfo {author} {\bibfnamefont {R.}~\bibnamefont {{Gomel}}}, \bibinfo {author} {\bibfnamefont {A.}~\bibnamefont {{Gomez}}}, \bibinfo {author} {\bibfnamefont {J.}~\bibnamefont {{Gonz{\'a}lez-N{\'u}{\~n}ez}}}, \bibinfo {author} {\bibfnamefont {I.}~\bibnamefont {{Gonz{\'a}lez-Santamar{\'\i}a}}}, \bibinfo {author}
  {\bibfnamefont {J.~J.}\ \bibnamefont {{Gonz{\'a}lez-Vidal}}}, \bibinfo {author} {\bibfnamefont {M.}~\bibnamefont {{Granvik}}}, \bibinfo {author} {\bibfnamefont {P.}~\bibnamefont {{Guillout}}}, \bibinfo {author} {\bibfnamefont {J.}~\bibnamefont {{Guiraud}}}, \bibinfo {author} {\bibfnamefont {R.}~\bibnamefont {{Guti{\'e}rrez-S{\'a}nchez}}}, \bibinfo {author} {\bibfnamefont {L.~P.}\ \bibnamefont {{Guy}}}, \bibinfo {author} {\bibfnamefont {D.}~\bibnamefont {{Hatzidimitriou}}}, \bibinfo {author} {\bibfnamefont {M.}~\bibnamefont {{Hauser}}}, \bibinfo {author} {\bibfnamefont {M.}~\bibnamefont {{Haywood}}}, \bibinfo {author} {\bibfnamefont {A.}~\bibnamefont {{Helmer}}}, \bibinfo {author} {\bibfnamefont {A.}~\bibnamefont {{Helmi}}}, \bibinfo {author} {\bibfnamefont {T.}~\bibnamefont {{Hilger}}}, \bibinfo {author} {\bibfnamefont {M.~H.}\ \bibnamefont {{Sarmiento}}}, \bibinfo {author} {\bibfnamefont {S.~L.}\ \bibnamefont {{Hidalgo}}}, \bibinfo {author} {\bibfnamefont {N.}~\bibnamefont {{H{\l}adczuk}}}, \bibinfo
  {author} {\bibfnamefont {D.}~\bibnamefont {{Hobbs}}}, \bibinfo {author} {\bibfnamefont {G.}~\bibnamefont {{Holland}}}, \bibinfo {author} {\bibfnamefont {H.~E.}\ \bibnamefont {{Huckle}}}, \bibinfo {author} {\bibfnamefont {K.}~\bibnamefont {{Jardine}}}, \bibinfo {author} {\bibfnamefont {G.}~\bibnamefont {{Jasniewicz}}}, \bibinfo {author} {\bibfnamefont {A.}~\bibnamefont {{Jean-Antoine Piccolo}}}, \bibinfo {author} {\bibfnamefont {{\'O}.}~\bibnamefont {{Jim{\'e}nez-Arranz}}}, \bibinfo {author} {\bibfnamefont {J.}~\bibnamefont {{Juaristi Campillo}}}, \bibinfo {author} {\bibfnamefont {F.}~\bibnamefont {{Julbe}}}, \bibinfo {author} {\bibfnamefont {L.}~\bibnamefont {{Karbevska}}}, \bibinfo {author} {\bibfnamefont {P.}~\bibnamefont {{Kervella}}}, \bibinfo {author} {\bibfnamefont {S.}~\bibnamefont {{Khanna}}}, \bibinfo {author} {\bibfnamefont {G.}~\bibnamefont {{Kordopatis}}}, \bibinfo {author} {\bibfnamefont {{\'A}.}~\bibnamefont {{K{\'o}sp{\'a}l}}}, \bibinfo {author} {\bibfnamefont {Z.}~\bibnamefont
  {{Kostrzewa-Rutkowska}}}, \bibinfo {author} {\bibfnamefont {K.}~\bibnamefont {{Kruszy{\'n}ska}}}, \bibinfo {author} {\bibfnamefont {M.}~\bibnamefont {{Kun}}}, \bibinfo {author} {\bibfnamefont {P.}~\bibnamefont {{Laizeau}}}, \bibinfo {author} {\bibfnamefont {S.}~\bibnamefont {{Lambert}}}, \bibinfo {author} {\bibfnamefont {A.~F.}\ \bibnamefont {{Lanza}}}, \bibinfo {author} {\bibfnamefont {Y.}~\bibnamefont {{Lasne}}}, \bibinfo {author} {\bibfnamefont {J.~F.}\ \bibnamefont {{Le Campion}}}, \bibinfo {author} {\bibfnamefont {Y.}~\bibnamefont {{Lebreton}}}, \bibinfo {author} {\bibfnamefont {T.}~\bibnamefont {{Lebzelter}}}, \bibinfo {author} {\bibfnamefont {S.}~\bibnamefont {{Leccia}}}, \bibinfo {author} {\bibfnamefont {N.}~\bibnamefont {{Leclerc}}}, \bibinfo {author} {\bibfnamefont {I.}~\bibnamefont {{Lecoeur-Taibi}}}, \bibinfo {author} {\bibfnamefont {S.}~\bibnamefont {{Liao}}}, \bibinfo {author} {\bibfnamefont {E.~L.}\ \bibnamefont {{Licata}}}, \bibinfo {author} {\bibfnamefont {H.~E.~P.}\ \bibnamefont
  {{Lindstr{\o}m}}}, \bibinfo {author} {\bibfnamefont {T.~A.}\ \bibnamefont {{Lister}}}, \bibinfo {author} {\bibfnamefont {E.}~\bibnamefont {{Livanou}}}, \bibinfo {author} {\bibfnamefont {A.}~\bibnamefont {{Lorca}}}, \bibinfo {author} {\bibfnamefont {C.}~\bibnamefont {{Loup}}}, \bibinfo {author} {\bibfnamefont {P.}~\bibnamefont {{Madrero Pardo}}}, \bibinfo {author} {\bibfnamefont {A.}~\bibnamefont {{Magdaleno Romeo}}}, \bibinfo {author} {\bibfnamefont {S.}~\bibnamefont {{Managau}}}, \bibinfo {author} {\bibfnamefont {R.~G.}\ \bibnamefont {{Mann}}}, \bibinfo {author} {\bibfnamefont {M.}~\bibnamefont {{Manteiga}}}, \bibinfo {author} {\bibfnamefont {J.~M.}\ \bibnamefont {{Marchant}}}, \bibinfo {author} {\bibfnamefont {M.}~\bibnamefont {{Marconi}}}, \bibinfo {author} {\bibfnamefont {J.}~\bibnamefont {{Marcos}}}, \bibinfo {author} {\bibfnamefont {M.~M.~S.}\ \bibnamefont {{Marcos Santos}}}, \bibinfo {author} {\bibfnamefont {D.}~\bibnamefont {{Mar{\'\i}n Pina}}}, \bibinfo {author} {\bibfnamefont {S.}~\bibnamefont
  {{Marinoni}}}, \bibinfo {author} {\bibfnamefont {F.}~\bibnamefont {{Marocco}}}, \bibinfo {author} {\bibfnamefont {D.~J.}\ \bibnamefont {{Marshall}}}, \bibinfo {author} {\bibfnamefont {L.}~\bibnamefont {{Martin Polo}}}, \bibinfo {author} {\bibfnamefont {J.~M.}\ \bibnamefont {{Mart{\'\i}n-Fleitas}}}, \bibinfo {author} {\bibfnamefont {G.}~\bibnamefont {{Marton}}}, \bibinfo {author} {\bibfnamefont {N.}~\bibnamefont {{Mary}}}, \bibinfo {author} {\bibfnamefont {A.}~\bibnamefont {{Masip}}}, \bibinfo {author} {\bibfnamefont {D.}~\bibnamefont {{Massari}}}, \bibinfo {author} {\bibfnamefont {A.}~\bibnamefont {{Mastrobuono-Battisti}}}, \bibinfo {author} {\bibfnamefont {T.}~\bibnamefont {{Mazeh}}}, \bibinfo {author} {\bibfnamefont {P.~J.}\ \bibnamefont {{McMillan}}}, \bibinfo {author} {\bibfnamefont {S.}~\bibnamefont {{Messina}}}, \bibinfo {author} {\bibfnamefont {D.}~\bibnamefont {{Michalik}}}, \bibinfo {author} {\bibfnamefont {N.~R.}\ \bibnamefont {{Millar}}}, \bibinfo {author} {\bibfnamefont {A.}~\bibnamefont
  {{Mints}}}, \bibinfo {author} {\bibfnamefont {D.}~\bibnamefont {{Molina}}}, \bibinfo {author} {\bibfnamefont {R.}~\bibnamefont {{Molinaro}}}, \bibinfo {author} {\bibfnamefont {L.}~\bibnamefont {{Moln{\'a}r}}}, \bibinfo {author} {\bibfnamefont {G.}~\bibnamefont {{Monari}}}, \bibinfo {author} {\bibfnamefont {M.}~\bibnamefont {{Mongui{\'o}}}}, \bibinfo {author} {\bibfnamefont {P.}~\bibnamefont {{Montegriffo}}}, \bibinfo {author} {\bibfnamefont {A.}~\bibnamefont {{Montero}}}, \bibinfo {author} {\bibfnamefont {R.}~\bibnamefont {{Mor}}}, \bibinfo {author} {\bibfnamefont {A.}~\bibnamefont {{Mora}}}, \bibinfo {author} {\bibfnamefont {R.}~\bibnamefont {{Morbidelli}}}, \bibinfo {author} {\bibfnamefont {T.}~\bibnamefont {{Morel}}}, \bibinfo {author} {\bibfnamefont {D.}~\bibnamefont {{Morris}}}, \bibinfo {author} {\bibfnamefont {T.}~\bibnamefont {{Muraveva}}}, \bibinfo {author} {\bibfnamefont {C.~P.}\ \bibnamefont {{Murphy}}}, \bibinfo {author} {\bibfnamefont {I.}~\bibnamefont {{Musella}}}, \bibinfo {author}
  {\bibfnamefont {Z.}~\bibnamefont {{Nagy}}}, \bibinfo {author} {\bibfnamefont {L.}~\bibnamefont {{Noval}}}, \bibinfo {author} {\bibfnamefont {F.}~\bibnamefont {{Oca{\~n}a}}}, \bibinfo {author} {\bibfnamefont {A.}~\bibnamefont {{Ogden}}}, \bibinfo {author} {\bibfnamefont {C.}~\bibnamefont {{Ordenovic}}}, \bibinfo {author} {\bibfnamefont {J.~O.}\ \bibnamefont {{Osinde}}}, \bibinfo {author} {\bibfnamefont {C.}~\bibnamefont {{Pagani}}}, \bibinfo {author} {\bibfnamefont {I.}~\bibnamefont {{Pagano}}}, \bibinfo {author} {\bibfnamefont {L.}~\bibnamefont {{Palaversa}}}, \bibinfo {author} {\bibfnamefont {P.~A.}\ \bibnamefont {{Palicio}}}, \bibinfo {author} {\bibfnamefont {L.}~\bibnamefont {{Pallas-Quintela}}}, \bibinfo {author} {\bibfnamefont {A.}~\bibnamefont {{Panahi}}}, \bibinfo {author} {\bibfnamefont {S.}~\bibnamefont {{Payne-Wardenaar}}}, \bibinfo {author} {\bibfnamefont {X.}~\bibnamefont {{Pe{\~n}alosa Esteller}}}, \bibinfo {author} {\bibfnamefont {A.}~\bibnamefont {{Penttil{\"a}}}}, \bibinfo {author}
  {\bibfnamefont {B.}~\bibnamefont {{Pichon}}}, \bibinfo {author} {\bibfnamefont {A.~M.}\ \bibnamefont {{Piersimoni}}}, \bibinfo {author} {\bibfnamefont {F.~X.}\ \bibnamefont {{Pineau}}}, \bibinfo {author} {\bibfnamefont {E.}~\bibnamefont {{Plachy}}}, \bibinfo {author} {\bibfnamefont {G.}~\bibnamefont {{Plum}}}, \bibinfo {author} {\bibfnamefont {E.}~\bibnamefont {{Poggio}}}, \bibinfo {author} {\bibfnamefont {A.}~\bibnamefont {{Pr{\v{s}}a}}}, \bibinfo {author} {\bibfnamefont {L.}~\bibnamefont {{Pulone}}}, \bibinfo {author} {\bibfnamefont {E.}~\bibnamefont {{Racero}}}, \bibinfo {author} {\bibfnamefont {S.}~\bibnamefont {{Ragaini}}}, \bibinfo {author} {\bibfnamefont {M.}~\bibnamefont {{Rainer}}}, \bibinfo {author} {\bibfnamefont {C.~M.}\ \bibnamefont {{Raiteri}}}, \bibinfo {author} {\bibfnamefont {P.}~\bibnamefont {{Ramos}}}, \bibinfo {author} {\bibfnamefont {M.}~\bibnamefont {{Ramos-Lerate}}}, \bibinfo {author} {\bibfnamefont {P.}~\bibnamefont {{Re Fiorentin}}}, \bibinfo {author} {\bibfnamefont
  {S.}~\bibnamefont {{Regibo}}}, \bibinfo {author} {\bibfnamefont {P.~J.}\ \bibnamefont {{Richards}}}, \bibinfo {author} {\bibfnamefont {C.}~\bibnamefont {{Rios Diaz}}}, \bibinfo {author} {\bibfnamefont {V.}~\bibnamefont {{Ripepi}}}, \bibinfo {author} {\bibfnamefont {A.}~\bibnamefont {{Riva}}}, \bibinfo {author} {\bibfnamefont {H.~W.}\ \bibnamefont {{Rix}}}, \bibinfo {author} {\bibfnamefont {G.}~\bibnamefont {{Rixon}}}, \bibinfo {author} {\bibfnamefont {N.}~\bibnamefont {{Robichon}}}, \bibinfo {author} {\bibfnamefont {A.~C.}\ \bibnamefont {{Robin}}}, \bibinfo {author} {\bibfnamefont {C.}~\bibnamefont {{Robin}}}, \bibinfo {author} {\bibfnamefont {M.}~\bibnamefont {{Roelens}}}, \bibinfo {author} {\bibfnamefont {H.~R.~O.}\ \bibnamefont {{Rogues}}}, \bibinfo {author} {\bibfnamefont {L.}~\bibnamefont {{Rohrbasser}}}, \bibinfo {author} {\bibfnamefont {M.}~\bibnamefont {{Romero-G{\'o}mez}}}, \bibinfo {author} {\bibfnamefont {N.}~\bibnamefont {{Rowell}}}, \bibinfo {author} {\bibfnamefont {F.}~\bibnamefont {{Royer}}},
  \bibinfo {author} {\bibfnamefont {D.}~\bibnamefont {{Ruz Mieres}}}, \bibinfo {author} {\bibfnamefont {K.~A.}\ \bibnamefont {{Rybicki}}}, \bibinfo {author} {\bibfnamefont {G.}~\bibnamefont {{Sadowski}}}, \bibinfo {author} {\bibfnamefont {A.}~\bibnamefont {{S{\'a}ez N{\'u}{\~n}ez}}}, \bibinfo {author} {\bibfnamefont {A.}~\bibnamefont {{Sagrist{\`a} Sell{\'e}s}}}, \bibinfo {author} {\bibfnamefont {J.}~\bibnamefont {{Sahlmann}}}, \bibinfo {author} {\bibfnamefont {E.}~\bibnamefont {{Salguero}}}, \bibinfo {author} {\bibfnamefont {N.}~\bibnamefont {{Samaras}}}, \bibinfo {author} {\bibfnamefont {V.}~\bibnamefont {{Sanchez Gimenez}}}, \bibinfo {author} {\bibfnamefont {N.}~\bibnamefont {{Sanna}}}, \bibinfo {author} {\bibfnamefont {R.}~\bibnamefont {{Santove{\~n}a}}}, \bibinfo {author} {\bibfnamefont {M.}~\bibnamefont {{Sarasso}}}, \bibinfo {author} {\bibfnamefont {M.}~\bibnamefont {{Schultheis}}}, \bibinfo {author} {\bibfnamefont {E.}~\bibnamefont {{Sciacca}}}, \bibinfo {author} {\bibfnamefont {M.}~\bibnamefont
  {{Segol}}}, \bibinfo {author} {\bibfnamefont {J.~C.}\ \bibnamefont {{Segovia}}}, \bibinfo {author} {\bibfnamefont {D.}~\bibnamefont {{S{\'e}gransan}}}, \bibinfo {author} {\bibfnamefont {D.}~\bibnamefont {{Semeux}}}, \bibinfo {author} {\bibfnamefont {S.}~\bibnamefont {{Shahaf}}}, \bibinfo {author} {\bibfnamefont {H.~I.}\ \bibnamefont {{Siddiqui}}}, \bibinfo {author} {\bibfnamefont {A.}~\bibnamefont {{Siebert}}}, \bibinfo {author} {\bibfnamefont {L.}~\bibnamefont {{Siltala}}}, \bibinfo {author} {\bibfnamefont {A.}~\bibnamefont {{Silvelo}}}, \bibinfo {author} {\bibfnamefont {E.}~\bibnamefont {{Slezak}}}, \bibinfo {author} {\bibfnamefont {I.}~\bibnamefont {{Slezak}}}, \bibinfo {author} {\bibfnamefont {O.~N.}\ \bibnamefont {{Snaith}}}, \bibinfo {author} {\bibfnamefont {E.}~\bibnamefont {{Solano}}}, \bibinfo {author} {\bibfnamefont {F.}~\bibnamefont {{Solitro}}}, \bibinfo {author} {\bibfnamefont {D.}~\bibnamefont {{Souami}}}, \bibinfo {author} {\bibfnamefont {J.}~\bibnamefont {{Souchay}}}, \bibinfo {author}
  {\bibfnamefont {A.}~\bibnamefont {{Spagna}}}, \bibinfo {author} {\bibfnamefont {L.}~\bibnamefont {{Spina}}}, \bibinfo {author} {\bibfnamefont {F.}~\bibnamefont {{Spoto}}}, \bibinfo {author} {\bibfnamefont {I.~A.}\ \bibnamefont {{Steele}}}, \bibinfo {author} {\bibfnamefont {H.}~\bibnamefont {{Steidelm{\"u}ller}}}, \bibinfo {author} {\bibfnamefont {C.~A.}\ \bibnamefont {{Stephenson}}}, \bibinfo {author} {\bibfnamefont {M.}~\bibnamefont {{S{\"u}veges}}}, \bibinfo {author} {\bibfnamefont {J.}~\bibnamefont {{Surdej}}}, \bibinfo {author} {\bibfnamefont {L.}~\bibnamefont {{Szabados}}}, \bibinfo {author} {\bibfnamefont {E.}~\bibnamefont {{Szegedi-Elek}}}, \bibinfo {author} {\bibfnamefont {F.}~\bibnamefont {{Taris}}}, \bibinfo {author} {\bibfnamefont {M.~B.}\ \bibnamefont {{Taylor}}}, \bibinfo {author} {\bibfnamefont {R.}~\bibnamefont {{Teixeira}}}, \bibinfo {author} {\bibfnamefont {L.}~\bibnamefont {{Tolomei}}}, \bibinfo {author} {\bibfnamefont {N.}~\bibnamefont {{Tonello}}}, \bibinfo {author} {\bibfnamefont
  {F.}~\bibnamefont {{Torra}}}, \bibinfo {author} {\bibfnamefont {J.}~\bibnamefont {{Torra}}}, \bibinfo {author} {\bibfnamefont {G.}~\bibnamefont {{Torralba Elipe}}}, \bibinfo {author} {\bibfnamefont {M.}~\bibnamefont {{Trabucchi}}}, \bibinfo {author} {\bibfnamefont {A.~T.}\ \bibnamefont {{Tsounis}}}, \bibinfo {author} {\bibfnamefont {C.}~\bibnamefont {{Turon}}}, \bibinfo {author} {\bibfnamefont {A.}~\bibnamefont {{Ulla}}}, \bibinfo {author} {\bibfnamefont {N.}~\bibnamefont {{Unger}}}, \bibinfo {author} {\bibfnamefont {M.~V.}\ \bibnamefont {{Vaillant}}}, \bibinfo {author} {\bibfnamefont {E.}~\bibnamefont {{van Dillen}}}, \bibinfo {author} {\bibfnamefont {W.}~\bibnamefont {{van Reeven}}}, \bibinfo {author} {\bibfnamefont {O.}~\bibnamefont {{Vanel}}}, \bibinfo {author} {\bibfnamefont {A.}~\bibnamefont {{Vecchiato}}}, \bibinfo {author} {\bibfnamefont {Y.}~\bibnamefont {{Viala}}}, \bibinfo {author} {\bibfnamefont {D.}~\bibnamefont {{Vicente}}}, \bibinfo {author} {\bibfnamefont {S.}~\bibnamefont {{Voutsinas}}},
  \bibinfo {author} {\bibfnamefont {M.}~\bibnamefont {{Weiler}}}, \bibinfo {author} {\bibfnamefont {T.}~\bibnamefont {{Wevers}}}, \bibinfo {author} {\bibfnamefont {{\L}.}~\bibnamefont {{Wyrzykowski}}}, \bibinfo {author} {\bibfnamefont {A.}~\bibnamefont {{Yoldas}}}, \bibinfo {author} {\bibfnamefont {P.}~\bibnamefont {{Yvard}}}, \bibinfo {author} {\bibfnamefont {H.}~\bibnamefont {{Zhao}}}, \bibinfo {author} {\bibfnamefont {J.}~\bibnamefont {{Zorec}}}, \bibinfo {author} {\bibfnamefont {S.}~\bibnamefont {{Zucker}}},\ and\ \bibinfo {author} {\bibfnamefont {T.}~\bibnamefont {{Zwitter}}},\ }\href {https://doi.org/10.1051/0004-6361/202243800} {\bibfield  {journal} {\bibinfo  {journal} {\aap}\ }\textbf {\bibinfo {volume} {674}},\ \bibinfo {eid} {A39} (\bibinfo {year} {2023})},\ \Eprint {https://arxiv.org/abs/2206.05870} {arXiv:2206.05870 [astro-ph.SR]} \BibitemShut {NoStop}%
\bibitem [{\citenamefont {{Gaia Collaboration}}\ \emph {et~al.}(2018)\citenamefont {{Gaia Collaboration}}, \citenamefont {{Katz}}, \citenamefont {{Antoja}}, \citenamefont {{Romero-G{\'o}mez}}, \citenamefont {{Drimmel}}, \citenamefont {{Reyl{\'e}}}, \citenamefont {{Seabroke}}, \citenamefont {{Soubiran}}, \citenamefont {{Babusiaux}}, \citenamefont {{Di Matteo}}, \citenamefont {{Figueras}}, \citenamefont {{Poggio}}, \citenamefont {{Robin}}, \citenamefont {{Evans}}, \citenamefont {{Brown}}, \citenamefont {{Vallenari}}, \citenamefont {{Prusti}}, \citenamefont {{de Bruijne}}, \citenamefont {{Bailer-Jones}}, \citenamefont {{Biermann}}, \citenamefont {{Eyer}}, \citenamefont {{Jansen}}, \citenamefont {{Jordi}}, \citenamefont {{Klioner}}, \citenamefont {{Lammers}}, \citenamefont {{Lindegren}}, \citenamefont {{Luri}}, \citenamefont {{Mignard}}, \citenamefont {{Panem}}, \citenamefont {{Pourbaix}}, \citenamefont {{Randich}}, \citenamefont {{Sartoretti}}, \citenamefont {{Siddiqui}}, \citenamefont {{van Leeuwen}},
  \citenamefont {{Walton}}, \citenamefont {{Arenou}}, \citenamefont {{Bastian}}, \citenamefont {{Cropper}}, \citenamefont {{Lattanzi}}, \citenamefont {{Bakker}}, \citenamefont {{Cacciari}}, \citenamefont {{Casta n}}, \citenamefont {{Chaoul}}, \citenamefont {{Cheek}}, \citenamefont {{De Angeli}}, \citenamefont {{Fabricius}}, \citenamefont {{Guerra}}, \citenamefont {{Holl}}, \citenamefont {{Masana}}, \citenamefont {{Messineo}}, \citenamefont {{Mowlavi}}, \citenamefont {{Nienartowicz}}, \citenamefont {{Panuzzo}}, \citenamefont {{Portell}}, \citenamefont {{Riello}}, \citenamefont {{Tanga}}, \citenamefont {{Th{\'e}venin}}, \citenamefont {{Gracia-Abril}}, \citenamefont {{Comoretto}}, \citenamefont {{Garcia-Reinaldos}}, \citenamefont {{Teyssier}}, \citenamefont {{Altmann}}, \citenamefont {{Andrae}}, \citenamefont {{Audard}}, \citenamefont {{Bellas-Velidis}}, \citenamefont {{Benson}}, \citenamefont {{Berthier}}, \citenamefont {{Blomme}}, \citenamefont {{Burgess}}, \citenamefont {{Busso}}, \citenamefont {{Carry}},
  \citenamefont {{Cellino}}, \citenamefont {{Clementini}}, \citenamefont {{Clotet}}, \citenamefont {{Creevey}}, \citenamefont {{Davidson}}, \citenamefont {{De Ridder}}, \citenamefont {{Delchambre}}, \citenamefont {{Dell'Oro}}, \citenamefont {{Ducourant}}, \citenamefont {{Fern{\'a}ndez-Hern{\'a}ndez}}, \citenamefont {{Fouesneau}}, \citenamefont {{Fr{\'e}mat}}, \citenamefont {{Galluccio}}, \citenamefont {{Garc{\'\i}a-Torres}}, \citenamefont {{Gonz{\'a}lez-N{\'u}{\~n}ez}}, \citenamefont {{Gonz{\'a}lez-Vidal}}, \citenamefont {{Gosset}}, \citenamefont {{Guy}}, \citenamefont {{Halbwachs}}, \citenamefont {{Hambly}}, \citenamefont {{Harrison}}, \citenamefont {{Hern{\'a}ndez}}, \citenamefont {{Hestroffer}}, \citenamefont {{Hodgkin}}, \citenamefont {{Hutton}}, \citenamefont {{Jasniewicz}}, \citenamefont {{Jean-Antoine-Piccolo}}, \citenamefont {{Jordan}}, \citenamefont {{Korn}}, \citenamefont {{Krone-Martins}}, \citenamefont {{Lanzafame}}, \citenamefont {{Lebzelter}}, \citenamefont {{L{\"o}ffler}}, \citenamefont
  {{Manteiga}}, \citenamefont {{Marrese}}, \citenamefont {{Mart{\'\i}n-Fleitas}}, \citenamefont {{Moitinho}}, \citenamefont {{Mora}}, \citenamefont {{Muinonen}}, \citenamefont {{Osinde}}, \citenamefont {{Pancino}}, \citenamefont {{Pauwels}}, \citenamefont {{Petit}}, \citenamefont {{Recio-Blanco}}, \citenamefont {{Richards}}, \citenamefont {{Rimoldini}}, \citenamefont {{Sarro}}, \citenamefont {{Siopis}}, \citenamefont {{Smith}}, \citenamefont {{Sozzetti}}, \citenamefont {{S{\"u}veges}}, \citenamefont {{Torra}}, \citenamefont {{van Reeven}}, \citenamefont {{Abbas}}, \citenamefont {{Abreu Aramburu}}, \citenamefont {{Accart}}, \citenamefont {{Aerts}}, \citenamefont {{Altavilla}}, \citenamefont {{{\'A}lvarez}}, \citenamefont {{Alvarez}}, \citenamefont {{Alves}}, \citenamefont {{Anderson}}, \citenamefont {{Andrei}}, \citenamefont {{Anglada Varela}}, \citenamefont {{Antiche}}, \citenamefont {{Arcay}}, \citenamefont {{Astraatmadja}}, \citenamefont {{Bach}}, \citenamefont {{Baker}}, \citenamefont
  {{Balaguer-N{\'u}{\~n}ez}}, \citenamefont {{Balm}}, \citenamefont {{Barache}}, \citenamefont {{Barata}}, \citenamefont {{Barbato}}, \citenamefont {{Barblan}}, \citenamefont {{Barklem}}, \citenamefont {{Barrado}}, \citenamefont {{Barros}}, \citenamefont {{Barstow}}, \citenamefont {{Bartholom{\'e} Mu{\~n}oz}}, \citenamefont {{Bassilana}}, \citenamefont {{Becciani}}, \citenamefont {{Bellazzini}}, \citenamefont {{Berihuete}}, \citenamefont {{Bertone}}, \citenamefont {{Bianchi}}, \citenamefont {{Bienaym{\'e}}}, \citenamefont {{Blanco-Cuaresma}}, \citenamefont {{Boch}}, \citenamefont {{Boeche}}, \citenamefont {{Bombrun}}, \citenamefont {{Borrachero}}, \citenamefont {{Bossini}}, \citenamefont {{Bouquillon}}, \citenamefont {{Bourda}}, \citenamefont {{Bragaglia}}, \citenamefont {{Bramante}}, \citenamefont {{Breddels}}, \citenamefont {{Bressan}}, \citenamefont {{Brouillet}}, \citenamefont {{Br{\"u}semeister}}, \citenamefont {{Brugaletta}}, \citenamefont {{Bucciarelli}}, \citenamefont {{Burlacu}}, \citenamefont
  {{Busonero}}, \citenamefont {{Butkevich}}, \citenamefont {{Buzzi}}, \citenamefont {{Caffau}}, \citenamefont {{Cancelliere}}, \citenamefont {{Cannizzaro}}, \citenamefont {{Cantat-Gaudin}}, \citenamefont {{Carballo}}, \citenamefont {{Carlucci}}, \citenamefont {{Carrasco}}, \citenamefont {{Casamiquela}}, \citenamefont {{Castellani}}, \citenamefont {{Castro-Ginard}}, \citenamefont {{Charlot}}, \citenamefont {{Chemin}}, \citenamefont {{Chiavassa}}, \citenamefont {{Cocozza}}, \citenamefont {{Costigan}}, \citenamefont {{Cowell}}, \citenamefont {{Crifo}}, \citenamefont {{Crosta}}, \citenamefont {{Crowley}}, \citenamefont {{Cuypers}}, \citenamefont {{Dafonte}}, \citenamefont {{Damerdji}}, \citenamefont {{Dapergolas}}, \citenamefont {{David}}, \citenamefont {{David}}, \citenamefont {{de Laverny}}, \citenamefont {{De Luise}}, \citenamefont {{De March}}, \citenamefont {{de Souza}}, \citenamefont {{de Torres}}, \citenamefont {{Debosscher}}, \citenamefont {{del Pozo}}, \citenamefont {{Delbo}}, \citenamefont {{Delgado}},
  \citenamefont {{Delgado}}, \citenamefont {{Diakite}}, \citenamefont {{Diener}}, \citenamefont {{Distefano}}, \citenamefont {{Dolding}}, \citenamefont {{Drazinos}}, \citenamefont {{Dur{\'a}n}}, \citenamefont {{Edvardsson}}, \citenamefont {{Enke}}, \citenamefont {{Eriksson}}, \citenamefont {{Esquej}}, \citenamefont {{Eynard Bontemps}}, \citenamefont {{Fabre}}, \citenamefont {{Fabrizio}}, \citenamefont {{Faigler}}, \citenamefont {{Falc a}}, \citenamefont {{Farr{\`a}s Casas}}, \citenamefont {{Federici}}, \citenamefont {{Fedorets}}, \citenamefont {{Fernique}}, \citenamefont {{Filippi}}, \citenamefont {{Findeisen}}, \citenamefont {{Fonti}}, \citenamefont {{Fraile}}, \citenamefont {{Fraser}}, \citenamefont {{Fr{\'e}zouls}}, \citenamefont {{Gai}}, \citenamefont {{Galleti}}, \citenamefont {{Garabato}}, \citenamefont {{Garc{\'\i}a-Sedano}}, \citenamefont {{Garofalo}}, \citenamefont {{Garralda}}, \citenamefont {{Gavel}}, \citenamefont {{Gavras}}, \citenamefont {{Gerssen}}, \citenamefont {{Geyer}}, \citenamefont
  {{Giacobbe}}, \citenamefont {{Gilmore}}, \citenamefont {{Girona}}, \citenamefont {{Giuffrida}}, \citenamefont {{Glass}}, \citenamefont {{Gomes}}, \citenamefont {{Granvik}}, \citenamefont {{Gueguen}}, \citenamefont {{Guerrier}}, \citenamefont {{Guiraud}}, \citenamefont {{Guti{\'e}}}, \citenamefont {{Haigron}}, \citenamefont {{Hatzidimitriou}}, \citenamefont {{Hauser}}, \citenamefont {{Haywood}}, \citenamefont {{Heiter}}, \citenamefont {{Helmi}}, \citenamefont {{Heu}}, \citenamefont {{Hilger}}, \citenamefont {{Hobbs}}, \citenamefont {{Hofmann}}, \citenamefont {{Holland}}, \citenamefont {{Huckle}}, \citenamefont {{Hypki}}, \citenamefont {{Icardi}}, \citenamefont {{Jan{\ss}en}}, \citenamefont {{Jevardat de Fombelle}}, \citenamefont {{Jonker}}, \citenamefont {{Juh{\'a}sz}}, \citenamefont {{Julbe}}, \citenamefont {{Karampelas}}, \citenamefont {{Kewley}}, \citenamefont {{Klar}}, \citenamefont {{Kochoska}}, \citenamefont {{Kohley}}, \citenamefont {{Kolenberg}}, \citenamefont {{Kontizas}}, \citenamefont
  {{Kontizas}}, \citenamefont {{Koposov}}, \citenamefont {{Kordopatis}}, \citenamefont {{Kostrzewa-Rutkowska}}, \citenamefont {{Koubsky}}, \citenamefont {{Lambert}}, \citenamefont {{Lanza}}, \citenamefont {{Lasne}}, \citenamefont {{Lavigne}}, \citenamefont {{Le Fustec}}, \citenamefont {{Le Poncin-Lafitte}}, \citenamefont {{Lebreton}}, \citenamefont {{Leccia}}, \citenamefont {{Leclerc}}, \citenamefont {{Lecoeur-Taibi}}, \citenamefont {{Lenhardt}}, \citenamefont {{Leroux}}, \citenamefont {{Liao}}, \citenamefont {{Licata}}, \citenamefont {{Lindstr{\o}m}}, \citenamefont {{Lister}}, \citenamefont {{Livanou}}, \citenamefont {{Lobel}}, \citenamefont {{L{\'o}pez}}, \citenamefont {{Managau}}, \citenamefont {{Mann}}, \citenamefont {{Mantelet}}, \citenamefont {{Marchal}}, \citenamefont {{Marchant}}, \citenamefont {{Marconi}}, \citenamefont {{Marinoni}}, \citenamefont {{Marschalk{\'o}}}, \citenamefont {{Marshall}}, \citenamefont {{Martino}}, \citenamefont {{Marton}}, \citenamefont {{Mary}}, \citenamefont {{Massari}},
  \citenamefont {{Matijevi{\v{c}}}}, \citenamefont {{Mazeh}}, \citenamefont {{McMillan}}, \citenamefont {{Messina}}, \citenamefont {{Michalik}}, \citenamefont {{Millar}}, \citenamefont {{Molina}}, \citenamefont {{Molinaro}}, \citenamefont {{Moln{\'a}r}}, \citenamefont {{Montegriffo}}, \citenamefont {{Mor}}, \citenamefont {{Morbidelli}}, \citenamefont {{Morel}}, \citenamefont {{Morris}}, \citenamefont {{Mulone}}, \citenamefont {{Muraveva}}, \citenamefont {{Musella}}, \citenamefont {{Nelemans}}, \citenamefont {{Nicastro}}, \citenamefont {{Noval}}, \citenamefont {{O'Mullane}}, \citenamefont {{Ord{\'e}novic}}, \citenamefont {{Ord{\'o}{\~n}ez-Blanco}}, \citenamefont {{Osborne}}, \citenamefont {{Pagani}}, \citenamefont {{Pagano}}, \citenamefont {{Pailler}}, \citenamefont {{Palacin}}, \citenamefont {{Palaversa}}, \citenamefont {{Panahi}}, \citenamefont {{Pawlak}}, \citenamefont {{Piersimoni}}, \citenamefont {{Pineau}}, \citenamefont {{Plachy}}, \citenamefont {{Plum}}, \citenamefont {{Poujoulet}}, \citenamefont
  {{Pr{\v{s}}a}}, \citenamefont {{Pulone}}, \citenamefont {{Racero}}, \citenamefont {{Ragaini}}, \citenamefont {{Rambaux}}, \citenamefont {{Ramos-Lerate}}, \citenamefont {{Regibo}}, \citenamefont {{Riclet}}, \citenamefont {{Ripepi}}, \citenamefont {{Riva}}, \citenamefont {{Rivard}}, \citenamefont {{Rixon}}, \citenamefont {{Roegiers}}, \citenamefont {{Roelens}}, \citenamefont {{Rowell}}, \citenamefont {{Royer}}, \citenamefont {{Ruiz-Dern}}, \citenamefont {{Sadowski}}, \citenamefont {{Sagrist{\`a} Sell{\'e}s}}, \citenamefont {{Sahlmann}}, \citenamefont {{Salgado}}, \citenamefont {{Salguero}}, \citenamefont {{Sanna}}, \citenamefont {{Santana-Ros}}, \citenamefont {{Sarasso}}, \citenamefont {{Savietto}}, \citenamefont {{Schultheis}}, \citenamefont {{Sciacca}}, \citenamefont {{Segol}}, \citenamefont {{Segovia}}, \citenamefont {{S{\'e}gransan}}, \citenamefont {{Shih}}, \citenamefont {{Siltala}}, \citenamefont {{Silva}}, \citenamefont {{Smart}}, \citenamefont {{Smith}}, \citenamefont {{Solano}}, \citenamefont
  {{Solitro}}, \citenamefont {{Sordo}}, \citenamefont {{Soria Nieto}}, \citenamefont {{Souchay}}, \citenamefont {{Spagna}}, \citenamefont {{Spoto}}, \citenamefont {{Stampa}}, \citenamefont {{Steele}}, \citenamefont {{Steidelm{\"u}ller}}, \citenamefont {{Stephenson}}, \citenamefont {{Stoev}}, \citenamefont {{Suess}}, \citenamefont {{Surdej}}, \citenamefont {{Szabados}}, \citenamefont {{Szegedi-Elek}}, \citenamefont {{Tapiador}}, \citenamefont {{Taris}}, \citenamefont {{Tauran}}, \citenamefont {{Taylor}}, \citenamefont {{Teixeira}}, \citenamefont {{Terrett}}, \citenamefont {{Teyssandier}}, \citenamefont {{Thuillot}}, \citenamefont {{Titarenko}}, \citenamefont {{Torra Clotet}}, \citenamefont {{Turon}}, \citenamefont {{Ulla}}, \citenamefont {{Utrilla}}, \citenamefont {{Uzzi}}, \citenamefont {{Vaillant}}, \citenamefont {{Valentini}}, \citenamefont {{Valette}}, \citenamefont {{van Elteren}}, \citenamefont {{Van Hemelryck}}, \citenamefont {{van Leeuwen}}, \citenamefont {{Vaschetto}}, \citenamefont {{Vecchiato}},
  \citenamefont {{Veljanoski}}, \citenamefont {{Viala}}, \citenamefont {{Vicente}}, \citenamefont {{Vogt}}, \citenamefont {{von Essen}}, \citenamefont {{Voss}}, \citenamefont {{Votruba}}, \citenamefont {{Voutsinas}}, \citenamefont {{Walmsley}}, \citenamefont {{Weiler}}, \citenamefont {{Wertz}}, \citenamefont {{Wevers}}, \citenamefont {{Wyrzykowski}}, \citenamefont {{Yoldas}}, \citenamefont {{{\v{Z}}erjal}}, \citenamefont {{Ziaeepour}}, \citenamefont {{Zorec}}, \citenamefont {{Zschocke}}, \citenamefont {{Zucker}}, \citenamefont {{Zurbach}},\ and\ \citenamefont {{Zwitter}}}]{GaiaDR2_maps}%
  \BibitemOpen
  \bibfield  {author} {\bibinfo {author} {\bibnamefont {{Gaia Collaboration}}}, \bibinfo {author} {\bibfnamefont {D.}~\bibnamefont {{Katz}}}, \bibinfo {author} {\bibfnamefont {T.}~\bibnamefont {{Antoja}}}, \bibinfo {author} {\bibfnamefont {M.}~\bibnamefont {{Romero-G{\'o}mez}}}, \bibinfo {author} {\bibfnamefont {R.}~\bibnamefont {{Drimmel}}}, \bibinfo {author} {\bibfnamefont {C.}~\bibnamefont {{Reyl{\'e}}}}, \bibinfo {author} {\bibfnamefont {G.~M.}\ \bibnamefont {{Seabroke}}}, \bibinfo {author} {\bibfnamefont {C.}~\bibnamefont {{Soubiran}}}, \bibinfo {author} {\bibfnamefont {C.}~\bibnamefont {{Babusiaux}}}, \bibinfo {author} {\bibfnamefont {P.}~\bibnamefont {{Di Matteo}}}, \bibinfo {author} {\bibfnamefont {F.}~\bibnamefont {{Figueras}}}, \bibinfo {author} {\bibfnamefont {E.}~\bibnamefont {{Poggio}}}, \bibinfo {author} {\bibfnamefont {A.~C.}\ \bibnamefont {{Robin}}}, \bibinfo {author} {\bibfnamefont {D.~W.}\ \bibnamefont {{Evans}}}, \bibinfo {author} {\bibfnamefont {A.~G.~A.}\ \bibnamefont {{Brown}}}, \bibinfo
  {author} {\bibfnamefont {A.}~\bibnamefont {{Vallenari}}}, \bibinfo {author} {\bibfnamefont {T.}~\bibnamefont {{Prusti}}}, \bibinfo {author} {\bibfnamefont {J.~H.~J.}\ \bibnamefont {{de Bruijne}}}, \bibinfo {author} {\bibfnamefont {C.~A.~L.}\ \bibnamefont {{Bailer-Jones}}}, \bibinfo {author} {\bibfnamefont {M.}~\bibnamefont {{Biermann}}}, \bibinfo {author} {\bibfnamefont {L.}~\bibnamefont {{Eyer}}}, \bibinfo {author} {\bibfnamefont {F.}~\bibnamefont {{Jansen}}}, \bibinfo {author} {\bibfnamefont {C.}~\bibnamefont {{Jordi}}}, \bibinfo {author} {\bibfnamefont {S.~A.}\ \bibnamefont {{Klioner}}}, \bibinfo {author} {\bibfnamefont {U.}~\bibnamefont {{Lammers}}}, \bibinfo {author} {\bibfnamefont {L.}~\bibnamefont {{Lindegren}}}, \bibinfo {author} {\bibfnamefont {X.}~\bibnamefont {{Luri}}}, \bibinfo {author} {\bibfnamefont {F.}~\bibnamefont {{Mignard}}}, \bibinfo {author} {\bibfnamefont {C.}~\bibnamefont {{Panem}}}, \bibinfo {author} {\bibfnamefont {D.}~\bibnamefont {{Pourbaix}}}, \bibinfo {author} {\bibfnamefont
  {S.}~\bibnamefont {{Randich}}}, \bibinfo {author} {\bibfnamefont {P.}~\bibnamefont {{Sartoretti}}}, \bibinfo {author} {\bibfnamefont {H.~I.}\ \bibnamefont {{Siddiqui}}}, \bibinfo {author} {\bibfnamefont {F.}~\bibnamefont {{van Leeuwen}}}, \bibinfo {author} {\bibfnamefont {N.~A.}\ \bibnamefont {{Walton}}}, \bibinfo {author} {\bibfnamefont {F.}~\bibnamefont {{Arenou}}}, \bibinfo {author} {\bibfnamefont {U.}~\bibnamefont {{Bastian}}}, \bibinfo {author} {\bibfnamefont {M.}~\bibnamefont {{Cropper}}}, \bibinfo {author} {\bibfnamefont {M.~G.}\ \bibnamefont {{Lattanzi}}}, \bibinfo {author} {\bibfnamefont {J.}~\bibnamefont {{Bakker}}}, \bibinfo {author} {\bibfnamefont {C.}~\bibnamefont {{Cacciari}}}, \bibinfo {author} {\bibfnamefont {J.}~\bibnamefont {{Casta n}}}, \bibinfo {author} {\bibfnamefont {L.}~\bibnamefont {{Chaoul}}}, \bibinfo {author} {\bibfnamefont {N.}~\bibnamefont {{Cheek}}}, \bibinfo {author} {\bibfnamefont {F.}~\bibnamefont {{De Angeli}}}, \bibinfo {author} {\bibfnamefont {C.}~\bibnamefont
  {{Fabricius}}}, \bibinfo {author} {\bibfnamefont {R.}~\bibnamefont {{Guerra}}}, \bibinfo {author} {\bibfnamefont {B.}~\bibnamefont {{Holl}}}, \bibinfo {author} {\bibfnamefont {E.}~\bibnamefont {{Masana}}}, \bibinfo {author} {\bibfnamefont {R.}~\bibnamefont {{Messineo}}}, \bibinfo {author} {\bibfnamefont {N.}~\bibnamefont {{Mowlavi}}}, \bibinfo {author} {\bibfnamefont {K.}~\bibnamefont {{Nienartowicz}}}, \bibinfo {author} {\bibfnamefont {P.}~\bibnamefont {{Panuzzo}}}, \bibinfo {author} {\bibfnamefont {J.}~\bibnamefont {{Portell}}}, \bibinfo {author} {\bibfnamefont {M.}~\bibnamefont {{Riello}}}, \bibinfo {author} {\bibfnamefont {P.}~\bibnamefont {{Tanga}}}, \bibinfo {author} {\bibfnamefont {F.}~\bibnamefont {{Th{\'e}venin}}}, \bibinfo {author} {\bibfnamefont {G.}~\bibnamefont {{Gracia-Abril}}}, \bibinfo {author} {\bibfnamefont {G.}~\bibnamefont {{Comoretto}}}, \bibinfo {author} {\bibfnamefont {M.}~\bibnamefont {{Garcia-Reinaldos}}}, \bibinfo {author} {\bibfnamefont {D.}~\bibnamefont {{Teyssier}}}, \bibinfo
  {author} {\bibfnamefont {M.}~\bibnamefont {{Altmann}}}, \bibinfo {author} {\bibfnamefont {R.}~\bibnamefont {{Andrae}}}, \bibinfo {author} {\bibfnamefont {M.}~\bibnamefont {{Audard}}}, \bibinfo {author} {\bibfnamefont {I.}~\bibnamefont {{Bellas-Velidis}}}, \bibinfo {author} {\bibfnamefont {K.}~\bibnamefont {{Benson}}}, \bibinfo {author} {\bibfnamefont {J.}~\bibnamefont {{Berthier}}}, \bibinfo {author} {\bibfnamefont {R.}~\bibnamefont {{Blomme}}}, \bibinfo {author} {\bibfnamefont {P.}~\bibnamefont {{Burgess}}}, \bibinfo {author} {\bibfnamefont {G.}~\bibnamefont {{Busso}}}, \bibinfo {author} {\bibfnamefont {B.}~\bibnamefont {{Carry}}}, \bibinfo {author} {\bibfnamefont {A.}~\bibnamefont {{Cellino}}}, \bibinfo {author} {\bibfnamefont {G.}~\bibnamefont {{Clementini}}}, \bibinfo {author} {\bibfnamefont {M.}~\bibnamefont {{Clotet}}}, \bibinfo {author} {\bibfnamefont {O.}~\bibnamefont {{Creevey}}}, \bibinfo {author} {\bibfnamefont {M.}~\bibnamefont {{Davidson}}}, \bibinfo {author} {\bibfnamefont {J.}~\bibnamefont
  {{De Ridder}}}, \bibinfo {author} {\bibfnamefont {L.}~\bibnamefont {{Delchambre}}}, \bibinfo {author} {\bibfnamefont {A.}~\bibnamefont {{Dell'Oro}}}, \bibinfo {author} {\bibfnamefont {C.}~\bibnamefont {{Ducourant}}}, \bibinfo {author} {\bibfnamefont {J.}~\bibnamefont {{Fern{\'a}ndez-Hern{\'a}ndez}}}, \bibinfo {author} {\bibfnamefont {M.}~\bibnamefont {{Fouesneau}}}, \bibinfo {author} {\bibfnamefont {Y.}~\bibnamefont {{Fr{\'e}mat}}}, \bibinfo {author} {\bibfnamefont {L.}~\bibnamefont {{Galluccio}}}, \bibinfo {author} {\bibfnamefont {M.}~\bibnamefont {{Garc{\'\i}a-Torres}}}, \bibinfo {author} {\bibfnamefont {J.}~\bibnamefont {{Gonz{\'a}lez-N{\'u}{\~n}ez}}}, \bibinfo {author} {\bibfnamefont {J.~J.}\ \bibnamefont {{Gonz{\'a}lez-Vidal}}}, \bibinfo {author} {\bibfnamefont {E.}~\bibnamefont {{Gosset}}}, \bibinfo {author} {\bibfnamefont {L.~P.}\ \bibnamefont {{Guy}}}, \bibinfo {author} {\bibfnamefont {J.~L.}\ \bibnamefont {{Halbwachs}}}, \bibinfo {author} {\bibfnamefont {N.~C.}\ \bibnamefont {{Hambly}}}, \bibinfo
  {author} {\bibfnamefont {D.~L.}\ \bibnamefont {{Harrison}}}, \bibinfo {author} {\bibfnamefont {J.}~\bibnamefont {{Hern{\'a}ndez}}}, \bibinfo {author} {\bibfnamefont {D.}~\bibnamefont {{Hestroffer}}}, \bibinfo {author} {\bibfnamefont {S.~T.}\ \bibnamefont {{Hodgkin}}}, \bibinfo {author} {\bibfnamefont {A.}~\bibnamefont {{Hutton}}}, \bibinfo {author} {\bibfnamefont {G.}~\bibnamefont {{Jasniewicz}}}, \bibinfo {author} {\bibfnamefont {A.}~\bibnamefont {{Jean-Antoine-Piccolo}}}, \bibinfo {author} {\bibfnamefont {S.}~\bibnamefont {{Jordan}}}, \bibinfo {author} {\bibfnamefont {A.~J.}\ \bibnamefont {{Korn}}}, \bibinfo {author} {\bibfnamefont {A.}~\bibnamefont {{Krone-Martins}}}, \bibinfo {author} {\bibfnamefont {A.~C.}\ \bibnamefont {{Lanzafame}}}, \bibinfo {author} {\bibfnamefont {T.}~\bibnamefont {{Lebzelter}}}, \bibinfo {author} {\bibfnamefont {W.}~\bibnamefont {{L{\"o}ffler}}}, \bibinfo {author} {\bibfnamefont {M.}~\bibnamefont {{Manteiga}}}, \bibinfo {author} {\bibfnamefont {P.~M.}\ \bibnamefont {{Marrese}}},
  \bibinfo {author} {\bibfnamefont {J.~M.}\ \bibnamefont {{Mart{\'\i}n-Fleitas}}}, \bibinfo {author} {\bibfnamefont {A.}~\bibnamefont {{Moitinho}}}, \bibinfo {author} {\bibfnamefont {A.}~\bibnamefont {{Mora}}}, \bibinfo {author} {\bibfnamefont {K.}~\bibnamefont {{Muinonen}}}, \bibinfo {author} {\bibfnamefont {J.}~\bibnamefont {{Osinde}}}, \bibinfo {author} {\bibfnamefont {E.}~\bibnamefont {{Pancino}}}, \bibinfo {author} {\bibfnamefont {T.}~\bibnamefont {{Pauwels}}}, \bibinfo {author} {\bibfnamefont {J.~M.}\ \bibnamefont {{Petit}}}, \bibinfo {author} {\bibfnamefont {A.}~\bibnamefont {{Recio-Blanco}}}, \bibinfo {author} {\bibfnamefont {P.~J.}\ \bibnamefont {{Richards}}}, \bibinfo {author} {\bibfnamefont {L.}~\bibnamefont {{Rimoldini}}}, \bibinfo {author} {\bibfnamefont {L.~M.}\ \bibnamefont {{Sarro}}}, \bibinfo {author} {\bibfnamefont {C.}~\bibnamefont {{Siopis}}}, \bibinfo {author} {\bibfnamefont {M.}~\bibnamefont {{Smith}}}, \bibinfo {author} {\bibfnamefont {A.}~\bibnamefont {{Sozzetti}}}, \bibinfo {author}
  {\bibfnamefont {M.}~\bibnamefont {{S{\"u}veges}}}, \bibinfo {author} {\bibfnamefont {J.}~\bibnamefont {{Torra}}}, \bibinfo {author} {\bibfnamefont {W.}~\bibnamefont {{van Reeven}}}, \bibinfo {author} {\bibfnamefont {U.}~\bibnamefont {{Abbas}}}, \bibinfo {author} {\bibfnamefont {A.}~\bibnamefont {{Abreu Aramburu}}}, \bibinfo {author} {\bibfnamefont {S.}~\bibnamefont {{Accart}}}, \bibinfo {author} {\bibfnamefont {C.}~\bibnamefont {{Aerts}}}, \bibinfo {author} {\bibfnamefont {G.}~\bibnamefont {{Altavilla}}}, \bibinfo {author} {\bibfnamefont {M.~A.}\ \bibnamefont {{{\'A}lvarez}}}, \bibinfo {author} {\bibfnamefont {R.}~\bibnamefont {{Alvarez}}}, \bibinfo {author} {\bibfnamefont {J.}~\bibnamefont {{Alves}}}, \bibinfo {author} {\bibfnamefont {R.~I.}\ \bibnamefont {{Anderson}}}, \bibinfo {author} {\bibfnamefont {A.~H.}\ \bibnamefont {{Andrei}}}, \bibinfo {author} {\bibfnamefont {E.}~\bibnamefont {{Anglada Varela}}}, \bibinfo {author} {\bibfnamefont {E.}~\bibnamefont {{Antiche}}}, \bibinfo {author} {\bibfnamefont
  {B.}~\bibnamefont {{Arcay}}}, \bibinfo {author} {\bibfnamefont {T.~L.}\ \bibnamefont {{Astraatmadja}}}, \bibinfo {author} {\bibfnamefont {N.}~\bibnamefont {{Bach}}}, \bibinfo {author} {\bibfnamefont {S.~G.}\ \bibnamefont {{Baker}}}, \bibinfo {author} {\bibfnamefont {L.}~\bibnamefont {{Balaguer-N{\'u}{\~n}ez}}}, \bibinfo {author} {\bibfnamefont {P.}~\bibnamefont {{Balm}}}, \bibinfo {author} {\bibfnamefont {C.}~\bibnamefont {{Barache}}}, \bibinfo {author} {\bibfnamefont {C.}~\bibnamefont {{Barata}}}, \bibinfo {author} {\bibfnamefont {D.}~\bibnamefont {{Barbato}}}, \bibinfo {author} {\bibfnamefont {F.}~\bibnamefont {{Barblan}}}, \bibinfo {author} {\bibfnamefont {P.~S.}\ \bibnamefont {{Barklem}}}, \bibinfo {author} {\bibfnamefont {D.}~\bibnamefont {{Barrado}}}, \bibinfo {author} {\bibfnamefont {M.}~\bibnamefont {{Barros}}}, \bibinfo {author} {\bibfnamefont {M.~A.}\ \bibnamefont {{Barstow}}}, \bibinfo {author} {\bibfnamefont {L.}~\bibnamefont {{Bartholom{\'e} Mu{\~n}oz}}}, \bibinfo {author} {\bibfnamefont
  {J.~L.}\ \bibnamefont {{Bassilana}}}, \bibinfo {author} {\bibfnamefont {U.}~\bibnamefont {{Becciani}}}, \bibinfo {author} {\bibfnamefont {M.}~\bibnamefont {{Bellazzini}}}, \bibinfo {author} {\bibfnamefont {A.}~\bibnamefont {{Berihuete}}}, \bibinfo {author} {\bibfnamefont {S.}~\bibnamefont {{Bertone}}}, \bibinfo {author} {\bibfnamefont {L.}~\bibnamefont {{Bianchi}}}, \bibinfo {author} {\bibfnamefont {O.}~\bibnamefont {{Bienaym{\'e}}}}, \bibinfo {author} {\bibfnamefont {S.}~\bibnamefont {{Blanco-Cuaresma}}}, \bibinfo {author} {\bibfnamefont {T.}~\bibnamefont {{Boch}}}, \bibinfo {author} {\bibfnamefont {C.}~\bibnamefont {{Boeche}}}, \bibinfo {author} {\bibfnamefont {A.}~\bibnamefont {{Bombrun}}}, \bibinfo {author} {\bibfnamefont {R.}~\bibnamefont {{Borrachero}}}, \bibinfo {author} {\bibfnamefont {D.}~\bibnamefont {{Bossini}}}, \bibinfo {author} {\bibfnamefont {S.}~\bibnamefont {{Bouquillon}}}, \bibinfo {author} {\bibfnamefont {G.}~\bibnamefont {{Bourda}}}, \bibinfo {author} {\bibfnamefont {A.}~\bibnamefont
  {{Bragaglia}}}, \bibinfo {author} {\bibfnamefont {L.}~\bibnamefont {{Bramante}}}, \bibinfo {author} {\bibfnamefont {M.~A.}\ \bibnamefont {{Breddels}}}, \bibinfo {author} {\bibfnamefont {A.}~\bibnamefont {{Bressan}}}, \bibinfo {author} {\bibfnamefont {N.}~\bibnamefont {{Brouillet}}}, \bibinfo {author} {\bibfnamefont {T.}~\bibnamefont {{Br{\"u}semeister}}}, \bibinfo {author} {\bibfnamefont {E.}~\bibnamefont {{Brugaletta}}}, \bibinfo {author} {\bibfnamefont {B.}~\bibnamefont {{Bucciarelli}}}, \bibinfo {author} {\bibfnamefont {A.}~\bibnamefont {{Burlacu}}}, \bibinfo {author} {\bibfnamefont {D.}~\bibnamefont {{Busonero}}}, \bibinfo {author} {\bibfnamefont {A.~G.}\ \bibnamefont {{Butkevich}}}, \bibinfo {author} {\bibfnamefont {R.}~\bibnamefont {{Buzzi}}}, \bibinfo {author} {\bibfnamefont {E.}~\bibnamefont {{Caffau}}}, \bibinfo {author} {\bibfnamefont {R.}~\bibnamefont {{Cancelliere}}}, \bibinfo {author} {\bibfnamefont {G.}~\bibnamefont {{Cannizzaro}}}, \bibinfo {author} {\bibfnamefont {T.}~\bibnamefont
  {{Cantat-Gaudin}}}, \bibinfo {author} {\bibfnamefont {R.}~\bibnamefont {{Carballo}}}, \bibinfo {author} {\bibfnamefont {T.}~\bibnamefont {{Carlucci}}}, \bibinfo {author} {\bibfnamefont {J.~M.}\ \bibnamefont {{Carrasco}}}, \bibinfo {author} {\bibfnamefont {L.}~\bibnamefont {{Casamiquela}}}, \bibinfo {author} {\bibfnamefont {M.}~\bibnamefont {{Castellani}}}, \bibinfo {author} {\bibfnamefont {A.}~\bibnamefont {{Castro-Ginard}}}, \bibinfo {author} {\bibfnamefont {P.}~\bibnamefont {{Charlot}}}, \bibinfo {author} {\bibfnamefont {L.}~\bibnamefont {{Chemin}}}, \bibinfo {author} {\bibfnamefont {A.}~\bibnamefont {{Chiavassa}}}, \bibinfo {author} {\bibfnamefont {G.}~\bibnamefont {{Cocozza}}}, \bibinfo {author} {\bibfnamefont {G.}~\bibnamefont {{Costigan}}}, \bibinfo {author} {\bibfnamefont {S.}~\bibnamefont {{Cowell}}}, \bibinfo {author} {\bibfnamefont {F.}~\bibnamefont {{Crifo}}}, \bibinfo {author} {\bibfnamefont {M.}~\bibnamefont {{Crosta}}}, \bibinfo {author} {\bibfnamefont {C.}~\bibnamefont {{Crowley}}}, \bibinfo
  {author} {\bibfnamefont {J.}~\bibnamefont {{Cuypers}}}, \bibinfo {author} {\bibfnamefont {C.}~\bibnamefont {{Dafonte}}}, \bibinfo {author} {\bibfnamefont {Y.}~\bibnamefont {{Damerdji}}}, \bibinfo {author} {\bibfnamefont {A.}~\bibnamefont {{Dapergolas}}}, \bibinfo {author} {\bibfnamefont {P.}~\bibnamefont {{David}}}, \bibinfo {author} {\bibfnamefont {M.}~\bibnamefont {{David}}}, \bibinfo {author} {\bibfnamefont {P.}~\bibnamefont {{de Laverny}}}, \bibinfo {author} {\bibfnamefont {F.}~\bibnamefont {{De Luise}}}, \bibinfo {author} {\bibfnamefont {R.}~\bibnamefont {{De March}}}, \bibinfo {author} {\bibfnamefont {R.}~\bibnamefont {{de Souza}}}, \bibinfo {author} {\bibfnamefont {A.}~\bibnamefont {{de Torres}}}, \bibinfo {author} {\bibfnamefont {J.}~\bibnamefont {{Debosscher}}}, \bibinfo {author} {\bibfnamefont {E.}~\bibnamefont {{del Pozo}}}, \bibinfo {author} {\bibfnamefont {M.}~\bibnamefont {{Delbo}}}, \bibinfo {author} {\bibfnamefont {A.}~\bibnamefont {{Delgado}}}, \bibinfo {author} {\bibfnamefont {H.~E.}\
  \bibnamefont {{Delgado}}}, \bibinfo {author} {\bibfnamefont {S.}~\bibnamefont {{Diakite}}}, \bibinfo {author} {\bibfnamefont {C.}~\bibnamefont {{Diener}}}, \bibinfo {author} {\bibfnamefont {E.}~\bibnamefont {{Distefano}}}, \bibinfo {author} {\bibfnamefont {C.}~\bibnamefont {{Dolding}}}, \bibinfo {author} {\bibfnamefont {P.}~\bibnamefont {{Drazinos}}}, \bibinfo {author} {\bibfnamefont {J.}~\bibnamefont {{Dur{\'a}n}}}, \bibinfo {author} {\bibfnamefont {B.}~\bibnamefont {{Edvardsson}}}, \bibinfo {author} {\bibfnamefont {H.}~\bibnamefont {{Enke}}}, \bibinfo {author} {\bibfnamefont {K.}~\bibnamefont {{Eriksson}}}, \bibinfo {author} {\bibfnamefont {P.}~\bibnamefont {{Esquej}}}, \bibinfo {author} {\bibfnamefont {G.}~\bibnamefont {{Eynard Bontemps}}}, \bibinfo {author} {\bibfnamefont {C.}~\bibnamefont {{Fabre}}}, \bibinfo {author} {\bibfnamefont {M.}~\bibnamefont {{Fabrizio}}}, \bibinfo {author} {\bibfnamefont {S.}~\bibnamefont {{Faigler}}}, \bibinfo {author} {\bibfnamefont {A.~J.}\ \bibnamefont {{Falc a}}},
  \bibinfo {author} {\bibfnamefont {M.}~\bibnamefont {{Farr{\`a}s Casas}}}, \bibinfo {author} {\bibfnamefont {L.}~\bibnamefont {{Federici}}}, \bibinfo {author} {\bibfnamefont {G.}~\bibnamefont {{Fedorets}}}, \bibinfo {author} {\bibfnamefont {P.}~\bibnamefont {{Fernique}}}, \bibinfo {author} {\bibfnamefont {F.}~\bibnamefont {{Filippi}}}, \bibinfo {author} {\bibfnamefont {K.}~\bibnamefont {{Findeisen}}}, \bibinfo {author} {\bibfnamefont {A.}~\bibnamefont {{Fonti}}}, \bibinfo {author} {\bibfnamefont {E.}~\bibnamefont {{Fraile}}}, \bibinfo {author} {\bibfnamefont {M.}~\bibnamefont {{Fraser}}}, \bibinfo {author} {\bibfnamefont {B.}~\bibnamefont {{Fr{\'e}zouls}}}, \bibinfo {author} {\bibfnamefont {M.}~\bibnamefont {{Gai}}}, \bibinfo {author} {\bibfnamefont {S.}~\bibnamefont {{Galleti}}}, \bibinfo {author} {\bibfnamefont {D.}~\bibnamefont {{Garabato}}}, \bibinfo {author} {\bibfnamefont {F.}~\bibnamefont {{Garc{\'\i}a-Sedano}}}, \bibinfo {author} {\bibfnamefont {A.}~\bibnamefont {{Garofalo}}}, \bibinfo {author}
  {\bibfnamefont {N.}~\bibnamefont {{Garralda}}}, \bibinfo {author} {\bibfnamefont {A.}~\bibnamefont {{Gavel}}}, \bibinfo {author} {\bibfnamefont {P.}~\bibnamefont {{Gavras}}}, \bibinfo {author} {\bibfnamefont {J.}~\bibnamefont {{Gerssen}}}, \bibinfo {author} {\bibfnamefont {R.}~\bibnamefont {{Geyer}}}, \bibinfo {author} {\bibfnamefont {P.}~\bibnamefont {{Giacobbe}}}, \bibinfo {author} {\bibfnamefont {G.}~\bibnamefont {{Gilmore}}}, \bibinfo {author} {\bibfnamefont {S.}~\bibnamefont {{Girona}}}, \bibinfo {author} {\bibfnamefont {G.}~\bibnamefont {{Giuffrida}}}, \bibinfo {author} {\bibfnamefont {F.}~\bibnamefont {{Glass}}}, \bibinfo {author} {\bibfnamefont {M.}~\bibnamefont {{Gomes}}}, \bibinfo {author} {\bibfnamefont {M.}~\bibnamefont {{Granvik}}}, \bibinfo {author} {\bibfnamefont {A.}~\bibnamefont {{Gueguen}}}, \bibinfo {author} {\bibfnamefont {A.}~\bibnamefont {{Guerrier}}}, \bibinfo {author} {\bibfnamefont {J.}~\bibnamefont {{Guiraud}}}, \bibinfo {author} {\bibfnamefont {R.}~\bibnamefont {{Guti{\'e}}}},
  \bibinfo {author} {\bibfnamefont {R.}~\bibnamefont {{Haigron}}}, \bibinfo {author} {\bibfnamefont {D.}~\bibnamefont {{Hatzidimitriou}}}, \bibinfo {author} {\bibfnamefont {M.}~\bibnamefont {{Hauser}}}, \bibinfo {author} {\bibfnamefont {M.}~\bibnamefont {{Haywood}}}, \bibinfo {author} {\bibfnamefont {U.}~\bibnamefont {{Heiter}}}, \bibinfo {author} {\bibfnamefont {A.}~\bibnamefont {{Helmi}}}, \bibinfo {author} {\bibfnamefont {J.}~\bibnamefont {{Heu}}}, \bibinfo {author} {\bibfnamefont {T.}~\bibnamefont {{Hilger}}}, \bibinfo {author} {\bibfnamefont {D.}~\bibnamefont {{Hobbs}}}, \bibinfo {author} {\bibfnamefont {W.}~\bibnamefont {{Hofmann}}}, \bibinfo {author} {\bibfnamefont {G.}~\bibnamefont {{Holland}}}, \bibinfo {author} {\bibfnamefont {H.~E.}\ \bibnamefont {{Huckle}}}, \bibinfo {author} {\bibfnamefont {A.}~\bibnamefont {{Hypki}}}, \bibinfo {author} {\bibfnamefont {V.}~\bibnamefont {{Icardi}}}, \bibinfo {author} {\bibfnamefont {K.}~\bibnamefont {{Jan{\ss}en}}}, \bibinfo {author} {\bibfnamefont
  {G.}~\bibnamefont {{Jevardat de Fombelle}}}, \bibinfo {author} {\bibfnamefont {P.~G.}\ \bibnamefont {{Jonker}}}, \bibinfo {author} {\bibfnamefont {{\'A}.~L.}\ \bibnamefont {{Juh{\'a}sz}}}, \bibinfo {author} {\bibfnamefont {F.}~\bibnamefont {{Julbe}}}, \bibinfo {author} {\bibfnamefont {A.}~\bibnamefont {{Karampelas}}}, \bibinfo {author} {\bibfnamefont {A.}~\bibnamefont {{Kewley}}}, \bibinfo {author} {\bibfnamefont {J.}~\bibnamefont {{Klar}}}, \bibinfo {author} {\bibfnamefont {A.}~\bibnamefont {{Kochoska}}}, \bibinfo {author} {\bibfnamefont {R.}~\bibnamefont {{Kohley}}}, \bibinfo {author} {\bibfnamefont {K.}~\bibnamefont {{Kolenberg}}}, \bibinfo {author} {\bibfnamefont {M.}~\bibnamefont {{Kontizas}}}, \bibinfo {author} {\bibfnamefont {E.}~\bibnamefont {{Kontizas}}}, \bibinfo {author} {\bibfnamefont {S.~E.}\ \bibnamefont {{Koposov}}}, \bibinfo {author} {\bibfnamefont {G.}~\bibnamefont {{Kordopatis}}}, \bibinfo {author} {\bibfnamefont {Z.}~\bibnamefont {{Kostrzewa-Rutkowska}}}, \bibinfo {author} {\bibfnamefont
  {P.}~\bibnamefont {{Koubsky}}}, \bibinfo {author} {\bibfnamefont {S.}~\bibnamefont {{Lambert}}}, \bibinfo {author} {\bibfnamefont {A.~F.}\ \bibnamefont {{Lanza}}}, \bibinfo {author} {\bibfnamefont {Y.}~\bibnamefont {{Lasne}}}, \bibinfo {author} {\bibfnamefont {J.~B.}\ \bibnamefont {{Lavigne}}}, \bibinfo {author} {\bibfnamefont {Y.}~\bibnamefont {{Le Fustec}}}, \bibinfo {author} {\bibfnamefont {C.}~\bibnamefont {{Le Poncin-Lafitte}}}, \bibinfo {author} {\bibfnamefont {Y.}~\bibnamefont {{Lebreton}}}, \bibinfo {author} {\bibfnamefont {S.}~\bibnamefont {{Leccia}}}, \bibinfo {author} {\bibfnamefont {N.}~\bibnamefont {{Leclerc}}}, \bibinfo {author} {\bibfnamefont {I.}~\bibnamefont {{Lecoeur-Taibi}}}, \bibinfo {author} {\bibfnamefont {H.}~\bibnamefont {{Lenhardt}}}, \bibinfo {author} {\bibfnamefont {F.}~\bibnamefont {{Leroux}}}, \bibinfo {author} {\bibfnamefont {S.}~\bibnamefont {{Liao}}}, \bibinfo {author} {\bibfnamefont {E.}~\bibnamefont {{Licata}}}, \bibinfo {author} {\bibfnamefont {H.~E.~P.}\ \bibnamefont
  {{Lindstr{\o}m}}}, \bibinfo {author} {\bibfnamefont {T.~A.}\ \bibnamefont {{Lister}}}, \bibinfo {author} {\bibfnamefont {E.}~\bibnamefont {{Livanou}}}, \bibinfo {author} {\bibfnamefont {A.}~\bibnamefont {{Lobel}}}, \bibinfo {author} {\bibfnamefont {M.}~\bibnamefont {{L{\'o}pez}}}, \bibinfo {author} {\bibfnamefont {S.}~\bibnamefont {{Managau}}}, \bibinfo {author} {\bibfnamefont {R.~G.}\ \bibnamefont {{Mann}}}, \bibinfo {author} {\bibfnamefont {G.}~\bibnamefont {{Mantelet}}}, \bibinfo {author} {\bibfnamefont {O.}~\bibnamefont {{Marchal}}}, \bibinfo {author} {\bibfnamefont {J.~M.}\ \bibnamefont {{Marchant}}}, \bibinfo {author} {\bibfnamefont {M.}~\bibnamefont {{Marconi}}}, \bibinfo {author} {\bibfnamefont {S.}~\bibnamefont {{Marinoni}}}, \bibinfo {author} {\bibfnamefont {G.}~\bibnamefont {{Marschalk{\'o}}}}, \bibinfo {author} {\bibfnamefont {D.~J.}\ \bibnamefont {{Marshall}}}, \bibinfo {author} {\bibfnamefont {M.}~\bibnamefont {{Martino}}}, \bibinfo {author} {\bibfnamefont {G.}~\bibnamefont {{Marton}}},
  \bibinfo {author} {\bibfnamefont {N.}~\bibnamefont {{Mary}}}, \bibinfo {author} {\bibfnamefont {D.}~\bibnamefont {{Massari}}}, \bibinfo {author} {\bibfnamefont {G.}~\bibnamefont {{Matijevi{\v{c}}}}}, \bibinfo {author} {\bibfnamefont {T.}~\bibnamefont {{Mazeh}}}, \bibinfo {author} {\bibfnamefont {P.~J.}\ \bibnamefont {{McMillan}}}, \bibinfo {author} {\bibfnamefont {S.}~\bibnamefont {{Messina}}}, \bibinfo {author} {\bibfnamefont {D.}~\bibnamefont {{Michalik}}}, \bibinfo {author} {\bibfnamefont {N.~R.}\ \bibnamefont {{Millar}}}, \bibinfo {author} {\bibfnamefont {D.}~\bibnamefont {{Molina}}}, \bibinfo {author} {\bibfnamefont {R.}~\bibnamefont {{Molinaro}}}, \bibinfo {author} {\bibfnamefont {L.}~\bibnamefont {{Moln{\'a}r}}}, \bibinfo {author} {\bibfnamefont {P.}~\bibnamefont {{Montegriffo}}}, \bibinfo {author} {\bibfnamefont {R.}~\bibnamefont {{Mor}}}, \bibinfo {author} {\bibfnamefont {R.}~\bibnamefont {{Morbidelli}}}, \bibinfo {author} {\bibfnamefont {T.}~\bibnamefont {{Morel}}}, \bibinfo {author}
  {\bibfnamefont {D.}~\bibnamefont {{Morris}}}, \bibinfo {author} {\bibfnamefont {A.~F.}\ \bibnamefont {{Mulone}}}, \bibinfo {author} {\bibfnamefont {T.}~\bibnamefont {{Muraveva}}}, \bibinfo {author} {\bibfnamefont {I.}~\bibnamefont {{Musella}}}, \bibinfo {author} {\bibfnamefont {G.}~\bibnamefont {{Nelemans}}}, \bibinfo {author} {\bibfnamefont {L.}~\bibnamefont {{Nicastro}}}, \bibinfo {author} {\bibfnamefont {L.}~\bibnamefont {{Noval}}}, \bibinfo {author} {\bibfnamefont {W.}~\bibnamefont {{O'Mullane}}}, \bibinfo {author} {\bibfnamefont {C.}~\bibnamefont {{Ord{\'e}novic}}}, \bibinfo {author} {\bibfnamefont {D.}~\bibnamefont {{Ord{\'o}{\~n}ez-Blanco}}}, \bibinfo {author} {\bibfnamefont {P.}~\bibnamefont {{Osborne}}}, \bibinfo {author} {\bibfnamefont {C.}~\bibnamefont {{Pagani}}}, \bibinfo {author} {\bibfnamefont {I.}~\bibnamefont {{Pagano}}}, \bibinfo {author} {\bibfnamefont {F.}~\bibnamefont {{Pailler}}}, \bibinfo {author} {\bibfnamefont {H.}~\bibnamefont {{Palacin}}}, \bibinfo {author} {\bibfnamefont
  {L.}~\bibnamefont {{Palaversa}}}, \bibinfo {author} {\bibfnamefont {A.}~\bibnamefont {{Panahi}}}, \bibinfo {author} {\bibfnamefont {M.}~\bibnamefont {{Pawlak}}}, \bibinfo {author} {\bibfnamefont {A.~M.}\ \bibnamefont {{Piersimoni}}}, \bibinfo {author} {\bibfnamefont {F.~X.}\ \bibnamefont {{Pineau}}}, \bibinfo {author} {\bibfnamefont {E.}~\bibnamefont {{Plachy}}}, \bibinfo {author} {\bibfnamefont {G.}~\bibnamefont {{Plum}}}, \bibinfo {author} {\bibfnamefont {E.}~\bibnamefont {{Poujoulet}}}, \bibinfo {author} {\bibfnamefont {A.}~\bibnamefont {{Pr{\v{s}}a}}}, \bibinfo {author} {\bibfnamefont {L.}~\bibnamefont {{Pulone}}}, \bibinfo {author} {\bibfnamefont {E.}~\bibnamefont {{Racero}}}, \bibinfo {author} {\bibfnamefont {S.}~\bibnamefont {{Ragaini}}}, \bibinfo {author} {\bibfnamefont {N.}~\bibnamefont {{Rambaux}}}, \bibinfo {author} {\bibfnamefont {M.}~\bibnamefont {{Ramos-Lerate}}}, \bibinfo {author} {\bibfnamefont {S.}~\bibnamefont {{Regibo}}}, \bibinfo {author} {\bibfnamefont {F.}~\bibnamefont {{Riclet}}},
  \bibinfo {author} {\bibfnamefont {V.}~\bibnamefont {{Ripepi}}}, \bibinfo {author} {\bibfnamefont {A.}~\bibnamefont {{Riva}}}, \bibinfo {author} {\bibfnamefont {A.}~\bibnamefont {{Rivard}}}, \bibinfo {author} {\bibfnamefont {G.}~\bibnamefont {{Rixon}}}, \bibinfo {author} {\bibfnamefont {T.}~\bibnamefont {{Roegiers}}}, \bibinfo {author} {\bibfnamefont {M.}~\bibnamefont {{Roelens}}}, \bibinfo {author} {\bibfnamefont {N.}~\bibnamefont {{Rowell}}}, \bibinfo {author} {\bibfnamefont {F.}~\bibnamefont {{Royer}}}, \bibinfo {author} {\bibfnamefont {L.}~\bibnamefont {{Ruiz-Dern}}}, \bibinfo {author} {\bibfnamefont {G.}~\bibnamefont {{Sadowski}}}, \bibinfo {author} {\bibfnamefont {T.}~\bibnamefont {{Sagrist{\`a} Sell{\'e}s}}}, \bibinfo {author} {\bibfnamefont {J.}~\bibnamefont {{Sahlmann}}}, \bibinfo {author} {\bibfnamefont {J.}~\bibnamefont {{Salgado}}}, \bibinfo {author} {\bibfnamefont {E.}~\bibnamefont {{Salguero}}}, \bibinfo {author} {\bibfnamefont {N.}~\bibnamefont {{Sanna}}}, \bibinfo {author} {\bibfnamefont
  {T.}~\bibnamefont {{Santana-Ros}}}, \bibinfo {author} {\bibfnamefont {M.}~\bibnamefont {{Sarasso}}}, \bibinfo {author} {\bibfnamefont {H.}~\bibnamefont {{Savietto}}}, \bibinfo {author} {\bibfnamefont {M.}~\bibnamefont {{Schultheis}}}, \bibinfo {author} {\bibfnamefont {E.}~\bibnamefont {{Sciacca}}}, \bibinfo {author} {\bibfnamefont {M.}~\bibnamefont {{Segol}}}, \bibinfo {author} {\bibfnamefont {J.~C.}\ \bibnamefont {{Segovia}}}, \bibinfo {author} {\bibfnamefont {D.}~\bibnamefont {{S{\'e}gransan}}}, \bibinfo {author} {\bibfnamefont {I.~C.}\ \bibnamefont {{Shih}}}, \bibinfo {author} {\bibfnamefont {L.}~\bibnamefont {{Siltala}}}, \bibinfo {author} {\bibfnamefont {A.~F.}\ \bibnamefont {{Silva}}}, \bibinfo {author} {\bibfnamefont {R.~L.}\ \bibnamefont {{Smart}}}, \bibinfo {author} {\bibfnamefont {K.~W.}\ \bibnamefont {{Smith}}}, \bibinfo {author} {\bibfnamefont {E.}~\bibnamefont {{Solano}}}, \bibinfo {author} {\bibfnamefont {F.}~\bibnamefont {{Solitro}}}, \bibinfo {author} {\bibfnamefont {R.}~\bibnamefont
  {{Sordo}}}, \bibinfo {author} {\bibfnamefont {S.}~\bibnamefont {{Soria Nieto}}}, \bibinfo {author} {\bibfnamefont {J.}~\bibnamefont {{Souchay}}}, \bibinfo {author} {\bibfnamefont {A.}~\bibnamefont {{Spagna}}}, \bibinfo {author} {\bibfnamefont {F.}~\bibnamefont {{Spoto}}}, \bibinfo {author} {\bibfnamefont {U.}~\bibnamefont {{Stampa}}}, \bibinfo {author} {\bibfnamefont {I.~A.}\ \bibnamefont {{Steele}}}, \bibinfo {author} {\bibfnamefont {H.}~\bibnamefont {{Steidelm{\"u}ller}}}, \bibinfo {author} {\bibfnamefont {C.~A.}\ \bibnamefont {{Stephenson}}}, \bibinfo {author} {\bibfnamefont {H.}~\bibnamefont {{Stoev}}}, \bibinfo {author} {\bibfnamefont {F.~F.}\ \bibnamefont {{Suess}}}, \bibinfo {author} {\bibfnamefont {J.}~\bibnamefont {{Surdej}}}, \bibinfo {author} {\bibfnamefont {L.}~\bibnamefont {{Szabados}}}, \bibinfo {author} {\bibfnamefont {E.}~\bibnamefont {{Szegedi-Elek}}}, \bibinfo {author} {\bibfnamefont {D.}~\bibnamefont {{Tapiador}}}, \bibinfo {author} {\bibfnamefont {F.}~\bibnamefont {{Taris}}}, \bibinfo
  {author} {\bibfnamefont {G.}~\bibnamefont {{Tauran}}}, \bibinfo {author} {\bibfnamefont {M.~B.}\ \bibnamefont {{Taylor}}}, \bibinfo {author} {\bibfnamefont {R.}~\bibnamefont {{Teixeira}}}, \bibinfo {author} {\bibfnamefont {D.}~\bibnamefont {{Terrett}}}, \bibinfo {author} {\bibfnamefont {P.}~\bibnamefont {{Teyssandier}}}, \bibinfo {author} {\bibfnamefont {W.}~\bibnamefont {{Thuillot}}}, \bibinfo {author} {\bibfnamefont {A.}~\bibnamefont {{Titarenko}}}, \bibinfo {author} {\bibfnamefont {F.}~\bibnamefont {{Torra Clotet}}}, \bibinfo {author} {\bibfnamefont {C.}~\bibnamefont {{Turon}}}, \bibinfo {author} {\bibfnamefont {A.}~\bibnamefont {{Ulla}}}, \bibinfo {author} {\bibfnamefont {E.}~\bibnamefont {{Utrilla}}}, \bibinfo {author} {\bibfnamefont {S.}~\bibnamefont {{Uzzi}}}, \bibinfo {author} {\bibfnamefont {M.}~\bibnamefont {{Vaillant}}}, \bibinfo {author} {\bibfnamefont {G.}~\bibnamefont {{Valentini}}}, \bibinfo {author} {\bibfnamefont {V.}~\bibnamefont {{Valette}}}, \bibinfo {author} {\bibfnamefont
  {A.}~\bibnamefont {{van Elteren}}}, \bibinfo {author} {\bibfnamefont {E.}~\bibnamefont {{Van Hemelryck}}}, \bibinfo {author} {\bibfnamefont {M.}~\bibnamefont {{van Leeuwen}}}, \bibinfo {author} {\bibfnamefont {M.}~\bibnamefont {{Vaschetto}}}, \bibinfo {author} {\bibfnamefont {A.}~\bibnamefont {{Vecchiato}}}, \bibinfo {author} {\bibfnamefont {J.}~\bibnamefont {{Veljanoski}}}, \bibinfo {author} {\bibfnamefont {Y.}~\bibnamefont {{Viala}}}, \bibinfo {author} {\bibfnamefont {D.}~\bibnamefont {{Vicente}}}, \bibinfo {author} {\bibfnamefont {S.}~\bibnamefont {{Vogt}}}, \bibinfo {author} {\bibfnamefont {C.}~\bibnamefont {{von Essen}}}, \bibinfo {author} {\bibfnamefont {H.}~\bibnamefont {{Voss}}}, \bibinfo {author} {\bibfnamefont {V.}~\bibnamefont {{Votruba}}}, \bibinfo {author} {\bibfnamefont {S.}~\bibnamefont {{Voutsinas}}}, \bibinfo {author} {\bibfnamefont {G.}~\bibnamefont {{Walmsley}}}, \bibinfo {author} {\bibfnamefont {M.}~\bibnamefont {{Weiler}}}, \bibinfo {author} {\bibfnamefont {O.}~\bibnamefont {{Wertz}}},
  \bibinfo {author} {\bibfnamefont {T.}~\bibnamefont {{Wevers}}}, \bibinfo {author} {\bibfnamefont {{\L}.}~\bibnamefont {{Wyrzykowski}}}, \bibinfo {author} {\bibfnamefont {A.}~\bibnamefont {{Yoldas}}}, \bibinfo {author} {\bibfnamefont {M.}~\bibnamefont {{{\v{Z}}erjal}}}, \bibinfo {author} {\bibfnamefont {H.}~\bibnamefont {{Ziaeepour}}}, \bibinfo {author} {\bibfnamefont {J.}~\bibnamefont {{Zorec}}}, \bibinfo {author} {\bibfnamefont {S.}~\bibnamefont {{Zschocke}}}, \bibinfo {author} {\bibfnamefont {S.}~\bibnamefont {{Zucker}}}, \bibinfo {author} {\bibfnamefont {C.}~\bibnamefont {{Zurbach}}},\ and\ \bibinfo {author} {\bibfnamefont {T.}~\bibnamefont {{Zwitter}}},\ }\href {https://doi.org/10.1051/0004-6361/201832865} {\bibfield  {journal} {\bibinfo  {journal} {\aap}\ }\textbf {\bibinfo {volume} {616}},\ \bibinfo {eid} {A11} (\bibinfo {year} {2018})},\ \Eprint {https://arxiv.org/abs/1804.09380} {arXiv:1804.09380 [astro-ph.GA]} \BibitemShut {NoStop}%
\bibitem [{\citenamefont {{P{\~o}der}}\ \emph {et~al.}(2023)\citenamefont {{P{\~o}der}}, \citenamefont {{Benito}}, \citenamefont {{Pata}}, \citenamefont {{Kipper}}, \citenamefont {{Ramler}}, \citenamefont {{H{\"u}tsi}}, \citenamefont {{Kolka}},\ and\ \citenamefont {{Thomas}}}]{Poder2023}%
  \BibitemOpen
  \bibfield  {author} {\bibinfo {author} {\bibfnamefont {S.}~\bibnamefont {{P{\~o}der}}}, \bibinfo {author} {\bibfnamefont {M.}~\bibnamefont {{Benito}}}, \bibinfo {author} {\bibfnamefont {J.}~\bibnamefont {{Pata}}}, \bibinfo {author} {\bibfnamefont {R.}~\bibnamefont {{Kipper}}}, \bibinfo {author} {\bibfnamefont {H.}~\bibnamefont {{Ramler}}}, \bibinfo {author} {\bibfnamefont {G.}~\bibnamefont {{H{\"u}tsi}}}, \bibinfo {author} {\bibfnamefont {I.}~\bibnamefont {{Kolka}}},\ and\ \bibinfo {author} {\bibfnamefont {G.~F.}\ \bibnamefont {{Thomas}}},\ }\href {https://doi.org/10.1051/0004-6361/202346474} {\bibfield  {journal} {\bibinfo  {journal} {\aap}\ }\textbf {\bibinfo {volume} {676}},\ \bibinfo {eid} {A134} (\bibinfo {year} {2023})},\ \Eprint {https://arxiv.org/abs/2309.02895} {arXiv:2309.02895 [astro-ph.GA]} \BibitemShut {NoStop}%
\bibitem [{\citenamefont {{McKee}}\ \emph {et~al.}(2015)\citenamefont {{McKee}}, \citenamefont {{Parravano}},\ and\ \citenamefont {{Hollenbach}}}]{McKee2015}%
  \BibitemOpen
  \bibfield  {author} {\bibinfo {author} {\bibfnamefont {C.~F.}\ \bibnamefont {{McKee}}}, \bibinfo {author} {\bibfnamefont {A.}~\bibnamefont {{Parravano}}},\ and\ \bibinfo {author} {\bibfnamefont {D.~J.}\ \bibnamefont {{Hollenbach}}},\ }\href {https://doi.org/10.1088/0004-637X/814/1/13} {\bibfield  {journal} {\bibinfo  {journal} {\apj}\ }\textbf {\bibinfo {volume} {814}},\ \bibinfo {eid} {13} (\bibinfo {year} {2015})},\ \Eprint {https://arxiv.org/abs/1509.05334} {arXiv:1509.05334 [astro-ph.GA]} \BibitemShut {NoStop}%
\bibitem [{\citenamefont {{Guo}}\ \emph {et~al.}(2020)\citenamefont {{Guo}}, \citenamefont {{Liu}}, \citenamefont {{Mao}}, \citenamefont {{Xue}}, \citenamefont {{Long}},\ and\ \citenamefont {{Zhang}}}]{Guo2020}%
  \BibitemOpen
  \bibfield  {author} {\bibinfo {author} {\bibfnamefont {R.}~\bibnamefont {{Guo}}}, \bibinfo {author} {\bibfnamefont {C.}~\bibnamefont {{Liu}}}, \bibinfo {author} {\bibfnamefont {S.}~\bibnamefont {{Mao}}}, \bibinfo {author} {\bibfnamefont {X.-X.}\ \bibnamefont {{Xue}}}, \bibinfo {author} {\bibfnamefont {R.~J.}\ \bibnamefont {{Long}}},\ and\ \bibinfo {author} {\bibfnamefont {L.}~\bibnamefont {{Zhang}}},\ }\href {https://doi.org/10.1093/mnras/staa1483} {\bibfield  {journal} {\bibinfo  {journal} {\mnras}\ }\textbf {\bibinfo {volume} {495}},\ \bibinfo {pages} {4828} (\bibinfo {year} {2020})},\ \Eprint {https://arxiv.org/abs/2005.12018} {arXiv:2005.12018 [astro-ph.GA]} \BibitemShut {NoStop}%
\bibitem [{\citenamefont {{Bienaym{\'e}}}\ \emph {et~al.}(2014)\citenamefont {{Bienaym{\'e}}}, \citenamefont {{Famaey}}, \citenamefont {{Siebert}}, \citenamefont {{Freeman}}, \citenamefont {{Gibson}}, \citenamefont {{Gilmore}}, \citenamefont {{Grebel}}, \citenamefont {{Bland-Hawthorn}}, \citenamefont {{Kordopatis}}, \citenamefont {{Munari}}, \citenamefont {{Navarro}}, \citenamefont {{Parker}}, \citenamefont {{Reid}}, \citenamefont {{Seabroke}}, \citenamefont {{Siviero}}, \citenamefont {{Steinmetz}}, \citenamefont {{Watson}}, \citenamefont {{Wyse}},\ and\ \citenamefont {{Zwitter}}}]{Bienayme2014}%
  \BibitemOpen
  \bibfield  {author} {\bibinfo {author} {\bibfnamefont {O.}~\bibnamefont {{Bienaym{\'e}}}}, \bibinfo {author} {\bibfnamefont {B.}~\bibnamefont {{Famaey}}}, \bibinfo {author} {\bibfnamefont {A.}~\bibnamefont {{Siebert}}}, \bibinfo {author} {\bibfnamefont {K.~C.}\ \bibnamefont {{Freeman}}}, \bibinfo {author} {\bibfnamefont {B.~K.}\ \bibnamefont {{Gibson}}}, \bibinfo {author} {\bibfnamefont {G.}~\bibnamefont {{Gilmore}}}, \bibinfo {author} {\bibfnamefont {E.~K.}\ \bibnamefont {{Grebel}}}, \bibinfo {author} {\bibfnamefont {J.}~\bibnamefont {{Bland-Hawthorn}}}, \bibinfo {author} {\bibfnamefont {G.}~\bibnamefont {{Kordopatis}}}, \bibinfo {author} {\bibfnamefont {U.}~\bibnamefont {{Munari}}}, \bibinfo {author} {\bibfnamefont {J.~F.}\ \bibnamefont {{Navarro}}}, \bibinfo {author} {\bibfnamefont {Q.}~\bibnamefont {{Parker}}}, \bibinfo {author} {\bibfnamefont {W.}~\bibnamefont {{Reid}}}, \bibinfo {author} {\bibfnamefont {G.~M.}\ \bibnamefont {{Seabroke}}}, \bibinfo {author} {\bibfnamefont {A.}~\bibnamefont
  {{Siviero}}}, \bibinfo {author} {\bibfnamefont {M.}~\bibnamefont {{Steinmetz}}}, \bibinfo {author} {\bibfnamefont {F.}~\bibnamefont {{Watson}}}, \bibinfo {author} {\bibfnamefont {R.~F.~G.}\ \bibnamefont {{Wyse}}},\ and\ \bibinfo {author} {\bibfnamefont {T.}~\bibnamefont {{Zwitter}}},\ }\href {https://doi.org/10.1051/0004-6361/201424478} {\bibfield  {journal} {\bibinfo  {journal} {\aap}\ }\textbf {\bibinfo {volume} {571}},\ \bibinfo {eid} {A92} (\bibinfo {year} {2014})},\ \Eprint {https://arxiv.org/abs/1406.6896} {arXiv:1406.6896 [astro-ph.GA]} \BibitemShut {NoStop}%
\bibitem [{\citenamefont {{Buch}}\ \emph {et~al.}(2019)\citenamefont {{Buch}}, \citenamefont {{Leung}},\ and\ \citenamefont {{Fan}}}]{Buch2019}%
  \BibitemOpen
  \bibfield  {author} {\bibinfo {author} {\bibfnamefont {J.}~\bibnamefont {{Buch}}}, \bibinfo {author} {\bibfnamefont {J.~S.~C.}\ \bibnamefont {{Leung}}},\ and\ \bibinfo {author} {\bibfnamefont {J.}~\bibnamefont {{Fan}}},\ }\href {https://doi.org/10.1088/1475-7516/2019/04/026} {\bibfield  {journal} {\bibinfo  {journal} {\jcap}\ }\textbf {\bibinfo {volume} {2019}},\ \bibinfo {eid} {026} (\bibinfo {year} {2019})},\ \Eprint {https://arxiv.org/abs/1808.05603} {arXiv:1808.05603 [astro-ph.GA]} \BibitemShut {NoStop}%
\bibitem [{\citenamefont {{Salomon}}\ \emph {et~al.}(2020)\citenamefont {{Salomon}}, \citenamefont {{Bienaym{\'e}}}, \citenamefont {{Reyl{\'e}}}, \citenamefont {{Robin}},\ and\ \citenamefont {{Famaey}}}]{Salomon2020}%
  \BibitemOpen
  \bibfield  {author} {\bibinfo {author} {\bibfnamefont {J.-B.}\ \bibnamefont {{Salomon}}}, \bibinfo {author} {\bibfnamefont {O.}~\bibnamefont {{Bienaym{\'e}}}}, \bibinfo {author} {\bibfnamefont {C.}~\bibnamefont {{Reyl{\'e}}}}, \bibinfo {author} {\bibfnamefont {A.~C.}\ \bibnamefont {{Robin}}},\ and\ \bibinfo {author} {\bibfnamefont {B.}~\bibnamefont {{Famaey}}},\ }\href {https://doi.org/10.1051/0004-6361/202038535} {\bibfield  {journal} {\bibinfo  {journal} {\aap}\ }\textbf {\bibinfo {volume} {643}},\ \bibinfo {eid} {A75} (\bibinfo {year} {2020})},\ \Eprint {https://arxiv.org/abs/2009.04495} {arXiv:2009.04495 [astro-ph.GA]} \BibitemShut {NoStop}%
\bibitem [{\citenamefont {{Widmark}}(2019)}]{Widmark2019}%
  \BibitemOpen
  \bibfield  {author} {\bibinfo {author} {\bibfnamefont {A.}~\bibnamefont {{Widmark}}},\ }\href {https://doi.org/10.1051/0004-6361/201834718} {\bibfield  {journal} {\bibinfo  {journal} {\aap}\ }\textbf {\bibinfo {volume} {623}},\ \bibinfo {eid} {A30} (\bibinfo {year} {2019})},\ \Eprint {https://arxiv.org/abs/1811.07911} {arXiv:1811.07911 [astro-ph.GA]} \BibitemShut {NoStop}%
\bibitem [{\citenamefont {{Haines}}\ \emph {et~al.}(2019)\citenamefont {{Haines}}, \citenamefont {{D'Onghia}}, \citenamefont {{Famaey}}, \citenamefont {{Laporte}},\ and\ \citenamefont {{Hernquist}}}]{Haines2019}%
  \BibitemOpen
  \bibfield  {author} {\bibinfo {author} {\bibfnamefont {T.}~\bibnamefont {{Haines}}}, \bibinfo {author} {\bibfnamefont {E.}~\bibnamefont {{D'Onghia}}}, \bibinfo {author} {\bibfnamefont {B.}~\bibnamefont {{Famaey}}}, \bibinfo {author} {\bibfnamefont {C.}~\bibnamefont {{Laporte}}},\ and\ \bibinfo {author} {\bibfnamefont {L.}~\bibnamefont {{Hernquist}}},\ }\href {https://doi.org/10.3847/2041-8213/ab25f3} {\bibfield  {journal} {\bibinfo  {journal} {\apjl}\ }\textbf {\bibinfo {volume} {879}},\ \bibinfo {eid} {L15} (\bibinfo {year} {2019})},\ \Eprint {https://arxiv.org/abs/1903.00607} {arXiv:1903.00607 [astro-ph.GA]} \BibitemShut {NoStop}%
\bibitem [{\citenamefont {{Lim}}\ \emph {et~al.}(2023)\citenamefont {{Lim}}, \citenamefont {{Putney}}, \citenamefont {{Buckley}},\ and\ \citenamefont {{Shih}}}]{Lim2023}%
  \BibitemOpen
  \bibfield  {author} {\bibinfo {author} {\bibfnamefont {S.~H.}\ \bibnamefont {{Lim}}}, \bibinfo {author} {\bibfnamefont {E.}~\bibnamefont {{Putney}}}, \bibinfo {author} {\bibfnamefont {M.~R.}\ \bibnamefont {{Buckley}}},\ and\ \bibinfo {author} {\bibfnamefont {D.}~\bibnamefont {{Shih}}},\ }\href {https://doi.org/10.48550/arXiv.2305.13358} {\bibfield  {journal} {\bibinfo  {journal} {arXiv e-prints}\ ,\ \bibinfo {eid} {arXiv:2305.13358}} (\bibinfo {year} {2023})},\ \Eprint {https://arxiv.org/abs/2305.13358} {arXiv:2305.13358 [astro-ph.GA]} \BibitemShut {NoStop}%
\bibitem [{\citenamefont {Rasmussen}\ and\ \citenamefont {Williams}(2005)}]{RasmussenWilliams2005}%
  \BibitemOpen
  \bibfield  {author} {\bibinfo {author} {\bibfnamefont {C.~E.}\ \bibnamefont {Rasmussen}}\ and\ \bibinfo {author} {\bibfnamefont {C.~K.~I.}\ \bibnamefont {Williams}},\ }\href {https://doi.org/10.7551/mitpress/3206.001.0001} {\emph {\bibinfo {title} {{Gaussian Processes for Machine Learning}}}}\ (\bibinfo  {publisher} {The MIT Press},\ \bibinfo {year} {2005})\BibitemShut {NoStop}%
\bibitem [{\citenamefont {Harris}\ \emph {et~al.}(2020)\citenamefont {Harris}, \citenamefont {Millman}, \citenamefont {van~der Walt}, \citenamefont {Gommers}, \citenamefont {Virtanen}, \citenamefont {Cournapeau}, \citenamefont {Wieser}, \citenamefont {Taylor}, \citenamefont {Berg}, \citenamefont {Smith}, \citenamefont {Kern}, \citenamefont {Picus}, \citenamefont {Hoyer}, \citenamefont {van Kerkwijk}, \citenamefont {Brett}, \citenamefont {Haldane}, \citenamefont {del R{\'{i}}o}, \citenamefont {Wiebe}, \citenamefont {Peterson}, \citenamefont {G{\'{e}}rard-Marchant}, \citenamefont {Sheppard}, \citenamefont {Reddy}, \citenamefont {Weckesser}, \citenamefont {Abbasi}, \citenamefont {Gohlke},\ and\ \citenamefont {Oliphant}}]{numpy}%
  \BibitemOpen
  \bibfield  {author} {\bibinfo {author} {\bibfnamefont {C.~R.}\ \bibnamefont {Harris}}, \bibinfo {author} {\bibfnamefont {K.~J.}\ \bibnamefont {Millman}}, \bibinfo {author} {\bibfnamefont {S.~J.}\ \bibnamefont {van~der Walt}}, \bibinfo {author} {\bibfnamefont {R.}~\bibnamefont {Gommers}}, \bibinfo {author} {\bibfnamefont {P.}~\bibnamefont {Virtanen}}, \bibinfo {author} {\bibfnamefont {D.}~\bibnamefont {Cournapeau}}, \bibinfo {author} {\bibfnamefont {E.}~\bibnamefont {Wieser}}, \bibinfo {author} {\bibfnamefont {J.}~\bibnamefont {Taylor}}, \bibinfo {author} {\bibfnamefont {S.}~\bibnamefont {Berg}}, \bibinfo {author} {\bibfnamefont {N.~J.}\ \bibnamefont {Smith}}, \bibinfo {author} {\bibfnamefont {R.}~\bibnamefont {Kern}}, \bibinfo {author} {\bibfnamefont {M.}~\bibnamefont {Picus}}, \bibinfo {author} {\bibfnamefont {S.}~\bibnamefont {Hoyer}}, \bibinfo {author} {\bibfnamefont {M.~H.}\ \bibnamefont {van Kerkwijk}}, \bibinfo {author} {\bibfnamefont {M.}~\bibnamefont {Brett}}, \bibinfo {author} {\bibfnamefont
  {A.}~\bibnamefont {Haldane}}, \bibinfo {author} {\bibfnamefont {J.~F.}\ \bibnamefont {del R{\'{i}}o}}, \bibinfo {author} {\bibfnamefont {M.}~\bibnamefont {Wiebe}}, \bibinfo {author} {\bibfnamefont {P.}~\bibnamefont {Peterson}}, \bibinfo {author} {\bibfnamefont {P.}~\bibnamefont {G{\'{e}}rard-Marchant}}, \bibinfo {author} {\bibfnamefont {K.}~\bibnamefont {Sheppard}}, \bibinfo {author} {\bibfnamefont {T.}~\bibnamefont {Reddy}}, \bibinfo {author} {\bibfnamefont {W.}~\bibnamefont {Weckesser}}, \bibinfo {author} {\bibfnamefont {H.}~\bibnamefont {Abbasi}}, \bibinfo {author} {\bibfnamefont {C.}~\bibnamefont {Gohlke}},\ and\ \bibinfo {author} {\bibfnamefont {T.~E.}\ \bibnamefont {Oliphant}},\ }\href {https://doi.org/10.1038/s41586-020-2649-2} {\bibfield  {journal} {\bibinfo  {journal} {Nature}\ }\textbf {\bibinfo {volume} {585}},\ \bibinfo {pages} {357} (\bibinfo {year} {2020})}\BibitemShut {NoStop}%
\bibitem [{\citenamefont {Pedregosa}\ \emph {et~al.}(2011)\citenamefont {Pedregosa}, \citenamefont {Varoquaux}, \citenamefont {Gramfort}, \citenamefont {Michel}, \citenamefont {Thirion}, \citenamefont {Grisel}, \citenamefont {Blondel}, \citenamefont {Prettenhofer}, \citenamefont {Weiss}, \citenamefont {Dubourg}, \citenamefont {Vanderplas}, \citenamefont {Passos}, \citenamefont {Cournapeau}, \citenamefont {Brucher}, \citenamefont {Perrot},\ and\ \citenamefont {{{\'E}}douard Duchesnay}}]{scikitlearn}%
  \BibitemOpen
  \bibfield  {author} {\bibinfo {author} {\bibfnamefont {F.}~\bibnamefont {Pedregosa}}, \bibinfo {author} {\bibfnamefont {G.}~\bibnamefont {Varoquaux}}, \bibinfo {author} {\bibfnamefont {A.}~\bibnamefont {Gramfort}}, \bibinfo {author} {\bibfnamefont {V.}~\bibnamefont {Michel}}, \bibinfo {author} {\bibfnamefont {B.}~\bibnamefont {Thirion}}, \bibinfo {author} {\bibfnamefont {O.}~\bibnamefont {Grisel}}, \bibinfo {author} {\bibfnamefont {M.}~\bibnamefont {Blondel}}, \bibinfo {author} {\bibfnamefont {P.}~\bibnamefont {Prettenhofer}}, \bibinfo {author} {\bibfnamefont {R.}~\bibnamefont {Weiss}}, \bibinfo {author} {\bibfnamefont {V.}~\bibnamefont {Dubourg}}, \bibinfo {author} {\bibfnamefont {J.}~\bibnamefont {Vanderplas}}, \bibinfo {author} {\bibfnamefont {A.}~\bibnamefont {Passos}}, \bibinfo {author} {\bibfnamefont {D.}~\bibnamefont {Cournapeau}}, \bibinfo {author} {\bibfnamefont {M.}~\bibnamefont {Brucher}}, \bibinfo {author} {\bibfnamefont {M.}~\bibnamefont {Perrot}},\ and\ \bibinfo {author} {\bibnamefont
  {{{\'E}}douard Duchesnay}},\ }\href {http://jmlr.org/papers/v12/pedregosa11a.html} {\bibfield  {journal} {\bibinfo  {journal} {Journal of Machine Learning Research}\ }\textbf {\bibinfo {volume} {12}},\ \bibinfo {pages} {2825} (\bibinfo {year} {2011})}\BibitemShut {NoStop}%
\bibitem [{\citenamefont {{Astropy Collaboration}}\ \emph {et~al.}(2022)\citenamefont {{Astropy Collaboration}}, \citenamefont {{Price-Whelan}}, \citenamefont {{Lim}}, \citenamefont {{Earl}}, \citenamefont {{Starkman}}, \citenamefont {{Bradley}}, \citenamefont {{Shupe}}, \citenamefont {{Patil}}, \citenamefont {{Corrales}}, \citenamefont {{Brasseur}}, \citenamefont {{N{\"o}the}}, \citenamefont {{Donath}}, \citenamefont {{Tollerud}}, \citenamefont {{Morris}}, \citenamefont {{Ginsburg}}, \citenamefont {{Vaher}}, \citenamefont {{Weaver}}, \citenamefont {{Tocknell}}, \citenamefont {{Jamieson}}, \citenamefont {{van Kerkwijk}}, \citenamefont {{Robitaille}}, \citenamefont {{Merry}}, \citenamefont {{Bachetti}}, \citenamefont {{G{\"u}nther}}, \citenamefont {{Aldcroft}}, \citenamefont {{Alvarado-Montes}}, \citenamefont {{Archibald}}, \citenamefont {{B{\'o}di}}, \citenamefont {{Bapat}}, \citenamefont {{Barentsen}}, \citenamefont {{Baz{\'a}n}}, \citenamefont {{Biswas}}, \citenamefont {{Boquien}}, \citenamefont {{Burke}},
  \citenamefont {{Cara}}, \citenamefont {{Cara}}, \citenamefont {{Conroy}}, \citenamefont {{Conseil}}, \citenamefont {{Craig}}, \citenamefont {{Cross}}, \citenamefont {{Cruz}}, \citenamefont {{D'Eugenio}}, \citenamefont {{Dencheva}}, \citenamefont {{Devillepoix}}, \citenamefont {{Dietrich}}, \citenamefont {{Eigenbrot}}, \citenamefont {{Erben}}, \citenamefont {{Ferreira}}, \citenamefont {{Foreman-Mackey}}, \citenamefont {{Fox}}, \citenamefont {{Freij}}, \citenamefont {{Garg}}, \citenamefont {{Geda}}, \citenamefont {{Glattly}}, \citenamefont {{Gondhalekar}}, \citenamefont {{Gordon}}, \citenamefont {{Grant}}, \citenamefont {{Greenfield}}, \citenamefont {{Groener}}, \citenamefont {{Guest}}, \citenamefont {{Gurovich}}, \citenamefont {{Handberg}}, \citenamefont {{Hart}}, \citenamefont {{Hatfield-Dodds}}, \citenamefont {{Homeier}}, \citenamefont {{Hosseinzadeh}}, \citenamefont {{Jenness}}, \citenamefont {{Jones}}, \citenamefont {{Joseph}}, \citenamefont {{Kalmbach}}, \citenamefont {{Karamehmetoglu}}, \citenamefont
  {{Ka{\l}uszy{\'n}ski}}, \citenamefont {{Kelley}}, \citenamefont {{Kern}}, \citenamefont {{Kerzendorf}}, \citenamefont {{Koch}}, \citenamefont {{Kulumani}}, \citenamefont {{Lee}}, \citenamefont {{Ly}}, \citenamefont {{Ma}}, \citenamefont {{MacBride}}, \citenamefont {{Maljaars}}, \citenamefont {{Muna}}, \citenamefont {{Murphy}}, \citenamefont {{Norman}}, \citenamefont {{O'Steen}}, \citenamefont {{Oman}}, \citenamefont {{Pacifici}}, \citenamefont {{Pascual}}, \citenamefont {{Pascual-Granado}}, \citenamefont {{Patil}}, \citenamefont {{Perren}}, \citenamefont {{Pickering}}, \citenamefont {{Rastogi}}, \citenamefont {{Roulston}}, \citenamefont {{Ryan}}, \citenamefont {{Rykoff}}, \citenamefont {{Sabater}}, \citenamefont {{Sakurikar}}, \citenamefont {{Salgado}}, \citenamefont {{Sanghi}}, \citenamefont {{Saunders}}, \citenamefont {{Savchenko}}, \citenamefont {{Schwardt}}, \citenamefont {{Seifert-Eckert}}, \citenamefont {{Shih}}, \citenamefont {{Jain}}, \citenamefont {{Shukla}}, \citenamefont {{Sick}}, \citenamefont
  {{Simpson}}, \citenamefont {{Singanamalla}}, \citenamefont {{Singer}}, \citenamefont {{Singhal}}, \citenamefont {{Sinha}}, \citenamefont {{Sip{\H{o}}cz}}, \citenamefont {{Spitler}}, \citenamefont {{Stansby}}, \citenamefont {{Streicher}}, \citenamefont {{{\v{S}}umak}}, \citenamefont {{Swinbank}}, \citenamefont {{Taranu}}, \citenamefont {{Tewary}}, \citenamefont {{Tremblay}}, \citenamefont {{de Val-Borro}}, \citenamefont {{Van Kooten}}, \citenamefont {{Vasovi{\'c}}}, \citenamefont {{Verma}}, \citenamefont {{de Miranda Cardoso}}, \citenamefont {{Williams}}, \citenamefont {{Wilson}}, \citenamefont {{Winkel}}, \citenamefont {{Wood-Vasey}}, \citenamefont {{Xue}}, \citenamefont {{Yoachim}}, \citenamefont {{Zhang}}, \citenamefont {{Zonca}},\ and\ \citenamefont {{Astropy Project Contributors}}}]{astropy}%
  \BibitemOpen
  \bibfield  {author} {\bibinfo {author} {\bibnamefont {{Astropy Collaboration}}}, \bibinfo {author} {\bibfnamefont {A.~M.}\ \bibnamefont {{Price-Whelan}}}, \bibinfo {author} {\bibfnamefont {P.~L.}\ \bibnamefont {{Lim}}}, \bibinfo {author} {\bibfnamefont {N.}~\bibnamefont {{Earl}}}, \bibinfo {author} {\bibfnamefont {N.}~\bibnamefont {{Starkman}}}, \bibinfo {author} {\bibfnamefont {L.}~\bibnamefont {{Bradley}}}, \bibinfo {author} {\bibfnamefont {D.~L.}\ \bibnamefont {{Shupe}}}, \bibinfo {author} {\bibfnamefont {A.~A.}\ \bibnamefont {{Patil}}}, \bibinfo {author} {\bibfnamefont {L.}~\bibnamefont {{Corrales}}}, \bibinfo {author} {\bibfnamefont {C.~E.}\ \bibnamefont {{Brasseur}}}, \bibinfo {author} {\bibfnamefont {M.}~\bibnamefont {{N{\"o}the}}}, \bibinfo {author} {\bibfnamefont {A.}~\bibnamefont {{Donath}}}, \bibinfo {author} {\bibfnamefont {E.}~\bibnamefont {{Tollerud}}}, \bibinfo {author} {\bibfnamefont {B.~M.}\ \bibnamefont {{Morris}}}, \bibinfo {author} {\bibfnamefont {A.}~\bibnamefont {{Ginsburg}}}, \bibinfo
  {author} {\bibfnamefont {E.}~\bibnamefont {{Vaher}}}, \bibinfo {author} {\bibfnamefont {B.~A.}\ \bibnamefont {{Weaver}}}, \bibinfo {author} {\bibfnamefont {J.}~\bibnamefont {{Tocknell}}}, \bibinfo {author} {\bibfnamefont {W.}~\bibnamefont {{Jamieson}}}, \bibinfo {author} {\bibfnamefont {M.~H.}\ \bibnamefont {{van Kerkwijk}}}, \bibinfo {author} {\bibfnamefont {T.~P.}\ \bibnamefont {{Robitaille}}}, \bibinfo {author} {\bibfnamefont {B.}~\bibnamefont {{Merry}}}, \bibinfo {author} {\bibfnamefont {M.}~\bibnamefont {{Bachetti}}}, \bibinfo {author} {\bibfnamefont {H.~M.}\ \bibnamefont {{G{\"u}nther}}}, \bibinfo {author} {\bibfnamefont {T.~L.}\ \bibnamefont {{Aldcroft}}}, \bibinfo {author} {\bibfnamefont {J.~A.}\ \bibnamefont {{Alvarado-Montes}}}, \bibinfo {author} {\bibfnamefont {A.~M.}\ \bibnamefont {{Archibald}}}, \bibinfo {author} {\bibfnamefont {A.}~\bibnamefont {{B{\'o}di}}}, \bibinfo {author} {\bibfnamefont {S.}~\bibnamefont {{Bapat}}}, \bibinfo {author} {\bibfnamefont {G.}~\bibnamefont {{Barentsen}}},
  \bibinfo {author} {\bibfnamefont {J.}~\bibnamefont {{Baz{\'a}n}}}, \bibinfo {author} {\bibfnamefont {M.}~\bibnamefont {{Biswas}}}, \bibinfo {author} {\bibfnamefont {M.}~\bibnamefont {{Boquien}}}, \bibinfo {author} {\bibfnamefont {D.~J.}\ \bibnamefont {{Burke}}}, \bibinfo {author} {\bibfnamefont {D.}~\bibnamefont {{Cara}}}, \bibinfo {author} {\bibfnamefont {M.}~\bibnamefont {{Cara}}}, \bibinfo {author} {\bibfnamefont {K.~E.}\ \bibnamefont {{Conroy}}}, \bibinfo {author} {\bibfnamefont {S.}~\bibnamefont {{Conseil}}}, \bibinfo {author} {\bibfnamefont {M.~W.}\ \bibnamefont {{Craig}}}, \bibinfo {author} {\bibfnamefont {R.~M.}\ \bibnamefont {{Cross}}}, \bibinfo {author} {\bibfnamefont {K.~L.}\ \bibnamefont {{Cruz}}}, \bibinfo {author} {\bibfnamefont {F.}~\bibnamefont {{D'Eugenio}}}, \bibinfo {author} {\bibfnamefont {N.}~\bibnamefont {{Dencheva}}}, \bibinfo {author} {\bibfnamefont {H.~A.~R.}\ \bibnamefont {{Devillepoix}}}, \bibinfo {author} {\bibfnamefont {J.~P.}\ \bibnamefont {{Dietrich}}}, \bibinfo {author}
  {\bibfnamefont {A.~D.}\ \bibnamefont {{Eigenbrot}}}, \bibinfo {author} {\bibfnamefont {T.}~\bibnamefont {{Erben}}}, \bibinfo {author} {\bibfnamefont {L.}~\bibnamefont {{Ferreira}}}, \bibinfo {author} {\bibfnamefont {D.}~\bibnamefont {{Foreman-Mackey}}}, \bibinfo {author} {\bibfnamefont {R.}~\bibnamefont {{Fox}}}, \bibinfo {author} {\bibfnamefont {N.}~\bibnamefont {{Freij}}}, \bibinfo {author} {\bibfnamefont {S.}~\bibnamefont {{Garg}}}, \bibinfo {author} {\bibfnamefont {R.}~\bibnamefont {{Geda}}}, \bibinfo {author} {\bibfnamefont {L.}~\bibnamefont {{Glattly}}}, \bibinfo {author} {\bibfnamefont {Y.}~\bibnamefont {{Gondhalekar}}}, \bibinfo {author} {\bibfnamefont {K.~D.}\ \bibnamefont {{Gordon}}}, \bibinfo {author} {\bibfnamefont {D.}~\bibnamefont {{Grant}}}, \bibinfo {author} {\bibfnamefont {P.}~\bibnamefont {{Greenfield}}}, \bibinfo {author} {\bibfnamefont {A.~M.}\ \bibnamefont {{Groener}}}, \bibinfo {author} {\bibfnamefont {S.}~\bibnamefont {{Guest}}}, \bibinfo {author} {\bibfnamefont {S.}~\bibnamefont
  {{Gurovich}}}, \bibinfo {author} {\bibfnamefont {R.}~\bibnamefont {{Handberg}}}, \bibinfo {author} {\bibfnamefont {A.}~\bibnamefont {{Hart}}}, \bibinfo {author} {\bibfnamefont {Z.}~\bibnamefont {{Hatfield-Dodds}}}, \bibinfo {author} {\bibfnamefont {D.}~\bibnamefont {{Homeier}}}, \bibinfo {author} {\bibfnamefont {G.}~\bibnamefont {{Hosseinzadeh}}}, \bibinfo {author} {\bibfnamefont {T.}~\bibnamefont {{Jenness}}}, \bibinfo {author} {\bibfnamefont {C.~K.}\ \bibnamefont {{Jones}}}, \bibinfo {author} {\bibfnamefont {P.}~\bibnamefont {{Joseph}}}, \bibinfo {author} {\bibfnamefont {J.~B.}\ \bibnamefont {{Kalmbach}}}, \bibinfo {author} {\bibfnamefont {E.}~\bibnamefont {{Karamehmetoglu}}}, \bibinfo {author} {\bibfnamefont {M.}~\bibnamefont {{Ka{\l}uszy{\'n}ski}}}, \bibinfo {author} {\bibfnamefont {M.~S.~P.}\ \bibnamefont {{Kelley}}}, \bibinfo {author} {\bibfnamefont {N.}~\bibnamefont {{Kern}}}, \bibinfo {author} {\bibfnamefont {W.~E.}\ \bibnamefont {{Kerzendorf}}}, \bibinfo {author} {\bibfnamefont {E.~W.}\
  \bibnamefont {{Koch}}}, \bibinfo {author} {\bibfnamefont {S.}~\bibnamefont {{Kulumani}}}, \bibinfo {author} {\bibfnamefont {A.}~\bibnamefont {{Lee}}}, \bibinfo {author} {\bibfnamefont {C.}~\bibnamefont {{Ly}}}, \bibinfo {author} {\bibfnamefont {Z.}~\bibnamefont {{Ma}}}, \bibinfo {author} {\bibfnamefont {C.}~\bibnamefont {{MacBride}}}, \bibinfo {author} {\bibfnamefont {J.~M.}\ \bibnamefont {{Maljaars}}}, \bibinfo {author} {\bibfnamefont {D.}~\bibnamefont {{Muna}}}, \bibinfo {author} {\bibfnamefont {N.~A.}\ \bibnamefont {{Murphy}}}, \bibinfo {author} {\bibfnamefont {H.}~\bibnamefont {{Norman}}}, \bibinfo {author} {\bibfnamefont {R.}~\bibnamefont {{O'Steen}}}, \bibinfo {author} {\bibfnamefont {K.~A.}\ \bibnamefont {{Oman}}}, \bibinfo {author} {\bibfnamefont {C.}~\bibnamefont {{Pacifici}}}, \bibinfo {author} {\bibfnamefont {S.}~\bibnamefont {{Pascual}}}, \bibinfo {author} {\bibfnamefont {J.}~\bibnamefont {{Pascual-Granado}}}, \bibinfo {author} {\bibfnamefont {R.~R.}\ \bibnamefont {{Patil}}}, \bibinfo {author}
  {\bibfnamefont {G.~I.}\ \bibnamefont {{Perren}}}, \bibinfo {author} {\bibfnamefont {T.~E.}\ \bibnamefont {{Pickering}}}, \bibinfo {author} {\bibfnamefont {T.}~\bibnamefont {{Rastogi}}}, \bibinfo {author} {\bibfnamefont {B.~R.}\ \bibnamefont {{Roulston}}}, \bibinfo {author} {\bibfnamefont {D.~F.}\ \bibnamefont {{Ryan}}}, \bibinfo {author} {\bibfnamefont {E.~S.}\ \bibnamefont {{Rykoff}}}, \bibinfo {author} {\bibfnamefont {J.}~\bibnamefont {{Sabater}}}, \bibinfo {author} {\bibfnamefont {P.}~\bibnamefont {{Sakurikar}}}, \bibinfo {author} {\bibfnamefont {J.}~\bibnamefont {{Salgado}}}, \bibinfo {author} {\bibfnamefont {A.}~\bibnamefont {{Sanghi}}}, \bibinfo {author} {\bibfnamefont {N.}~\bibnamefont {{Saunders}}}, \bibinfo {author} {\bibfnamefont {V.}~\bibnamefont {{Savchenko}}}, \bibinfo {author} {\bibfnamefont {L.}~\bibnamefont {{Schwardt}}}, \bibinfo {author} {\bibfnamefont {M.}~\bibnamefont {{Seifert-Eckert}}}, \bibinfo {author} {\bibfnamefont {A.~Y.}\ \bibnamefont {{Shih}}}, \bibinfo {author} {\bibfnamefont
  {A.~S.}\ \bibnamefont {{Jain}}}, \bibinfo {author} {\bibfnamefont {G.}~\bibnamefont {{Shukla}}}, \bibinfo {author} {\bibfnamefont {J.}~\bibnamefont {{Sick}}}, \bibinfo {author} {\bibfnamefont {C.}~\bibnamefont {{Simpson}}}, \bibinfo {author} {\bibfnamefont {S.}~\bibnamefont {{Singanamalla}}}, \bibinfo {author} {\bibfnamefont {L.~P.}\ \bibnamefont {{Singer}}}, \bibinfo {author} {\bibfnamefont {J.}~\bibnamefont {{Singhal}}}, \bibinfo {author} {\bibfnamefont {M.}~\bibnamefont {{Sinha}}}, \bibinfo {author} {\bibfnamefont {B.~M.}\ \bibnamefont {{Sip{\H{o}}cz}}}, \bibinfo {author} {\bibfnamefont {L.~R.}\ \bibnamefont {{Spitler}}}, \bibinfo {author} {\bibfnamefont {D.}~\bibnamefont {{Stansby}}}, \bibinfo {author} {\bibfnamefont {O.}~\bibnamefont {{Streicher}}}, \bibinfo {author} {\bibfnamefont {J.}~\bibnamefont {{{\v{S}}umak}}}, \bibinfo {author} {\bibfnamefont {J.~D.}\ \bibnamefont {{Swinbank}}}, \bibinfo {author} {\bibfnamefont {D.~S.}\ \bibnamefont {{Taranu}}}, \bibinfo {author} {\bibfnamefont {N.}~\bibnamefont
  {{Tewary}}}, \bibinfo {author} {\bibfnamefont {G.~R.}\ \bibnamefont {{Tremblay}}}, \bibinfo {author} {\bibfnamefont {M.}~\bibnamefont {{de Val-Borro}}}, \bibinfo {author} {\bibfnamefont {S.~J.}\ \bibnamefont {{Van Kooten}}}, \bibinfo {author} {\bibfnamefont {Z.}~\bibnamefont {{Vasovi{\'c}}}}, \bibinfo {author} {\bibfnamefont {S.}~\bibnamefont {{Verma}}}, \bibinfo {author} {\bibfnamefont {J.~V.}\ \bibnamefont {{de Miranda Cardoso}}}, \bibinfo {author} {\bibfnamefont {P.~K.~G.}\ \bibnamefont {{Williams}}}, \bibinfo {author} {\bibfnamefont {T.~J.}\ \bibnamefont {{Wilson}}}, \bibinfo {author} {\bibfnamefont {B.}~\bibnamefont {{Winkel}}}, \bibinfo {author} {\bibfnamefont {W.~M.}\ \bibnamefont {{Wood-Vasey}}}, \bibinfo {author} {\bibfnamefont {R.}~\bibnamefont {{Xue}}}, \bibinfo {author} {\bibfnamefont {P.}~\bibnamefont {{Yoachim}}}, \bibinfo {author} {\bibfnamefont {C.}~\bibnamefont {{Zhang}}}, \bibinfo {author} {\bibfnamefont {A.}~\bibnamefont {{Zonca}}},\ and\ \bibinfo {author} {\bibnamefont {{Astropy Project
  Contributors}}},\ }\href {https://doi.org/10.3847/1538-4357/ac7c74} {\bibfield  {journal} {\bibinfo  {journal} {\apj}\ }\textbf {\bibinfo {volume} {935}},\ \bibinfo {eid} {167} (\bibinfo {year} {2022})},\ \Eprint {https://arxiv.org/abs/2206.14220} {arXiv:2206.14220 [astro-ph.IM]} \BibitemShut {NoStop}%
\bibitem [{\citenamefont {Hunter}(2007)}]{matplotlib}%
  \BibitemOpen
  \bibfield  {author} {\bibinfo {author} {\bibfnamefont {J.~D.}\ \bibnamefont {Hunter}},\ }\href {https://doi.org/10.1109/MCSE.2007.55} {\bibfield  {journal} {\bibinfo  {journal} {Computing in Science \& Engineering}\ }\textbf {\bibinfo {volume} {9}},\ \bibinfo {pages} {90} (\bibinfo {year} {2007})}\BibitemShut {NoStop}%
\bibitem [{\citenamefont {{Foreman-Mackey}}\ \emph {et~al.}(2013)\citenamefont {{Foreman-Mackey}}, \citenamefont {{Hogg}}, \citenamefont {{Lang}},\ and\ \citenamefont {{Goodman}}}]{emcee}%
  \BibitemOpen
  \bibfield  {author} {\bibinfo {author} {\bibfnamefont {D.}~\bibnamefont {{Foreman-Mackey}}}, \bibinfo {author} {\bibfnamefont {D.~W.}\ \bibnamefont {{Hogg}}}, \bibinfo {author} {\bibfnamefont {D.}~\bibnamefont {{Lang}}},\ and\ \bibinfo {author} {\bibfnamefont {J.}~\bibnamefont {{Goodman}}},\ }\href {https://doi.org/10.1086/670067} {\bibfield  {journal} {\bibinfo  {journal} {\pasp}\ }\textbf {\bibinfo {volume} {125}},\ \bibinfo {pages} {306} (\bibinfo {year} {2013})},\ \Eprint {https://arxiv.org/abs/1202.3665} {arXiv:1202.3665 [astro-ph.IM]} \BibitemShut {NoStop}%
\bibitem [{\citenamefont {{Bobylev}}\ and\ \citenamefont {{Bajkova}}(2010)}]{BobylevBajkova2010}%
  \BibitemOpen
  \bibfield  {author} {\bibinfo {author} {\bibfnamefont {V.~V.}\ \bibnamefont {{Bobylev}}}\ and\ \bibinfo {author} {\bibfnamefont {A.~T.}\ \bibnamefont {{Bajkova}}},\ }\href {https://doi.org/10.1111/j.1365-2966.2010.17244.x} {\bibfield  {journal} {\bibinfo  {journal} {\mnras}\ }\textbf {\bibinfo {volume} {408}},\ \bibinfo {pages} {1788} (\bibinfo {year} {2010})},\ \Eprint {https://arxiv.org/abs/1006.5152} {arXiv:1006.5152 [astro-ph.GA]} \BibitemShut {NoStop}%
\bibitem [{\citenamefont {{Branham}}(2010)}]{Branham2010}%
  \BibitemOpen
  \bibfield  {author} {\bibinfo {author} {\bibfnamefont {J.}~\bibnamefont {{Branham}}, \bibfnamefont {Richard~L.}},\ }\href {https://doi.org/10.1111/j.1365-2966.2010.17389.x} {\bibfield  {journal} {\bibinfo  {journal} {\mnras}\ }\textbf {\bibinfo {volume} {409}},\ \bibinfo {pages} {1269} (\bibinfo {year} {2010})}\BibitemShut {NoStop}%
\bibitem [{\citenamefont {{Shen}}\ and\ \citenamefont {{Zhang}}(2010)}]{ShenZhang2010}%
  \BibitemOpen
  \bibfield  {author} {\bibinfo {author} {\bibfnamefont {M.}~\bibnamefont {{Shen}}}\ and\ \bibinfo {author} {\bibfnamefont {H.}~\bibnamefont {{Zhang}}},\ }\href {https://doi.org/10.1016/j.chinastron.2009.12.008} {\bibfield  {journal} {\bibinfo  {journal} {\caa}\ }\textbf {\bibinfo {volume} {34}},\ \bibinfo {pages} {89} (\bibinfo {year} {2010})}\BibitemShut {NoStop}%
\bibitem [{\citenamefont {{Bobylev}}\ and\ \citenamefont {{Bajkova}}(2011)}]{BobylevBajkova2011}%
  \BibitemOpen
  \bibfield  {author} {\bibinfo {author} {\bibfnamefont {V.~V.}\ \bibnamefont {{Bobylev}}}\ and\ \bibinfo {author} {\bibfnamefont {A.~T.}\ \bibnamefont {{Bajkova}}},\ }\href {https://doi.org/10.1134/S0320010811080018} {\bibfield  {journal} {\bibinfo  {journal} {Astronomy Letters}\ }\textbf {\bibinfo {volume} {37}},\ \bibinfo {pages} {526} (\bibinfo {year} {2011})},\ \Eprint {https://arxiv.org/abs/1106.5849} {arXiv:1106.5849 [astro-ph.GA]} \BibitemShut {NoStop}%
\bibitem [{\citenamefont {{Branham}}(2011)}]{Branham2011}%
  \BibitemOpen
  \bibfield  {author} {\bibinfo {author} {\bibfnamefont {J.}~\bibnamefont {{Branham}}, \bibfnamefont {R.~L.}},\ }\href@noop {} {\bibfield  {journal} {\bibinfo  {journal} {\rmxaa}\ }\textbf {\bibinfo {volume} {47}},\ \bibinfo {pages} {197} (\bibinfo {year} {2011})}\BibitemShut {NoStop}%
\bibitem [{\citenamefont {{Siebert}}\ \emph {et~al.}(2011)\citenamefont {{Siebert}}, \citenamefont {{Famaey}}, \citenamefont {{Minchev}}, \citenamefont {{Seabroke}}, \citenamefont {{Binney}}, \citenamefont {{Burnett}}, \citenamefont {{Freeman}}, \citenamefont {{Williams}}, \citenamefont {{Bienaym{\'e}}}, \citenamefont {{Bland-Hawthorn}}, \citenamefont {{Campbell}}, \citenamefont {{Fulbright}}, \citenamefont {{Gibson}}, \citenamefont {{Gilmore}}, \citenamefont {{Grebel}}, \citenamefont {{Helmi}}, \citenamefont {{Munari}}, \citenamefont {{Navarro}}, \citenamefont {{Parker}}, \citenamefont {{Reid}}, \citenamefont {{Siviero}}, \citenamefont {{Steinmetz}}, \citenamefont {{Watson}}, \citenamefont {{Wyse}},\ and\ \citenamefont {{Zwitter}}}]{Siebert2011}%
  \BibitemOpen
  \bibfield  {author} {\bibinfo {author} {\bibfnamefont {A.}~\bibnamefont {{Siebert}}}, \bibinfo {author} {\bibfnamefont {B.}~\bibnamefont {{Famaey}}}, \bibinfo {author} {\bibfnamefont {I.}~\bibnamefont {{Minchev}}}, \bibinfo {author} {\bibfnamefont {G.~M.}\ \bibnamefont {{Seabroke}}}, \bibinfo {author} {\bibfnamefont {J.}~\bibnamefont {{Binney}}}, \bibinfo {author} {\bibfnamefont {B.}~\bibnamefont {{Burnett}}}, \bibinfo {author} {\bibfnamefont {K.~C.}\ \bibnamefont {{Freeman}}}, \bibinfo {author} {\bibfnamefont {M.}~\bibnamefont {{Williams}}}, \bibinfo {author} {\bibfnamefont {O.}~\bibnamefont {{Bienaym{\'e}}}}, \bibinfo {author} {\bibfnamefont {J.}~\bibnamefont {{Bland-Hawthorn}}}, \bibinfo {author} {\bibfnamefont {R.}~\bibnamefont {{Campbell}}}, \bibinfo {author} {\bibfnamefont {J.~P.}\ \bibnamefont {{Fulbright}}}, \bibinfo {author} {\bibfnamefont {B.~K.}\ \bibnamefont {{Gibson}}}, \bibinfo {author} {\bibfnamefont {G.}~\bibnamefont {{Gilmore}}}, \bibinfo {author} {\bibfnamefont {E.~K.}\ \bibnamefont
  {{Grebel}}}, \bibinfo {author} {\bibfnamefont {A.}~\bibnamefont {{Helmi}}}, \bibinfo {author} {\bibfnamefont {U.}~\bibnamefont {{Munari}}}, \bibinfo {author} {\bibfnamefont {J.~F.}\ \bibnamefont {{Navarro}}}, \bibinfo {author} {\bibfnamefont {Q.~A.}\ \bibnamefont {{Parker}}}, \bibinfo {author} {\bibfnamefont {W.~A.}\ \bibnamefont {{Reid}}}, \bibinfo {author} {\bibfnamefont {A.}~\bibnamefont {{Siviero}}}, \bibinfo {author} {\bibfnamefont {M.}~\bibnamefont {{Steinmetz}}}, \bibinfo {author} {\bibfnamefont {F.}~\bibnamefont {{Watson}}}, \bibinfo {author} {\bibfnamefont {R.~F.~G.}\ \bibnamefont {{Wyse}}},\ and\ \bibinfo {author} {\bibfnamefont {T.}~\bibnamefont {{Zwitter}}},\ }\href {https://doi.org/10.1111/j.1365-2966.2010.18037.x} {\bibfield  {journal} {\bibinfo  {journal} {\mnras}\ }\textbf {\bibinfo {volume} {412}},\ \bibinfo {pages} {2026} (\bibinfo {year} {2011})},\ \Eprint {https://arxiv.org/abs/1011.4092} {arXiv:1011.4092 [astro-ph.GA]} \BibitemShut {NoStop}%
\bibitem [{\citenamefont {{Stepanishchev}}\ and\ \citenamefont {{Bobylev}}(2011)}]{StepanishchevBobylev2011}%
  \BibitemOpen
  \bibfield  {author} {\bibinfo {author} {\bibfnamefont {A.~S.}\ \bibnamefont {{Stepanishchev}}}\ and\ \bibinfo {author} {\bibfnamefont {V.~V.}\ \bibnamefont {{Bobylev}}},\ }\href {https://doi.org/10.1134/S1063773711030054} {\bibfield  {journal} {\bibinfo  {journal} {Astronomy Letters}\ }\textbf {\bibinfo {volume} {37}},\ \bibinfo {pages} {254} (\bibinfo {year} {2011})}\BibitemShut {NoStop}%
\bibitem [{\citenamefont {{Stepanishchev}}\ and\ \citenamefont {{Bobylev}}(2013)}]{StepanishchevBobylev2013}%
  \BibitemOpen
  \bibfield  {author} {\bibinfo {author} {\bibfnamefont {A.~S.}\ \bibnamefont {{Stepanishchev}}}\ and\ \bibinfo {author} {\bibfnamefont {V.~V.}\ \bibnamefont {{Bobylev}}},\ }\href {https://doi.org/10.1134/S1063773713030067} {\bibfield  {journal} {\bibinfo  {journal} {Astronomy Letters}\ }\textbf {\bibinfo {volume} {39}},\ \bibinfo {pages} {185} (\bibinfo {year} {2013})},\ \Eprint {https://arxiv.org/abs/1302.1457} {arXiv:1302.1457 [astro-ph.GA]} \BibitemShut {NoStop}%
\bibitem [{\citenamefont {{Branham}}(2014)}]{Branham2014}%
  \BibitemOpen
  \bibfield  {author} {\bibinfo {author} {\bibfnamefont {R.~L.}\ \bibnamefont {{Branham}}},\ }\href {https://doi.org/10.1007/s10509-014-2005-9} {\bibfield  {journal} {\bibinfo  {journal} {\apss}\ }\textbf {\bibinfo {volume} {353}},\ \bibinfo {pages} {179} (\bibinfo {year} {2014})}\BibitemShut {NoStop}%
\bibitem [{\citenamefont {{Vityazev}}\ and\ \citenamefont {{Tsvetkov}}(2014)}]{VityazevTsvetkov2014}%
  \BibitemOpen
  \bibfield  {author} {\bibinfo {author} {\bibfnamefont {V.~V.}\ \bibnamefont {{Vityazev}}}\ and\ \bibinfo {author} {\bibfnamefont {A.~S.}\ \bibnamefont {{Tsvetkov}}},\ }\href {https://doi.org/10.1093/mnras/stu953} {\bibfield  {journal} {\bibinfo  {journal} {\mnras}\ }\textbf {\bibinfo {volume} {442}},\ \bibinfo {pages} {1249} (\bibinfo {year} {2014})}\BibitemShut {NoStop}%
\bibitem [{\citenamefont {{Bobylev}}\ and\ \citenamefont {{Bajkova}}(2016)}]{BobylevBajkova2016}%
  \BibitemOpen
  \bibfield  {author} {\bibinfo {author} {\bibfnamefont {V.~V.}\ \bibnamefont {{Bobylev}}}\ and\ \bibinfo {author} {\bibfnamefont {A.~T.}\ \bibnamefont {{Bajkova}}},\ }\href {https://doi.org/10.1134/S1063773716020018} {\bibfield  {journal} {\bibinfo  {journal} {Astronomy Letters}\ }\textbf {\bibinfo {volume} {42}},\ \bibinfo {pages} {90} (\bibinfo {year} {2016})},\ \Eprint {https://arxiv.org/abs/1512.08100} {arXiv:1512.08100 [astro-ph.GA]} \BibitemShut {NoStop}%
\bibitem [{\citenamefont {{Bobylev}}\ and\ \citenamefont {{Bajkova}}(2017)}]{BobylevBajkova2017}%
  \BibitemOpen
  \bibfield  {author} {\bibinfo {author} {\bibfnamefont {V.~V.}\ \bibnamefont {{Bobylev}}}\ and\ \bibinfo {author} {\bibfnamefont {A.~T.}\ \bibnamefont {{Bajkova}}},\ }\href {https://doi.org/10.1134/S1063773717030021} {\bibfield  {journal} {\bibinfo  {journal} {Astronomy Letters}\ }\textbf {\bibinfo {volume} {43}},\ \bibinfo {pages} {159} (\bibinfo {year} {2017})},\ \Eprint {https://arxiv.org/abs/1611.00794} {arXiv:1611.00794 [astro-ph.GA]} \BibitemShut {NoStop}%
\bibitem [{\citenamefont {{Vityazev}}\ \emph {et~al.}(2017)\citenamefont {{Vityazev}}, \citenamefont {{Tsvetkov}}, \citenamefont {{Bobylev}},\ and\ \citenamefont {{Bajkova}}}]{Vityazev2017}%
  \BibitemOpen
  \bibfield  {author} {\bibinfo {author} {\bibfnamefont {V.~V.}\ \bibnamefont {{Vityazev}}}, \bibinfo {author} {\bibfnamefont {A.~S.}\ \bibnamefont {{Tsvetkov}}}, \bibinfo {author} {\bibfnamefont {V.~V.}\ \bibnamefont {{Bobylev}}},\ and\ \bibinfo {author} {\bibfnamefont {A.~T.}\ \bibnamefont {{Bajkova}}},\ }\href {https://doi.org/10.1007/s10511-017-9499-0} {\bibfield  {journal} {\bibinfo  {journal} {Astrophysics}\ }\textbf {\bibinfo {volume} {60}},\ \bibinfo {pages} {462} (\bibinfo {year} {2017})},\ \Eprint {https://arxiv.org/abs/1711.01476} {arXiv:1711.01476 [astro-ph.GA]} \BibitemShut {NoStop}%
\bibitem [{\citenamefont {{Bobylev}}\ and\ \citenamefont {{Bajkova}}(2018)}]{BobylevBajkova2018}%
  \BibitemOpen
  \bibfield  {author} {\bibinfo {author} {\bibfnamefont {V.~V.}\ \bibnamefont {{Bobylev}}}\ and\ \bibinfo {author} {\bibfnamefont {A.~T.}\ \bibnamefont {{Bajkova}}},\ }\href {https://doi.org/10.1134/S1063773718020020} {\bibfield  {journal} {\bibinfo  {journal} {Astronomy Letters}\ }\textbf {\bibinfo {volume} {44}},\ \bibinfo {pages} {184} (\bibinfo {year} {2018})},\ \Eprint {https://arxiv.org/abs/1801.07431} {arXiv:1801.07431 [astro-ph.GA]} \BibitemShut {NoStop}%
\bibitem [{\citenamefont {{Vickers}}\ and\ \citenamefont {{Smith}}(2018)}]{VickersSmith2018}%
  \BibitemOpen
  \bibfield  {author} {\bibinfo {author} {\bibfnamefont {J.~J.}\ \bibnamefont {{Vickers}}}\ and\ \bibinfo {author} {\bibfnamefont {M.~C.}\ \bibnamefont {{Smith}}},\ }\href {https://doi.org/10.3847/1538-4357/aac323} {\bibfield  {journal} {\bibinfo  {journal} {\apj}\ }\textbf {\bibinfo {volume} {860}},\ \bibinfo {eid} {91} (\bibinfo {year} {2018})},\ \Eprint {https://arxiv.org/abs/1805.02332} {arXiv:1805.02332 [astro-ph.GA]} \BibitemShut {NoStop}%
\bibitem [{\citenamefont {{Vityazev}}\ \emph {et~al.}(2018)\citenamefont {{Vityazev}}, \citenamefont {{Popov}}, \citenamefont {{Tsvetkov}}, \citenamefont {{Petrov}}, \citenamefont {{Trofimov}},\ and\ \citenamefont {{Kiyaev}}}]{Vityazev2018}%
  \BibitemOpen
  \bibfield  {author} {\bibinfo {author} {\bibfnamefont {V.~V.}\ \bibnamefont {{Vityazev}}}, \bibinfo {author} {\bibfnamefont {A.~V.}\ \bibnamefont {{Popov}}}, \bibinfo {author} {\bibfnamefont {A.~S.}\ \bibnamefont {{Tsvetkov}}}, \bibinfo {author} {\bibfnamefont {S.~D.}\ \bibnamefont {{Petrov}}}, \bibinfo {author} {\bibfnamefont {D.~A.}\ \bibnamefont {{Trofimov}}},\ and\ \bibinfo {author} {\bibfnamefont {V.~I.}\ \bibnamefont {{Kiyaev}}},\ }\href {https://doi.org/10.1134/S1063773718030040} {\bibfield  {journal} {\bibinfo  {journal} {Astronomy Letters}\ }\textbf {\bibinfo {volume} {44}},\ \bibinfo {pages} {236} (\bibinfo {year} {2018})}\BibitemShut {NoStop}%
\bibitem [{\citenamefont {{Krisanova}}\ \emph {et~al.}(2020)\citenamefont {{Krisanova}}, \citenamefont {{Bobylev}},\ and\ \citenamefont {{Bajkova}}}]{Krisanova2020}%
  \BibitemOpen
  \bibfield  {author} {\bibinfo {author} {\bibfnamefont {O.~I.}\ \bibnamefont {{Krisanova}}}, \bibinfo {author} {\bibfnamefont {V.~V.}\ \bibnamefont {{Bobylev}}},\ and\ \bibinfo {author} {\bibfnamefont {A.~T.}\ \bibnamefont {{Bajkova}}},\ }\href {https://doi.org/10.1134/S1063773720060067} {\bibfield  {journal} {\bibinfo  {journal} {Astronomy Letters}\ }\textbf {\bibinfo {volume} {46}},\ \bibinfo {pages} {370} (\bibinfo {year} {2020})},\ \Eprint {https://arxiv.org/abs/2008.10981} {arXiv:2008.10981 [astro-ph.GA]} \BibitemShut {NoStop}%
\bibitem [{\citenamefont {{Nouh}}\ and\ \citenamefont {{Elsanhoury}}(2020)}]{NouhElsanhoury2020}%
  \BibitemOpen
  \bibfield  {author} {\bibinfo {author} {\bibfnamefont {M.~I.}\ \bibnamefont {{Nouh}}}\ and\ \bibinfo {author} {\bibfnamefont {W.~H.}\ \bibnamefont {{Elsanhoury}}},\ }\href {https://doi.org/10.1007/s10511-020-09624-5} {\bibfield  {journal} {\bibinfo  {journal} {Astrophysics}\ }\textbf {\bibinfo {volume} {63}},\ \bibinfo {pages} {179} (\bibinfo {year} {2020})},\ \Eprint {https://arxiv.org/abs/1905.07614} {arXiv:1905.07614 [astro-ph.SR]} \BibitemShut {NoStop}%
\bibitem [{\citenamefont {{Wang}}\ \emph {et~al.}(2021)\citenamefont {{Wang}}, \citenamefont {{Zhang}}, \citenamefont {{Huang}}, \citenamefont {{Chen}}, \citenamefont {{Wang}},\ and\ \citenamefont {{Wang}}}]{Wang2021}%
  \BibitemOpen
  \bibfield  {author} {\bibinfo {author} {\bibfnamefont {F.}~\bibnamefont {{Wang}}}, \bibinfo {author} {\bibfnamefont {H.~W.}\ \bibnamefont {{Zhang}}}, \bibinfo {author} {\bibfnamefont {Y.}~\bibnamefont {{Huang}}}, \bibinfo {author} {\bibfnamefont {B.~Q.}\ \bibnamefont {{Chen}}}, \bibinfo {author} {\bibfnamefont {H.~F.}\ \bibnamefont {{Wang}}},\ and\ \bibinfo {author} {\bibfnamefont {C.}~\bibnamefont {{Wang}}},\ }\href {https://doi.org/10.1093/mnras/stab848} {\bibfield  {journal} {\bibinfo  {journal} {\mnras}\ }\textbf {\bibinfo {volume} {504}},\ \bibinfo {pages} {199} (\bibinfo {year} {2021})},\ \Eprint {https://arxiv.org/abs/2103.10232} {arXiv:2103.10232 [astro-ph.GA]} \BibitemShut {NoStop}%
\bibitem [{\citenamefont {{Nelson}}\ and\ \citenamefont {{Widrow}}(2022)}]{NelsonWidrow2022}%
  \BibitemOpen
  \bibfield  {author} {\bibinfo {author} {\bibfnamefont {P.}~\bibnamefont {{Nelson}}}\ and\ \bibinfo {author} {\bibfnamefont {L.~M.}\ \bibnamefont {{Widrow}}},\ }\href {https://doi.org/10.1093/mnras/stac2594} {\bibfield  {journal} {\bibinfo  {journal} {\mnras}\ }\textbf {\bibinfo {volume} {516}},\ \bibinfo {pages} {5429} (\bibinfo {year} {2022})},\ \Eprint {https://arxiv.org/abs/2206.04627} {arXiv:2206.04627 [astro-ph.GA]} \BibitemShut {NoStop}%
\bibitem [{\citenamefont {{Guo}}\ and\ \citenamefont {{Qi}}(2023)}]{GuoQi2023}%
  \BibitemOpen
  \bibfield  {author} {\bibinfo {author} {\bibfnamefont {S.}~\bibnamefont {{Guo}}}\ and\ \bibinfo {author} {\bibfnamefont {Z.}~\bibnamefont {{Qi}}},\ }\href {https://doi.org/10.3390/universe9060252} {\bibfield  {journal} {\bibinfo  {journal} {Universe}\ }\textbf {\bibinfo {volume} {9}},\ \bibinfo {eid} {252} (\bibinfo {year} {2023})}\BibitemShut {NoStop}%
\bibitem [{\citenamefont {{Moni Bidin}}\ \emph {et~al.}(2012)\citenamefont {{Moni Bidin}}, \citenamefont {{Carraro}}, \citenamefont {{M{\'e}ndez}},\ and\ \citenamefont {{Smith}}}]{MoniBidin2012}%
  \BibitemOpen
  \bibfield  {author} {\bibinfo {author} {\bibfnamefont {C.}~\bibnamefont {{Moni Bidin}}}, \bibinfo {author} {\bibfnamefont {G.}~\bibnamefont {{Carraro}}}, \bibinfo {author} {\bibfnamefont {R.~A.}\ \bibnamefont {{M{\'e}ndez}}},\ and\ \bibinfo {author} {\bibfnamefont {R.}~\bibnamefont {{Smith}}},\ }\href {https://doi.org/10.1088/0004-637X/751/1/30} {\bibfield  {journal} {\bibinfo  {journal} {\apj}\ }\textbf {\bibinfo {volume} {751}},\ \bibinfo {eid} {30} (\bibinfo {year} {2012})},\ \Eprint {https://arxiv.org/abs/1204.3924} {arXiv:1204.3924 [astro-ph.GA]} \BibitemShut {NoStop}%
\bibitem [{\citenamefont {{Bovy}}\ and\ \citenamefont {{Tremaine}}(2012)}]{BovyTremaine2012}%
  \BibitemOpen
  \bibfield  {author} {\bibinfo {author} {\bibfnamefont {J.}~\bibnamefont {{Bovy}}}\ and\ \bibinfo {author} {\bibfnamefont {S.}~\bibnamefont {{Tremaine}}},\ }\href {https://doi.org/10.1088/0004-637X/756/1/89} {\bibfield  {journal} {\bibinfo  {journal} {\apj}\ }\textbf {\bibinfo {volume} {756}},\ \bibinfo {eid} {89} (\bibinfo {year} {2012})},\ \Eprint {https://arxiv.org/abs/1205.4033} {arXiv:1205.4033 [astro-ph.GA]} \BibitemShut {NoStop}%
\bibitem [{\citenamefont {{Garbari}}\ \emph {et~al.}(2012)\citenamefont {{Garbari}}, \citenamefont {{Liu}}, \citenamefont {{Read}},\ and\ \citenamefont {{Lake}}}]{Garbari2012}%
  \BibitemOpen
  \bibfield  {author} {\bibinfo {author} {\bibfnamefont {S.}~\bibnamefont {{Garbari}}}, \bibinfo {author} {\bibfnamefont {C.}~\bibnamefont {{Liu}}}, \bibinfo {author} {\bibfnamefont {J.~I.}\ \bibnamefont {{Read}}},\ and\ \bibinfo {author} {\bibfnamefont {G.}~\bibnamefont {{Lake}}},\ }\href {https://doi.org/10.1111/j.1365-2966.2012.21608.x} {\bibfield  {journal} {\bibinfo  {journal} {\mnras}\ }\textbf {\bibinfo {volume} {425}},\ \bibinfo {pages} {1445} (\bibinfo {year} {2012})},\ \Eprint {https://arxiv.org/abs/1206.0015} {arXiv:1206.0015 [astro-ph.GA]} \BibitemShut {NoStop}%
\bibitem [{\citenamefont {{Smith}}\ \emph {et~al.}(2012)\citenamefont {{Smith}}, \citenamefont {{Whiteoak}},\ and\ \citenamefont {{Evans}}}]{Smith2012}%
  \BibitemOpen
  \bibfield  {author} {\bibinfo {author} {\bibfnamefont {M.~C.}\ \bibnamefont {{Smith}}}, \bibinfo {author} {\bibfnamefont {S.~H.}\ \bibnamefont {{Whiteoak}}},\ and\ \bibinfo {author} {\bibfnamefont {N.~W.}\ \bibnamefont {{Evans}}},\ }\href {https://doi.org/10.1088/0004-637X/746/2/181} {\bibfield  {journal} {\bibinfo  {journal} {\apj}\ }\textbf {\bibinfo {volume} {746}},\ \bibinfo {eid} {181} (\bibinfo {year} {2012})},\ \Eprint {https://arxiv.org/abs/1111.6920} {arXiv:1111.6920 [astro-ph.GA]} \BibitemShut {NoStop}%
\bibitem [{\citenamefont {{Zhang}}\ \emph {et~al.}(2013)\citenamefont {{Zhang}}, \citenamefont {{Rix}}, \citenamefont {{van de Ven}}, \citenamefont {{Bovy}}, \citenamefont {{Liu}},\ and\ \citenamefont {{Zhao}}}]{Zhang2013}%
  \BibitemOpen
  \bibfield  {author} {\bibinfo {author} {\bibfnamefont {L.}~\bibnamefont {{Zhang}}}, \bibinfo {author} {\bibfnamefont {H.-W.}\ \bibnamefont {{Rix}}}, \bibinfo {author} {\bibfnamefont {G.}~\bibnamefont {{van de Ven}}}, \bibinfo {author} {\bibfnamefont {J.}~\bibnamefont {{Bovy}}}, \bibinfo {author} {\bibfnamefont {C.}~\bibnamefont {{Liu}}},\ and\ \bibinfo {author} {\bibfnamefont {G.}~\bibnamefont {{Zhao}}},\ }\href {https://doi.org/10.1088/0004-637X/772/2/108} {\bibfield  {journal} {\bibinfo  {journal} {\apj}\ }\textbf {\bibinfo {volume} {772}},\ \bibinfo {eid} {108} (\bibinfo {year} {2013})},\ \Eprint {https://arxiv.org/abs/1209.0256} {arXiv:1209.0256 [astro-ph.GA]} \BibitemShut {NoStop}%
\bibitem [{\citenamefont {{Xia}}\ \emph {et~al.}(2016)\citenamefont {{Xia}}, \citenamefont {{Liu}}, \citenamefont {{Mao}}, \citenamefont {{Song}}, \citenamefont {{Zhang}}, \citenamefont {{Long}}, \citenamefont {{Zhang}}, \citenamefont {{Hou}}, \citenamefont {{Wang}},\ and\ \citenamefont {{Wu}}}]{Xia2016}%
  \BibitemOpen
  \bibfield  {author} {\bibinfo {author} {\bibfnamefont {Q.}~\bibnamefont {{Xia}}}, \bibinfo {author} {\bibfnamefont {C.}~\bibnamefont {{Liu}}}, \bibinfo {author} {\bibfnamefont {S.}~\bibnamefont {{Mao}}}, \bibinfo {author} {\bibfnamefont {Y.}~\bibnamefont {{Song}}}, \bibinfo {author} {\bibfnamefont {L.}~\bibnamefont {{Zhang}}}, \bibinfo {author} {\bibfnamefont {R.~J.}\ \bibnamefont {{Long}}}, \bibinfo {author} {\bibfnamefont {Y.}~\bibnamefont {{Zhang}}}, \bibinfo {author} {\bibfnamefont {Y.}~\bibnamefont {{Hou}}}, \bibinfo {author} {\bibfnamefont {Y.}~\bibnamefont {{Wang}}},\ and\ \bibinfo {author} {\bibfnamefont {Y.}~\bibnamefont {{Wu}}},\ }\href {https://doi.org/10.1093/mnras/stw565} {\bibfield  {journal} {\bibinfo  {journal} {\mnras}\ }\textbf {\bibinfo {volume} {458}},\ \bibinfo {pages} {3839} (\bibinfo {year} {2016})},\ \Eprint {https://arxiv.org/abs/1510.06810} {arXiv:1510.06810 [astro-ph.GA]} \BibitemShut {NoStop}%
\bibitem [{\citenamefont {{Hagen}}\ and\ \citenamefont {{Helmi}}(2018)}]{Hagen2018}%
  \BibitemOpen
  \bibfield  {author} {\bibinfo {author} {\bibfnamefont {J.~H.~J.}\ \bibnamefont {{Hagen}}}\ and\ \bibinfo {author} {\bibfnamefont {A.}~\bibnamefont {{Helmi}}},\ }\href {https://doi.org/10.1051/0004-6361/201832903} {\bibfield  {journal} {\bibinfo  {journal} {\aap}\ }\textbf {\bibinfo {volume} {615}},\ \bibinfo {eid} {A99} (\bibinfo {year} {2018})},\ \Eprint {https://arxiv.org/abs/1802.09291} {arXiv:1802.09291 [astro-ph.GA]} \BibitemShut {NoStop}%
\bibitem [{\citenamefont {{Sivertsson}}\ \emph {et~al.}(2018)\citenamefont {{Sivertsson}}, \citenamefont {{Silverwood}}, \citenamefont {{Read}}, \citenamefont {{Bertone}},\ and\ \citenamefont {{Steger}}}]{Sivertsson2018}%
  \BibitemOpen
  \bibfield  {author} {\bibinfo {author} {\bibfnamefont {S.}~\bibnamefont {{Sivertsson}}}, \bibinfo {author} {\bibfnamefont {H.}~\bibnamefont {{Silverwood}}}, \bibinfo {author} {\bibfnamefont {J.~I.}\ \bibnamefont {{Read}}}, \bibinfo {author} {\bibfnamefont {G.}~\bibnamefont {{Bertone}}},\ and\ \bibinfo {author} {\bibfnamefont {P.}~\bibnamefont {{Steger}}},\ }\href {https://doi.org/10.1093/mnras/sty977} {\bibfield  {journal} {\bibinfo  {journal} {\mnras}\ }\textbf {\bibinfo {volume} {478}},\ \bibinfo {pages} {1677} (\bibinfo {year} {2018})},\ \Eprint {https://arxiv.org/abs/1708.07836} {arXiv:1708.07836 [astro-ph.GA]} \BibitemShut {NoStop}%
\bibitem [{\citenamefont {{Wardana}}\ \emph {et~al.}(2020)\citenamefont {{Wardana}}, \citenamefont {{Wulandari}}, \citenamefont {{Sulistiyowati}},\ and\ \citenamefont {{Khatami}}}]{Wardana2020}%
  \BibitemOpen
  \bibfield  {author} {\bibinfo {author} {\bibfnamefont {M.~D.}\ \bibnamefont {{Wardana}}}, \bibinfo {author} {\bibfnamefont {H.}~\bibnamefont {{Wulandari}}}, \bibinfo {author} {\bibnamefont {{Sulistiyowati}}},\ and\ \bibinfo {author} {\bibfnamefont {A.~H.}\ \bibnamefont {{Khatami}}},\ }in\ \href {https://doi.org/10.1051/epjconf/202024004002} {\emph {\bibinfo {booktitle} {European Physical Journal Web of Conferences}}},\ \bibinfo {series} {European Physical Journal Web of Conferences}, Vol.\ \bibinfo {volume} {240}\ (\bibinfo {year} {2020})\ p.\ \bibinfo {pages} {04002}\BibitemShut {NoStop}%
\bibitem [{\citenamefont {{Widmark}}\ \emph {et~al.}(2021)\citenamefont {{Widmark}}, \citenamefont {{Laporte}}, \citenamefont {{de Salas}},\ and\ \citenamefont {{Monari}}}]{Widmark2021}%
  \BibitemOpen
  \bibfield  {author} {\bibinfo {author} {\bibfnamefont {A.}~\bibnamefont {{Widmark}}}, \bibinfo {author} {\bibfnamefont {C.~F.~P.}\ \bibnamefont {{Laporte}}}, \bibinfo {author} {\bibfnamefont {P.~F.}\ \bibnamefont {{de Salas}}},\ and\ \bibinfo {author} {\bibfnamefont {G.}~\bibnamefont {{Monari}}},\ }\href {https://doi.org/10.1051/0004-6361/202141466} {\bibfield  {journal} {\bibinfo  {journal} {\aap}\ }\textbf {\bibinfo {volume} {653}},\ \bibinfo {eid} {A86} (\bibinfo {year} {2021})},\ \Eprint {https://arxiv.org/abs/2105.14030} {arXiv:2105.14030 [astro-ph.GA]} \BibitemShut {NoStop}%
\bibitem [{\citenamefont {{Guo}}\ \emph {et~al.}(2024)\citenamefont {{Guo}}, \citenamefont {{Li}}, \citenamefont {{Shen}}, \citenamefont {{Mao}},\ and\ \citenamefont {{Liu}}}]{Guo2024}%
  \BibitemOpen
  \bibfield  {author} {\bibinfo {author} {\bibfnamefont {R.}~\bibnamefont {{Guo}}}, \bibinfo {author} {\bibfnamefont {Z.-Y.}\ \bibnamefont {{Li}}}, \bibinfo {author} {\bibfnamefont {J.}~\bibnamefont {{Shen}}}, \bibinfo {author} {\bibfnamefont {S.}~\bibnamefont {{Mao}}},\ and\ \bibinfo {author} {\bibfnamefont {C.}~\bibnamefont {{Liu}}},\ }\href {https://doi.org/10.3847/1538-4357/ad037b} {\bibfield  {journal} {\bibinfo  {journal} {\apj}\ }\textbf {\bibinfo {volume} {960}},\ \bibinfo {eid} {133} (\bibinfo {year} {2024})},\ \Eprint {https://arxiv.org/abs/2310.10225} {arXiv:2310.10225 [astro-ph.GA]} \BibitemShut {NoStop}%
\bibitem [{\citenamefont {{Staudt}}\ \emph {et~al.}(2024)\citenamefont {{Staudt}}, \citenamefont {{Bullock}}, \citenamefont {{Boylan-Kolchin}}, \citenamefont {{Wetzel}},\ and\ \citenamefont {{Ou}}}]{Staudt2024}%
  \BibitemOpen
  \bibfield  {author} {\bibinfo {author} {\bibfnamefont {P.~G.}\ \bibnamefont {{Staudt}}}, \bibinfo {author} {\bibfnamefont {J.~S.}\ \bibnamefont {{Bullock}}}, \bibinfo {author} {\bibfnamefont {M.}~\bibnamefont {{Boylan-Kolchin}}}, \bibinfo {author} {\bibfnamefont {A.}~\bibnamefont {{Wetzel}}},\ and\ \bibinfo {author} {\bibfnamefont {X.}~\bibnamefont {{Ou}}},\ }\href {https://doi.org/10.48550/arXiv.2403.04122} {\bibfield  {journal} {\bibinfo  {journal} {arXiv e-prints}\ ,\ \bibinfo {eid} {arXiv:2403.04122}} (\bibinfo {year} {2024})},\ \Eprint {https://arxiv.org/abs/2403.04122} {arXiv:2403.04122 [astro-ph.GA]} \BibitemShut {NoStop}%
\bibitem [{\citenamefont {{Hu}}\ \emph {et~al.}(2020)\citenamefont {{Hu}}, \citenamefont {{Kramer}}, \citenamefont {{Wex}}, \citenamefont {{Champion}},\ and\ \citenamefont {{Kehl}}}]{Hu2020}%
  \BibitemOpen
  \bibfield  {author} {\bibinfo {author} {\bibfnamefont {H.}~\bibnamefont {{Hu}}}, \bibinfo {author} {\bibfnamefont {M.}~\bibnamefont {{Kramer}}}, \bibinfo {author} {\bibfnamefont {N.}~\bibnamefont {{Wex}}}, \bibinfo {author} {\bibfnamefont {D.~J.}\ \bibnamefont {{Champion}}},\ and\ \bibinfo {author} {\bibfnamefont {M.~S.}\ \bibnamefont {{Kehl}}},\ }\href {https://doi.org/10.1093/mnras/staa2107} {\bibfield  {journal} {\bibinfo  {journal} {\mnras}\ }\textbf {\bibinfo {volume} {497}},\ \bibinfo {pages} {3118} (\bibinfo {year} {2020})},\ \Eprint {https://arxiv.org/abs/2007.07725} {arXiv:2007.07725 [astro-ph.SR]} \BibitemShut {NoStop}%
\bibitem [{\citenamefont {{Lorimer}}\ and\ \citenamefont {{Kramer}}(2012)}]{HandbookPulsarAstro}%
  \BibitemOpen
  \bibfield  {author} {\bibinfo {author} {\bibfnamefont {D.~R.}\ \bibnamefont {{Lorimer}}}\ and\ \bibinfo {author} {\bibfnamefont {M.}~\bibnamefont {{Kramer}}},\ }\href@noop {} {\emph {\bibinfo {title} {{Handbook of Pulsar Astronomy}}}}\ (\bibinfo {year} {2012})\BibitemShut {NoStop}%
\end{thebibliography}%

\appendix 

\section{Oort Constant and Dark Matter Density References and Data}

In Table \ref{tab:oort_figure_data} we provide the data and references used to make Figure \ref{fig:oort_constants}.

\begin{table*}[]
    \centering
    \begin{tabular}{|llll|} \hline \hline
        $A$ & $B$ & Reference & Comments \\
        (km/s/kpc) & (km/s/kpc) & &  \\ \hline 
        17.8 $\pm$ 0.8 & -13.2 $\pm$ 1.5 & \cite{BobylevBajkova2010} & Masers \\
        14.85 $\pm$ 7.47 & -10.85 $\pm$ 6.63 & \cite{Branham2010} & Hipparcos F Giants \\
        17.42 $\pm$ 1.17 & -12.46 $\pm$ 0.86 & \cite{ShenZhang2010} & Hipparcos Cepheids, 0.2 kpc $<r<$ 3.0 kpc \\
        18.17 $\pm$ 1.14 & -11.91 $\pm$ 0.85 & & Above, 0.5 kpc $<r<$ 3.0 kpc \\
        17.9 $\pm$ 0.5 & -13.6 $\pm$ 1.0 & \cite{BobylevBajkova2011} & OB III Stars \\
        14.05 $\pm$ 3.28 & -9.30 $\pm$ 2.87 & \cite{Branham2011} & Hipparcos G III Stars \\
        9.2 $\pm$ 0.5 & -17.4 $\pm$ 0.5 & \cite{Siebert2011} & RAVE \\
        16.9 $\pm$ 1.2 & -13.5 $\pm$ 1.4 & \cite{StepanishchevBobylev2011} & Masers \\
        16.7 $\pm$ 0.6 & -12.0 $\pm$ 1.0 & \cite{StepanishchevBobylev2013} & Masers \\
        16.00 $\pm$ 0.36 & -14.17 $\pm$ 0.28 & \cite{Branham2014} & Hipparcos OB Stars \\
        14.2 $\pm$ 0.2 & -10.0 $\pm$ 0.2 & \cite{VityazevTsvetkov2014} & UCAC4 11th Mag \\
        9.1 $\pm$ 0.1 & -10.9 $\pm$ 0.1 & & Above, UCAC4 16th Mag \\
        12.1 $\pm$ 0.7 & -10.6 $\pm$ 0.5 & & Above, PPMXL 11th Mag \\
        10.3 $\pm$ 0.2 & -12.1 $\pm$ 0.2 & & Above, PPMXL 17th Mag \\
        12.2 $\pm$ 0.4 & -5.8 $\pm$ 0.3 & & Above, XPM 11th Mag \\
        7.0 $\pm$ 0.1 & 10.0 $\pm$ 0.1 & & Above, XPM 17th Mag \\
        17.12 $\pm$ 0.45 & -11.60 $\pm$ 0.77 & \cite{BobylevBajkova2016} & RAVE, $r<250$ pc \\
        16.53 $\pm$ 0.52 & -10.82 $\pm$ 0.93 & \cite{BobylevBajkova2017} & Gaia DR1 OB Stars \\
        17.77 $\pm$ 0.46 & -13.76 $\pm$ 0.71 & & Above, but different sample \\
        15.3 $\pm$ 0.4 & -11.9 $\pm$ 0.4 & \cite{Bovy2017} & Gaia DR1, $r<230$ pc \\
        15.57 $\pm$ 0.31 & -10.72 $\pm$ 0.5 & \cite{Vityazev2017} & RAVE5 $+$ TGAS \\
        15.07 $\pm$ 0.25 & -12.17 $\pm$ 0.39 & \cite{BobylevBajkova2018} & TGAS $+$ RAVE \\
        14.13 & -6.7 & \cite{VickersSmith2018} & TGAS-RAVE-LAMOST, Young Stars \\
        18.8 & -12.7 & & Above, Intermediate Age Stars \\
        17 & -2.1 & & Above, Old Stars \\
        16.29 $\pm$ 0.06 & -11.9 $\pm$ 0.05 & \cite{Vityazev2018} & TGAS MS Stars, $r<1.5$ kpc \\
        13.32 $\pm$ 0.09 & -12.71 $\pm$ 0.06 & & Above, TGAS RG Stars, $r<1.5$ kpc \\
        15.1 $\pm$ 0.1 & -13.4 $\pm$ 0.1 & \cite{Li2019} & Gaia DR2, $r<500$ pc \\
        15.73 $\pm$ 0.32 & -12.67 $\pm$ 0.34 & \cite{Krisanova2020} & Gaia DR2 \\
        15.6 $\pm$ 1.6 & -13.9 $\pm$ 1.8 & \cite{NouhElsanhoury2020} & SEGUE Halo RG Stars \\
        16.31 $\pm$ 0.89 & -11.99 $\pm$ 0.79 & \cite{Wang2021} & LAMOST A Stars, $r<600$ pc, $|z| < 100$ pc \\
        16.2 $\pm$ 0.2 & -11.7 $\pm$ 0.2 & \cite{NelsonWidrow2022} & Gaia DR2 ``(p)'' \\
        15.2 $\pm$ 0.8 & -12.4 $\pm$ 0.9 & & Above, ``(v)'' \\
        15.6 $\pm$ 1.6 & -15.8 $\pm$ 1.7 & \cite{GuoQi2023} & GCNS, Gaia EDR3 \\
        \hline \hline
    \end{tabular}
    \caption{Selected measurements of the Oort constants $A$ and $B$ since 2010.}
    \label{tab:oort_figure_data}
\end{table*}

In Table \ref{tab:dm_density} we provide the data and references used to make Figure \ref{fig:dm_density}.

\begin{table}[]
    \centering
    \begin{tabular}{|lr|} \hline \hline
        $\rho_{dm}$ & Reference \\
        (Msun pc$^{-3}$) &  \\ \hline 
        0.00062 $\pm$ 0.001 & \cite{MoniBidin2012} \\
        0.008 $\pm$ 0.003 & \cite{BovyTremaine2012} \\
        0.0087$^{+0.007}_{-0.002}$ & \cite{Garbari2012} \\
        0.005 & \cite{Smith2012} \\
        0.0065 $\pm$ 0.0023 & \cite{Zhang2013} \\
        0.006 $\pm$ 0.0018 & \cite{BovyRix2013} \\

        0.0143 $\pm$ 0.0011 & \cite{Bienayme2014} \\
        0.013 $\pm$ 0.003 & \cite{McKee2015} \\
        0.018 $\pm$ 0.0054 & \cite{Xia2016} \\
        0.018 $\pm$ 0.002 & \cite{Hagen2018} \\
        0.012 $\pm$ 0.002 & \cite{Sivertsson2018} \\
        0.016 $\pm$ 0.010 & \cite{Buch2019} \\
        0.0133$^{+0.0024}_{-0.0022}$ & \cite{Guo2020} \\
        0.0116 $\pm$ 0.0012 & \cite{Wardana2020} \\
        0.0085 $\pm$ 0.0039 & \cite{Widmark2021} \\
        -0.004$^{+0.05}_{-0.02}$ & \cite{Chakrabarti2021} \\
        0.012 $\pm$ 0.001 & \cite{Lim2023} \\
        0.0150 $\pm$ 0.0031 & \cite{Guo2024} \\
        0.011 $\pm$ 0.002 & \cite{Staudt2024} \\
        \hline \hline
    \end{tabular}
    \caption{Selected measurements of the dark matter density in the Galactic midplane since 2012.}
    \label{tab:dm_density}
\end{table}

\clearpage

\section{Individual Measurement vs. Fit Parameter Uncertainties}

\begin{figure}[h!]
    \centering
    \includegraphics[width=0.47\textwidth]{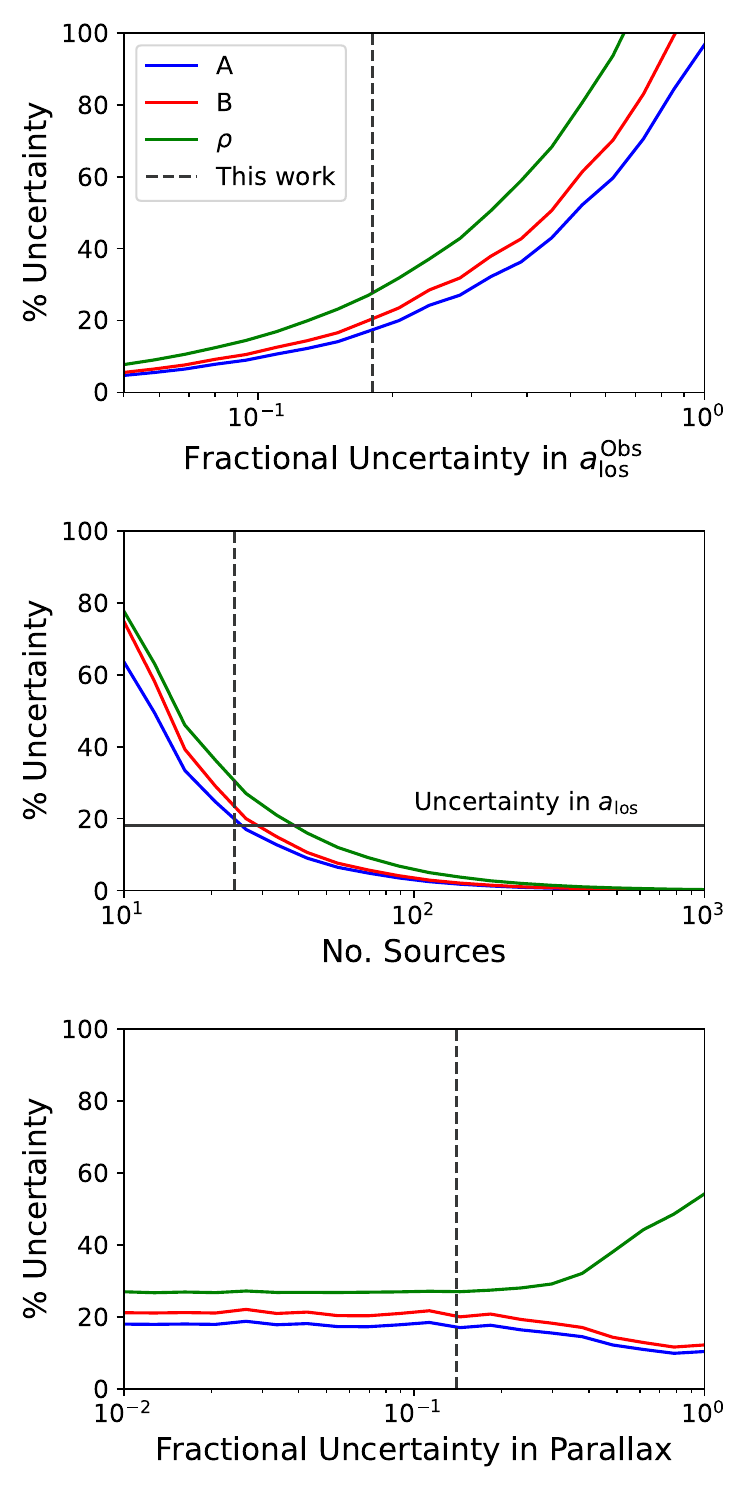}
    \caption{Fractional uncertainty in estimates of the Oort constants and Oort limit. Each row shows the dependence on the relative uncertainty of a given input, holding the other inputs constant at the mean values of the data in this work. Improving the uncertainty in the measured line-of-sight acceleration and increasing the number of sources has a substantial effect on the overall uncertainty in the estimated parameters. The uncertainty in parallax does not substantially impact the uncertainties in fit values.}
    \label{fig:errors_precision}
\end{figure}

Here we assess the impact of the various uncertainties in the pulsar timing measurements on the estimates of the Oort limit and Oort constants. Figure \ref{fig:errors_precision} shows the fractional uncertainty of each estimated value as a function of uncertainty in line-of-sight acceleration, uncertainty in parallax, and the number of sources. Each datapoint was obtained by simulating 1000 samples of $N$ pulsars from a random uniform distribution within 2 kpc of the Sun, and then injecting Gaussian errors in the observed line-of-sight acceleration and parallax for each source. The $a_\mathrm{los}$ data for each pulsar was taken from an $\alpha-\beta$ model, an $\alpha-\beta$ model was fit to the each random sample, and the uncertainty in each fit parameter was calculated from the MLE uncertainty procedure in Section \ref{sec:fits}. While uncertainty in line-of-sight acceleration and the number of sources strongly impact the uncertainty in the estimated values, the uncertainty in parallax does not strongly contribute to uncertainty in the estimated values.

\begin{figure}
    \centering
    \includegraphics[width=0.47\textwidth]{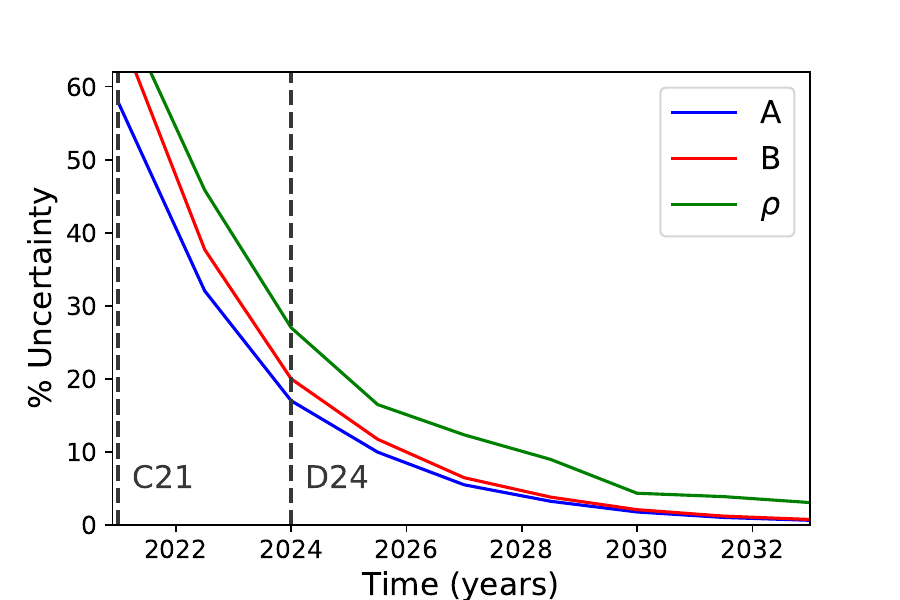}
    \caption{Projected uncertainty over time for the Oort constant and Oort limit measurements. This assumes that the number of sources and the uncertainties in their parallax and acceleration measurements will continue to improve at the same rate as they did between C21 and this work (D24). Within the next decade we expect that uncertainties on direct acceleration measurements of the Oort constants and Oort limit will be on the order of just a few percent.}
    \label{fig:time_errors}
\end{figure}

Note that this is a measure of the scatter in estimates of the Oort limit and Oort constants, not an assessment of error in the fit values; large parallax uncertainties ($>30\%$) cause $\alpha$ to be substantially overestimated in the fit models even though the uncertainties of the Oort constants appear to become smaller at large parallax uncertainties. Additionally, this analysis does not consider uncertainty in the velocity at the position of the Sun or the distance from the Sun to the center of the Galaxy; uncertainties in those measurements will increase the uncertainly in estimates of the Oort constants and Oort limit.

Note that it is possible to obtain estimates of fit parameters with uncertainties smaller than the actual measurement uncertainty in each individual measurement; this is shown in the middle panel of Figure \ref{fig:errors_precision}, where the uncertainty on the estimated values dips below the uncertainty in each $a_\mathrm{los}$ measurement between 20 and 40 sources. This is because these estimates are an aggregate measurement, similar to how the mean of several measurements has an uncertainty smaller than the uncertainty in each individual measurement. This indicates that even if the individual pulsar timing measurements have large uncertainties, it is important to include as many sources as possible in our datasets. 

Figure \ref{fig:time_errors} shows a projection of how the uncertainty in the Oort limit and Oort constants will change over time. This assumes that uncertainties in parallax measurements of each pulsar will halve every 4 years \citep[e.g.][]{Hu2020}, that uncertainty in $a_\mathrm{los}$ decreases as $T^{-5/3}$ where $T$ is the baseline of the pulsar timing data \citep{HandbookPulsarAstro}, and the number of pulsars doubles every 4 years (consistent with the increase in the number of usable sources between C21 and this work). Within the next ten years, we expect to be able to estimate the Oort limit and Oort constants to within a few percent using only direct acceleration measurements. 

Our method of computing the dark matter density in the midplane is limited by the uncertainties in the stellar and gas volume densities near the Sun. As it stands today, these values contribute an uncertainty of 0.006 M$_\odot$/pc$^{3}$ to our value of $\rho_{0,DM}.$ As the nominal value of $\rho_{0,DM}$ is expected to be somewhere on the order of $\sim$0.01 M$_\odot$/pc$^{3}$, this means that this method can only obtain a minimum relative uncertainty of 60\%, regardless of how well we can measure the Oort limit. Improvements will need to be made to the baryonic density estimates in order to fully utilize the projected Oort limit measurement from direct acceleration data. These improvements will probably be straightforward as next-generation observatories such as Rubin and Roman become active.

%
%
%


\end{document}